\documentclass[preprint,trackchanges]{aastex62}

\usepackage{amsmath}
\usepackage{natbib}
\hypersetup{linkcolor=cyan,citecolor=magenta,filecolor=yellow,urlcolor=blue}

\received{May 26, 2017}
\accepted{December 02, 2017}
\published{February 01, 2018}

\shorttitle{Color Evolution And Its Physical Implication}
\shortauthors{Li et al.}

\begin{document}

\title{A Large Catalog of Multiwavelength GRB Afterglows I: \\Color Evolution And Its Physical Implication}

\correspondingauthor{Liang Li}
\email{liang.li@fysik.su.se, yu.wang@icranet.org, lshao@hebtu.edu.cn, xfwu@pmo.ac.cn, hyf@nju.edu.cn}

\author[0000-0002-0786-7307]{Liang Li}
\affiliation{Department of Physics, KTH Royal Institute of Technology, SE-106 91 Stockholm, Sweden}
\affiliation{The Oskar Klein Centre for Cosmoparticle Physics}
\affiliation{Department of Physics, Stockholm University, SE-106 91 Stockholm, Sweden}
\affiliation{ICRANet, Piazza della Repubblica 10, I-65122 Pescara, Italy}
\affiliation{Purple Mountain Observatory, Chinese Academy of Sciences, Nanjing 210008, China}

\author{Yu Wang}
\affiliation{ICRANet, Piazza della Repubblica 10, I-65122 Pescara, Italy}
\affiliation{Dip. di Fisica and ICRA, Sapienza Universit¨¤ di Roma, Piazzale Aldo Moro 5, I-00185 Rome, Italy}

\author{Lang Shao}
\affiliation{Department of Space Sciences and Astronomy, Hebei Normal University, Shijiazhuang 050024, China}
\affiliation{Key Laboratory of Dark Matter and Space Astronomy, Purple Mountain Observatory, Chinese Academy of Sciences, Nanjing 210008, China}

\author{Xue-Feng Wu}
\affiliation{Purple Mountain Observatory, Chinese Academy of Sciences, Nanjing 210008, China}

\author{Yong-Feng Huang}
\affiliation{Department of Astronomy, Nanjing University, Nanjing 210093, Jiangsu, China}
\affiliation{Key Laboratory of Modern Astronomy and Astrophysics (Nanjing University), Ministry of Education, Nanjing 210093, China}

\author{Bing Zhang}
\affiliation{Department of Physics and Astronomy, University of Nevada, Las Vegas, NV 89154, USA}

\author{Felix Ryde}
\affiliation{Department of Physics, KTH Royal Institute of Technology, SE-106 91 Stockholm, Sweden}
\affiliation{The Oskar Klein Centre for Cosmoparticle Physics}

\author{Hoi-Fung Yu}
\affiliation{Department of Physics, KTH Royal Institute of Technology, SE-106 91 Stockholm, Sweden}
\affiliation{The Oskar Klein Centre for Cosmoparticle Physics}

\begin{abstract}

The spectrum of gamma-ray burst (GRB) afterglows can be studied with color indices. 
Here we present a large comprehensive catalog of 70 GRBs with multiwavelength optical transient data on which we perform a systematic study to find the temporal evolution of color indices.  
We categorize them into two samples based on how well the color indices are evaluated. 
The Golden sample includes 25 bursts mostly observed by GROND, and the Silver sample includes 45 bursts observed by other telescopes.
For the Golden sample, we find that 96\% of the color indices do not vary over time. However, the color indices do vary during short periods in most bursts.
The observed variations are consistent with effects of (i) the cooling frequency crossing the studied energy bands in a wind medium (43\%) and in a constant-density medium (30\%), (ii) early dust extinction (12\%), (iii) transition from reverse-shock to forward-shock emission (5\%), or (iv) an emergent SN emission (10\%). We also study the evolutionary properties of the mean color indices for different emission episodes.
We find that 86\% of the color indices in the 70 bursts show constancy between consecutive ones. The color index variations occur mainly during the late GRB-SN bump, the flare, and early reverse-shock emission components.  
We further perform a statistical analysis of various observational properties and model parameters (spectral index $\beta_{o}^{\rm CI}$, electron spectral indices $p^{\rm CI}$, etc.) using color indices.
Overall, we conclude that $\sim$ 90\% of colors are constant in time and can be accounted for by the simplest external forward-shock model, while the varying color indices call for more detailed modeling.

\end{abstract}

\section{Introduction} \label{sec:intro}

The first observation of gamma-ray burst (GRB) optical afterglow was that of the BeppoSAX-detected GRB~970228 \citep{1997Natur.386..686V}. Since the launch of the {\it Swift} satellite more than 10 yr ago \citep{2004ApJ...611.1005G}, many ground-based optical telescopes with increasing sensitivity have accumulated a rich collection of optical afterglows. In recent years, optical afterglows have been studied extensively either in terms of either their multiband light curves \citep[e.g.,][and references therein]{2002ApJ...571..779P,2006ApJ...637..873H,2006ApJ...637..889Z,2010ApJ...720.1513K,2011ApJ...734...96K,2012ApJ...758...27L,2013ApJ...774...13L,2015ApJS..219....9W,2017arXiv170103713R}, or their spectral energy distributions (SEDs) derived from simultaneous multiband photometry \citep[e.g.,][]{2004ApJ...608..846S,2007ApJ...661..787S,2006ApJ...641..993K,2010ApJ...720.1513K,2011ApJ...734...96K,2017arXiv170103713R}. The latter, if studied in a broad bandpass, provides tight constraints on the radiation mechanism and circumburst environment of GRBs.
Given the rapid progress made in the detection of afterglows in the {\it Swift} era, unforeseen temporal and spectral features were discovered,  which challenge our understanding of GRB afterglow physics \citep[][and references therein]{2007ChJAA...7....1Z,2009ARA&A..47..567G}. 

The temporal and spectral properties of optical transients can be studied by color indices (hereinafter CIs), defined as the magnitude difference between two filters.
They can be used to study the SED with a good temporal resolution, even when high-resolution spectra are not available.
For instance, CIs can resolve small variations in the spectral profiles of optical transients \citep{2001A&A...377..450S,2013EAS....61..271S}.
They are also helpful in identifying different radiation mechanisms or progenitors in a single burst. As an example, CIs were used as an indicator of the underlying Type Ib/c supernovae \citep[SNe;][]{2004A&A...427..901S}.

Theoretically, the conventional afterglow model has been discussed by many authors \citep[e.g.,][]{1997ApJ...476..232M,1998ApJ...497L..17S,2003MNRAS.341..263H}, and the relationship between temporal and spectral indices (the so-called closure relation) has been extensively reviewed in \cite{2004IJMPA..19.2385Z} and \cite{2013NewAR..57..141G}.
According to this model, which assumes that synchrotron radiation dominates the afterglow emission, the color index should not change with time.
Instead, if a change is observed, there may be several possible reasons \citep{Melandri:2017wl}, for instance:
(i) the cooling frequency crosses the studied energy bands \citep[e.g.,][]{2011A&A...535A..57F}. 
(ii) an additional emission component emerges, e.g., the SN counterpart observed in some GRBs, or the mixture of different emission stages, such as forward and reverse shock \citep{2003ApJ...597..455K} in the context of the fireball model \citep{2004RvMP...76.1143P}.
(iii) after being destroyed by the initial intense radiation, host extinction could be changed owing to dust photodestruction \citep{2002ApJ...580..261P, 2014MNRAS.440.1810M}. Some other possible reasons, such as a structured jet, could also produce the change in the CIs.

In this paper, we have made an extensive effort to collect publicly available optical and near-infrared photometric data. We restricted the study to 70 bursts with high-quality multiband observations, which were detected during both the pre-{\it Swift} and {\it Swift} eras.
\textcolor{orange}{Our goal is to study the temporal variability of CIs and explore its physical implications.}
In a following paper, we will present a detailed time-resolved analysis of afterglow spectra using X-ray and optical data, which will be used to study the extinction curve and explore the physics in the circumburst medium in the host galaxy (Li et al, in prep.).

This paper is organized as follows. The sample selection and data analysis are presented in Section 2.  
The statistical properties of CIs are summarized in Section 3.
The evolution of the CIs is presented in Section 4. 
The theoretical implications are discussed in Section 5.
Finally, conclusions are provided in Section 6.

\section{Sample Selection, Corrections, and Scaling}

We extensively compiled as many optical/near-IR (NIR) photometric data as possible, detected by ground-based telescopes from both pre-{\it Swift} and \emph{Swift}-era afterglows.
The bursts in our sample were observed before 2016\footnote{We note that the last burst satisfying our criteria is GRB 130925A.}.
We collected observational data from published papers and the Gamma-ray Burst Coordinates Networks Circular Service (GCN)\footnote{\url{https://gcn.gsfc.nasa.gov}}.
We initially selected GRBs with known redshift that have high-quality data in at least three optical bands. 
We then restricted the study to the GRBs that were observed in optical wavelength for at least 1 hr, facilitating the study of CIs temporal evolution. A total of 70 bursts satisfy these criteria (from 1997 February 28 to 2015 December). To obtain a trustworthy data set, the data sources should be well calibrated. We made corrections to the data as follows.

We studied the temporal properties of the CIs in the rest frame, $t_{\rm rest}=t_{\rm obs}/(1+z)$.
For some specific cases (GRB 970508, GRB 990510, GRB 990712, GRB 060614, GRB 060908, GRB 090426, GRB 091127, GRB 100621A, GRB 100814A, GRB 110918A, GRB 120422A, GRB 120729A, GRB 130702A, GRB 130925A), we subtracted the flux contribution at very late epochs that we interpret as coming from the host galaxy.
This contribution is determined by identifying constant flux values at a late time, $\sim 10^6$ s after the GRB trigger.

We calculated the Galactic extinction correction from the reddening map presented in \cite{2011ApJ...737..103S} for optical and NIR magnitudes, and we assume\footnote{$R_{\rm V}$ $\equiv$ $A_{\rm V}$/$E_{\rm B-V}$ is the total-to-selective extinction ratio, where the color excess $E_{\rm B-V}$ $\equiv$ $A_{\rm B}$-$A_{\rm V}$, is the difference between the extinction in B and V bands.} $R_{\rm v}$=3.1. 

The observed spectra are $k$-corrected, where $k$ is defined as $m_{\rm T}=m_{\rm O}-k$ \citep[e.g.,][]{1968ApJ...154...21O, 1997iagn.book.....P}, with $k=2.5(\beta_o-1)\log(1+z)$. Here $m_{\rm O}$ and $m_{\rm T}$ are the observed and true magnitudes, respectively. $\beta_o$ is the spectral index\footnote{We assume a typical value of 0.75 for those bursts whose spectral indices are not available, which are indicated by ellipsis dots in Table 1.} of the optical afterglow, and $z$ is the redshift.

\subsection{Corrections for Host Dust Extinction}

The extinction due to dust from the host galaxy can significantly affect the observations. For wavelength $\lambda_{i}$, the dust extinction is defined as
\begin{equation}
A(\lambda_{i})=m_{\rm O}(\lambda_{i})-m_{\rm T}(\lambda_{i}).
\label{eq:Alambda}
\end{equation}

The observed SED of the optical afterglow is modeled by an absorbed power law in frequency \citep{2006ApJ...641..993K}
\begin{equation}
F_{\nu} =F_{0}\nu^{-\beta_{o}}e^{-\tau(\nu_{\rm host})},
\label{eq:Host1}
\end{equation}
where
\begin{equation}
\tau(\nu_{\rm host}) =\frac{1}{1.086} A_{\rm V}\eta(\nu_{\rm host}),
\label{eq:Host2}
\end{equation}
in which $\beta_{o}$ is the intrinsic power-law slope of the SED, $F_{\rm 0}$ is the normalization constant, $\eta(\nu_{\rm host})=A^{\rm host}_{\lambda}/A^{\rm host}_{\rm V}$ is given by the extinction model assumed for the GRB host galaxy, and $A^{\rm host}_{\lambda}$ is the host galaxy extinction correction (known only with large uncertainties). 
We do not carry out spectral fitting in this paper. 
Instead, the host galaxy extinction $A^{\rm host}_{\rm V}$, the spectral index $\beta_{o}$, and the redshift $z$ are acquired from the literature, and their values are given in Table 1.

\begin{figure*}
\centering
\includegraphics[angle=0, scale=1.0]{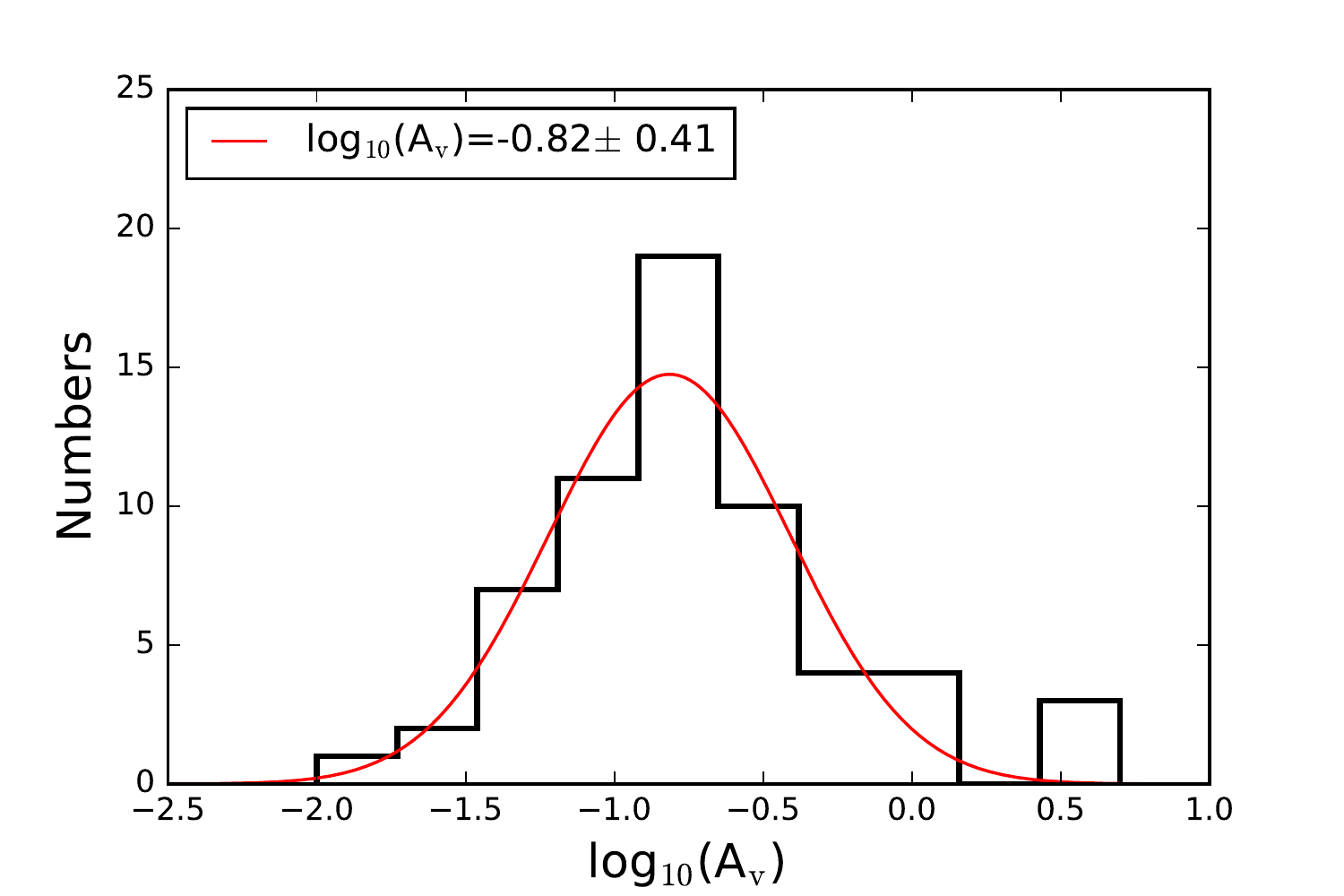}
\caption{Distribution of the host galaxy extinction in the V band, $A^{\rm host}_{\rm v}$, obtained from published papers. The best Gaussian fit gives the mean $A_{\rm v}^{\rm host}$ and the standard deviation such that log$_{10}$($A_{\rm v}^{\rm host})=-0.82\pm0.41$.}
\label{Av}
\end{figure*}

\begin{figure*}
\centering
\includegraphics[angle=0, scale=0.80]{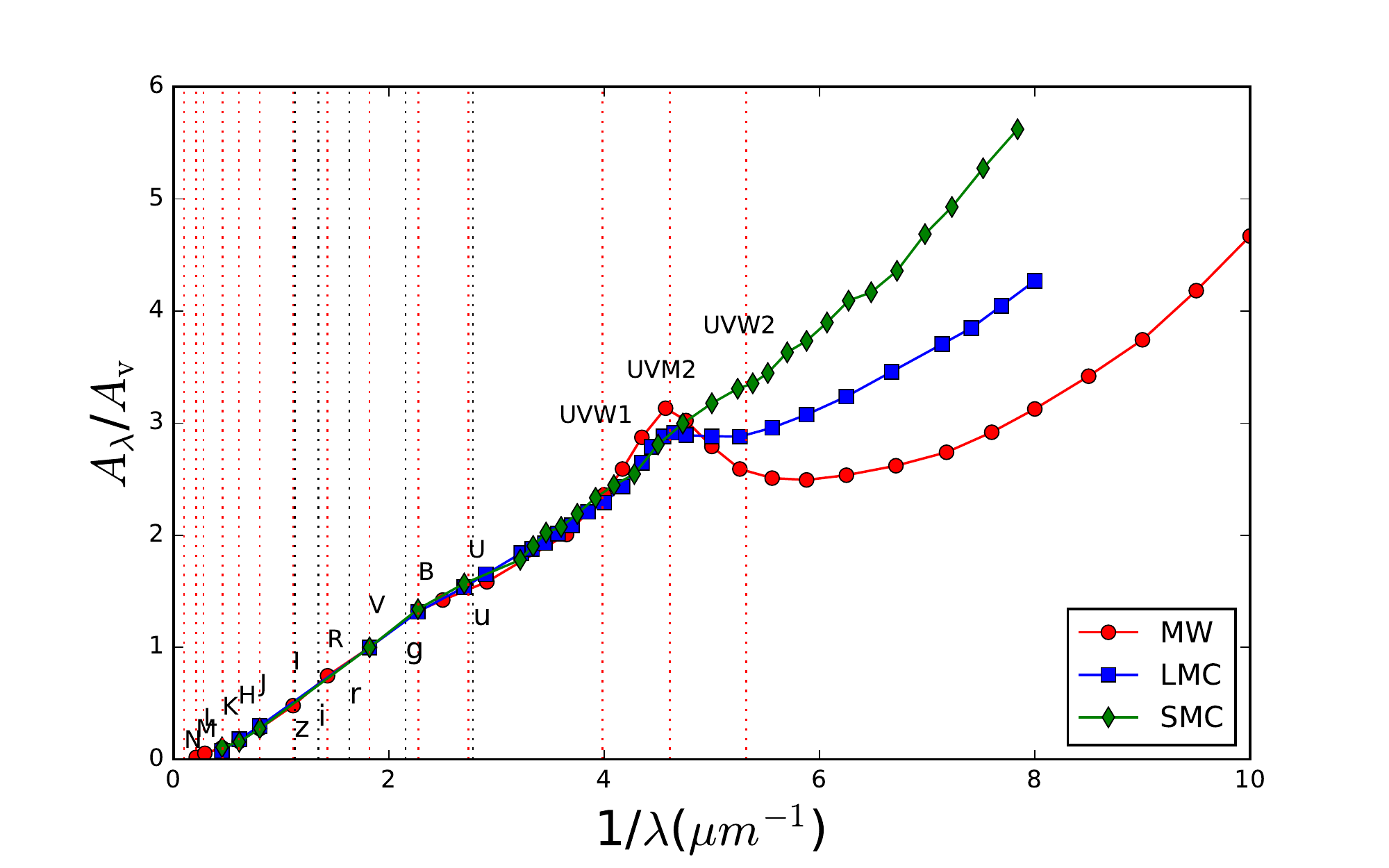}
\caption{Dust models for the host extinction reproduced from \cite{1992ApJ...395..130P}. The solid lines of different colors are for different models, while the black and red dashed lines correspond to the energy bands considered in this work.}
\label{HostCurves}
\end{figure*}

Specifically, $A^{\rm host}_{\rm V}$ is chosen by the following three steps.
First, we selected the values obtained by \cite{2006ApJ...641..993K} and \cite{2010ApJ...720.1513K} corresponding to the Small Magellanic Cloud (SMC) dust model. 
Second, when they are not available in \cite{2006ApJ...641..993K} and \cite{2010ApJ...720.1513K}, the extinction parameters are retrieved from other papers (given as a reference in Table 1) from the SMC dust model.
This model is our first choice since it is shown to best describe dust in GRB environments \citep{2010ApJ...720.1513K}.
When the information is not available for this model, the value of $A^{\rm host}_{\rm V}$ is obtained from either from the Milky Way (MW) or the large Magellanic Cloud (LMC) models.
Finally, for a few cases, when the value of $A^{\rm host}_{\rm V}$ is unavailable for any model, we use the value obtained by fitting a Gaussian to the extinction distribution (see Figure \ref{Av}), log$_{10}$($A_{\rm V}^{\rm host}$)=-0.82$\pm$0.41; here -0.82 and 0.41 are the mean value and standard deviation of the Gaussian fit\footnote{Hereinafter we adopt the same convention for all Gaussian fits.}, respectively. After obtaining the value of $A^{\rm host}_{\rm V}$, we derived the extinction values in any optical band, $A^{\rm host}_{\lambda}$, in the GRB host galaxies following \cite{1992ApJ...395..130P}. Figure \ref{HostCurves} illustrates the spectral behavior of $A^{\rm host}_{\lambda}/A^{\rm host}_{\rm V}$ for the three dust models and displays the position of the filters considered in this paper.

\subsection{Light-curve Fitting}

In this section, we describe the fits to the GRB light curves in our 70 bursts.
The purpose of this procedure is twofold.
First, it allows us to identify different emission phases.
Second, the results are used to interpolate the magnitude, in order to compute the CIs, when there are no simultaneous multi-wavelength observations available.

The observed magnitude $m$ is obtained from the flux $F$ as
\begin{equation}
m =-2.5 \rm log_{10} \left(\frac{\textit{F}}{\textit{F}_{0}} \right),
\label{eq:mag}
\end{equation}
where $F_{0}$ is the flux of an object with magnitude zero.
Note that in this paper we adopt the AB magnitude system.

\begin{figure*}
\centering
\includegraphics[angle=0, scale=0.80]{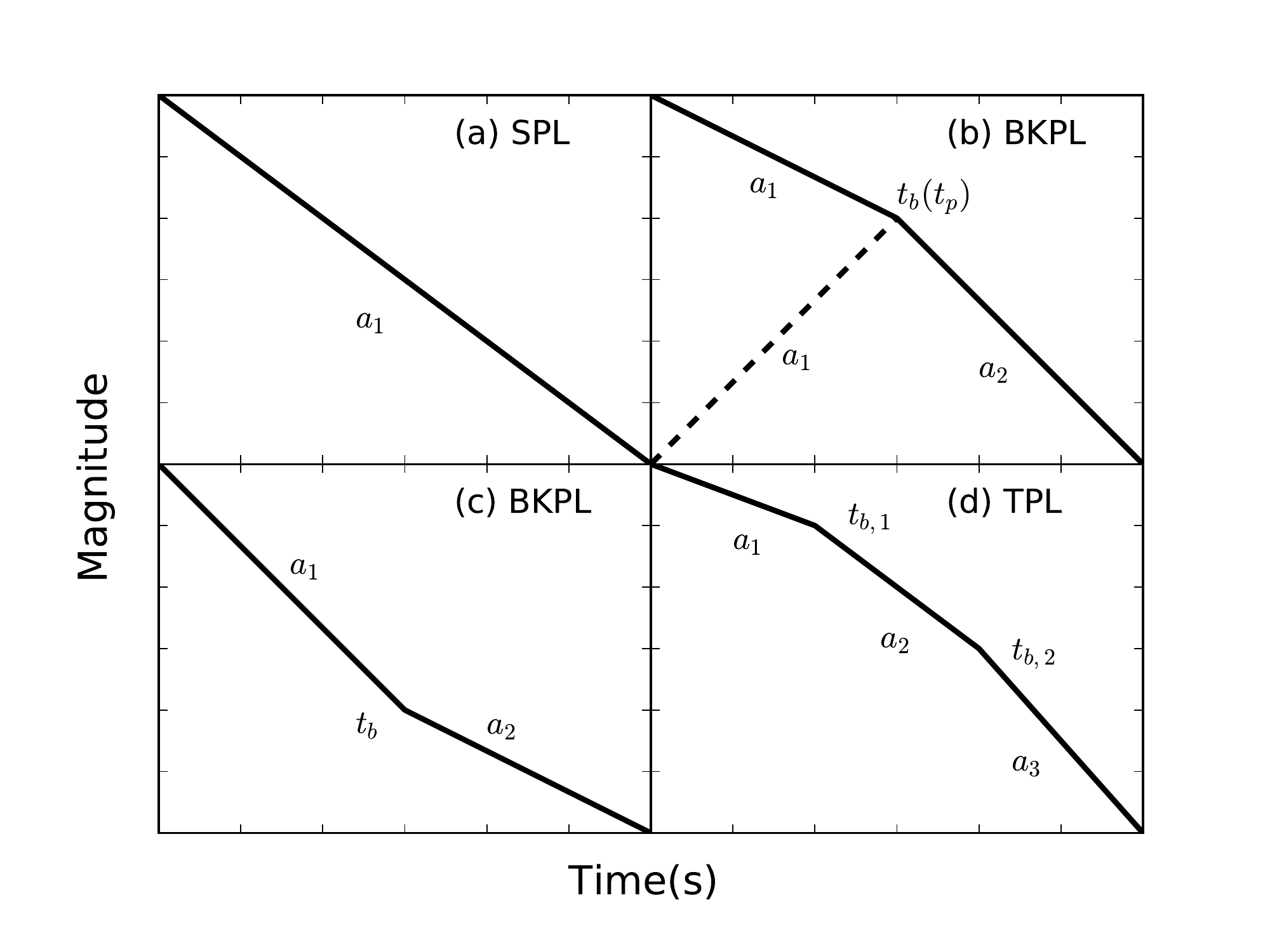}
\caption{Cartoons of the light-curve models. From top left to bottom right, the figure presents in turn (a) a single power-law model, a smoothly broken power-law with positive $\omega$, which can be adopted to fit (b) a break (solid line)/bump (dash line), and (c) negative $\omega$, and (d) finally a triple power-law.}
\label{Cartoon}
\end{figure*}

After taking into account all the correction factors (our Galaxy and the host galaxy extinction corrections, and the spectral $k$-correction), we fit the light curves in each band separately with a model of multiple components \citep{2012ApJ...758...27L}. 
The basic model is either a single power law (SPL) or a smoothly broken power law (BKPL) for the flux, resulting in\footnote{We apply these afterglow models to the light-curve fitting in magnitude (linear)- time (log) space.}:
\begin{equation}
\begin{split}
m_{\rm T}(t)=-2.5 \rm log_{10} (10^{-0.4m_{\rm c}}\, t^{-\alpha}
+10^{-0.4m_{\rm G}}+10^{-0.4m_{\rm K}}
+10^{-0.4m_{\rm host}}+10^{-0.4m_{\rm GF}}),
\end{split}
\label{eq:bkpl}
\end{equation}
\begin{equation}
\begin{split}
m_{\rm T}(t)=-2.5 \rm log_{10} \{10^{-0.4m_{\rm c}}[(t/t_{\rm b,1})^{\alpha_1\omega_{1}}
+(t/t_{\rm b,1})^{\alpha_2\omega_{1}}]^{-1/\omega_{1}}
+10^{-0.4m_{\rm G}}+10^{-0.4m_{\rm K}}+10^{-0.4m_{\rm host}}+10^{-0.4m_{\rm GF}} \},
\end{split}
\label{eq:pl}
\end{equation}
here $\alpha$, $\alpha_1$, and $\alpha_2$ are the temporal slopes, $t_{\rm b_{1}}$ is the break time, and $\omega_{1}$ describes the sharpness of the break (the smaller the value, the smoother the break). For most GRBs, we fix this last parameter to $\omega_{1}=3$. 
For a few cases (GRB 030226, GRB 050922C, GRB 061007, GRB 080603A) that have a smoother break, we fix it to $\omega_{1}=1$, which improves the fit. Equation (\ref{eq:bkpl}) can be used for both for the light curves with a break and those who exhibit an increasing behavior (e.g., afterglow onset indicated by the dashed line in Fig.\ref{Cartoon}b).
In addition, $m_{\rm c}$ is the magnitude constant value, $m_{\rm G}$ is the extinction magnitude for our Galaxy, $m_{\rm K}$ is the spectral $k$-correction, $m_{\rm host}$ is the extinction magnitude for the host galaxy, and $m_{\rm GF}$ is the possible contribution from the host galaxy at late time (if identified).
Note that $m_{\rm G}$, $m_{\rm K}$, and $m_{\rm host}$ are known before fitting.

We notice that for two bursts (GRB 061126 and GRB 080319B), the optical-afterglow light curves have a steeper decay slope $\alpha_{1}$ in the pre-break segment (which might originate from the reverse-shock emission) than the decay slope $\alpha_{2}$ in the post-break segment (possibly from the normal decay dominated by the external forward shock). However, due to the limitation of the BKPL model, it is impossible to fit the light curves whose pre-break slope is steeper than the post-break slope. In this case, we adopted the BKPL model function with a negative sharpness of the break $\omega$=-3 (Fig.\ref{Cartoon}c). 

A double-BKPL light curve (Fig.\ref{Cartoon}d) is also expected in some afterglow models. For example, it is theoretically expected that the afterglow light curves may have a shallow segment owing to energy injection at an early time, which then changes into a normal-decay segment when the energy injection is over, and finally steepens owing to a jet break \citep{2006ApJ...642..354Z}.
We therefore consider a smooth triple-power-law function (TPL) to fit the light curves \citep[e.g.,][]{2008ApJ...675..528L,2012ApJ...758...27L}:

\begin{equation}
\begin{split}
m_{\rm T}(t)=-2.5 \rm log_{10} \{((10^{-0.4m_{\rm c}}[(t/t_{\rm b,1})^{\alpha_1\omega_{1}} +(t/t_{\rm b,1})^{\alpha_2\omega_{1}}]^{-1/\omega_{1}})^{-1/\omega_{2}}\\
+((10^{-0.4m_{\rm c}}[(t/t_{\rm b,1})^{\alpha_1\omega_{1}}+(t/t_{\rm b,1})^{\alpha_2\omega_{1}}]^{-1/\omega_{1}})(t_{\rm b,2})(t/t_{\rm b,2})^{-\alpha_3}))^{-1/\omega_{2}})^{-1/\omega_{2}}\\
+10^{-0.4m_{\rm G}}+10^{-0.4m_{\rm K}}+10^{-0.4m_{\rm host}}+10^{-0.4m_{\rm GF}} \}.
\end{split}
\label{eq:tpl}
\end{equation}
where $\omega_2$ is the sharpness parameter at the second break time $t_{b,2}$.

The fits are performed with the IDL routine
"mpfitfun.pro"\footnote{\url{http://purl.com/net/mpfit}}\citep{2009ASPC..411..251M}, which uses the Levenberg-Marquardt algorithm to achieve minimization.
To select the best model, we compare the reduced $\chi^2$ values and choose the one that has a more reasonable value (close to 1).
For example, the $\chi^2_{\rm r}$ of GRB 990510 (R band) is 31/39 (a little less than 1) for the BKPL model and 169/36 (much larger than 1) for the SPL model. Therefore, we adopt the BKPL as the best model for GRB 990510.
Second, if both models have a $\chi^2_{\rm r}$ close to 1, our principle is to choose the simpler one (fewer parameters). For instance, for GRB 060908, a $\chi^2_{\rm r}$ is equal to 38/49 ($\sim$ 1) for the BKPL model and 52/46 ($\sim$ 1) for the SPL model. In this case, we choose the SPL model.

\begin{figure*}
\centering
\includegraphics[angle=0,scale=0.80]{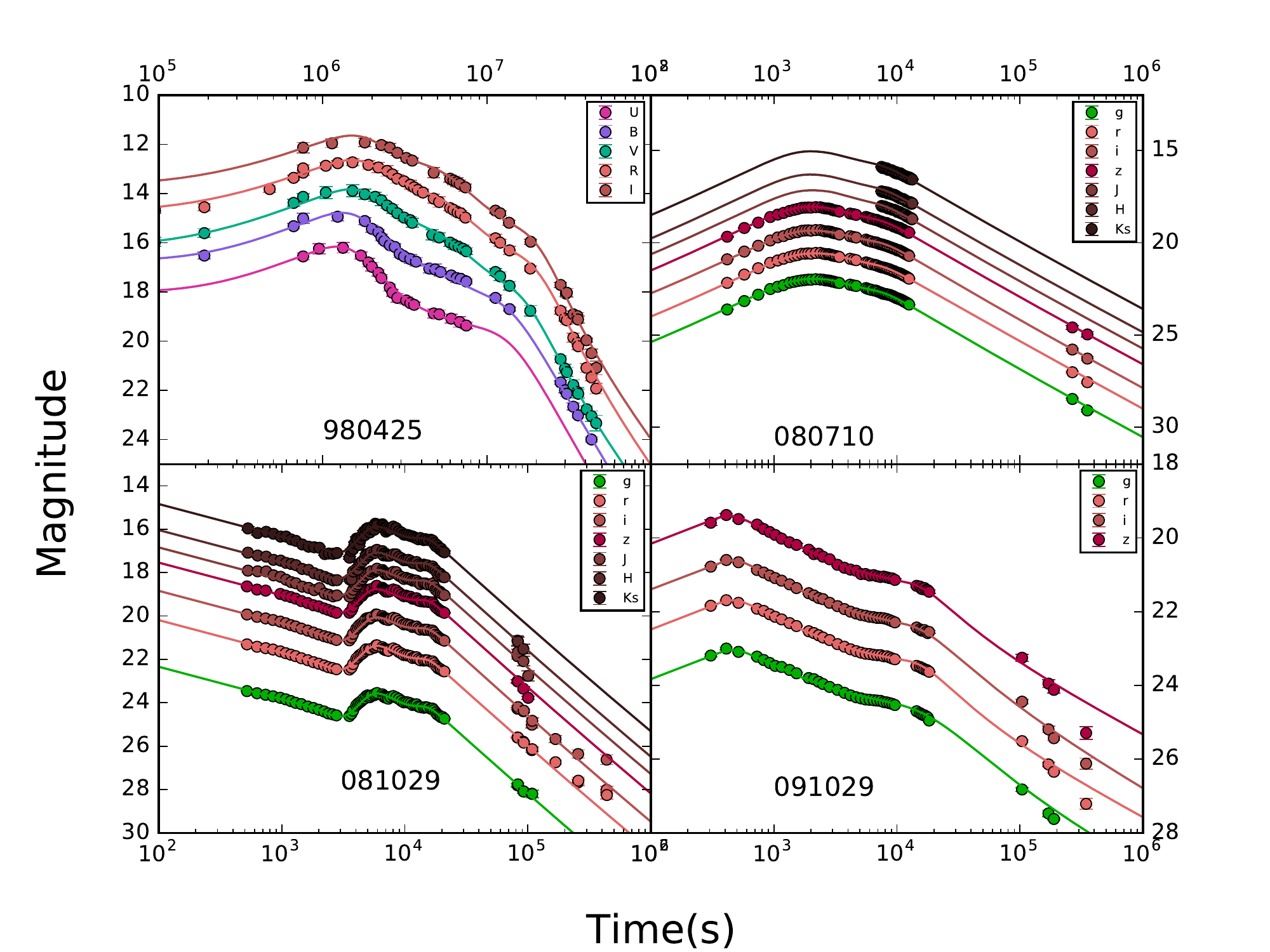}
\caption{Some examples of the best fitting of the multiband data for the optical-afterglow light curves.}
\label{Examples}
\end{figure*}

For multicomponent light curves (see \S 2.3), we introduce the minimum number of components by eye inspection of the temporal features. If the $\chi^2_{\rm r}$ is still much larger than 1, we continue to add more components and redo the fit, until the $\chi^2_{\rm r}$ becomes close to 1 \citep{2012ApJ...758...27L}.
For instance, the value of the $\chi^2_{\rm r}$ of GRB 071025 (J band) for the BKPL is 474/39 (much larger than 1), while it is 58/34 (close to 1) for the double-BKPL model. As a result, the double-BKPL model is selected as the best model for this burst. Examples of light-curve fitting with various models or their composition are shown in Figure \ref{Examples}.

\subsection{Component Identification}

Optical afterglows have complicated light curves with up to eight emission episodes \citep{2012ApJ...758...27L, 2015ApJ...805...13L, 2006ApJ...646..351L, 2013ApJ...774...13L, 2006A&A...451..821N, 2006ApJ...641..993K, 2010ApJ...720.1513K, 2011ApJ...734...96K, 2008MNRAS.387..497P, 2011MNRAS.414.3537P}. A synthetic optical-afterglow light curve is presented in \cite{2012ApJ...758...27L}. To study the spectral behaviors in these different emission phases, we investigate the temporal evolution of their mean CIs. Here we explain which components are considered and how they are identified based on their temporal and spectral features.

Ia: the prompt optical flares (Prompt-Optical);
in the very early time of some bursts when the prompt GRB emission is still going on, a highly variable optical emission component may be observed as in GRB 080319B\footnote{We discard the observations of the early optical-afterglow emission of GRB 080319B since it lacks multiband observations.}.
They are selected such that the observation time is smaller than $T_{90}$ and the temporal slope\footnote{Throughout the paper, the convention $F_{\nu}\propto \nu^{-\beta}t^{-\alpha}$ is adopted.} $\alpha$ $>$2.0.

Ib: the early optical flares (Reverse-Shock);
in a few cases, the early light curves have steep slopes, which likely indicate an early reverse-shock emission component, such as the early phase of GRB 061126. They were selected such that their typical temporal index $\alpha$ $\sim$ 1.7 and the peak time $t_{\rm p}$ is a few hundreds seconds.

II: an early shallow-decay component (Energy-Injection);
the afterglow light curve might show an initial shallow-decay segment followed by a normal-decay/post-jet-break segment, which is likely due to energy injection from a long-lasting spinning-down central engine or piling up of flare materials into the blast wave \citep{2007ApJ...670..565L, 2012ApJ...758...27L, 2015ApJ...805...13L}. This component is described as the energy injection phase.
They are identified by their temporal index $\alpha^{\rm Shallow}$$<$$\alpha^{\rm Normal}$ with a typical value $\alpha^{\rm Shallow}$ $\sim$ 0.5 and with a typical value of break time $t_{\rm b}$ $\sim$ 10$^{4}$ s. Here $\alpha^{\rm Shallow}$ is the temporal index during the shallow-decay segment and $\alpha^{\rm Normal}$ is the temporal index during the normal-decay phase.

III: the standard afterglow component (Onset/Normal-Decay);
the light curves sometimes have an early onset rising segment followed by a normal decay. In most of the cases, a lack of observations in the early time lead to only a single normal decay.
Afterglow onsets are identified with an early smooth bump with a typical value of peak time $t_{\rm p}$ of several hundreds of seconds, coupled to the following normal decay with $\alpha$ $\sim$ 1.2.

IV: the jet-break component (Jet-Break);
the light curves break into a steeper decay.
They are identified with $\alpha \sim p \sim 2.5$ \citep{2006ApJ...642..354Z} and a break time $\sim$ 10$^{5}$ s. Here {\it p} is the electron index.

V: the late optical flares (Flare);
the light curves have prompt-like flares during the afterglow phase when the prompt emission is turned off. They indicate the late-time activities of the central engine.
The late optical flares are characterized by a very sharp temporal index $\alpha$ $>$ 2.0.

VI: the late re-brightening bumps (Re-Bump);
the late bumps would emerge at late times, which is distinguished from the onset bump of the early afterglow.
Both are likely involved in the jet component that produces the re-brightening bump, which seems to be on-axis and independent of the prompt emission jet component \citep{2013ApJ...774...13L}.
They are described by a smooth bump around $t_{\rm p}$  $\sim 10^{5}$ seconds.

VII: the late SN bumps (SN-Bump);
in some cases, the optical transient light curves at late time show an SN bump.
They form a late smooth bump at $t_{\rm p}$  $\sim 10^{6}$ s.
In addition,  we checked that the GRB-SN associations were confirmed in the literature (see \S 5.2.3).

All the assigned emission components in our 70 bursts are presented in Table 2 and 3, and are marked with arrows in Figures \ref{GoldenLCs} and \ref{SilverLCs}.

\subsection{Definition of Color Index Variation}

Once the fits are obtained, one can derive the CIs. They are defined as the magnitude difference between two filters, 
\begin{equation}
\rm CI=m(\lambda_{2})-m(\lambda_{1}),
\label{eq:CI}
\end{equation}
where $\lambda_{2}<\lambda_{1}$ is required so that the smaller the CI, the bluer (or hotter) the optical afterglow.
In this paper, we uniquely focus on the CI between two adjacent bands\footnote{u-g, g-r, r-i, i-z for the SDSS photometric system and  UVW2-UVM2, UVM2-UVW1, UVW1-U, U-B, B-V, V-R, R-I, I-J, J-H, H-K for the commonly UBVRI photometric system.}. CI can be affected by optical extinction, which can be characterised by the color excess (CE), defined by the difference between the observed color index (CI$_{\rm O}$) and the intrinsic color index (CI$_{\rm T}$)
\begin{equation}
\rm CE=CI_{O}-CI_{T}=A(\lambda_{2})-A(\lambda_{1}).
\label{eq:CE}
\end{equation}

\subsubsection{Definition of the Golden and Silver Samples}

To compute the CIs, simultaneous observations or interpolated values are required. We also note that different telescopes have different photometric systems (i.e., different light filters and spectral response functions), which are either the common Johnson-Cousins UBVRI photometry system\footnote{UVW2(1880 {\AA}), UVM2(2170 {\AA}), UVW1(2510 {\AA}), U(3650 {\AA}), B(4400 {\AA}), V(5500 {\AA}), R(6588 {\AA}), I(8060 {\AA}), J(12350 {\AA}), H(16620 {\AA}), K(22000 {\AA})} or the Sloan Digital Sky Survey (SDSS) ugriz photometry system\footnote{u(3596 {\AA}), g(4586.9 {\AA}), r(6219.8 {\AA}), i(7640.7 {\AA}), z(8989.6 {\AA})}. Instead of converting one system into another with empirical correlations \cite[e.g.,][]{2006A&A...460..339J}, we group the bursts by photometric systems, which are studied separately. 

Therefore, we sort all GRBs into two samples based on the photometric system used for the observation. 
This also naturally separates bursts for which interpolation is required from those for which it is not.

\begin{figure*}
\includegraphics[angle=0,scale=0.40]{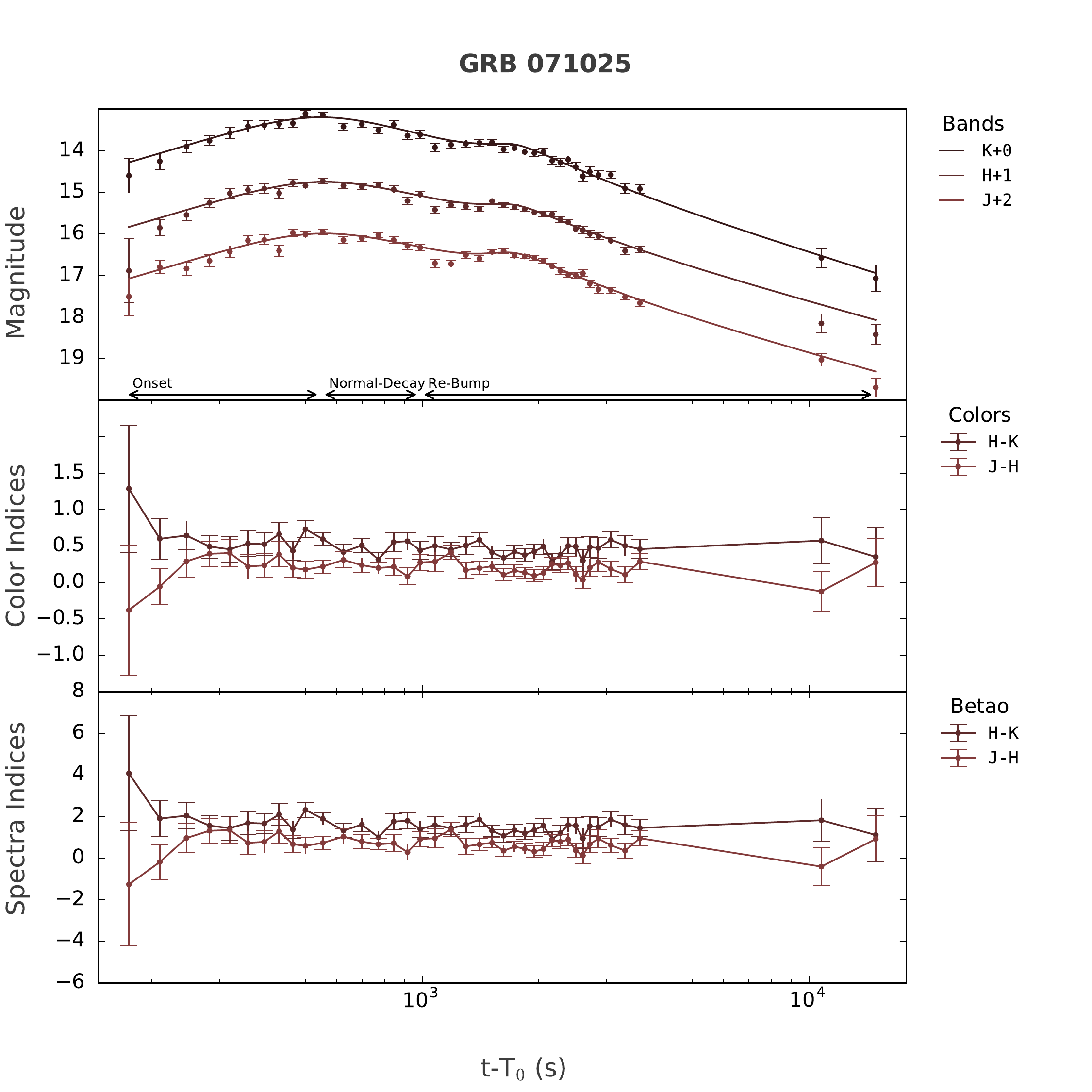}
\includegraphics[angle=0,scale=0.40]{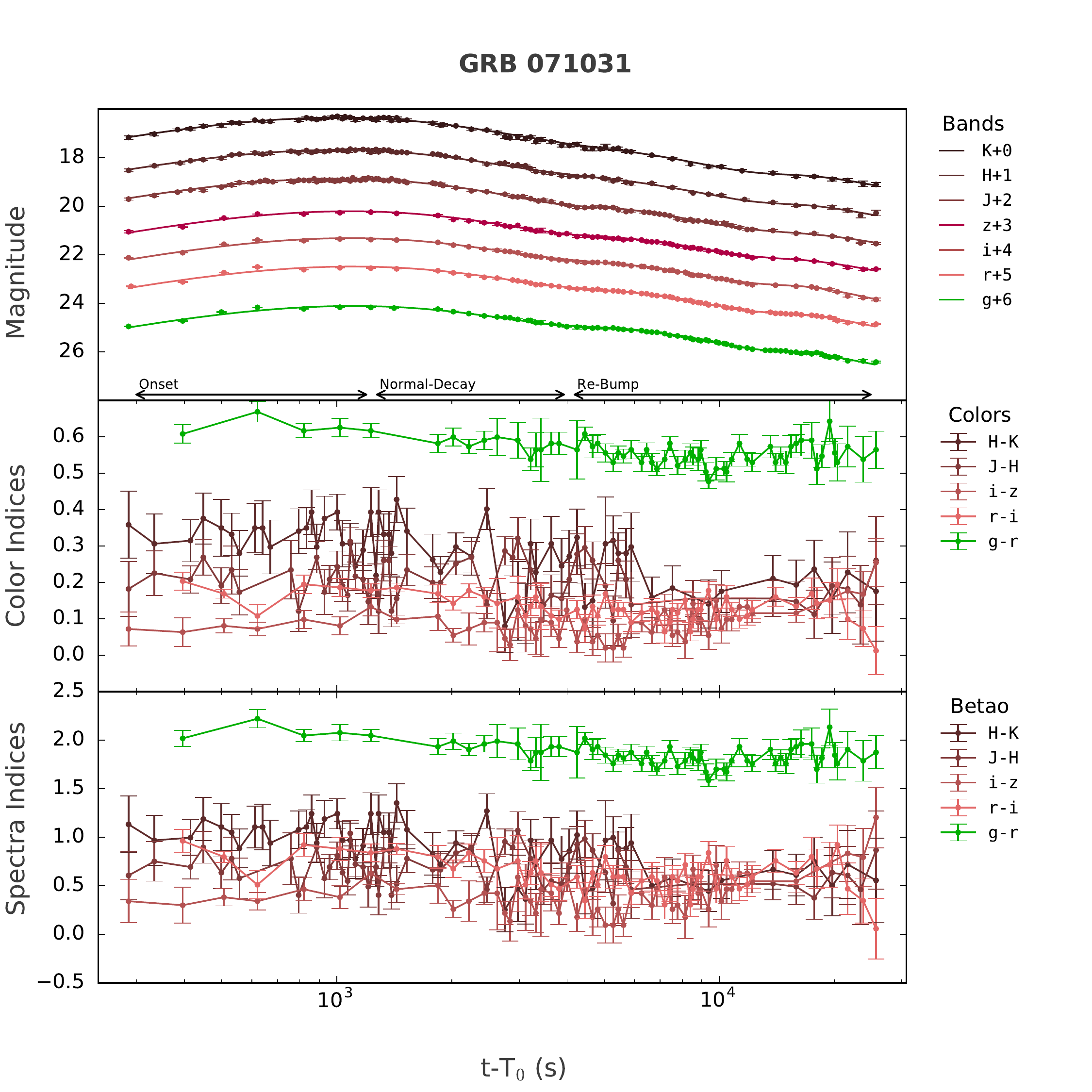}
\includegraphics[angle=0,scale=0.40]{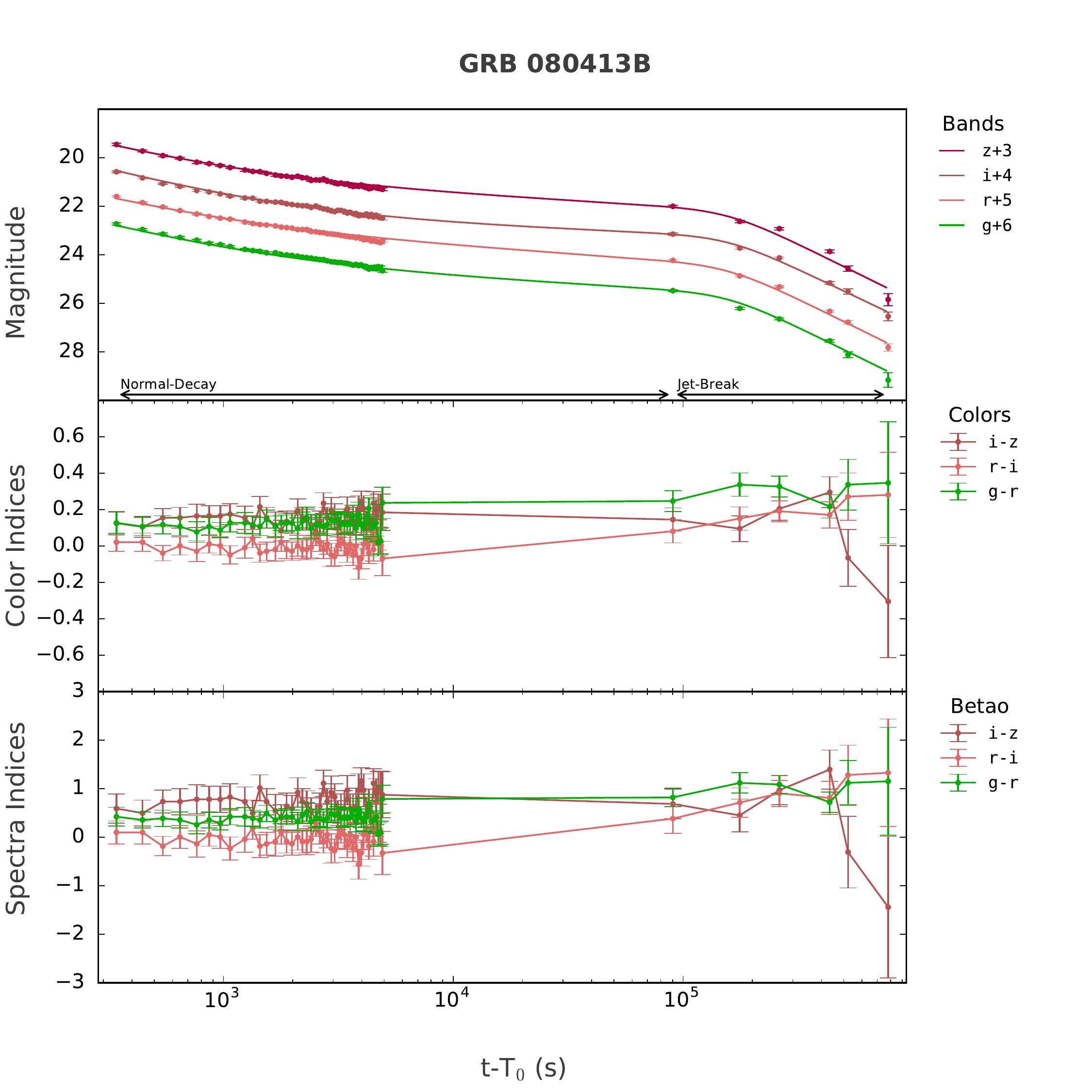}
\includegraphics[angle=0,scale=0.40]{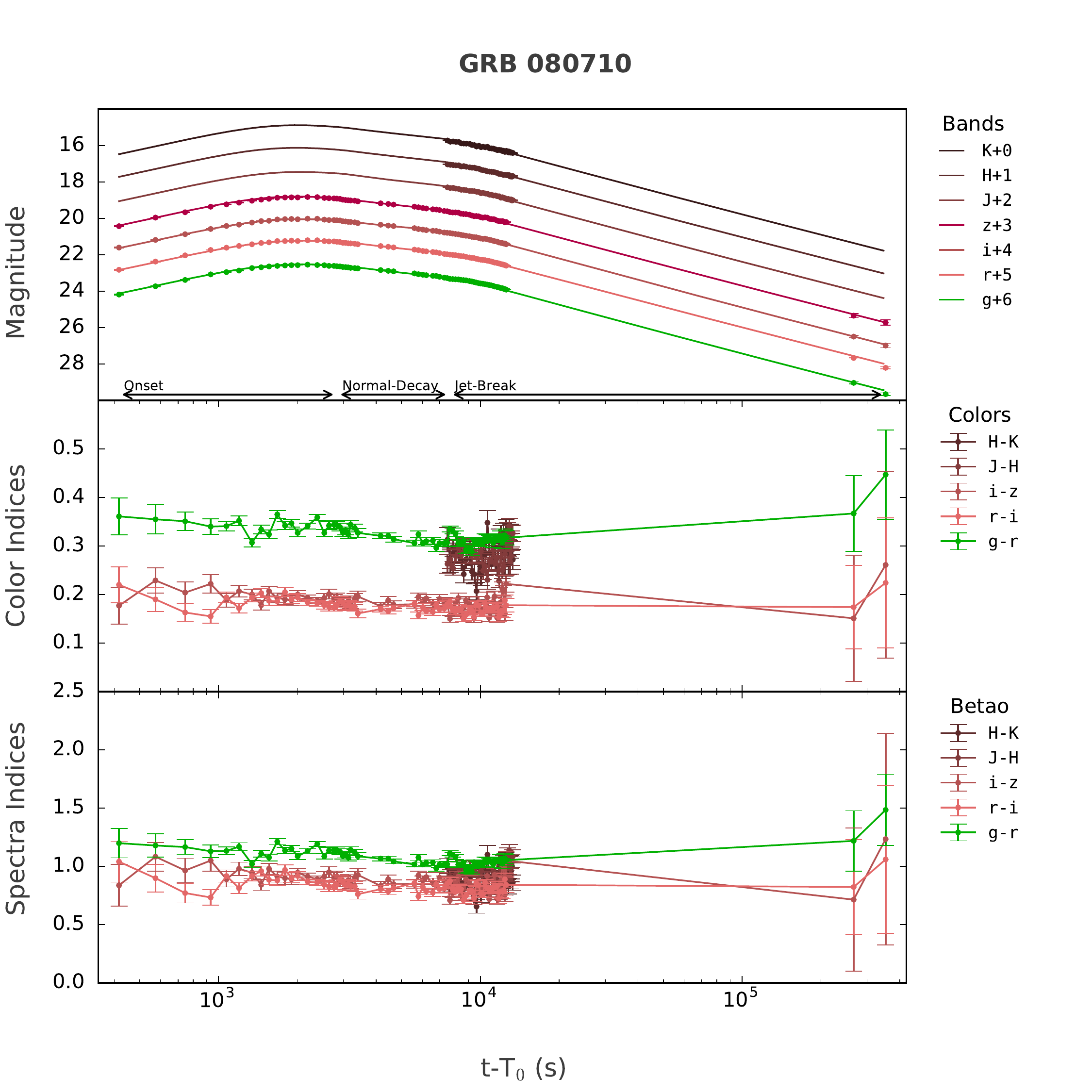}
\caption{Best multiband optical-afterglow light-curve fits and the time evolution of color indices and spectral indices for each burst for the Golden sample, in magnitude (linear)-time (logarithmic) scale space. The spectral indices are derived from the CI-$\beta_{o}$ correlation assuming a power-law spectral decay.}
\label{GoldenLCs}
\end{figure*}

\begin{figure*}
\includegraphics[angle=0,scale=0.40]{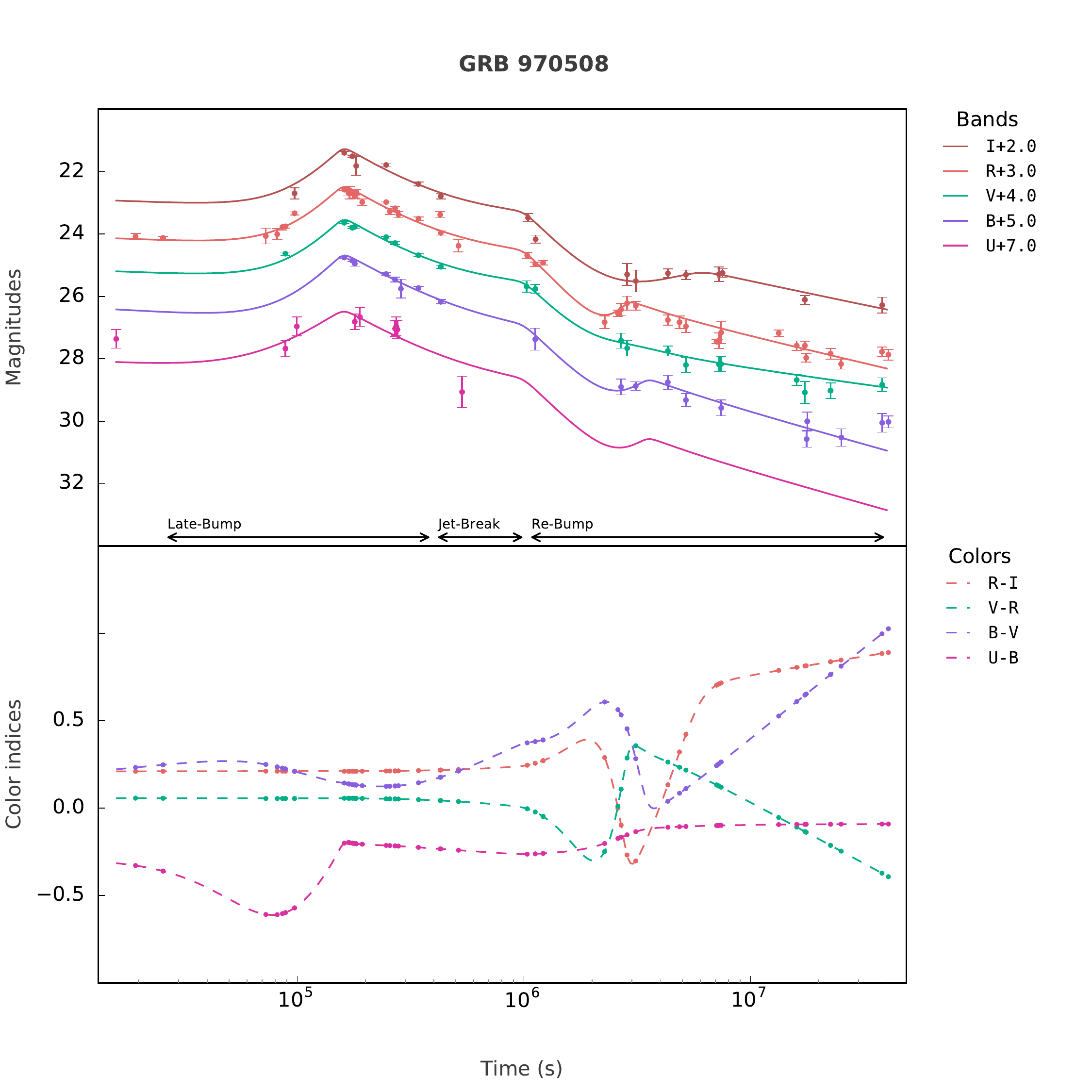}
\includegraphics[angle=0,scale=0.40]{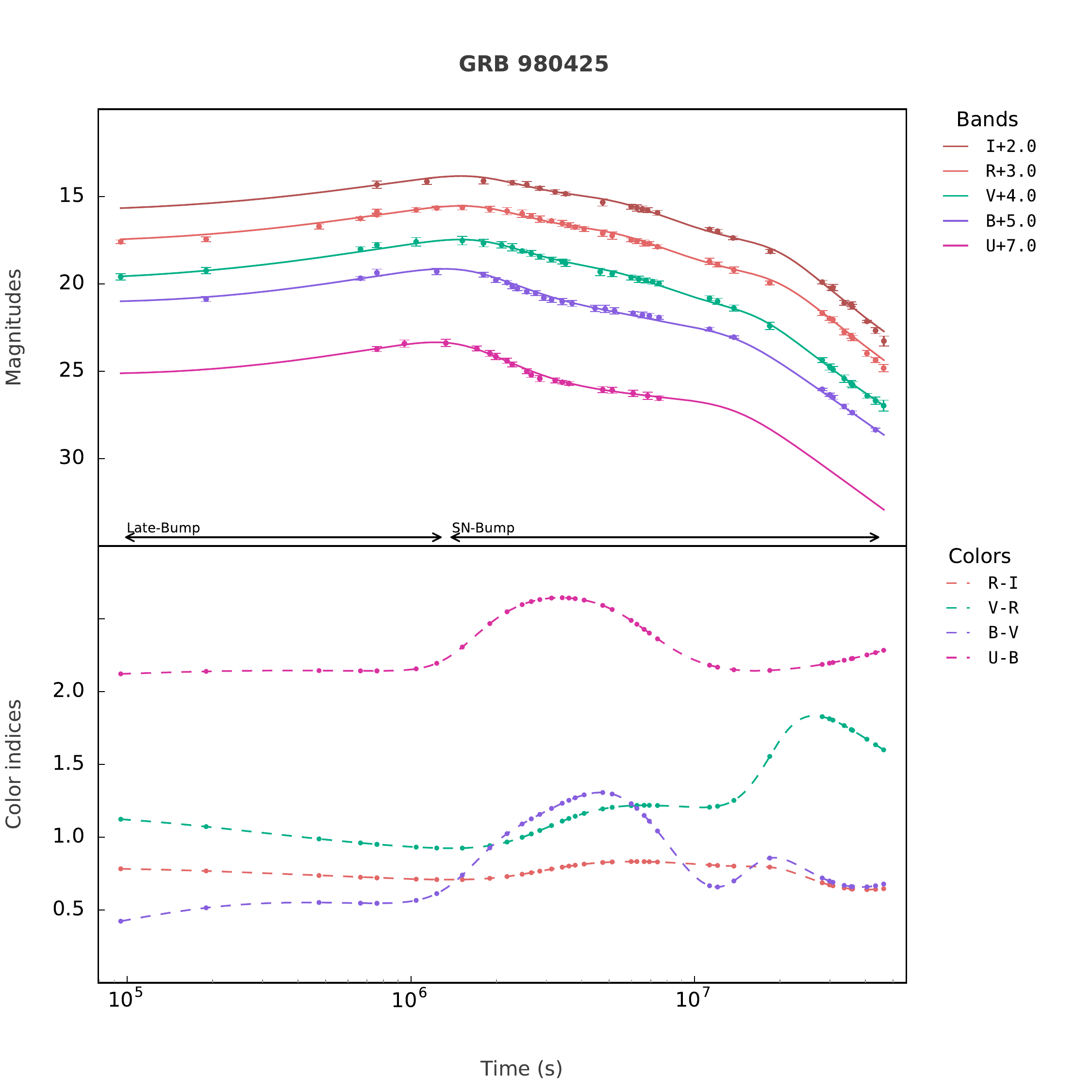}
\includegraphics[angle=0,scale=0.40]{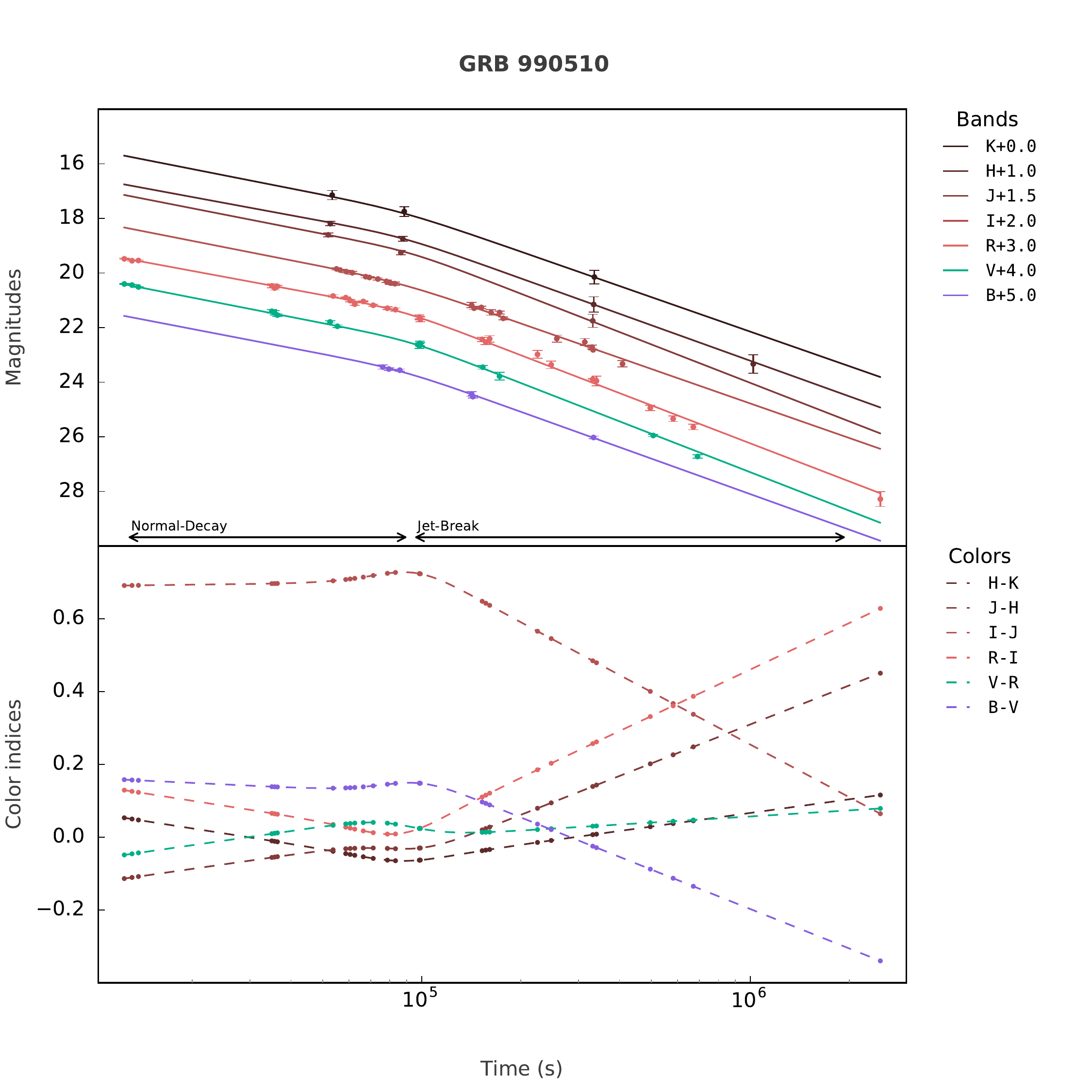}
\includegraphics[angle=0,scale=0.40]{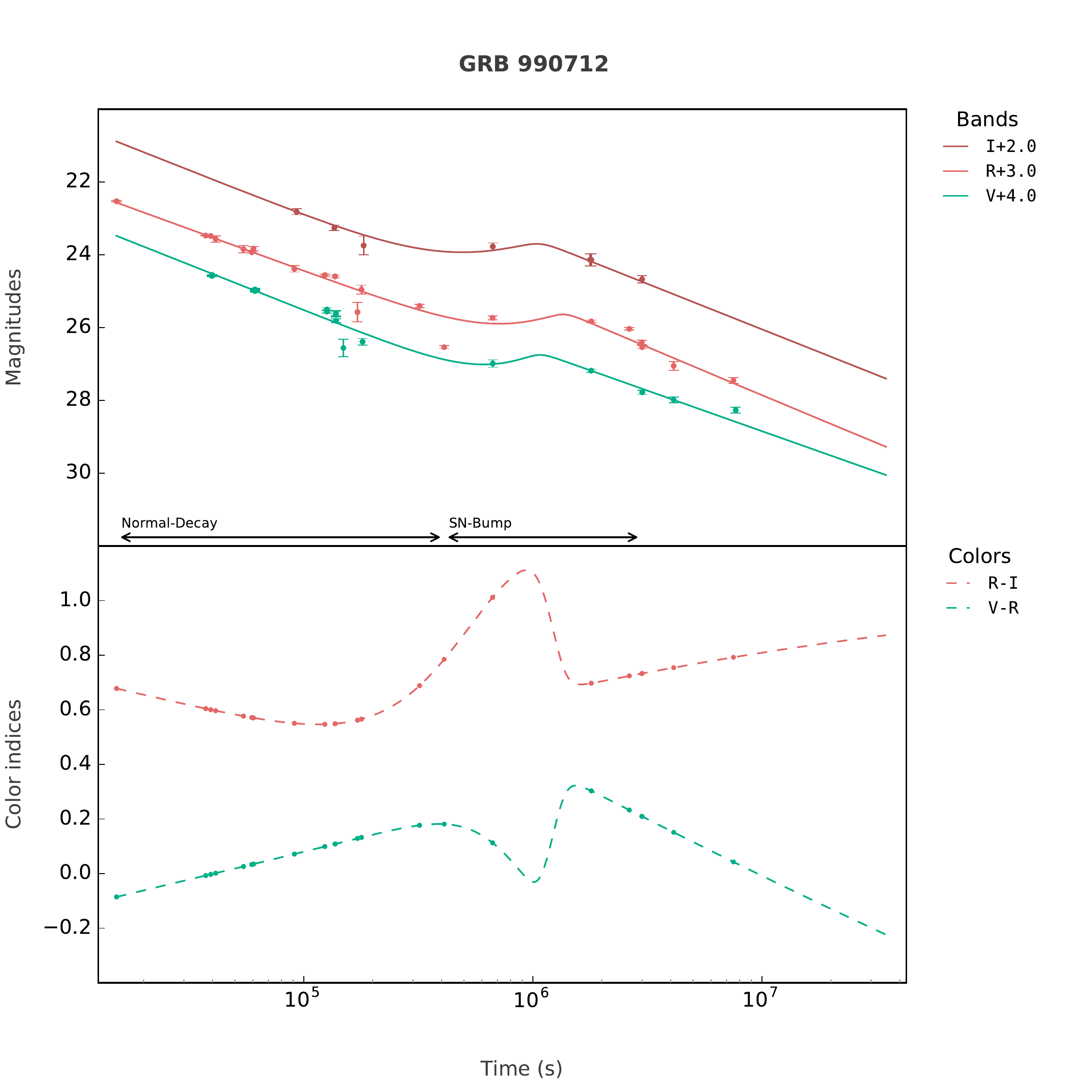}
\caption{The symbols are the same as Figure \ref{GoldenLCs}, but for the Silver sample.}
\label{SilverLCs}
\end{figure*}

\begin{itemize}
\item \textit{Golden sample}:

The 25 bursts have simultaneous observations in multiple filters in the optical/IR bands (g, r, i, z, J, H, Ks) for a wide time window. They are mostly observed by the Gamma-Ray Burst Optical/NIR Detector \citep[GROND;][]{2008PASP..120..405G}, with some events contributed from observations of the RAPid Telescopes for Optical Response \citep[RAPTOR;][]{2010SPIE.7737E..23W} and the Peters Automated IR Imaging Telescope \citep[PAIRITEL;][]{2006ASPC..351..751B}. These define our Golden sample. 
Obtaining the CI\footnote{We focus on five color indices (g-r, r-i, i-z, J-H and H-Ks) for our Golden sample and 10 color indices (UVW2-UVM2, UVM2-UVW1, UVW1-U, U-B, B-V, V-R, R-I, I-J, J-H and H-K) for our Silver sample.} for these bursts is straightforward.
The afterglow multicolor light curves and the time evolution of CIs are displayed in Figure \ref{GoldenLCs}. 

\item \textit{Silver sample}: 

The 45 remaining bursts in our sample do not have simultaneous observations in multiple filters but have instead multiple observations in different filters from different optical instruments at different times. 
For those bursts, we use the light-curve fits (\S 2.3) and interpolate the missing data points to obtain concurrent values in different filters (UVW2, UVM2, UVW1, U, B, V, R, I, J, H and K). 
This group of bursts defines the Silver sample. It has wider observed energy bands than the Golden sample. 
We try to include a maximum number of possible multiple data to represent the whole light curve.
The multicolor light curves\footnote{We uniformly added different numbers to make a fixed set of offsets to different bands. 
For the Golden sample: g: 6, r: 5, i: 4, z: 3, J: 2, H: 1, K: 0;
For the Silver sample: UVW2: 12, UVM2: 9, UVW1: 8, UVM1: 7, U: 7, B: 5, V: 4, R: 3, I: 2, J: 1.5, H: 1, K: 0, u: 14, g: 11, r: 10, i: 9, z: 8.} and the time evolution of CIs for each burst are shown in Figure \ref{SilverLCs}.

\end{itemize}

The most accurate CIs are obtained in cases where there are simultaneous observations in multiple filters. This is the case for the bursts in the Golden sample. For many of these bursts there are additional nonsimultaneous data as well, which are discarded in order to keep the sample pure. Three criteria are applied to consider whether to add the additional data sources into the Golden sample: (i) there is at least an order-of-magnitude extension in the duration of observation; (ii) the number of data points should be greater than (or at least equal to) that in the original data source; (iii) at least three new bands have good observational data. Most cases do not satisfy  these criteria. However, in two particular cases, GRB 090426 and GRB 130427A, there is a significant amount of additional data. We therefore use these cases to compare the results from the Golden and Silver samples.

For example, we add the R-band data to the multicolor light curves in the Golden sample for GRB 090426 (Figure \ref{GoldenLCs}).  
It provides a much earlier observation that can be compared to the GROND griz observation.

\subsubsection{Methods to Determine the Temporal Behavior of Color Indices}

We further define sets of CIs.   
The first one contains time-resolved data (Data Set I). 
The second set is made of component-wise averaged CIs (Data Set II).
Finally, the mean CIs over the entire burst duration give Data Set III.
Hereinafter all the analysis (such as the figures and the tables) will be marked with which sets of data are used.

According to \cite{2001A&A...377..450S}, CIs should not be derived from fits to the light curve. 
They claimed that any fits to the data would distort the CIs. 
The Golden sample gives directly CIs directly while the Silver sample CIs are obtained with a fit to the light curve. Studying the difference between Golden and Silver samples allows us to check the consistency between our results and that of \cite{2001A&A...377..450S}. 

To account for the missing data points, two methods can be employed:
(i) interpolation of the light curve from the fits (applied to the Silver sample),
and (ii) averaging of existing data points as in \cite{2001A&A...377..450S}. 
Interpolating missing data points results in large uncertainties but has the advantage of preserving a maximum number of simultaneous CIs. If the afterglow emission is smooth enough, the uncertainties should not be too large. On the other hand, averaging data results in more accurate estimation of the CIs, but strongly reduces their number. As pointed out by \cite{2001A&A...377..450S}, averaging is also not reliable when light curves have rapid changes. Interestingly, after comparing both methods, we find that the results are consistent. We note that our Set II, which is obtained by averaging the CIs on each afterglow component, is similar to averages in arbitrary time intervals, as in \cite{2001A&A...377..450S}, except that, in our paper, the time intervals are imposed by expected physical variations of the light curves.  

\begin{figure*}
\centering
\includegraphics[angle=0,scale=0.80]{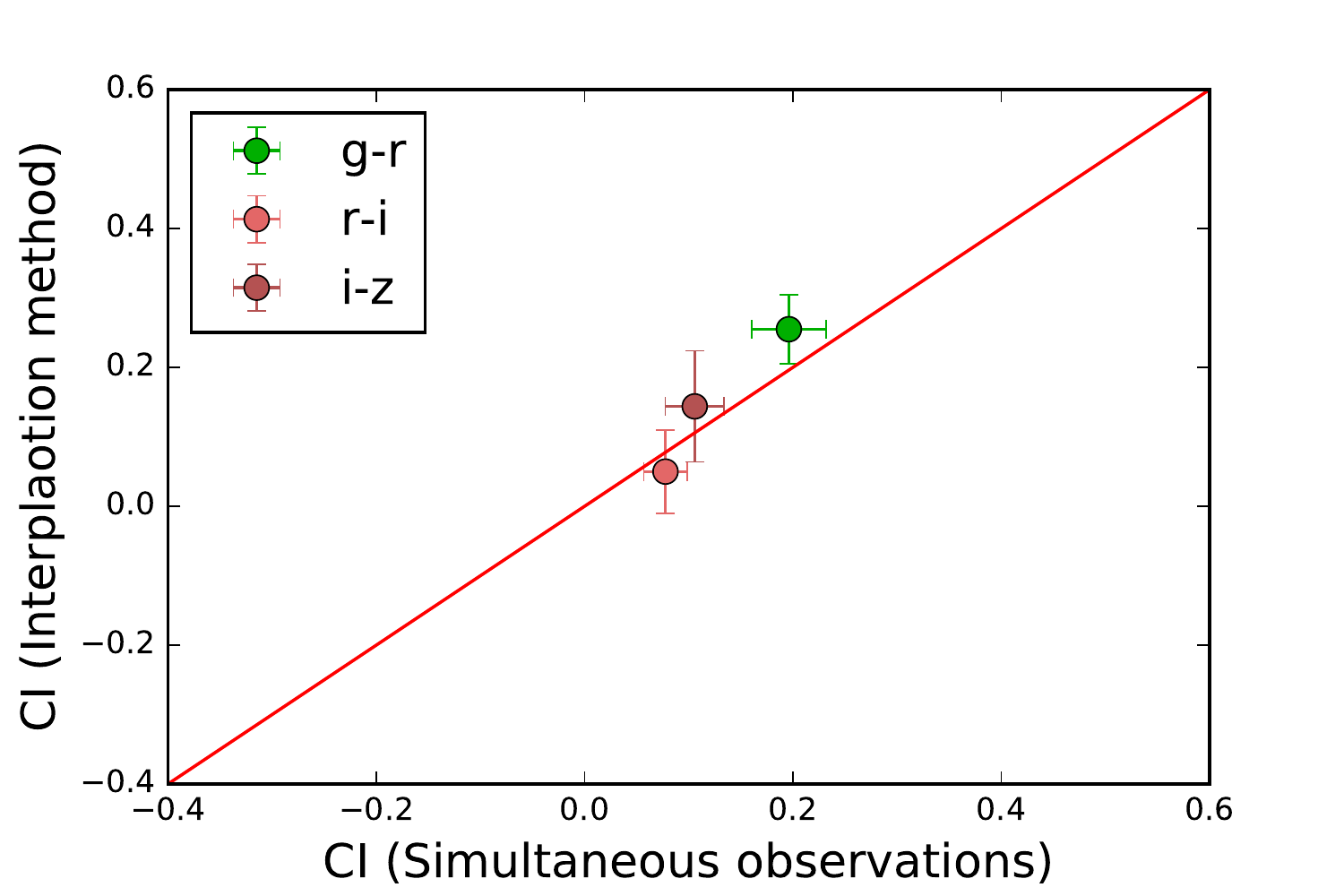}
\caption{Correlation of the CI between two different methods: Golden (CI is derived directly by simultaneous multiband observations) and Silver (interpolation method); data are derived from GRB 130427A. The equal line is represented by the solid curve.}
\label{MethodsChecking}
\end{figure*}

We investigate the correlation of the CIs from GRB 130427A using these two methods. This burst has a well-observed afterglow with a great number of data points. 
In Figure \ref{MethodsChecking}, we compare the CIs in the same observation duration between the Golden and the Silver samples. The CIs from the Golden sample are derived directly from RAPTOR \citep{2014Sci...343...38V} with simultaneous observations in multiple filters, while the CIs from the Silver sample are obtained by an interpolation method from the RATIR \citep{2017ApJ...837..116B}, which in general do not have simultaneous observation, compared with the GROND observation in the corresponding period. 
We find that the data cluster around the equal line, which indicates that the results of both methods are consistent with each other.

\subsubsection{Determination of Time Evolution}

To determine whether the CI changes with time, we consider the following quantitative analysis. First, we calculate the mean color index $\overline{m}$ and the standard deviation $\sigma_{\rm s}$ for each CI.
We do this for Data Set I and II defined in Section \S 2.4.2. If a significant variation exists for the $i$th CIs, the criteria $\mid m_{i}-\overline{m} \mid >\sigma_{m_{i}}+\sigma_{\rm s}$ should be met; here $\sigma_{m_{i}}$ is the error on the $i$th CI.
Note that both $\sigma_{m_{i}}$ and $\sigma_{\rm s}$ are positive. Therefore, we define two grades for the CI:
\begin{itemize}
\item Constant: the CI is consistent with the overall mean value.
\item Variable: the CI presents a significant variation.
\end{itemize}

\section{Statistical Properties of Color Indices}

\begin{figure*}
\centering
\includegraphics[angle=0, scale=0.80]{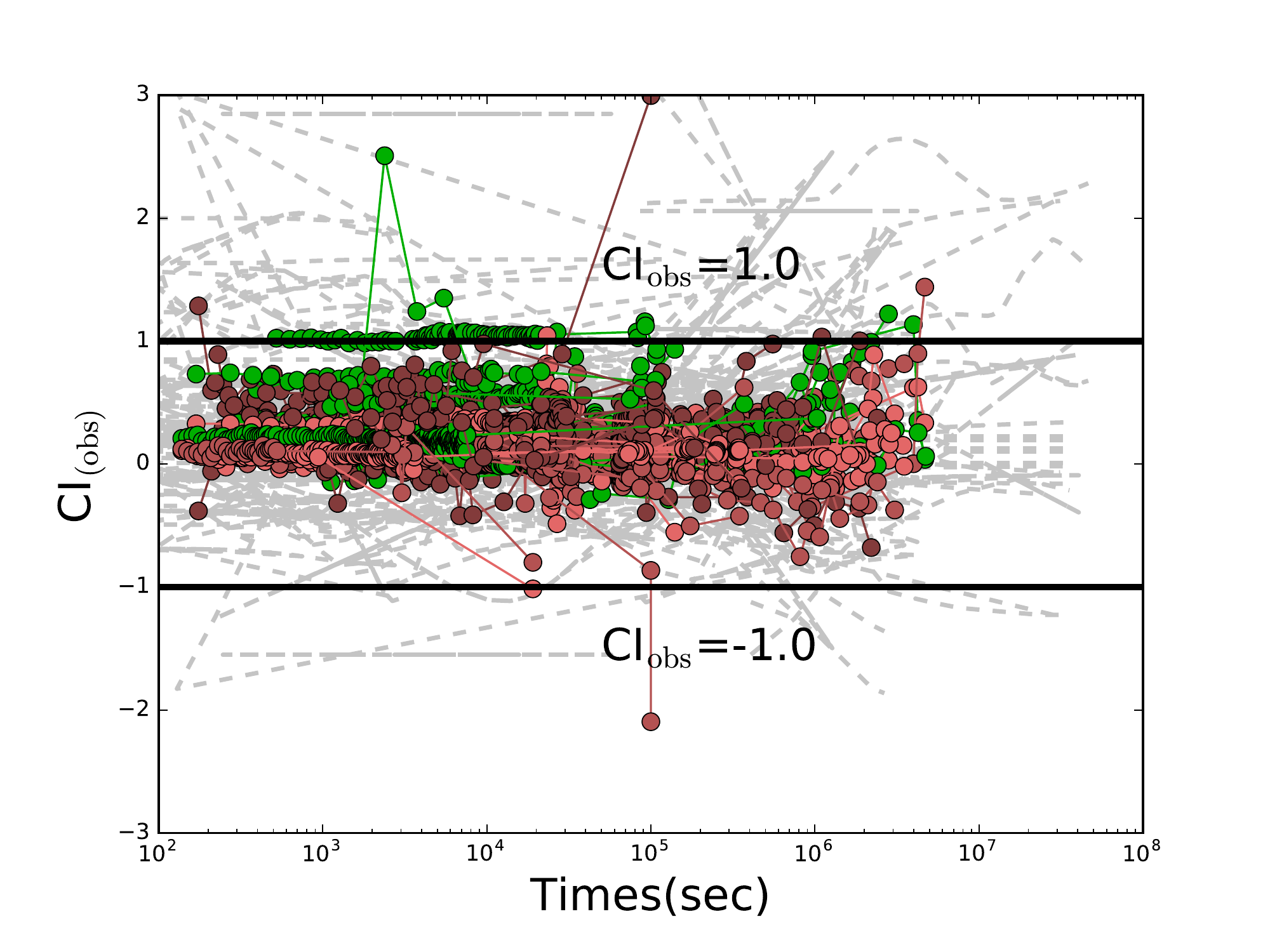}
\caption{Temporal evolution of all CIs for all bursts (Data Set I). The Silver sample is in grey, while each color represents a different combination of bands for the Golden sample. Two horizontal black lines show CI= -1 and 1.}
\label{TotalColors}
\end{figure*}

Table 4 summarizes the statistical properties of the CIs in our sample, which include the total number of data points, the distributed ranges, and their mean values.
In Figure \ref{TotalColors}, we present all CIs as functions of time in one panel to show the global behavior. We find that more than 90\% of the CIs distribute within [-1.0, 1.0] in the observed time intervals.

\begin{figure*}
\includegraphics[angle=0,scale=0.55]{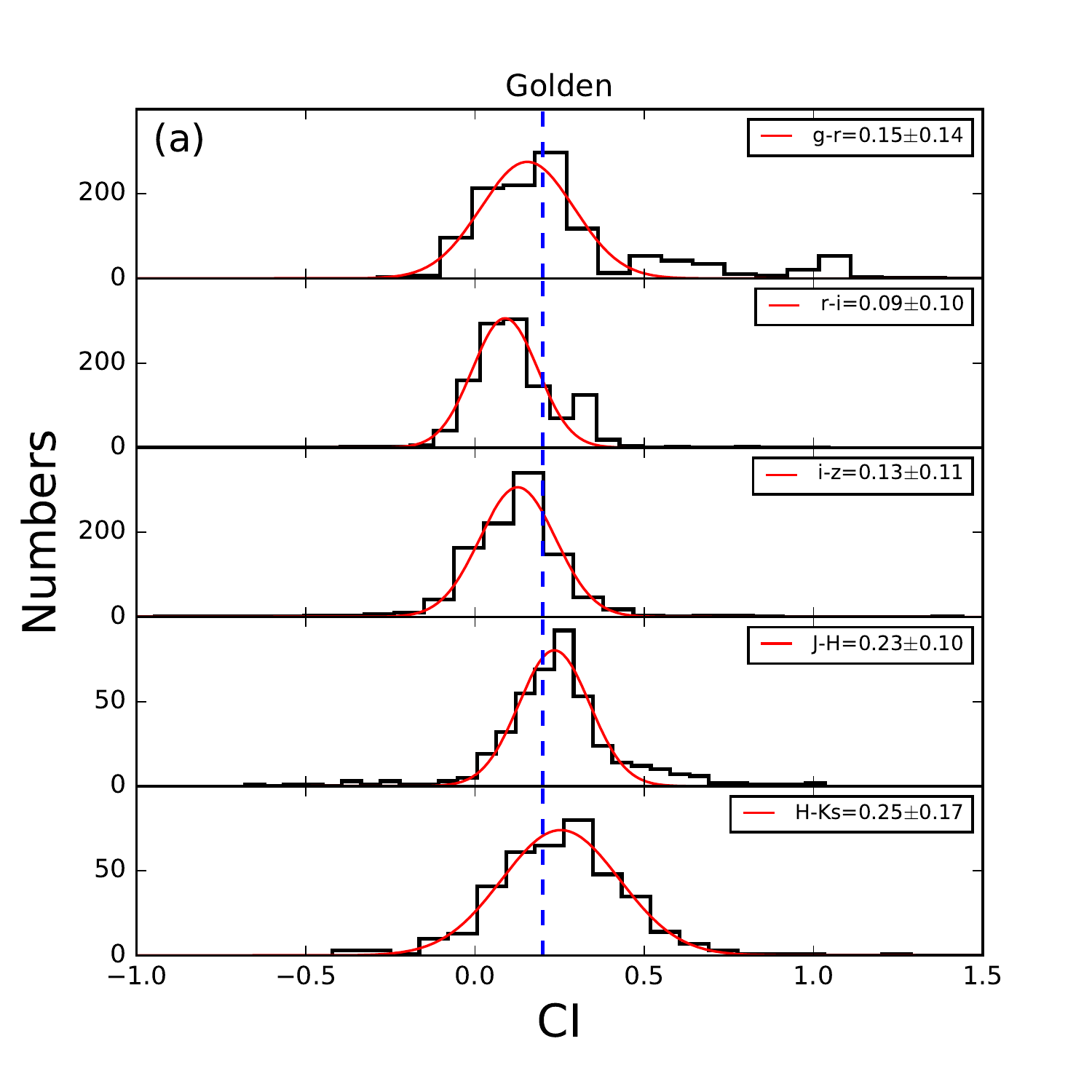}
\includegraphics[angle=0,scale=0.48]{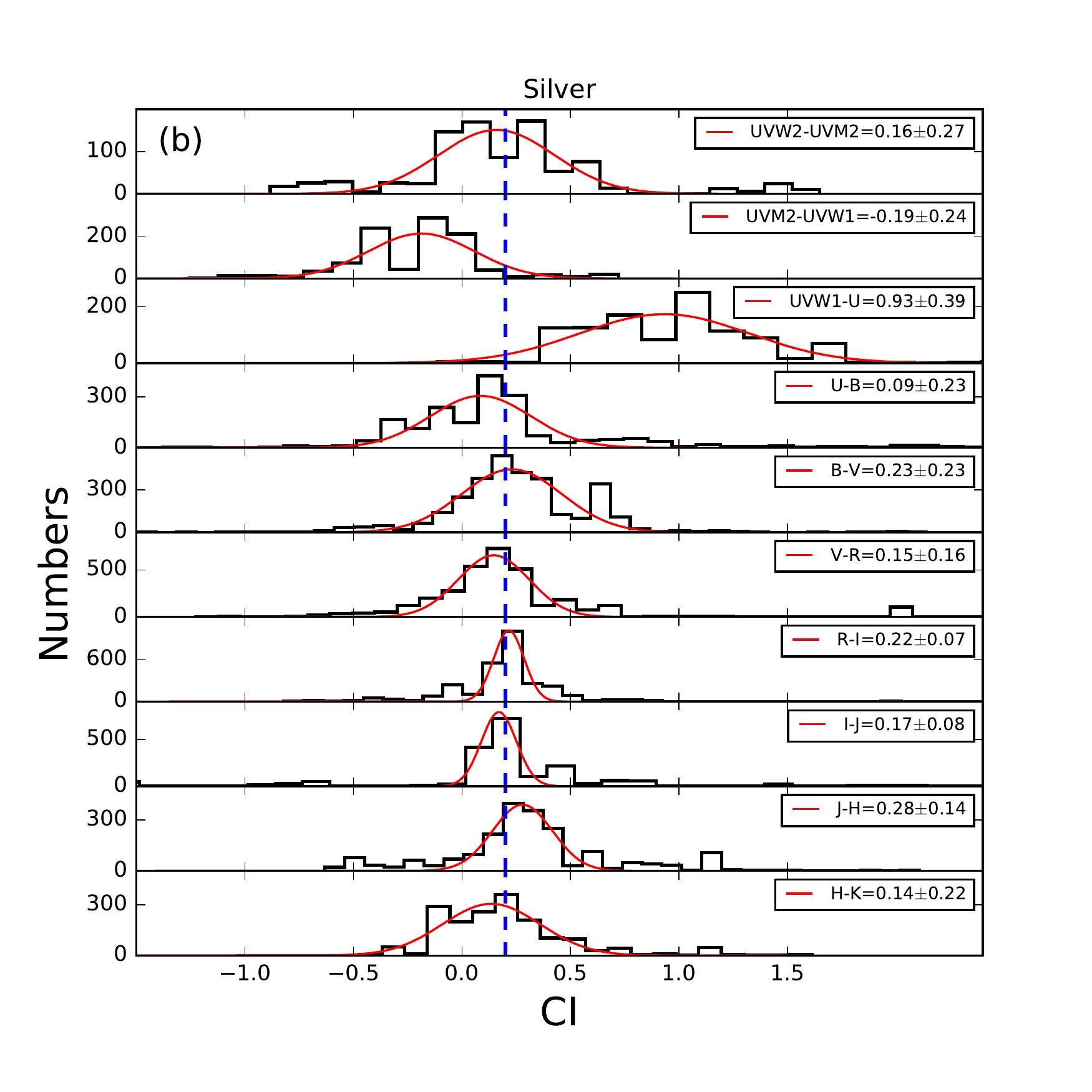}
\caption{Distributions of the color indices (Data Set I) for the (a) Golden and the (b) Silver sample. The red lines are the best Gaussian fits, whose results can be found in Table 4. The vertical blue dashed lines represent CIs equal to 0.2.}
\label{DisColors}
\end{figure*}

We analyze the distributions of the CIs (Data Set I) for both the Golden and Silver samples. 
They are all well fitted with Gaussian functions.
Figure \ref{DisColors} displays the distributions together with the best Gaussian fits and with the mean CI values, as well as their standard deviations.
For the Golden sample (Fig.\ref{DisColors}a), we have g-r=0.15$\pm$0.14, r-i=0.09$\pm$0.10, i-z=0.13$\pm$0.11, J-H=0.23$\pm$0.10 and H-Ks=0.25$\pm$0.17, respectively.
We find that the distributions\footnote{We note that these results are similar to the ones if only CIs with simultaneous observations for all five CIs are selected.} are consistent with each other with a mean value around 0.2. 
We note the presence of a tail at large values for the CI g-r.
For the Silver sample (Fig.\ref{DisColors}b), we obtained UVM2-UVW1=0.16$\pm$0.27, UVM2-UVW1=-0.19$\pm$0.24, UVW1-U=0.93$\pm$0.39, U-B=0.09$\pm$0.23, B-V=0.23$\pm$0.23, V-R=0.15$\pm$0.16, R-I=0.22$\pm$0.07, I-J=0.17$\pm$0.08, J-H=0.28$\pm$0.14 and H-K=0.14$\pm$0.22, respectively.
We find that the distributions are still consistent with each other with a mean value also around 0.2 for most of the CIs. This indicates that most of the CIs could be explained by the standard afterglow model with a single power-law spectral slope.

Note, however, that there are two inconsistent peaks in the Ultraviolet (UV) bands, with the UVM2-UVW1 CIs at -0.19 and UVW1-U CIs at 0.93. The difference between the distributions could originate from the method that the CIs are obtained for the Silver sample. Because of the fitting and interpolation procedures, they carry large uncertainties, not taken into account by the histograms.  
A second reason for this difference could be underestimation of the extinction corrections.
This is because the extinction curve in the host galaxy is largely uncertain\footnote{The uncertainty of the value stems from the method of obtaining the host extinction. If the extinction value of each energy band is constant, then the derived CIs will also be constant.} and therefore the corrected data could deviate from the actual values. This implies that after making the correction to the data, there could also be an additional reddening. Further investigation of the extinction property for each individual burst will be presented in a following paper (Li et al, in prep.). Stronger deviations are expected in high-energy bands in the Silver sample, tentatively explaining the possible reddening in those bands (Fig.\ref{DisColors}b). Also, very few GRB host galaxies show MW/LMC-like 2175 $\overset{\circ}{\rm A}$ features. However, the redshifts are distributed over a wide range of values, so even if such features exist, they would appear in different observer-frame bands. Therefore, the effect would be completely diluted. 

\begin{figure*}
\centering
\includegraphics[angle=0,scale=0.8]{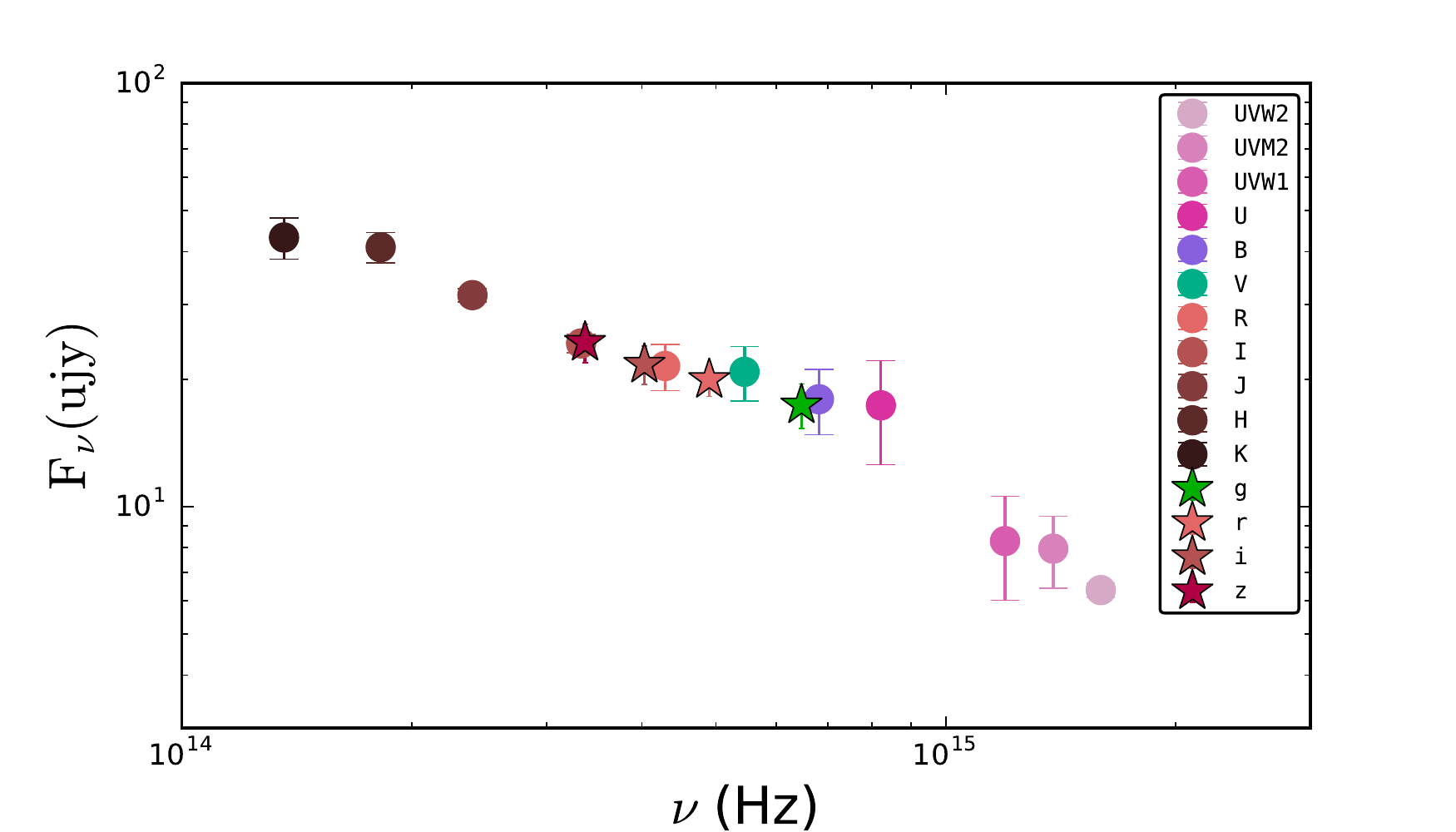}
\caption{SED of the afterglow, which derives from average values of CIs (Data Set III) and assumes a typical observation magnitudes for r band (Golden sample)/R band (Silver sample). Different colors represents different energy bands. The stars represent the Golden sample (SDSS griz system), while the circles represent the Silver sample (common UBVRI photometry system).}
\label{SED}
\end{figure*}

To test whether there exists an additional reddening, we analyze the SED, using the average CIs (Data Set III, regardless GRBs) and assuming a typical magnitude for the r-band (Golden sample)/R band (Silver sample). In Figure \ref{SED}, we show the SED of the afterglow from a wide-band observations. We note that the difference in the distributions between the Golden and Silver samples are negligible in the SED; thus, it is reasonable to analyze the SED of the Silver sample. However, we find that the SED is still not completely consistent with the theoretical expectations (a single power-law spectral slope). This implies that the traditional extinction models (LMC, SMC, MK, etc.) are not precise in fully describing the extinction character for some GRBs.

\begin{figure*}
\centering
\includegraphics[angle=0, scale=1.0]{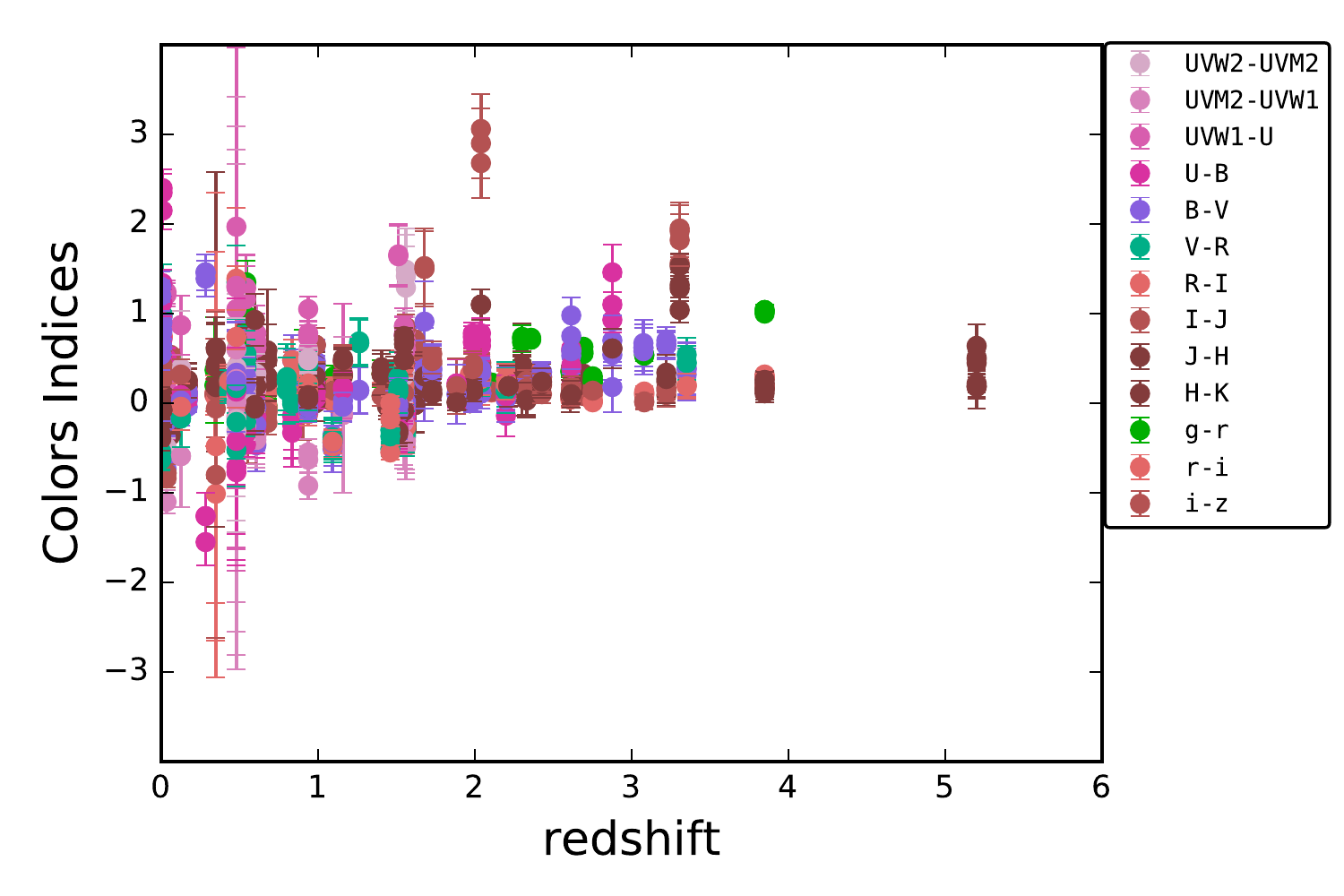}
\caption{Color indices (based on Data Set III) as a function of redshift. Different colors represent different bands.}
\label{RedshiftColors}
\end{figure*}

We conclude by estimating the influence of redshift on the CI. The redshifts of our bursts lie between \textit{z}=0.007 and 5.2. 
In Figure \ref{RedshiftColors}, we plot the CIs against the redshifts.
We obtain a Spearman correlation coefficient \textit{r}=0.21, corresponding to a chance probability $\hat{p}$= $\leq$ 10$^{-4}$. 
This indicates that the possible dependence of the CIs on the redshift is very weak.

\section{Temporal Evolution of the Color Indices}

\subsection{Time Evolution for Data Set I (Full Resolution Data)}

\begin{figure*}
\includegraphics[angle=0,scale=0.70]{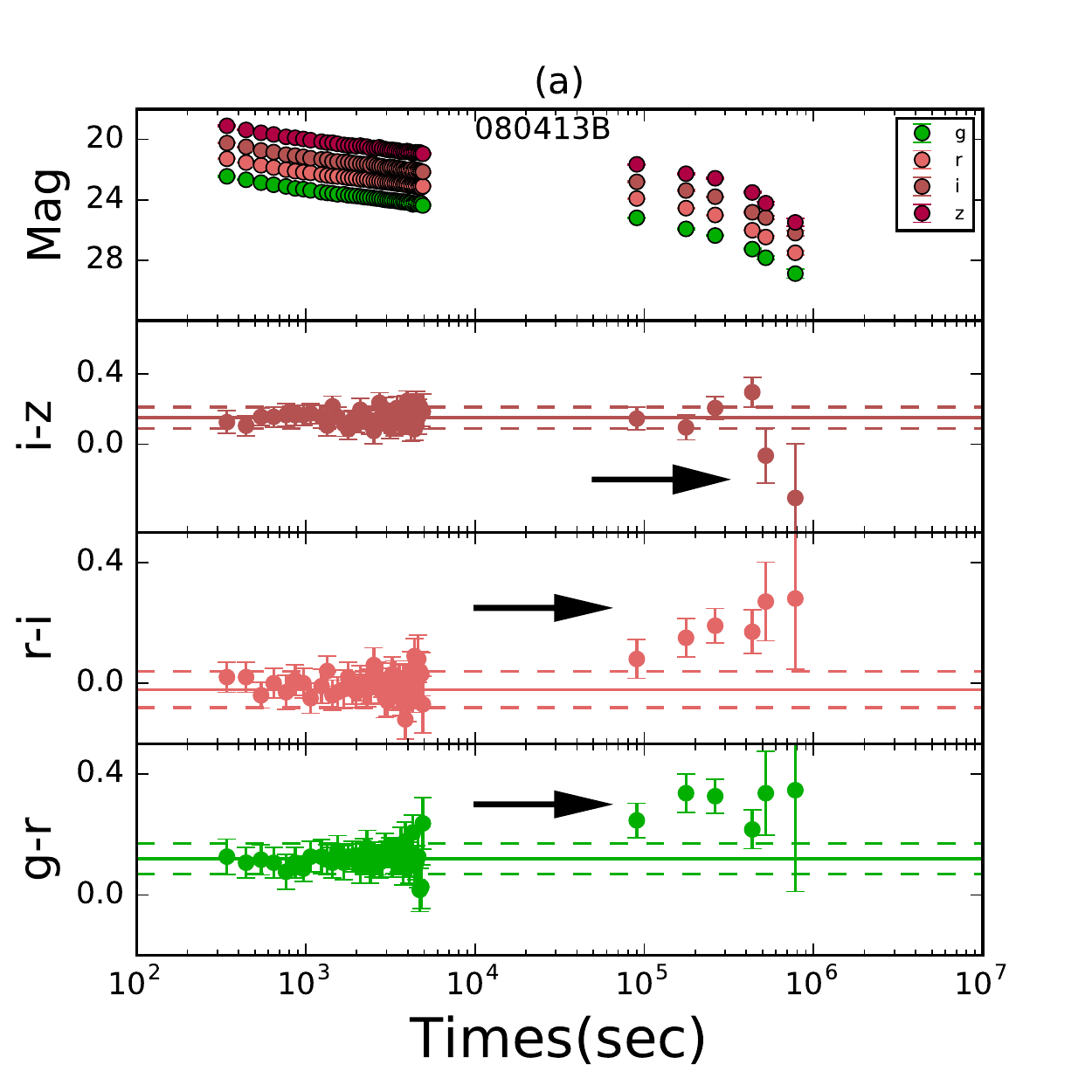}
\includegraphics[angle=0,scale=0.50]{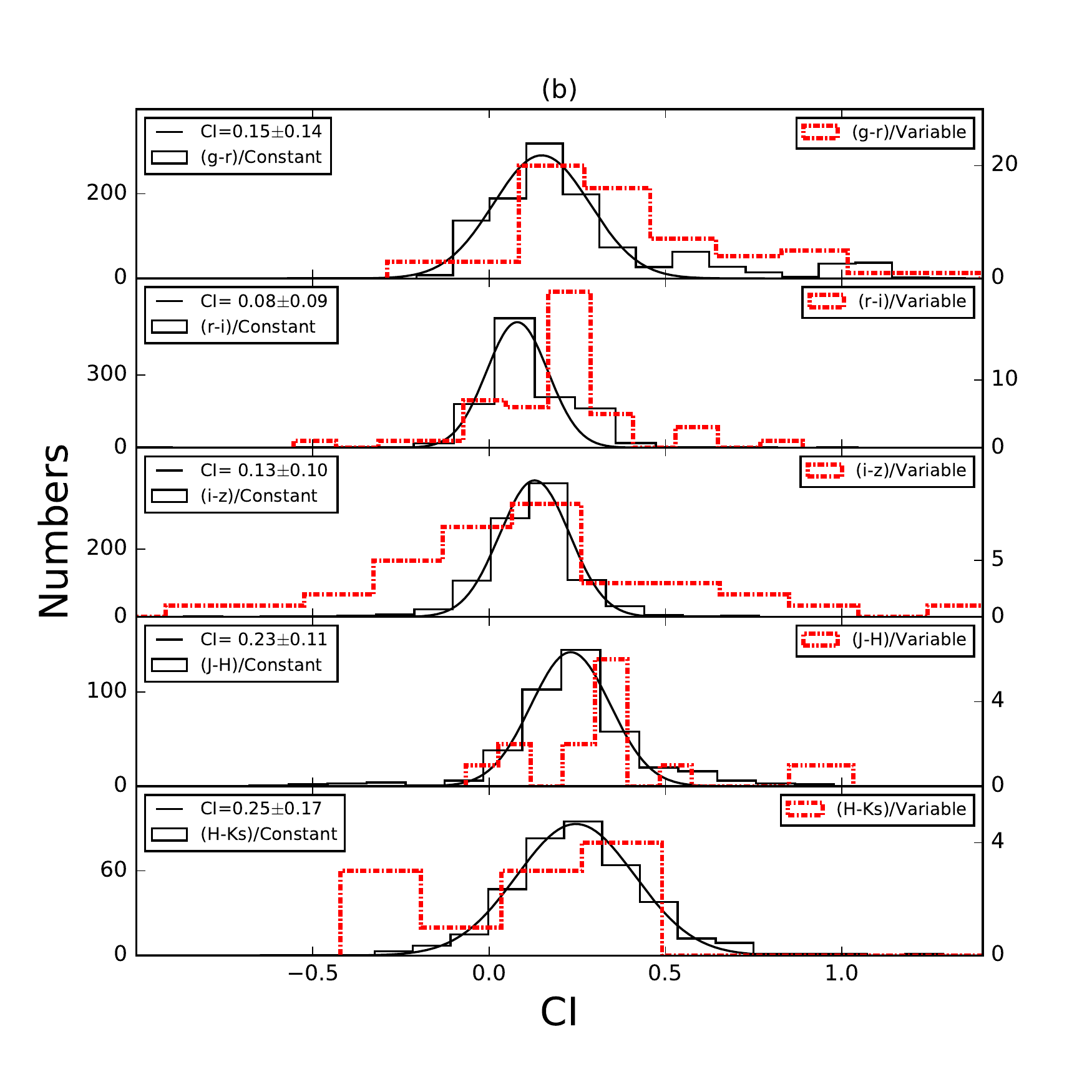}
\caption{Left: multiband light curves together with the color indices of GRB 080413B. Variable CIs are labeled with the arrow. Right: distribution of \textit{variable} (in red) and \textit{constant} (in gray) CIs. The best Gaussian fit of the \textit{constant} CI is represented by the gray curve, one has $g-r=0.15\pm0.14$, $r-i=0.08\pm0.09$, $i-z=0.13\pm0.10$, $J-H=0.23\pm0.11$ and $H-Ks= 0.25\pm0.17$, respectively.}
\label{GradeColors}
\end{figure*}

Following the definition of time evolution outlined in Section \S 2.4.3, we find that there are 20 out of 25 GRBs in the Golden sample that have at least more than two data points that are \textit{variable} CIs.
This indicates that the evolution of the CIs in the afterglow is very common.
GRB~080413B is taken as an example, and its CIs as functions of time are displayed in Figure~\ref{GradeColors}(a).  The \textit{variable} CIs are indicated by arrows. We note that the evolution of CIs tends to appear at a late time. 
In Figure~\ref{GradeColors}(b), we compare the distributions of \textit{constant} and \textit{variable} CIs. We see that the nonvarying CIs are distributed similarly from one color to the others. Fitting the histograms by a Gaussian function, one has g-r=0.15$\pm$0.14 (mean $\pm$ standard deviations), r-i=0.08$\pm$0.09, i-z=0.13$\pm$0.10, J-H=0.23$\pm$0.11 and H-Ks=0.25$\pm$0.17, respectively.

Instead, the group of varying CIs shows a broad distribution for different colors, possibly indicating different underlying phenomena.   
Table 5 summarizes the total number of \textit{constant}/\textit{variable} CIs.
In the 25 GRBs of the Golden sample, 4039 out of 4212 (95.9\%) values for all the five CIs in all the time bins are \textit{constant} while only 173 (4.1\%) are \textit{variable}.
This indicates that during most of the observed intervals, the CIs are constant, even for the 20 bursts that have a change (a small number of data points) in them.

\subsection{Time Evolution for Data Set II (Component-wise Average)}

We investigate the connection between the values of CIs and the afterglow light-curve components to determine their temporal evolution from one component to another. The components were identified in \S 2.4.
We calculate the mean value of CIs for each emission component\footnote{Note that if several similar emission phases exist at different times in a single burst, we merge them together and calculate the mean value. This is applied for GRB 021004 and GRB 071031 (with two late flares).}.
The results are presented in Table~2 for the Golden sample and in Table~3 for the Silver sample.

\begin{figure*}
\includegraphics[angle=0,scale=0.50]{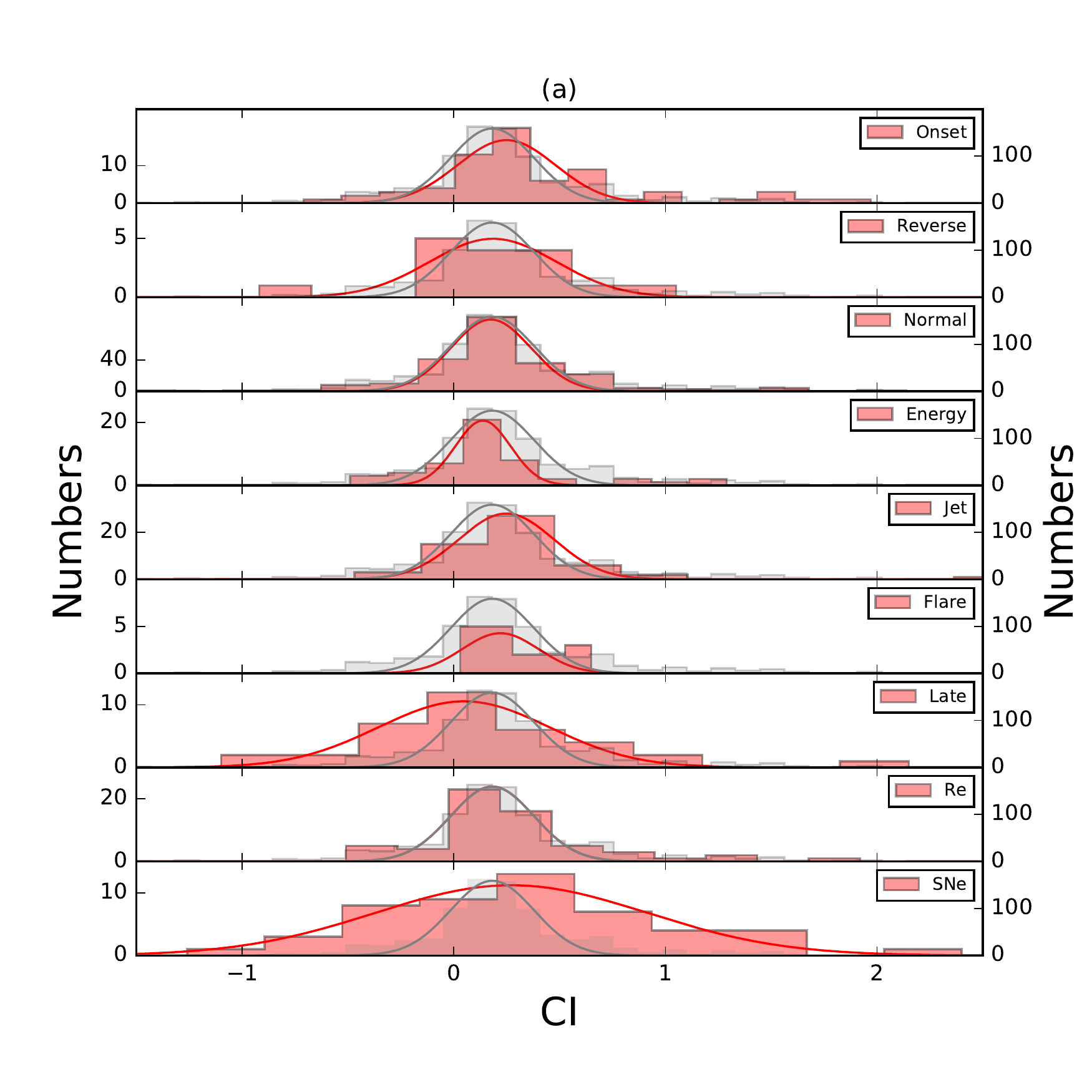}
\includegraphics[angle=0,scale=0.50]{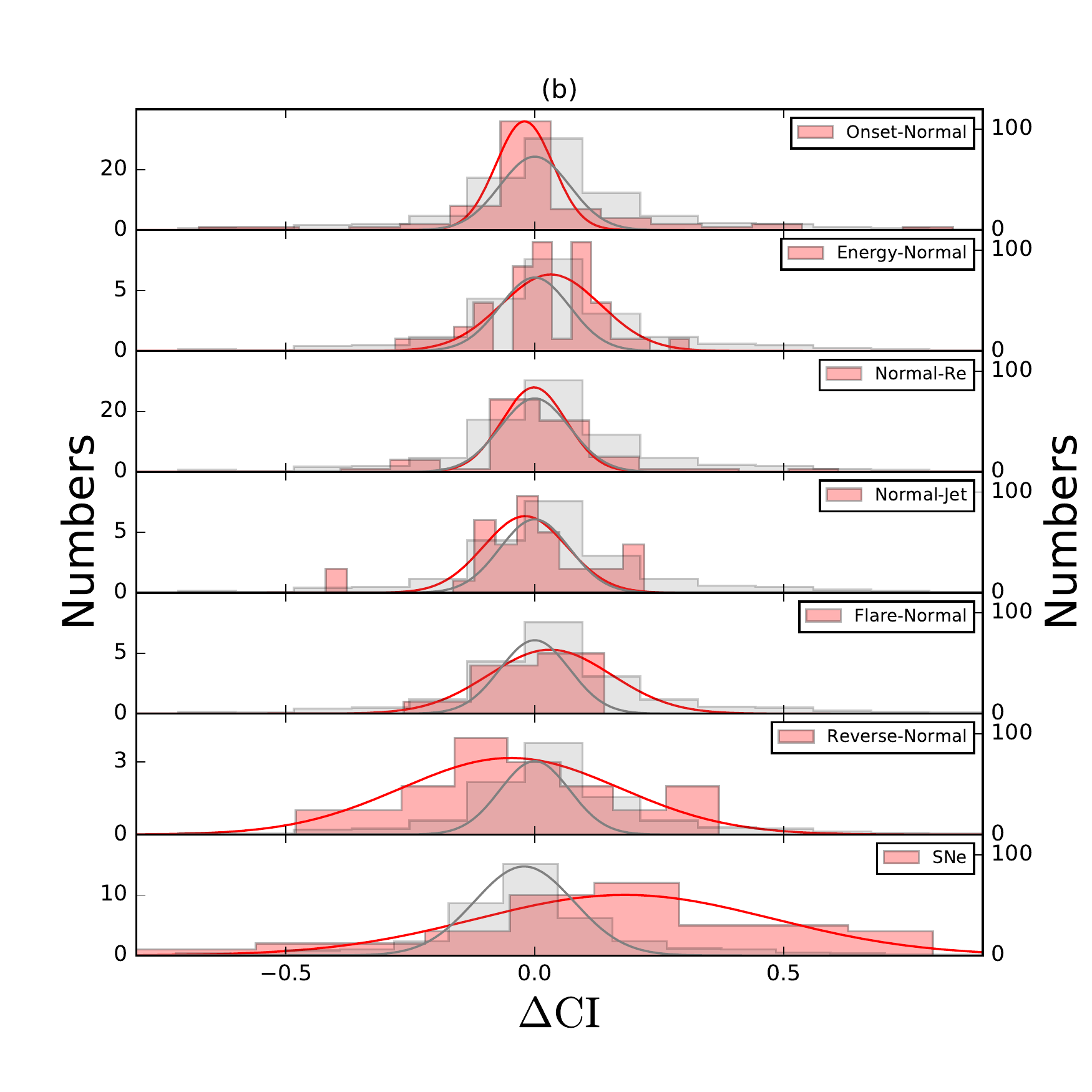}
\caption{Distributions of CI and $\Delta$CI for various emission components (in red), while the total distribution of CI or $\Delta$CI appears in grey.  
(a) Distributions of CI (Data Set II, regardless of the color and GRBs), together with the best Gaussian fits with Onset=0.25$\pm$0.23, Reverse=0.18$\pm$0.30, Normal=0.17$\pm$0.19, Energy=0.14$\pm$0.13, Jet=0.25$\pm$0.23, Flare=0.22$\pm$0.18, Late=0.05$\pm$0.40, Re=0.18$\pm$0.20, SNe=0.27$\pm$0.64, while the gray histograms are comparing with their global behaviors, and the gray lines are the best Guassian fits with Global=0.18$\pm$0.20.
(b) Distributions of $\Delta$CI (Data Set II, regardless of the color and of GRBs) between one component and its following one, together with the best Gaussian fit, whose results are given by Table 6.}
\label{DisComColors}
\end{figure*}

In Figure \ref{DisComColors}(a), we show the distribution of CIs among various optical emission components (Data Set II, regardless of the bursts and CIs) and the distribution of $\Delta$CI (results from Gaussian fits are summarized in Table 6) between one component and the following one (Figs. \ref{DisComColors}b). $\Delta \rm CI$ is defined by
\begin{equation}
\Delta \rm CI=CI(t_{2})-CI(t_{1}),
\label{eq:DetaCI}
\end{equation}
where $t_{2}>t_{1}$. 
This means that an increasing CI, with a red-to-blue color change, corresponds to $\Delta \rm CI>0$; in contrast, $\Delta \rm CI<0$ represents a decrease in CI with a blue-to-red color change.

We defined the global sample as the set of all CIs from both the Golden and Silver samples.  
We find that various emission components generally have similar distributions, with typical values $\sim$ 0.2, except the SN component, which presents a different $\Delta$CI distribution.
Comparing the CIs, we find that the distributions are consistent, though the number of data points is small. Note, however, that the distribution is very broad.

\begin{figure*}
\centering
\includegraphics[angle=0, scale=1.0]{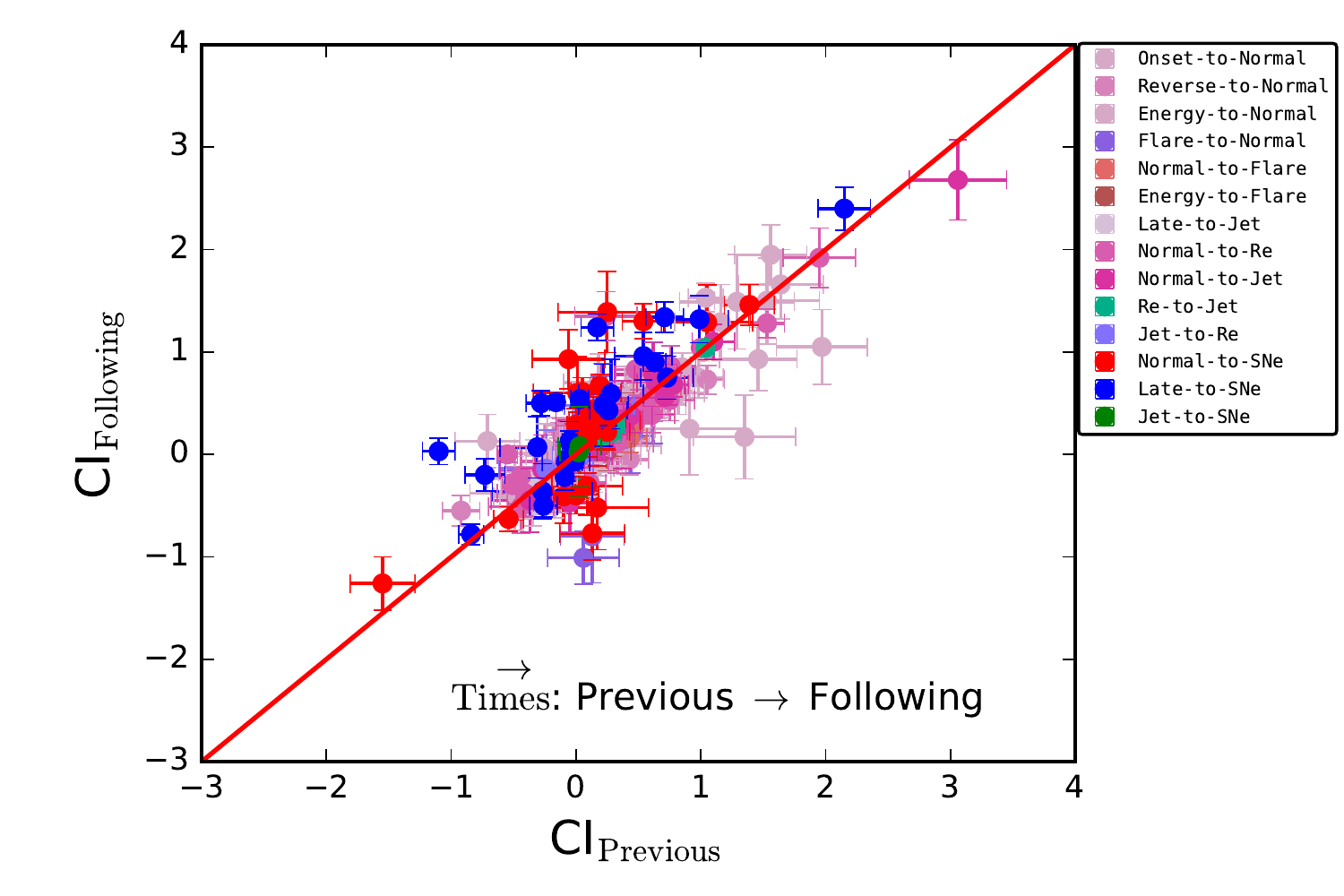}
\caption{CI correlation between one component and its following one (Data Set II for both the Golden and Silver samples). The equal line is represented by the continuous curve.}
\label{ComsComsColors}
\end{figure*}

We present in Figure \ref{ComsComsColors} the CI-CI correlations between consecutive components, regardless of the CIs. 
We compute the correlation index \textit{r} in all cases and find that it is systematically close to 1. The two smallest values are the transition to an SN component with r=0.57 and a flare component with r =0.60. 
In addition, a linear fit to the slope for most correlations is also close to 1. This indicates that the CIs do not have a significant change between consecutive components. 
The fit results, along with the Spearman correlation coefficient \textit{r} and the chance probability $\hat{p}$  are reported in Table 6. This table also gives the number of CIs per emission component and the distribution range of $\Delta$CI.

We find that 263 of 306 (85.9\%) pair components satisfy the criterion that the CI does not change. The large fraction of CI without significant time variation for both the Golden and Silver samples suggests that the emission mechanism of the various components in the optical transient may stay the same. This is consistent with the prediction of the external shock afterglow model, for which the emission is dominated by synchrotron radiation, provided that the observation bands do not cover the spectral breaks. 

Only 43 of the 306 (14.1\%) pair components have a significant difference.
For these pairs, we found that 18 include SN components, 5 include revese-shock components, and 5 include flares (see Table 5).
The Golden and the Silver samples show similar results.
In those pair components that include an SN, 36\% show variation.
Likewise, 20\% of the pair components including flares show variation.
This makes the SN and flare components the ones that exhibit most common variations.
This is consistent with the observational evidence that the optical transient spectrum typically changes during the associated SN and flares \citep[e.g.,][and references therein]{2001A&A...377..450S,2004A&A...427..901S,2006ApJ...642..354Z,2016ApJ...831..111M, 2016ApJ...825..107G, 2017ApJ...841L..15G}. 
We also found a few cases of CI variation for the jet break components.
This is note-worthy since a jet-break is believed to be purely dynamical. Therefore, CI evolution is not expected. 
Here, one should bare in mind the possibility that we could have identified the wrong component. For example, we could have identified a flare as a re-brightening bump if the flare is weak (GRB 021004, GRB 000301C, GRB 060707A). This is because they have similar light curves.
Another possibility is that the interpolation procedure for the Silver sample introduces uncertainties affecting the determination of the CI.

\section{Physical Implication}

\subsection{Constant Color Indices in the External Shock Model}

There is a natural connection between the CIs and the spectral indices.
Assuming that the SED of the afterglow in the considered bands is well described by an intrinsically single power law, the spectral flux density (erg cm$^{-2}$s$^{-1}$Hz$^{-1}$) for the afterglow is described by Equation (\ref{eq:Host1}).
The relation between the flux and the observed magnitude is given by Equation (\ref{eq:mag}), so we can write the CI, which is expressed from the flux density as
\begin{equation}
\rm CI=-2.5 \rm log_{10} \left(\frac{\textit{F}_{\nu_{2}}}{\textit{F}_{\nu_{1}}}\right)+C,
\label{eq:fluxcolors}
\end{equation}
where $C$=ZP$_{2}$-ZP$_{1}$, and ZP$_{i}$= 2.5log$_{10}$($F_{\nu_{i},0}$) is the zero-point of various bands $i$.
Therefore, the spectral index $\beta_{o}$ can be connected to the CI as
\begin{equation}
\rm CI=2.5\beta_{o} \rm log_{10} \left(\frac{\nu_{2}}{\nu_{1}}\right)
\label{eq:betaocolors}
\end{equation}

According to Equation (\ref{eq:betaocolors}), there exists a linear relation between CIs and spectral indices. Therefore, CIs provide a probe to investigate the spectral properties of the optical afterglow.

\begin{figure*}
\centering
\includegraphics[angle=0, scale=0.60]{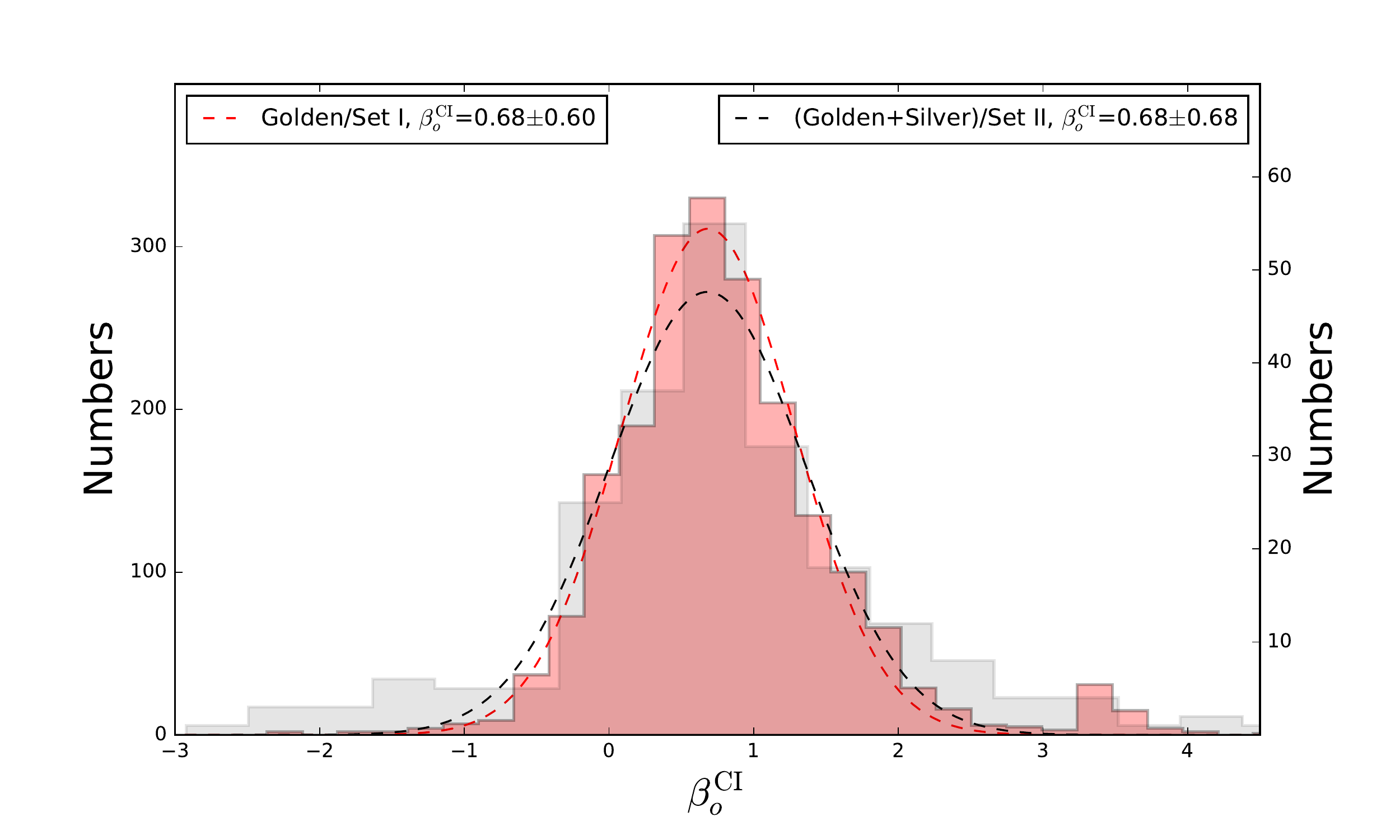}
\caption{Distributions of spectral indices $\beta_{o}^{\rm CI}$ and their best Gaussian fits (dashed line), $\beta_{o}^{\rm CI}$ are derived from the CI-$\beta_{o}$ relation (Eq \ref{eq:betaocolors}) using the \textit{constant} color indices of data Set I for the Golden sample and the Data Set II during the normal-decay phase for both the Golden and the Silver samples. We find $\beta_{o}^{\rm CI}$=0.68$\pm$0.60 for Data Set I and $\beta_{o}^{\rm CI}$=0.68$\pm$0.68 for Data Set II.}
\label{BetaoColors}
\end{figure*}

Using Equation (\ref{eq:betaocolors}), we derive the index $\beta_{o}^{\rm CI}$ using \textit{constant} CIs of Data Set I of the Golden sample (see Appendix Table 7) and the Data Set II of both the Golden and Silver samples during the normal-decay phase. We show the distribution of $\beta_{o}^{\rm CI}$ in Figure \ref{BetaoColors}. 
The distribution is well fitted by a Gaussian function, and $\beta_{o}^{\rm CI}$=0.68$\pm$0.60 for \textit{constant} CIs of Data Set I and $\beta_{o}^{\rm CI}$=0.68$\pm$0.68 for the Data Set II during the normal decay phase, which are consistent with each other.

The distributions are consistent with the predictions of the external shock model, with typical $\beta_{o}$ around 0.75 \citep{1997ApJ...476..232M,1998ApJ...497L..17S, 2006ApJ...642..354Z}. 
The spectral indices inconsistent with the external shock model are obtained from the high-energy bands, which could point toward a stronger reddening. 
We note that according to the external shock model, $\beta_{o}^{\rm CI}$ should cluster between 0.5 and 1.0. However, we found that a small fraction of data deviate from this range, which have large error bars, and thus they cannot accurately constrain the calculation of CI. Then this uncertainty propagates to the calculation of spectral index.

According to the standard synchrotron afterglow model, radiation is produced by electrons distributed with a power-law index \textit{p}. 
Therefore, the CIs can be derived from the following equations for both a constant-density interstellar medium (ISM) and a wind-like medium (wind):

(i) Fast cooling:
\begin{eqnarray}
\label{n} \rm CI=\left\{ \begin{array}{ll}
-2.5\rm log_{10} \left(\frac{\nu_{2}}{\nu_{1}}\right)^{-\frac{1}{2}}, & (\nu_{m}>\nu_{2}>\nu_{1}>\nu_{c}) \\
-2.5\rm log_{10} \left(\frac{\nu_{2}}{\nu_{1}}\right)^{-\frac{1}{2}}-2.5\rm log_{10} \left(\frac{\nu_{2}}{\nu_{m}}\right)^{-\frac{(p-1)}{2}}, & 
(\nu_{2}>\nu_{m}>\nu_{1}>\nu_{c})\\
-2.5\rm log_{10} \left(\frac{\nu_{2}}{\nu_{1}}\right)^{-\frac{p}{2}}, & 
(\nu_{2}>\nu_{1}>\nu_{m}>\nu_{c})\\
\end{array} \right.
\label{eq:FastCooling}
\end{eqnarray}

(ii) Slow cooling:
\begin{eqnarray}
\label{n} \rm CI=\left\{ \begin{array}{ll}
-2.5\rm log_{10} \left(\frac{\nu_{2}}{\nu_{1}}\right)^{-\frac{(p-1)}{2}}, &  
(\nu_{c}>\nu_{2}>\nu_{1}>\nu_{m}) \\
-2.5\rm log_{10} \left(\frac{\nu_{2}}{\nu_{1}}\right)^{-\frac{(p-1)}{2}}-2.5\rm log_{10} \left(\frac{\nu_{2}}{\nu_{c}}\right)^{-\frac{1}{2}}, & 
(\nu_{2}>\nu_{c}>\nu_{1}>\nu_{m})\\
-2.5\rm log_{10} \left(\frac{\nu_{2}}{\nu_{1}}\right)^{-\frac{p}{2}}, &  
(\nu_{2}>\nu_{1}>\nu_{c}>\nu_{m})\\
\end{array} \right.
\label{eq:SlowCooling}
\end{eqnarray}
Here $\nu_{c}$ is the cooling frequency, $\nu_{m}$ is the minimum frequency, and $\nu_{2}$ and $\nu_{1}$ are the frequencies of adjacent bands.
In the fast cooling regime, electrons at the injection Lorentz factor have time to lose their energy by emitting synchrotron radiation, while they do not in the slow cooling regime.  

\begin{figure*}
\centering
\includegraphics[angle=0,scale=1.0]{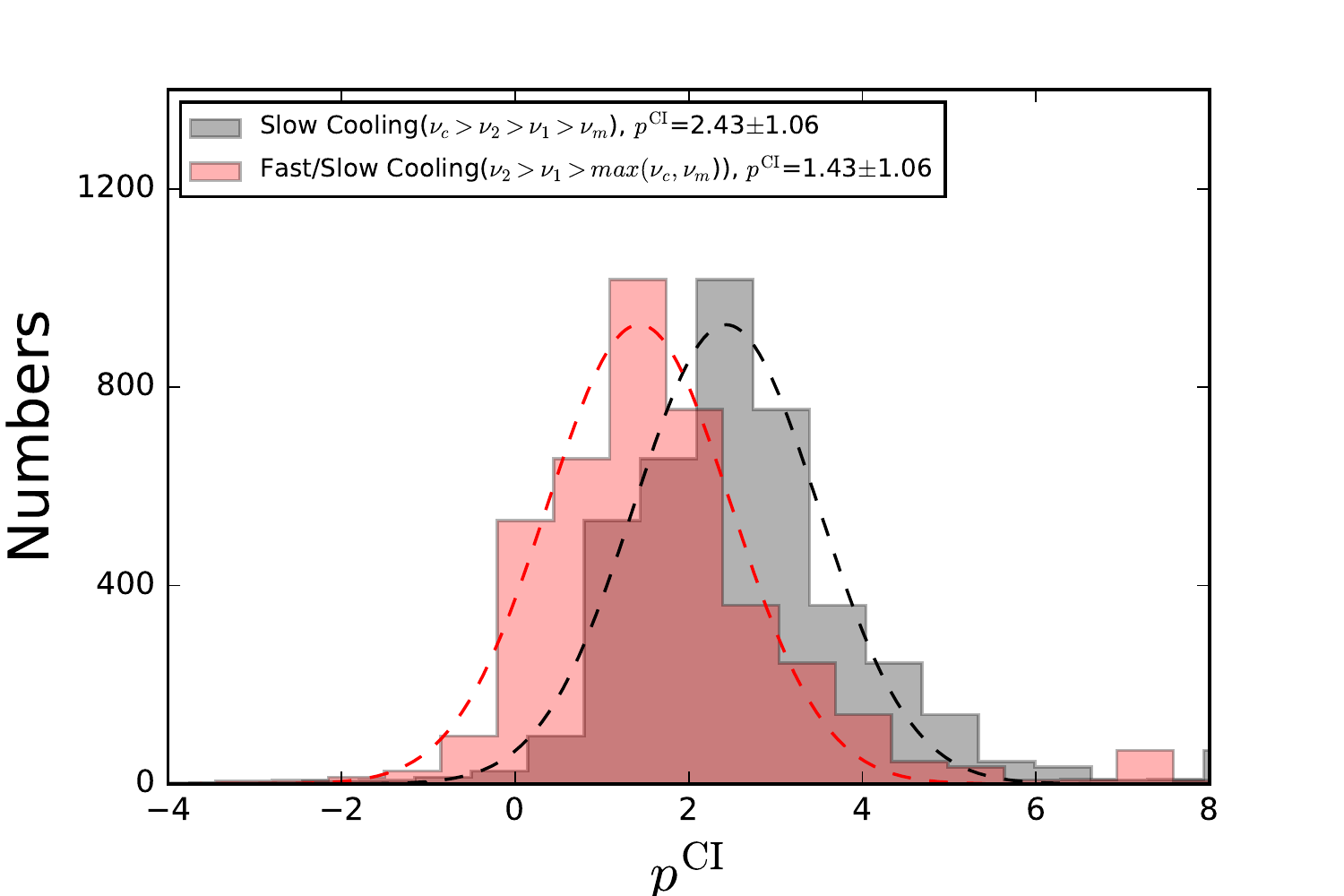}
\caption{Distributions of observed electron spectral indices $p^{\rm CI}$, derived from the \textit{constant} CI of the Golden sample (Data Set I). The dashed lines are the best Gaussian fits giving $p^{\rm CI}$=2.43$\pm$1.06 for the slow cooling case ($\nu_{c}>\nu_{2}>\nu_{1}>\nu_{m}$) and $p^{\rm CI}$=1.43$\pm$1.06 for the fast/slow cooling case ($\nu_{2}>\nu_{1}>max(\nu_{c},\nu_{m}$)).}
\label{PColors}
\end{figure*}

Next, we use the observed CIs (Data Set I of the Golden sample) to derive the electron power-law indices \textit{$p^{\rm CI}$} using the Equations (\ref{eq:FastCooling}-\ref{eq:SlowCooling}). In Figure \ref{PColors}, we present the distribution of \textit{$p^{\rm CI}$}, fit by a Gaussian. 
The mean values with their standard deviations are $p^{\rm CI}$=2.43$\pm$1.06 for slow cooling ($\nu_{c}>\nu_{2}>\nu_{1}>\nu_{m}$) and $p^{\rm CI}$=1.43$\pm$1.06 for fast/slow cooling ($\nu_{2}>\nu_{1}>max(\nu_{c},\nu_{m})$). 
We find that the slow cooling scenario is consistent with the theoretical predictions for relativistic shocks, \textit{$p$} $\sim$ 2.5. A wide range of the electron index is obtained, which is consistent with previous studies \citep{2006MNRAS.371.1441S,2008ApJ...675..528L}.  

\begin{figure*}
\centering
\includegraphics[angle=0,scale=0.70]{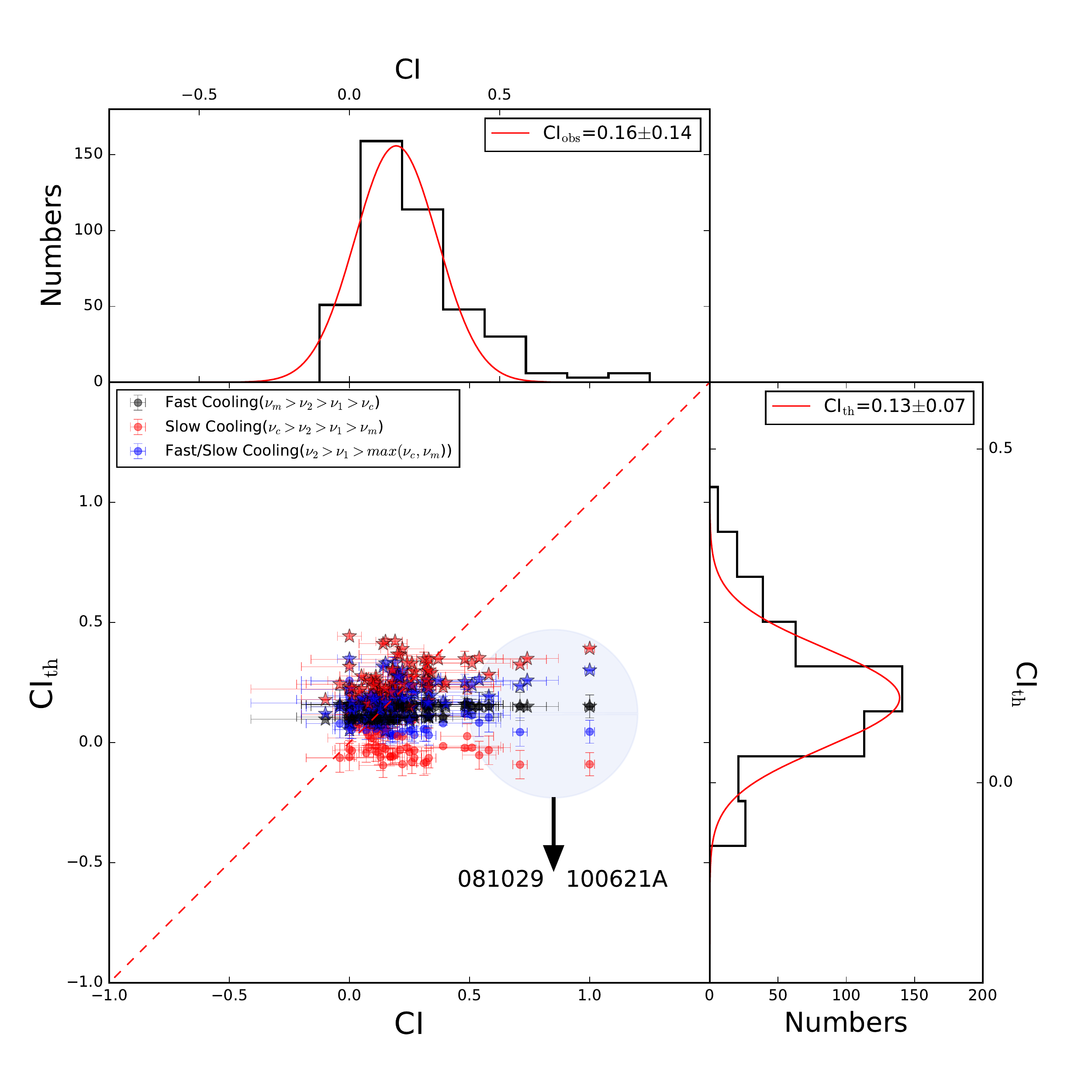}
\caption{Correlation with their distributions between theoretical and observed CI for all GRBs in the three different spectral regimes (Data Set III). The red dashed line is the equal line, and the red solid lines are the best Gaussian fits giving CI$_{\rm obs}$=0.16$\pm$0.14 for observed CI and CI$_{\rm th}$=0.13$\pm$0.07 and for theoretical CI.}
\label{CIthCIobs}
\end{figure*}

Next, we use the closure relations to compute the electron index $p$ from the temporal index $\alpha$ of the normal decay, which was obtained by fitting the light curve, or the spectral index $\beta_{o}$.
We then further estimate the expected value of the color indices CI$_{\rm th}$ (Data Set III of the Golden sample) using Eq.(\ref{eq:FastCooling}-\ref{eq:SlowCooling}).
Figure \ref{CIthCIobs} displays the correlation between the CI and CI$_{\rm th}$ for these spectral regimes. Their distributions are also shown with the best Gaussian fits CI$_{\rm obs}$=0.16$\pm$0.14 for observed CI and CI$_{\rm th}$=0.13$\pm$0.07 for theoretical CI.

We find that the data smoothly cluster around the equal line, and both observed CI and theoretical CI$_{\rm th}$ have a similar distributions, consistent with the afterglow synchrotron radiation model with an intrinsic power-law decay. 

We also find that a small part of CIs are far away from the equal line, which implies that these data are inconsistent with the model. These CIs are mainly derived from GRB 081029 and 100621A (Fig.\ref{CIthCIobs}). It is interesting that both GRB 081029 and GRB 100621A present a light curve with re-brightening bumps (see Figure \ref{GoldenLCs}). 
The late re-brightening bumps could come from different emission components. Overlapping bumps could alter the CIs estimation. 
However, we also find that CIs preceding the bump are still inconsistent with the model.
To investigate this further, we searched all the bursts in the Golden sample to find more cases presenting a clear late re-brightening bumps in the light curves. We find that, GRB 071025, GRB 091029, and GRB 100814A are consistent with the theoretical values and present a late re-brightening. 
The results show that the afterglow emission for both GRB 081029 and GRB 100621A is likely different from other bursts. 
 
\begin{figure*}
\includegraphics[angle=0,scale=0.45]{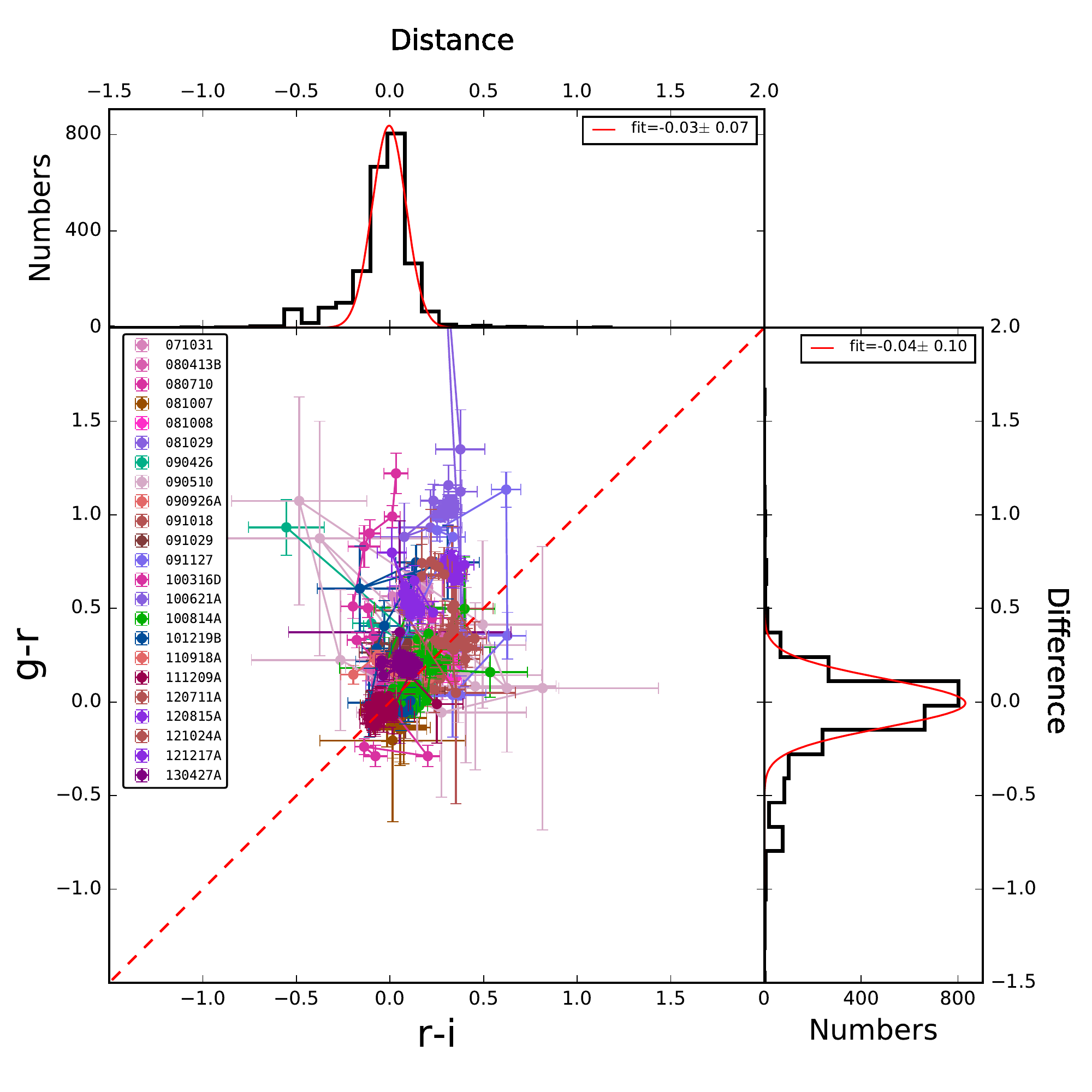}
\includegraphics[angle=0,scale=0.45]{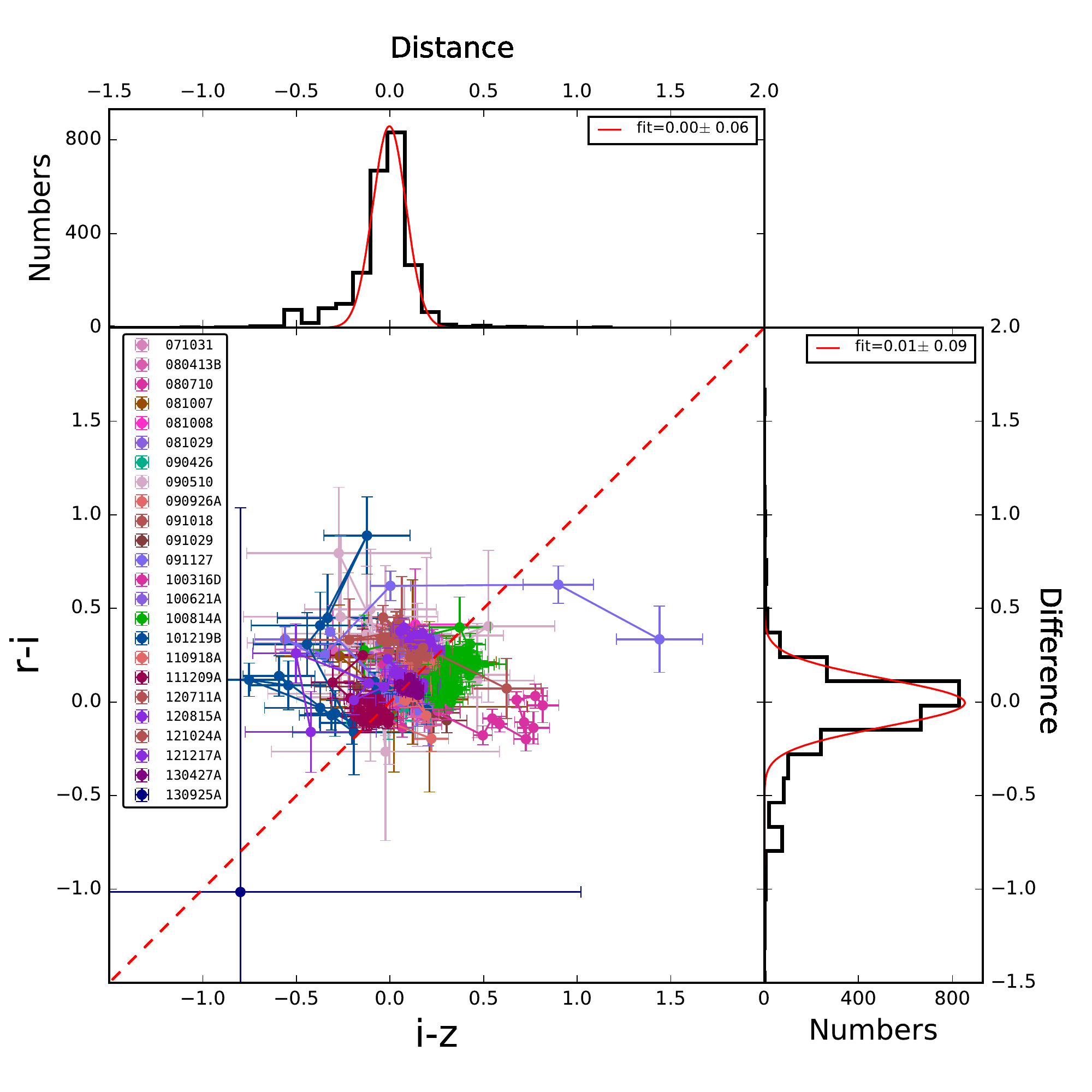}
\begin{center}
\centering
\includegraphics[angle=0,scale=0.45]{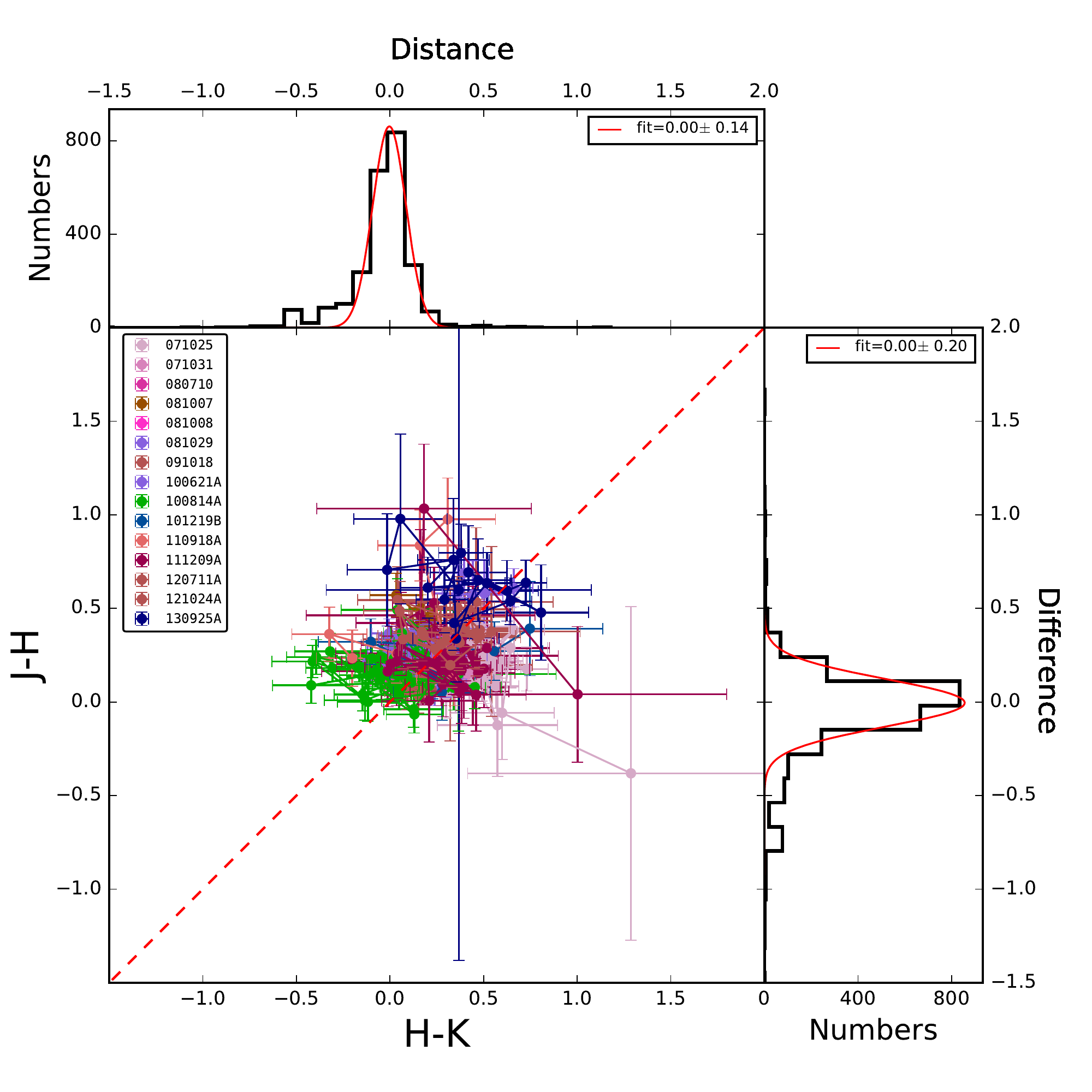}
\end{center}
\caption{Color-color diagrams with the distributions of distances between the data points to the equal line and the distributions of differences between two colors (Data Set I of the Golden sample). The red dashed lines are the equal lines, and the red solid lines are the best Gaussian fits.}
\label{ColorColors}
\end{figure*}

In Figure \ref{ColorColors}, we show color-color diagrams using the Golden sample (Data Set I) to investigate the evolution of CI shift.
We calculate the distance of data points to the equal line and the difference value between the two CIs. 
We find that most of the CIs cluster around the equal line and the typical difference value between two CIs is close to 0.
For the distances, one has (g-r)-(r-i)=-0.03$\pm$0.07, (r-i)-(i-z)=0.00$\pm$0.06 and (J-H)-(H-Ks)=0.00$\pm$0.14, respectively.
For their differences, one has (g-r)-(r-i)=-0.04$\pm$0.10, (r-i)-(i-z)=0.01$\pm$0.09, and (J-H)-(H-Ks)=0.00$\pm$0.20, respectively.

This result indicates that the CI shift is not significant, and the intrinsic reddening inside their host galaxies must be quite similar and relatively small for these events. This is also expected from the narrow distributions in wavelength of the ugriz filters. 
This result also implies that the afterglow spectral shapes are similar from one to another and can be described by a smooth power-law spectral decay, with no bumps or strong lines for these bands. These results are consistent with the finding of \cite{2004A&A...427..901S}.

\subsection{Possible Explanations For The Color Indices Variation}

Below we outline three possible explanations for the CI variation, and in \S 5.2.4 we compare the prediction to the data.

\subsubsection{The Cooling Frequency Crosses the Studied Energy Bands}

In the external shock synchrotron model, the cooling frequency, $\nu_c$, is likely to cross the optical bands at a time later than approximately 1 hr \citep{2014ApJ...780...82U}. This will cause a change in the spectral indices and thereby in the observed CIs.

According to Eqs. (\ref{eq:FastCooling})-(\ref{eq:SlowCooling}), if the cooling frequency crosses all the optical bands and the spectral regime from a stable phase transition to another stable phase, then the change in the CIs, $\Delta$CI, is given by

(i) Fast cooling:
\begin{eqnarray}
\label{n} \Delta \rm CI=\left\{ \begin{array}{ll}
-2.5\rm log_{10} \left(\frac{\nu_{2}}{\nu_{m}}\right)^{-\frac{(p-1)}{2}}, & 
(\rm from \ \nu_{2}>\nu_{m}>\nu_{1}>\nu_{c} \ \rm to \     \nu_{m}>\nu_{2}>\nu_{1}>\nu_{c}) \\
-2.5\rm log_{10} \left(\frac{\nu_{1}}{\nu_{m}}\right)^{-\frac{(p-1)}{2}}, & 
(\rm from \ \nu_{2}>\nu_{1}>\nu_{m}>\nu_{c} \ \rm to \   \nu_{2}>\nu_{m}>\nu_{1}>\nu_{c}) \\
-2.5\rm log_{10} \left(\frac{\nu_{2}}{\nu_{1}}\right)^{-\frac{(p-1)}{2}}, & 
(\rm from \ \nu_{2}>\nu_{1}>\nu_{m}>\nu_{c} \ \rm to \   \nu_{m}>\nu_{2}>\nu_{1}>\nu_{c}) \\

\end{array} \right.
\label{eq:detacolorsFastCooling}
\end{eqnarray}

(ii) Slow cooling:
\begin{eqnarray}
\label{n} \Delta \rm CI=\left\{ \begin{array}{ll}
-2.5\rm log_{10} \left(\frac{\nu_{2}}{\nu_{c}}\right)^{-\frac{1}{2}}, & 
(\rm from \ \nu_{2}>\nu_{c}>\nu_{1}>\nu_{m} \ \rm to \  \nu_{c}>\nu_{2}>\nu_{1}>\nu_{m})\\
-2.5\rm log_{10} \left(\frac{\nu_{1}}{\nu_{c}}\right)^{-\frac{1}{2}}, & 
(\rm from \ \nu_{2}>\nu_{1}>\nu_{c}>\nu_{m}  \ \rm to \  \nu_{2}>\nu_{c}>\nu_{1}>\nu_{m})\\
-2.5\rm log_{10} \left(\frac{\nu_{2}}{\nu_{1}}\right)^{-\frac{1}{2}}, &
(\rm from \ \nu_{2}>\nu_{1}>\nu_{c}>\nu_{m} \ \rm to \  \nu_{c}>\nu_{2}>\nu_{1}>\nu_{m})\\
\end{array} \right.
\label{eq:detacolorsSlowCooling}
\end{eqnarray}

In the constant environment model (the ISM model), the $\nu_c$ decreases with time \citep{1998ApJ...497L..17S}. The optical band is initially below $\nu_c$ (bluer spectrum). As $\nu_c$ decreases in time, the optical band becomes above $\nu_c$, and the spectrum becomes redder. This gives rise to a negative $\Delta$CI. On the other hand, in the wind model, $\nu_c$ increases with time \citep{1999ApJ...520L..29C}. The optical band is initially above $\nu_c$ (redder spectrum). As $\nu_c$ increases in time, the optical band becomes below $\nu_c$ (the spectrum becomes bluer). This gives a positive $\Delta$CI.

We note that the cooling break in the afterglow spectra may not be sharp  \citep{2014ApJ...780...82U}. The transition from one regime to another is rather smooth and requires a very long time, especially in the optical bands. For most of the cases analyzed in this paper the transition has not finished during the period investigated, so what we have tested is an upper limit of $\Delta$CI.

\subsubsection{Effects of Host Extinction}

The host extinction may also cause the CI change \citep[e.g.,][]{Waxman:2000er, 2014MNRAS.440.1810M}. After being eventually destroyed by the initial intense radiation, dust could reform at an early time near the burst region, changing the host extinction \citep{2002ApJ...580..261P}.  This leads to two temporal changes in the CI. 
During the early afterglow phase (Phase I), dust might be destroyed, and the CI can be described by a red-to-blue change, implying an increase in the CI with $\Delta \rm CI>0$. On the contrary, at a late time (Phase II), dust could reform and reddening will reappear, and the CI presents a blue-to-red change, which provides a decrease in the CI with $\Delta \rm CI<0$. We note that Phase II might not occur, since the photodestruction could be irreversible \citep{2002ApJ...580..261P,2003ApJ...585..775P,2003MNRAS.340..694L}.

\begin{figure*}
\begin{center}
\centering
\includegraphics[angle=0,scale=0.80]{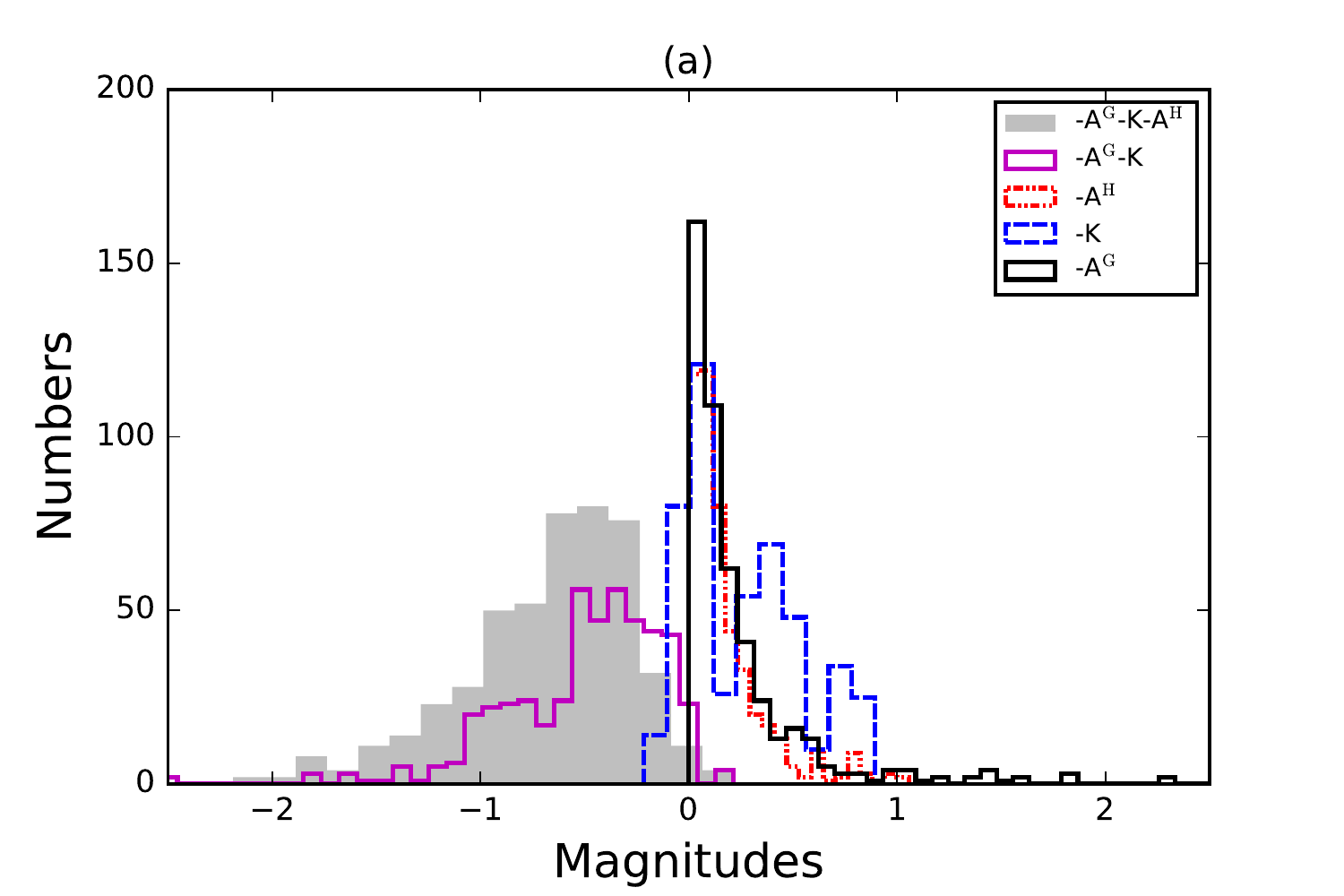}\\
\end{center}
\includegraphics[angle=0,scale=0.60]{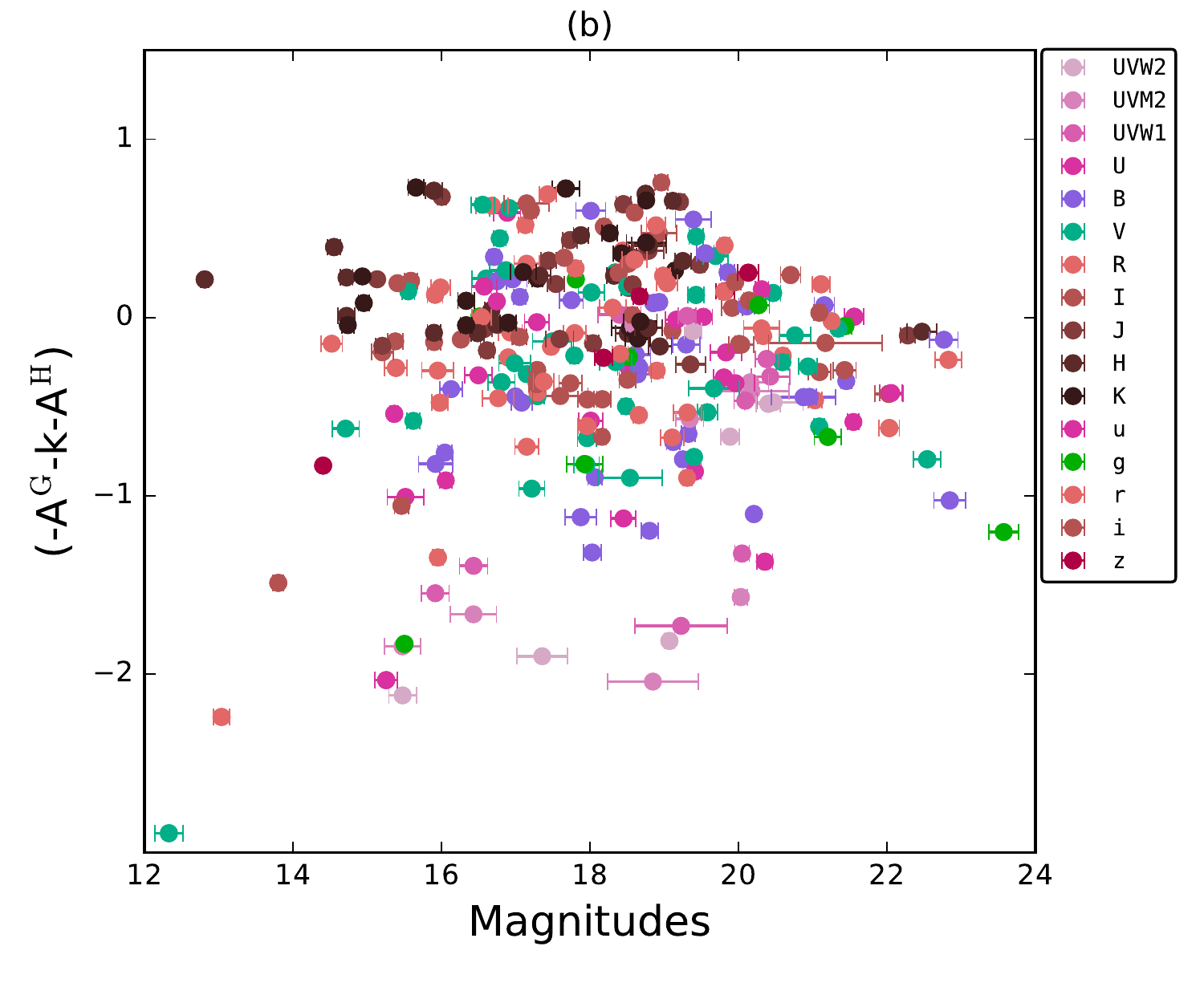}
\includegraphics[angle=0,scale=0.60]{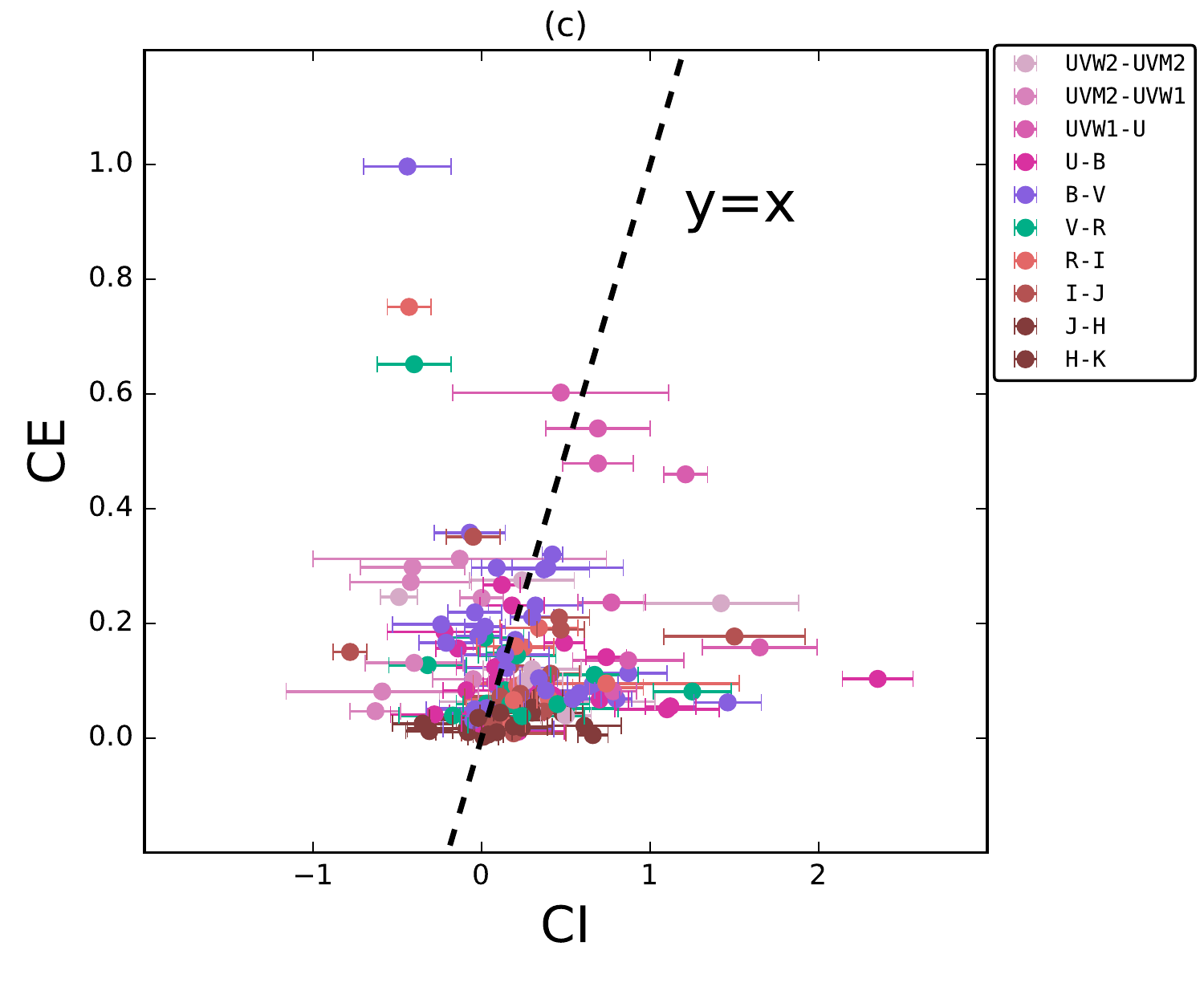}
\caption{Top: distributions of extinction and spectral $k$-correction factors (Data Set III). Bottom left: total correction factors compared against the magnitude in a given band. Bottom right: comparison between CI and CE. The dashed line is the equal line.}
\label{ExtinctionColors}
\end{figure*}

The true magnitudes in the optical bands are estimated through considerations of extinctions (in both in our Galaxy and the host galaxy) and the spectral $k$-correction. We display the distributions of all correction factors in Figure \ref{ExtinctionColors}. The factor that affects the achromatic energy band is given by (-$A^{\rm G}$-K-$A^{\rm H}$). This quantity is distributed in [-4.29, 0.22], and, as seen in Fig.\ref{ExtinctionColors}b, it is negligible compared to the brightness (Data Set III).

The factor affecting the CIs can be characterised by the color excess, CE, such that
\begin{equation}
\begin{split}
\rm CI_{T}=m_{O}(\lambda_{1})-A^{G}(\lambda_{1})-K-A^{H}(\lambda_{1}) \rm -\{m_{O}(\lambda_{2})-A^{G}(\lambda_{2})-K-A^{H}(\lambda_{2})\}\\
=\rm CI_{O}-\{A^{G}(\lambda_{1})-A^{G}(\lambda_{2})+A^{H}(\lambda_{1})-A^{H}(\lambda_{2})\}\\
=\rm CI_{O}-CE,
\end{split}
\label{eq:hostCE}
\end{equation}
where $A^{\rm H}$ is the magnitude of the host dust extinction and $A^{\rm G}$ is the magnitude of our Galaxy dust extinction.
Here the CE includes contributions from our Galaxy and the host galaxy, while the $k$-correction factor is the same for both bands.
Therefore CE=$\{A^{\rm G}(\lambda_{1})-A^{\rm G}(\lambda_{2})+A^{\rm H}(\lambda_{1})-A^{\rm H}(\lambda_{2})\}$ is the total correction, ranging from 0.00 to 0.99. 
Figure \ref{ExtinctionColors}(c) presents the CE as a function of CI. 
Since both the CI and CE have similar magnitudes, uncertainties on the corrections have large effects on the value of the CIs.
 
\subsubsection{Effects of the Supernova Emission Component}

GRB-associated SNe in some long bursts \citep{2006ARA&A..44..507W, 2003Natur.423..847H,2017AdAst2017E...5C} could produce an additional emission component embedded in the late afterglow synchrotron emission. This results in a CI variation \citep[e.g.,][]{2004A&A...427..901S,2004ApJ...609L..59G}, described as a red afterglow bump.
From the literature, 13 cases\footnote{GRB 980425/SN 1998bw \citep{1998Natur.395..670G}, GRB 030329/SN 2003dh \citep{2005AA...440..477R}, GRB 050525A/SN 2005nc \citep{2006ApJ...642L.103D}, GRB 060218/SN 2006aj \citep{2006Natur.442.1008C},GRB 081007/SN 2008hw \citep{2015AA...577A..44O}, GRB 091127/SN 2009nz \citep{2010ApJ...718L.150C}, GRB 100316D/SN 2010bh \citep{2011ApJ...740...41C},GRB 101219B/SN 2010ma \citep{2015AA...577A..44O}, GRB 120422A/SN 2012bz \citep{2012AA...547A..82M}, GRB 130215A/SN 2013ez \citep{2014AA...568A..19C}, GRB 130427A/SN 2013cq \citep{2013ApJ...776...98X}, 130702A/SN 2013dx \citep{2015AA...577A.116D}, GRB 130831A/SN 2013fu \citep{2014AA...568A..19C}.} of GRB-SN associations in our sample were reported.
We compare the CIs between the GRB-SN emission and their previous components, regardless of the nature of this previous component. We are only concerned here with the variation of the CI (Data Set II).
The results displayed by Figure \ref{ComsComsColors} show that a very high fraction (see Table 5) of CIs exhibit a significant variation, indicating a spectral change as expected.

\subsubsection{Statistical Analysis}

\begin{figure*}
\centering
\includegraphics[angle=0,scale=0.80]{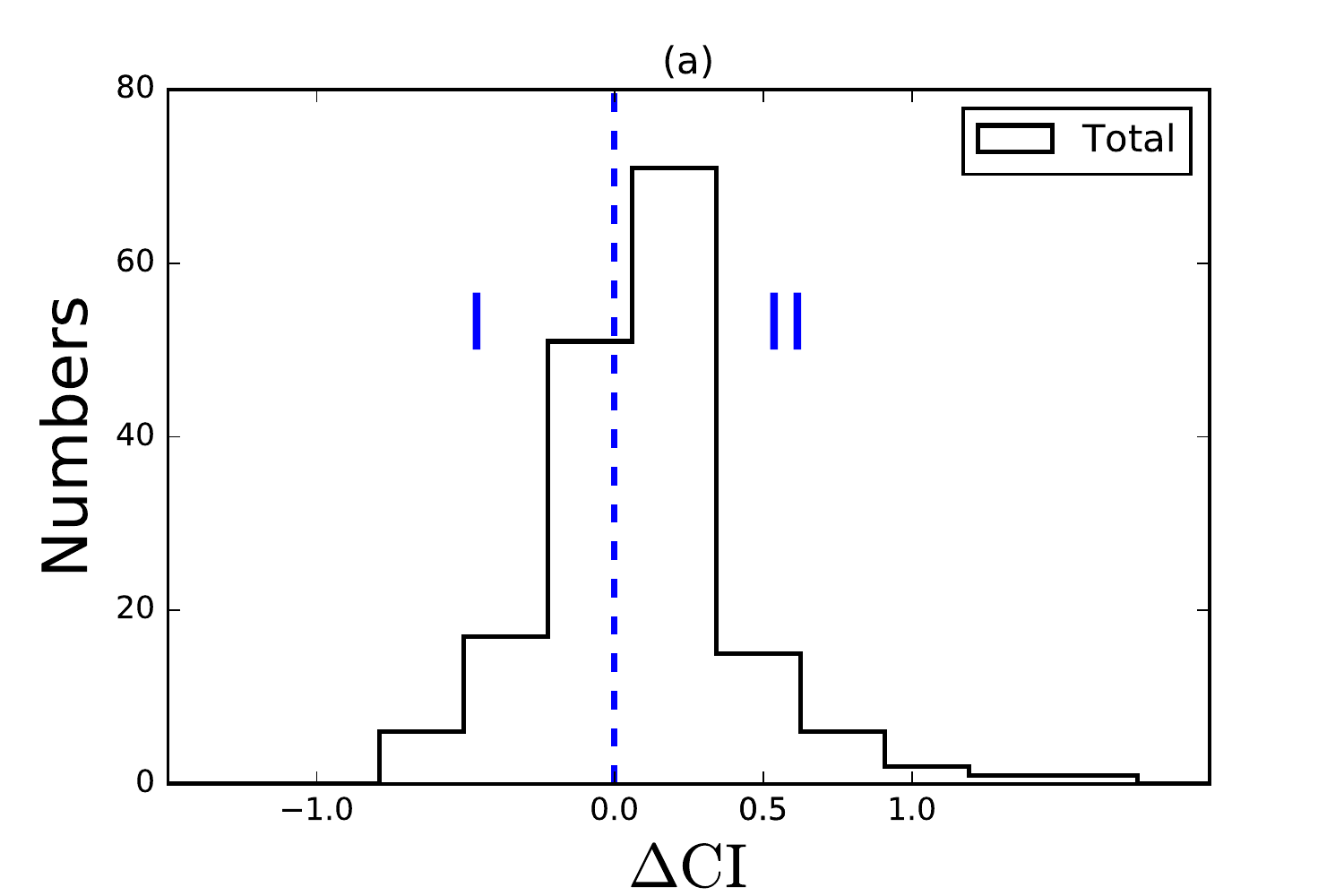}\\
\includegraphics[angle=0,scale=0.80]{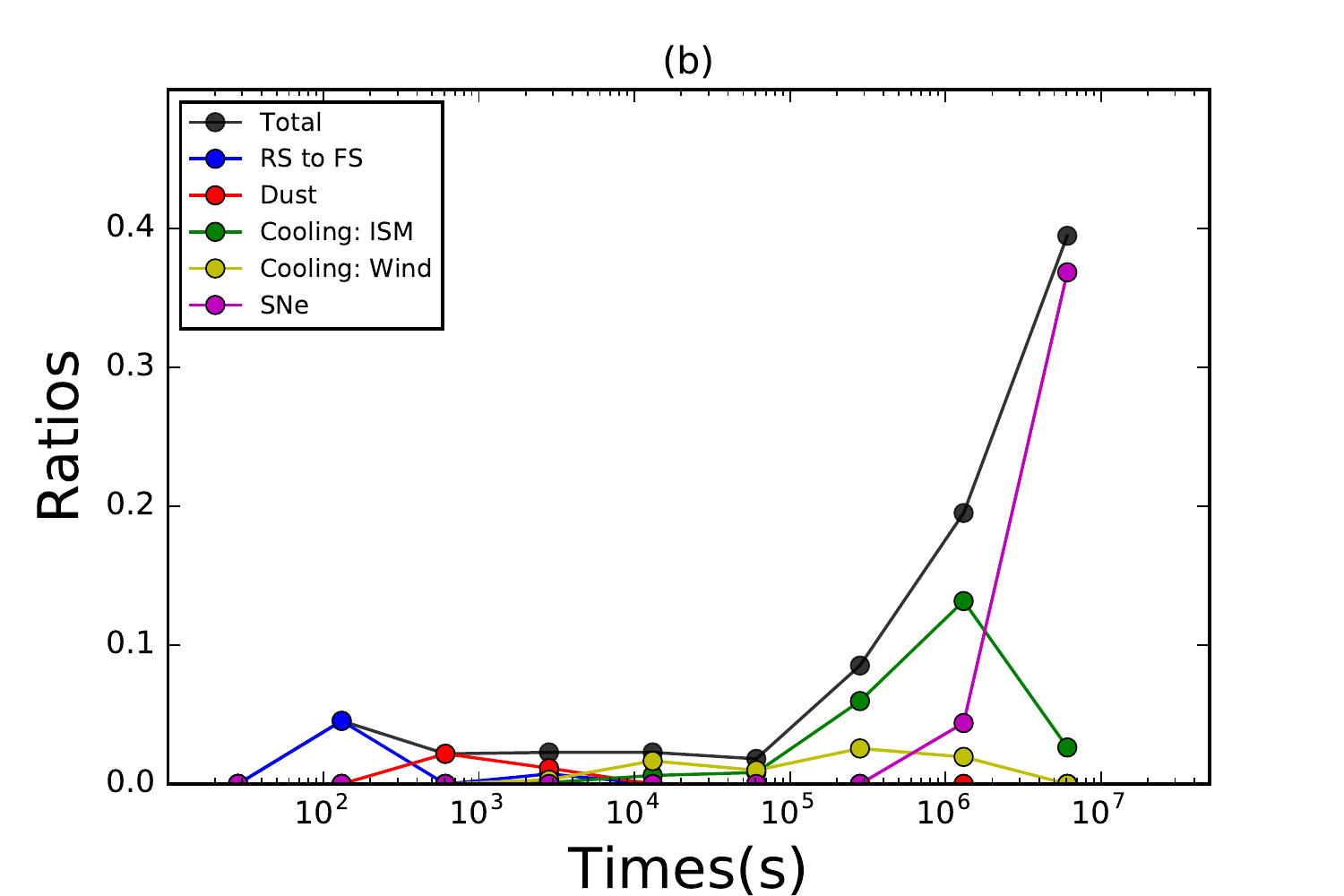}
\caption{(a) Distribution of $\Delta$CI for the Golden sample (Data Set I). The blue line separates positive and negative $\Delta$CI.
(b) Ratios of \textit{variable} CIs associated with a given phenomenon to the total number of \textit{variable} CIs in each time interval (see text for detailed explanations). Different colors represent different phenomena.}
\label{ThObsColors}
\end{figure*}

We perform a statistical analysis considering only {\it variable} CIs to explore possible physical origins. Figure \ref{ThObsColors}(a) shows the distribution of variable CIs.
If the observed variation is due to the crossing of the cooling spectral break,
then either positive or negative values of $\Delta$CIs are expected, depending on the density profile of the circumburst medium. As discussed above, a positive value of $\Delta$CIs is expected in the early afterglow phase (earlier than 1 hr) owing to dust destruction. Variable CIs for the SN are defined as the identification of a GRB-SN bump at a later time, which is confirmed in published papers (see \S 5.2.3).

Figure \ref{ThObsColors}(b) presents the temporal evolution of the ratio of varying CIs to the total number of varying CIs in a given time interval. We group the ratios depending, in turn, on (i) the identification of the SN component and whether the $\Delta$CIs are part of it and (ii) the sign of the $\Delta$CIs. 

Among the positive $\Delta$CIs (64 \% of all cases, labeled II in Figure \ref{ThObsColors}(a)) we find two peaks, at around 500s and $\sim 10^6$ s. The early peak is consistent with being due to dust extinction, corresponding to 12\% of the $\Delta$CIs. These cases are marked by the red line in the figure. The peak at $10^6$ s is consistent with effects of $\nu_c$ crossing the observed band in a wind environment, causing a red-to-blue change, i.e., a shallowing of the spectrum (yellow line in the figure), including 43\% of the variable CIs.

Likewise, among the negative $\Delta$CIs (36 \% of all cases, labelled I in Figure \ref{ThObsColors}(a)), there are two peaks, one at around 100 s, and one at $\sim 10^5-10^6$ s. The latter peak contains $\Delta$CIs that are consistent with effects of $\nu_c$ crossing the observed band in an ISM environment, causing a blue-to-red change, i.e., a steepening of the spectrum (green line in the figure), corresponding to 30\% of the cases. For the bursts that exhibit a color change at around 100s (blue curve in the figure), the most natural explanation is the transition from reverse-shock (RS) emission to forward-shock (FS) emission. Early on, the emission is likely dominated by the RS component and the optical band is below $\nu_{c,r}$ (bluer spectrum). At around 100 s, there is likely a transition from the RS to the FS emission (Type II light curve reported in \citealt{2003ApJ...595..950Z}). At this time, the FS may be already above $\nu_{c,f}$. This gives a redder color. For the wind model \citep{2003MNRAS.342.1131W, 2003ApJ...597..455K}, a similar situation is expected. This case includes $\sim$5\% of $\Delta$CIs.

Finally, the SN cases dominate the late-time variations (purple curve in the figure), the last one corresponding to 10\% of the variable CIs.

\section{Conclusion}\label{sec:Con}

We studied the multiband optical afterglows of 70 bursts with their CIs. The optical/NIR photometric data are not only based on {\it Swift}/UVOT but are also obtained from ground-based instruments, especially the GROND. This catalog of multiwavelength GRB afterglow data provides an opportunity for a statistical study.  
We corrected the data with the Galactic and host galaxy extinctions. After performing a spectral \textit{k}-correction, we divided the bursts into two samples depending on whether the different optical bands have a simultaneous observation. 
Our main results are summarized as follows:

\begin{itemize}

\item
A Golden sample includes 25 out of 70 GRBs with five CIs. The distributions of these CIs are approximately the same. There is no color shift between CIs, and in general they satisfy the afterglow model prediction of a single power-law spectral slope.
 
\item
A Silver sample includes 45 out of 70 GRBs with 10 CIs with wider energy ranges. The distributions of these CIs are consistent with the Golden sample in the corresponding energy bands. For most of these CIs, the typical value is around 0.2, and they also, in general, satisfy the afterglow model prediction of a single power-law spectral slope.
There are two significantly inconsistent peaks, mainly distributed in the UV bands, with UVM2-UVW1=-0.19 and UVW1-U=0.93. This is consistent with the theoretical prediction that the intrinsic reddening is more significant in the high-energy bands.

\item
After performing all the corrections for the data, the SED of the optical transient, which is derived from the average CI, presents a deviated single power-law spectral slope. This implies that the traditional extinction models could be not accurate enough to describe the extinction characteristics for some GRBs.

\item
Most CIs (95.9\%) are constant in the Golden sample (Data Set I).
They are tightly distributed with standard deviations of $\sim$ 0.2.
The variable CIs (4.1\%) are widely distributed, presenting evidence for several different emission mechanisms. We determined that around 30\% are due to the crossing of the cooling spectral break in an ISM medium, while 43\% are due to a wind-like medium, 10\% due to the SN emission, 12\% due to early dust extinction, and 5\% due to the transition from RS to FS emissions.

\item
Component-wise, we found that most cases of the varying CIs (Data Set II) are in the transition during the SNe, the Reversed-Shock, or the Flare components.

\item
We derived the correlations between CIs and spectral indices based on the standard afterglow synchrotron model and obtained $\beta_{o}^{\rm CI}$ (using the \textit{constant} CI of data Set I and Data Set II during the normal-decay phase) and $p^{\rm CI}$ (using the \textit{constant} CI of Data Set I). 
The typical values are $\beta_{o}^{\rm CI}$=0.68$\pm$0.60 (Data Set I) and $\beta_{o}^{\rm CI}$=0.68$\pm$0.68 (Data Set II). We found that these two distributions are consistent with each other. Moreover, $p^{\rm CI}$=2.43$\pm$1.06 for Slow cooling ($\nu_{c}>\nu_{2}>\nu_{1}>\nu_{m}$) and $p^{\rm CI}$=1.43$\pm$1.06 for Fast/Slow cooling ($\nu_{2}>\nu_{1}>max(\nu_{c},\nu_{m})$), which are consistent with the theoretical predictions for relativistic shocks. A wide range of $p$ values is obtained, which is consistent with previous findings. In turn, we derived the theoretical CIs using the $p$ values derive from the closure relation and compared them to the observed CIs. We found that they still clustered around the equal line. 

\item
We also investigated the overall behavior of the CIs:

(i) The CIs are independent of the redshift.

(ii) More that 90\% of CIs are distributed between -1.0 and 1.0. They have negligible variations compared to the decline of the brightness with time.

(iii) The dust extinction correction and the spectral $k$-correction can significantly affect the values of CIs and spectral indices.

\end{itemize}

We compiled a sample of multiband observations of optical transients to investigate the temporal evolution of the CIs. The variable CIs are clues to unveiling the origin and details of the optical transient and of GRBs in general, prompting for more prolonged and simultaneous observations.

\acknowledgments

We thank Damien B\'egu\'e, Yun-Feng Liang, Jin-Jun Geng, H\"usne Dereli, and Remo Ruffini for helpful discussions, and we are greatly indebted to the referee for an extremely careful reading of the manuscript and for helpful and valuable comments that greatly improved this paper.
This work is supported by the Swedish National Space Board, the Swedish Research Council, the National Basic Research Program of China (973 Program, grant no. 2014CB845800), and the National Natural Science Foundation of China (grant nos. 11725314, 11673068, 11473012, 11673068 and No. 11103083), and the Key Research Program of Frontier Sciences (QYZDB-SSW-SYS005), the Strategic Priority Research Program "Multi-waveband gravitational wave Universe"(grant no. XDB23000000).
L.L. acknowledges the support by the Erasmus Mundus Joint Doctorate Program via grant no. 2013-1471 from the EACEA of the European Commission. 
L.S. acknowledges the supported by the Joint NSFC-ISF Research Program (no. 11361140349), jointly funded by the National Natural Science Foundation of China and the Israel Science Foundation.
B.Z. acknowledges NASA NNX14AF85G for support.
F.R. is supported by the G\"oran Gustafsson Foundation for Research in Natural Sciences and Medicine.
Part of this work made use of our private Python library.

\clearpage
\startlongtable
\setlength{\tabcolsep}{0.35em}
\begin{deluxetable*}{cccccccccccccccccccccccccc}
\tabletypesize{\scriptsize}
\tablecaption{Properties of the GRB Sample with Multicolor Light Curves}
\tablehead{
\colhead{GRB}
&\colhead{Filters}
&\colhead{$\rm E_{B-V}^{G}$}
&\colhead{$\beta_{\rm o}$}
&\colhead{$\rm A_{v}^{host}$}
&\colhead{$z$}
&\colhead{References}\\
&
&\colhead{(Galaxy)}
&\colhead{(Spectral Index)}
&\colhead{(Host Galaxy)}
&\colhead{(Redshift)}
&\colhead{(for Data Sources)}
}
\startdata
970508&\textit{UBVRcIc}&0.034&1.11&0.38$\pm$0.11&0.8349&1,2,3,4,5,6\\
980425&\textit{UBVRcIc}&0.059&...&0.17$\pm$0.02&0.0085&7,8,9,10\\
990510&\textit{BVRIJHKs}&0.226&0.55&0.22$\pm$0.07&1.6187&11,12,13,14\\
990712&\textit{VRI}&0.035&0.99$\pm$0.02&0&0.4331&15,16,17\\
000301C&\textit{UBVRIJK}&0.053&0.70&0.12$\pm$0.06&2.0404&18,19,20,21,22,23\\
000926&\textit{UBVRI}&0.025&1.00$\pm$0.18&0.15$\pm$0.07&2.0387&24,25,26\\
021004&\textit{UBVRIJHK}&0.074&0.14&0.05&2.3304&27,28,29,30,31\\
030226&\textit{UBVRIJHK}&0.023&0.70$\pm$0.03&0.06$\pm$0.06&1.98691&32,33,34\\
\enddata
\end{deluxetable*}

\clearpage
\startlongtable
\begin{deluxetable*}{cccccccccc}
\tablecaption{Color Indices of the Golden Sample}
\tablewidth{0pt}
\tabletypesize{\scriptsize}
\tablehead{
\colhead{GRB}&
\colhead{Component\tablenotemark{a}}&
\colhead{Epoch\tablenotemark{b}}&
\colhead{Model\tablenotemark{c}}&
\colhead{$\rm g-r$}&
\colhead{$\rm r-i$}&
\colhead{$\rm i-z$}&
\colhead{$\rm J-H$}&
\colhead{$\rm H-Ks$}&\\
\cline{4-9}
&
&
\colhead{(ks)}&
&
\colhead{$\overline{\rm mag}$}&
\colhead{$\overline{\rm mag}$}&
\colhead{$\overline{\rm mag}$}&
\colhead{$\overline{\rm mag}$}&
\colhead{$\overline{\rm mag}$}&
}
\startdata
071025(0)\tablenotemark{d}&Global&0.17-14.9&...&...&...&...&0.19$\pm$0.14&0.51$\pm$0.15\\
071025(1)&Onset&0.17-0.6&BKPL&...&...&...&0.19$\pm$0.25&0.64$\pm$0.24\\
071025(2)&Normal-Decay&0.55-1.0&BKPL&...&...&...&0.22$\pm$0.10&0.49$\pm$0.11\\
071025(3)&Re-Bump&0.99-14.9&BKPL&...&...&...&0.18$\pm$0.11&0.45$\pm$0.13\\
071031(0)&Global&0.29-25.7&...&0.56$\pm$0.03&0.13$\pm$0.03&0.09$\pm$0.04&0.2$\pm$0.06&0.28$\pm$0.07\\
071031(1)&Onset&0.29-1.2&BKPL&0.63$\pm$0.02&0.17$\pm$0.02&0.09$\pm$0.03&0.21$\pm$0.05&0.33$\pm$0.06\\
071031(2)&Normal-Decay&1.23-4.0&BKPL&0.58$\pm$0.04&0.14$\pm$0.03&0.08$\pm$0.04&0.21$\pm$0.05&0.27$\pm$0.07\\
071031(3)&Re-Bump&4.05-25.7&BKPL&0.55$\pm$0.03&0.12$\pm$0.03&0.09$\pm$0.04&0.19$\pm$0.07&0.22$\pm$0.08\\
\enddata
\tablenotetext{a}{The identifications of various optical-afterglow emission components.}
\tablenotetext{b}{The time intervals for each component in the observer frame.}
\tablenotetext{c}{The models of the light-curve fits for each component.}
\tablenotetext{d}{The serial number, 0, is denoted as global, and other numbers are the components with time sequence.}
\end{deluxetable*}

\clearpage
\begin{longrotatetable}
\setlength{\tabcolsep}{0.35em}
\begin{deluxetable*}{cccccccccccccccccc}
\tabletypesize{\tiny}
\tablecaption{Color Indices of the Silver Sample}
\tablewidth{0pt}
\tablehead{
\colhead{GRB}&
\colhead{Component}&
\colhead{Epoch}&
\colhead{Model}&
&&&
\colhead{Color Indices}&\\
\cline{6-14}
&&
\colhead{(ks)}&
&
\colhead{$\overline{\rm UVW2-UVM2}$}&
\colhead{$\overline{\rm UVM2-UVW1}$}&
\colhead{$\overline{\rm UVW1-U}$}&
\colhead{$\overline{\rm U-B}$}&
\colhead{$\overline{\rm B-V}$}&
\colhead{$\overline{\rm V-R}$}&
\colhead{$\overline{\rm R-I}$}&
\colhead{$\overline{\rm I-J}$}&
\colhead{$\overline{\rm J-H}$}&
\colhead{$\overline{\rm H-K}$}&
}
\startdata
970508(0)&Global&25.57-40594.2&...&...&...&...&-0.23$\pm$0.38&0.32$\pm$0.28&0.02$\pm$0.23&0.34$\pm$0.23&...&...&...\\
970508(1)&Late-Bump&25.57-400.0&BKPL&...&...&...&-0.33$\pm$0.38&0.17$\pm$0.28&0.05$\pm$0.23&0.21$\pm$0.23&...&...&...\\
970508(2)&Jet-Break&400.0-1030.0&BKPL&...&...&...&-0.24$\pm$0.38&0.19$\pm$0.28&0.04$\pm$0.23&0.22$\pm$0.23&...&...&...\\
970508(3)&Re-Bump&1030.0-40594.2&BKPL&...&...&...&-0.14$\pm$0.38&0.48$\pm$0.28&-0.01$\pm$0.23&0.48$\pm$0.23&...&...&...\\
980425(0)&Global&94.85-46400.0&...&...&...&...&2.35$\pm$0.21&0.87$\pm$0.23&1.25$\pm$0.23&0.75$\pm$0.21&...&...&...\\
980425(1)&Late-Bump&94.85-1330.0&BKPL&...&...&...&2.15$\pm$0.21&0.54$\pm$0.23&0.99$\pm$0.23&0.73$\pm$0.21&...&...&...\\
980425(2)&SN-Bump&1330.0-46400.0&BKPL&...&...&...&2.40$\pm$0.21&0.96$\pm$0.23&1.32$\pm$0.23&0.75$\pm$0.21&...&...&...\\
990510(0)&Global&12.44-2000.0&...&...&...&...&...&0.09$\pm$0.09&...&...&0.64$\pm$0.17&...&-0.02$\pm$0.31\\
\enddata
\end{deluxetable*}
\end{longrotatetable}

\begin{deluxetable*}{cccccccccccccccccc}
\tabletypesize{\scriptsize}
\tablecaption{Statistics of Color Indices}
\tablehead{
\colhead{Color}&
\colhead{Range}&
\colhead{Mean}&
\colhead{Correlation}&
\colhead{$\beta_{o}^{\rm CI}$}&
\colhead{$\beta_{o}^{\rm CI}$}&\\
&
\colhead{(Total)}&
\colhead{(Gaussian Fitting)}&
\colhead{(CI-$\beta_{o}$)}&
\colhead{(CI=Mean)}&
\colhead{(CI=0.2)}&
}
\startdata
\hline
&Data Set I/Golden/Color Indices\tablenotemark{a}\\
\hline
g-r&(-0.29,2.51)&0.15$\pm$0.14&0.30$\beta_{o}$&0.50$\pm$0.47&0.67\\
r-i&(-1.01,1.04)&0.09$\pm$0.10&0.21$\beta_{o}$&0.43$\pm$0.48&0.95\\
i-z&(-2.10,1.44)&0.13$\pm$0.11&0.19$\beta_{o}$&0.68$\pm$0.58&1.05\\
J-H&(-0.68,1.03)&0.23$\pm$0.10&0.30$\beta_{o}$&0.77$\pm$0.33&0.67\\
H-Ks&(-0.42,3.00)&0.25$\pm$0.17&0.32$\beta_{o}$&0.78$\pm$0.53&0.63\\
\hline
&Data Set I/Silver/Color Indices\tablenotemark{a}\\
\hline
UVW2-UVM2&(-0.88,1.65)&0.16$\pm$0.27&0.16$\beta_{o}$&1.00$\pm$1.69&1.25\\
UVM2-UVW1&(-1.26,0.72)&-0.19$\pm$0.24&0.16$\beta_{o}$&-1.19$\pm$1.50&1.25\\
UVW1-U&(-0.11,3.02)&0.93$\pm$0.39&0.41$\beta_{o}$&2.27$\pm$0.95&0.49\\
U-B&(-1.83,2.65)&0.09$\pm$0.23&0.20$\beta_{o}$&0.45$\pm$1.15&1.00\\
B-V&(-1.50,2.14)&0.23$\pm$0.23&0.24$\beta_{o}$&0.96$\pm$0.96&0.83\\
V-R&(-1.23,2.9)&0.15$\pm$0.16&0.26$\beta_{o}$&0.58$\pm$0.62&0.77\\
R-I&(-0.82,2.85)&0.22$\pm$0.07&0.28$\beta_{o}$&0.79$\pm$0.25&0.71\\
I-J&(-1.86,3.15)&0.17$\pm$0.08&0.36$\beta_{o}$&0.47$\pm$0.22&0.56\\
J-H&(-0.63,2.11)&0.28$\pm$0.14&0.30$\beta_{o}$&0.93$\pm$0.47&0.67\\
H-K&(-0.47,1.61)&0.13$\pm$0.22&0.32$\beta_{o}$&0.41$\pm$0.69&0.63\\
\enddata
\tablenotetext{a}{Statistics of color indices regardless of GRBs and components.}
\end{deluxetable*}

\clearpage
\startlongtable
\begin{deluxetable*}{ccccc}
\tabletypesize{\scriptsize}
\tablecaption{Constant versus Variable Color Indices}
\tablehead{
\colhead{GRB/Color/Component}&
\colhead{Number/Fraction}&
\colhead{Number/Fraction}&
\colhead{Number}&\\
&
\colhead{(Constant)}&
\colhead{(Variation)}&
\colhead{(Total)}&
}
\startdata
\hline
&Data Set I/Golden/Color Indices\tablenotemark{a}\\
\hline
g-r&1135/94.9\%&61/5.1\%&1196\\
r-i&1135/95.9\%&48/4.1\%&1183\\
i-z&984/96.3\%&38/3.7\%&1022\\
J-H&408/96.7\%&14/3.3\%&422\\
H-K&377/96.9\%&12/3.1\%&389\\
\hline
Total Colors&4039/95.9\%&173/4.1\%&4212\\
\hline
&Data Set I/Golden/Components\tablenotemark{b}\\
\enddata
\tablenotetext{a}{Statistics regardless GRBs and components.}
\tablenotetext{b}{Statistics regardless GRBs and colors.}
\tablenotetext{c}{Statistics regardless colors and components.}
\end{deluxetable*}

\clearpage
\begin{deluxetable}{ccccccc}
\tabletypesize{\scriptsize}
\tablecaption{Results of Linear Correlation Analysis for Data Set II}
\tablehead{
\colhead{Component-to-Component}&
\colhead{Number}&
\colhead{$\Delta$CI}&
\colhead{$\Delta$CI}&
\colhead{{\it r}}&
\colhead{{\it p}}&\\
\colhead{(Previous-to-Following)}&
\colhead{(Total)}&
\colhead{(range)}&
\colhead{(Gaussian fits)}&
\colhead{(Spearman)}&
\colhead{(Spearman)}&
}
\startdata
Onset-to-Normal&68&[-1.18,0.84]&-0.02$\pm$0.06&0.85&$<10^{-4}$\\
Reverse-to-Normal&16&[-0.48,0.37]&-0.04$\pm$0.22&0.68&3.9$\times 10^{-3}$\\
Energy-to-Normal&42&[-0.28,0.31]&0.03$\pm$0.10&0.87&$<10^{-4}$\\
Flare-to-Normal&12&[-1.07,0.14]&0.03$\pm$0.12&0.77&3.6$\times 10^{-3}$\\
Normal-to-Flare&5&[-0.27,0.00]&...\tablenotemark{a}&0.80&$<10^{-4}$\\
Energy-to-Flare&8&[-0.22,0.05]&...&0.60&0.12\\
Late-to-Jet&4&[-0.01,0.09]&...&1.00&$<10^{-4}$\\
Normal-to-Re&56&[-0.39,1.11]&0.00$\pm$0.06&0.88&$<10^{-4}$\\
Normal-to-Jet&36&[-0.42,0.22]&-0.02$\pm$0.08&0.89&$<10^{-4}$\\
Re-to-Jet&5&[-0.07,0.02]&...&0.90&0.04\\
Jet-to-Re&4&[-0.05,0.29]&...&0.95&0.05\\
Normal-to-SN&24&[-0.90,1.14]&0.13$\pm$0.33&0.57&3.7$\times 10^{-3}$\\
Late-to-SN&24&[-0.24,1.13]&0.25$\pm$0.48&0.82&$<10^{-4}$\\
Jet-to-SN&2&[0.00,0.05]&...&...&...\\
\hline
Total&306&[-1.18,1.14]&-0.01$\pm$0.04&0.86&$<10^{-4}$\\
\hline
\enddata
\tablenotetext{a}{Small number of data points leading to unreliable fits.}
\end{deluxetable}

\clearpage
\thispagestyle{empty}
\setlength{\voffset}{-18mm}

\clearpage
\thispagestyle{empty}
\setlength{\voffset}{-18mm}

\clearpage
\thispagestyle{empty}
\setlength{\voffset}{-18mm}

\clearpage
\thispagestyle{empty}
\setlength{\voffset}{-18mm}

\clearpage
\thispagestyle{empty}
\setlength{\voffset}{-18mm}
\begin{figure*}
\includegraphics[angle=0,scale=0.40]{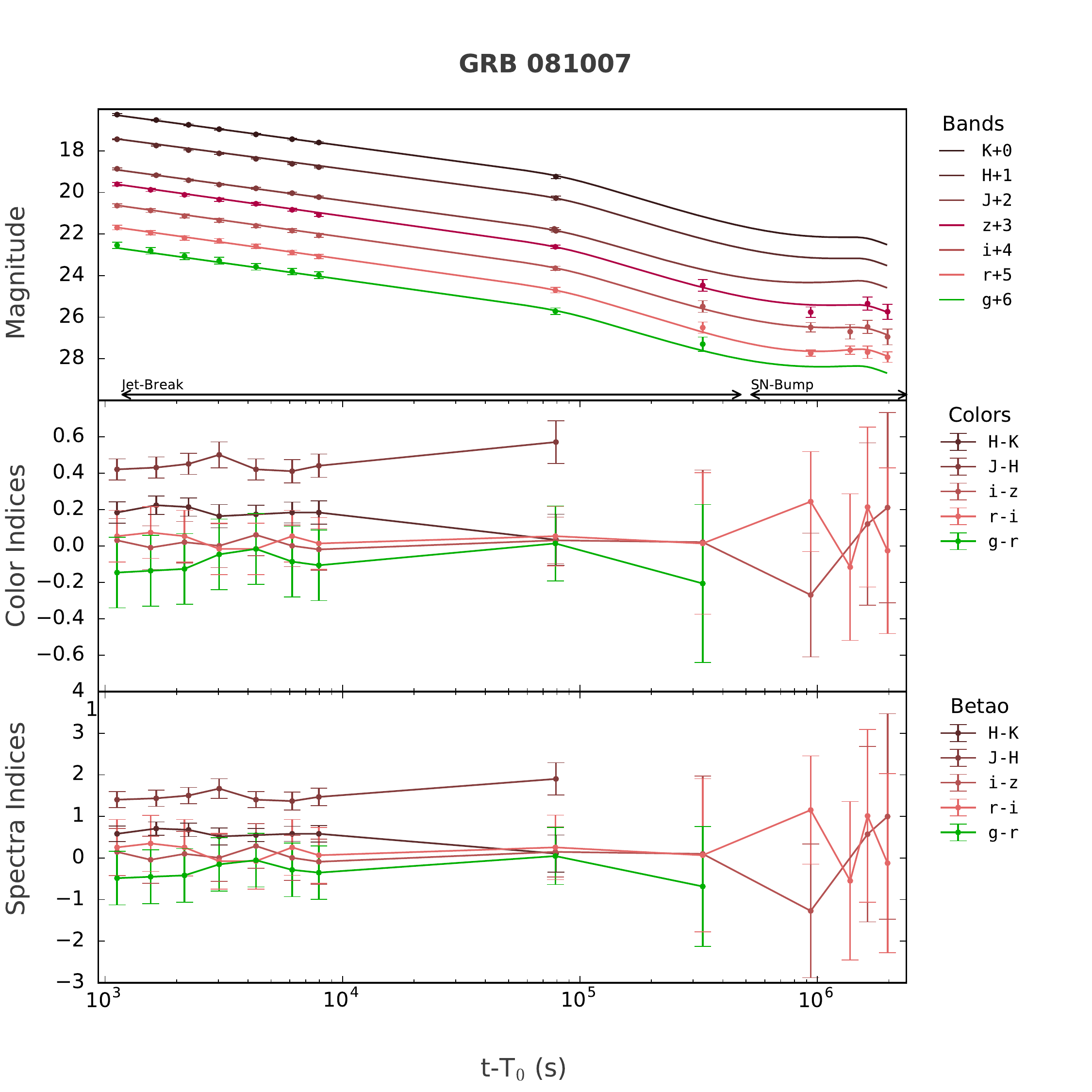}
\includegraphics[angle=0,scale=0.40]{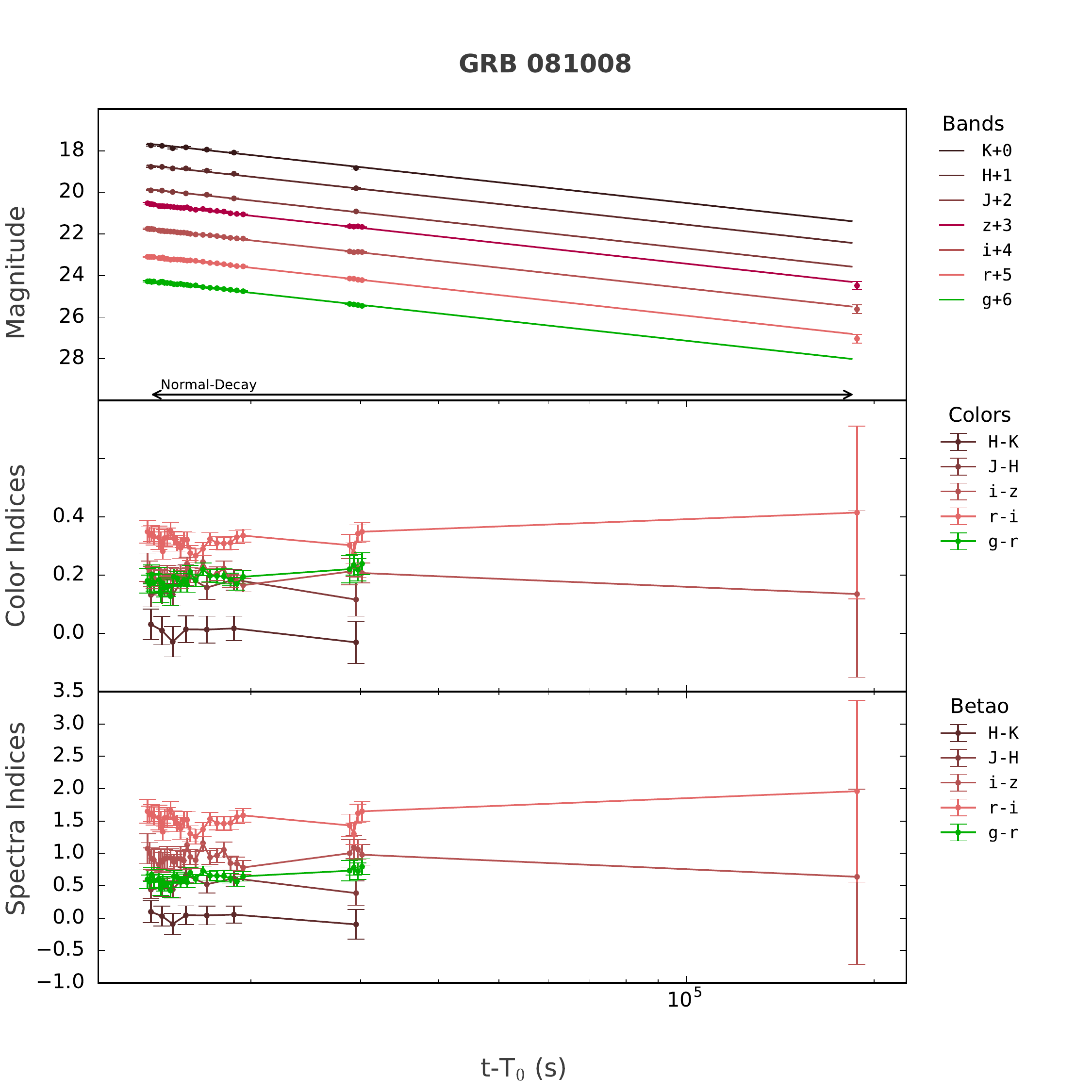}
\includegraphics[angle=0,scale=0.40]{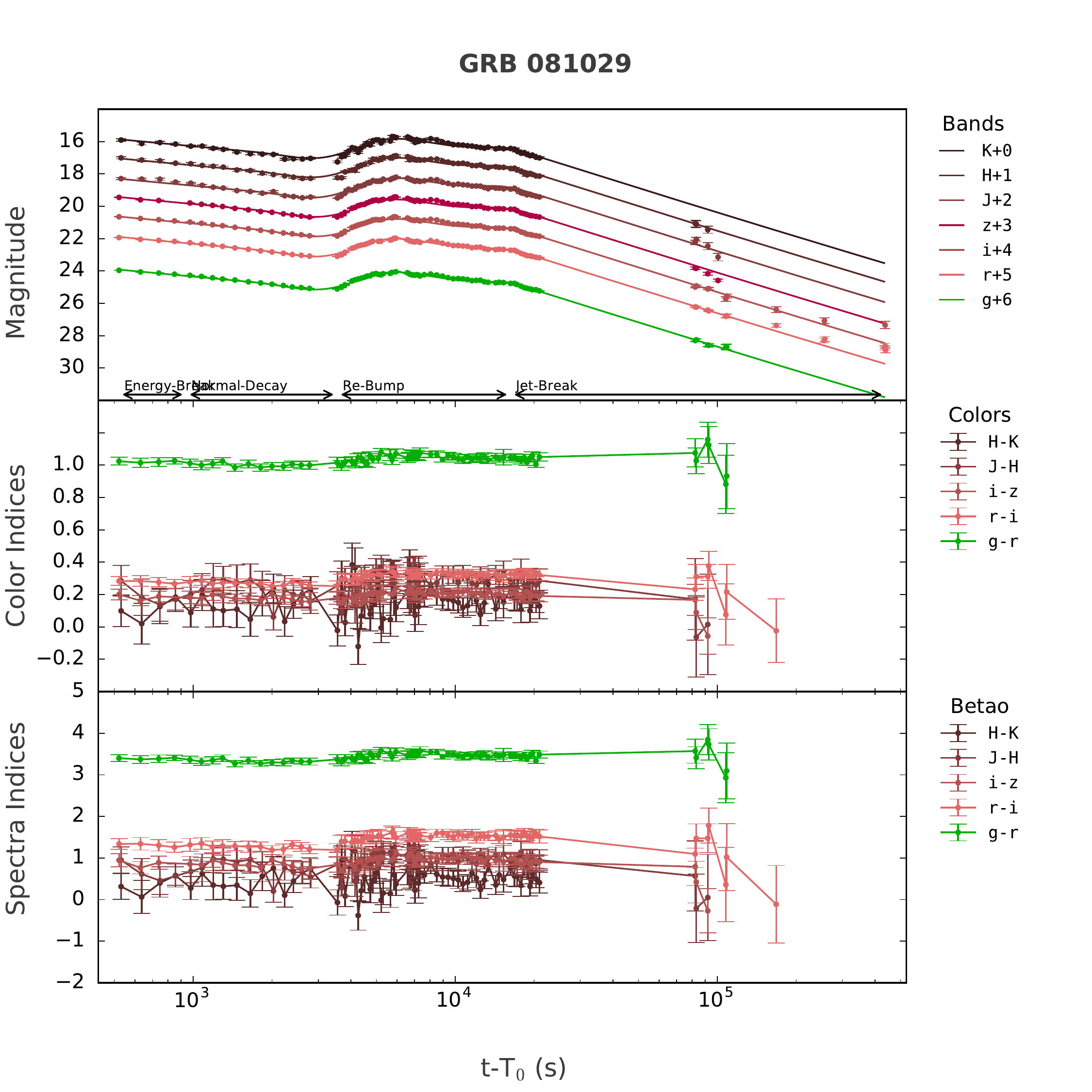}
\includegraphics[angle=0,scale=0.40]{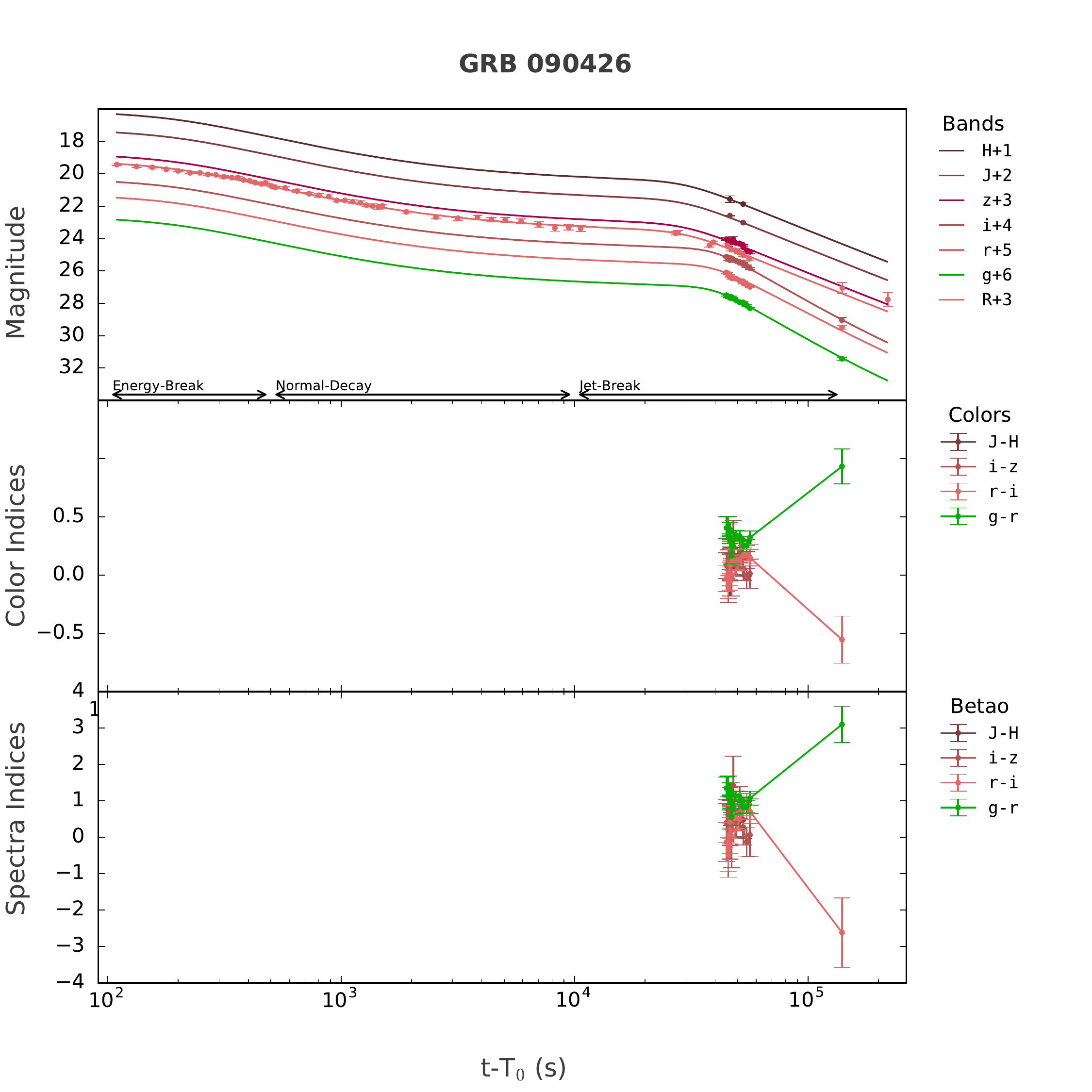}
\center{Fig. \ref{GoldenLCs}--- Continued}
\end{figure*}
\begin{figure*}
\includegraphics[angle=0,scale=0.40]{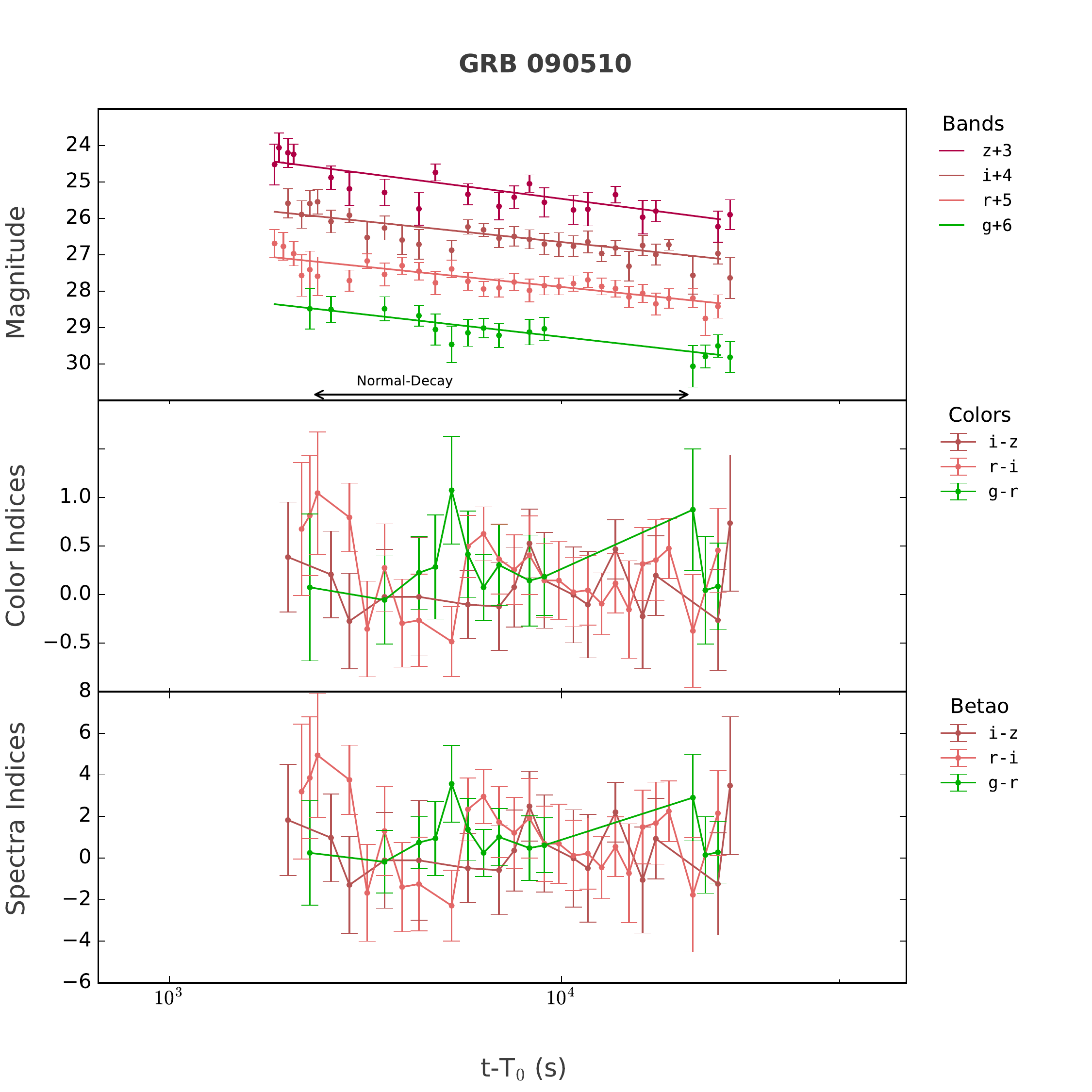}
\includegraphics[angle=0,scale=0.40]{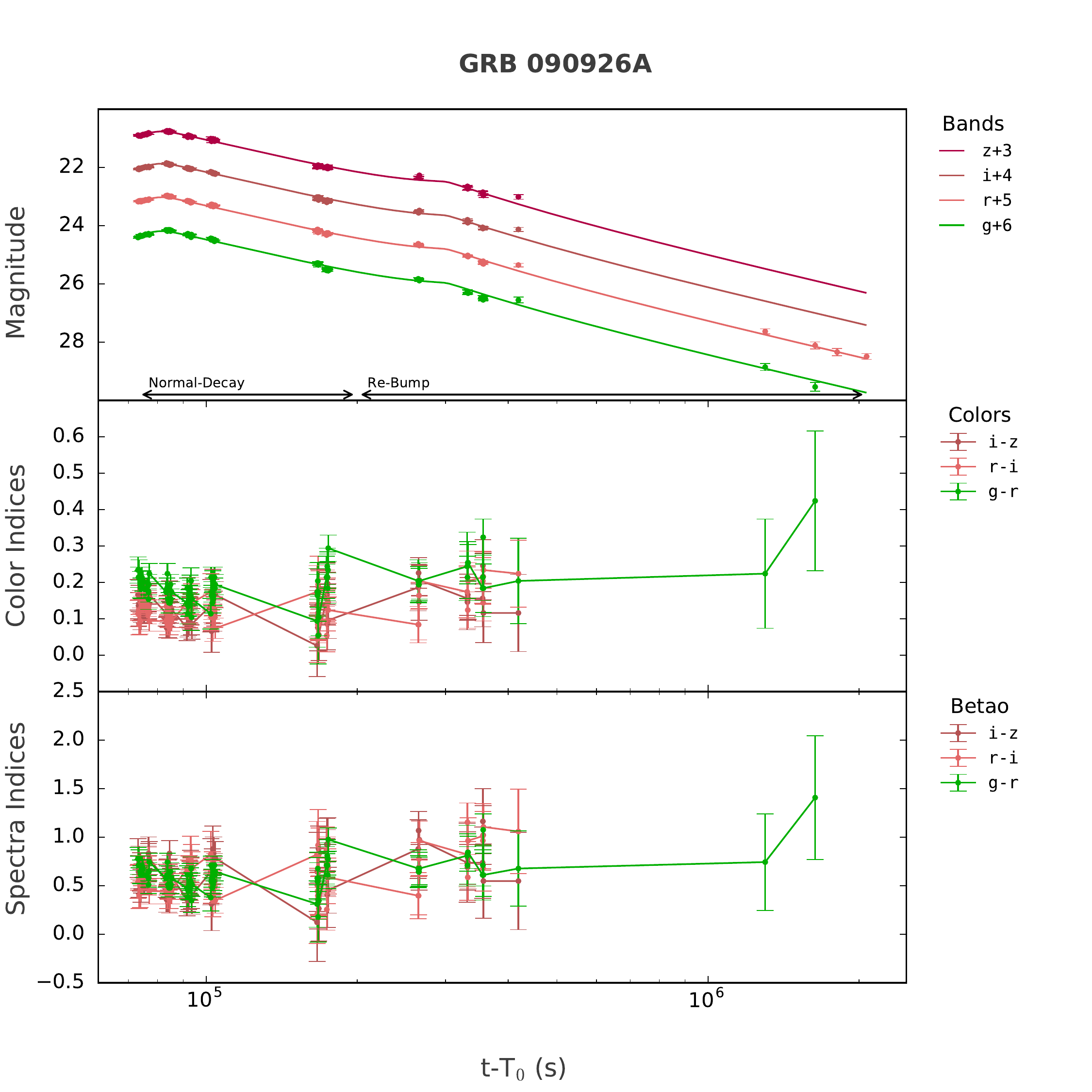}
\includegraphics[angle=0,scale=0.40]{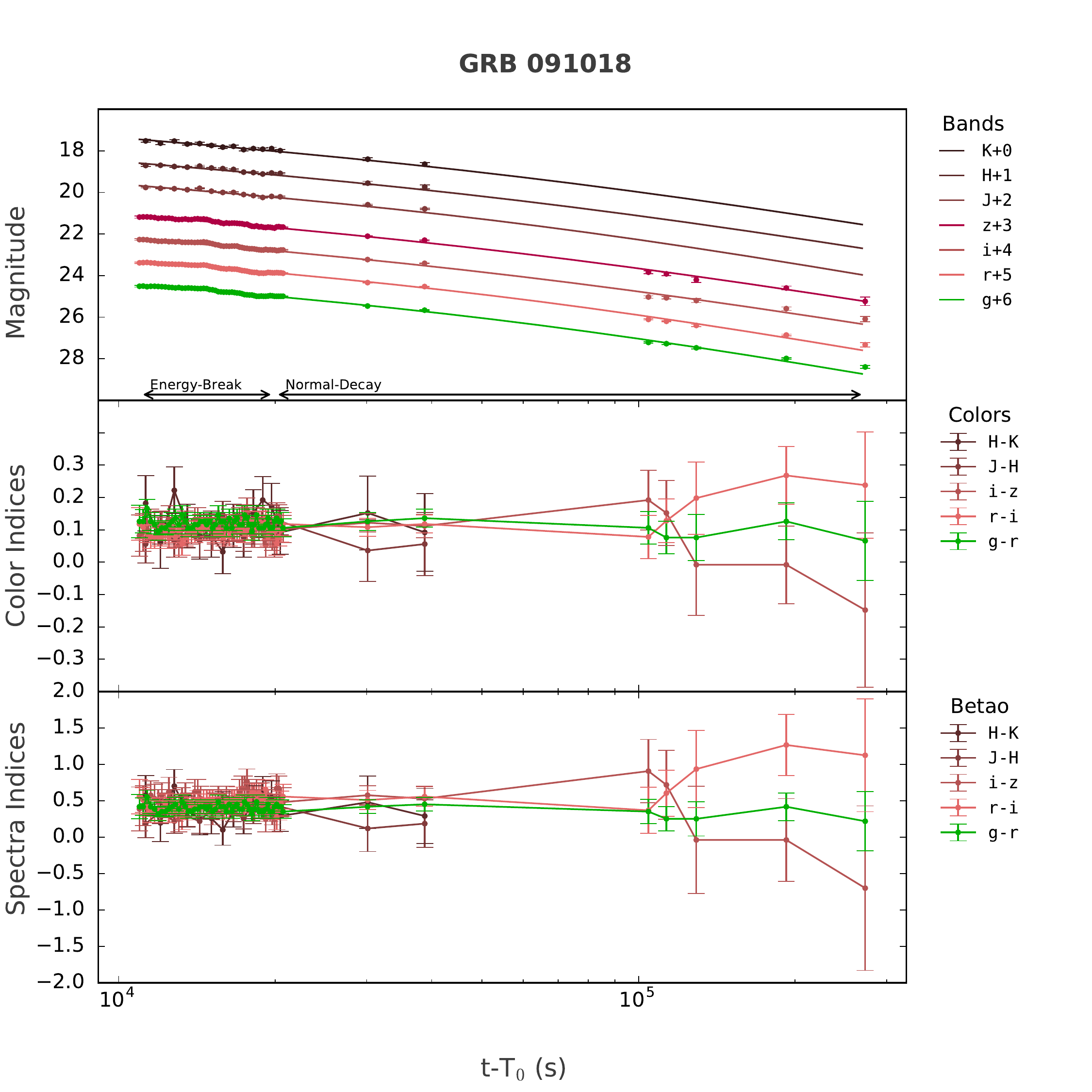}
\includegraphics[angle=0,scale=0.40]{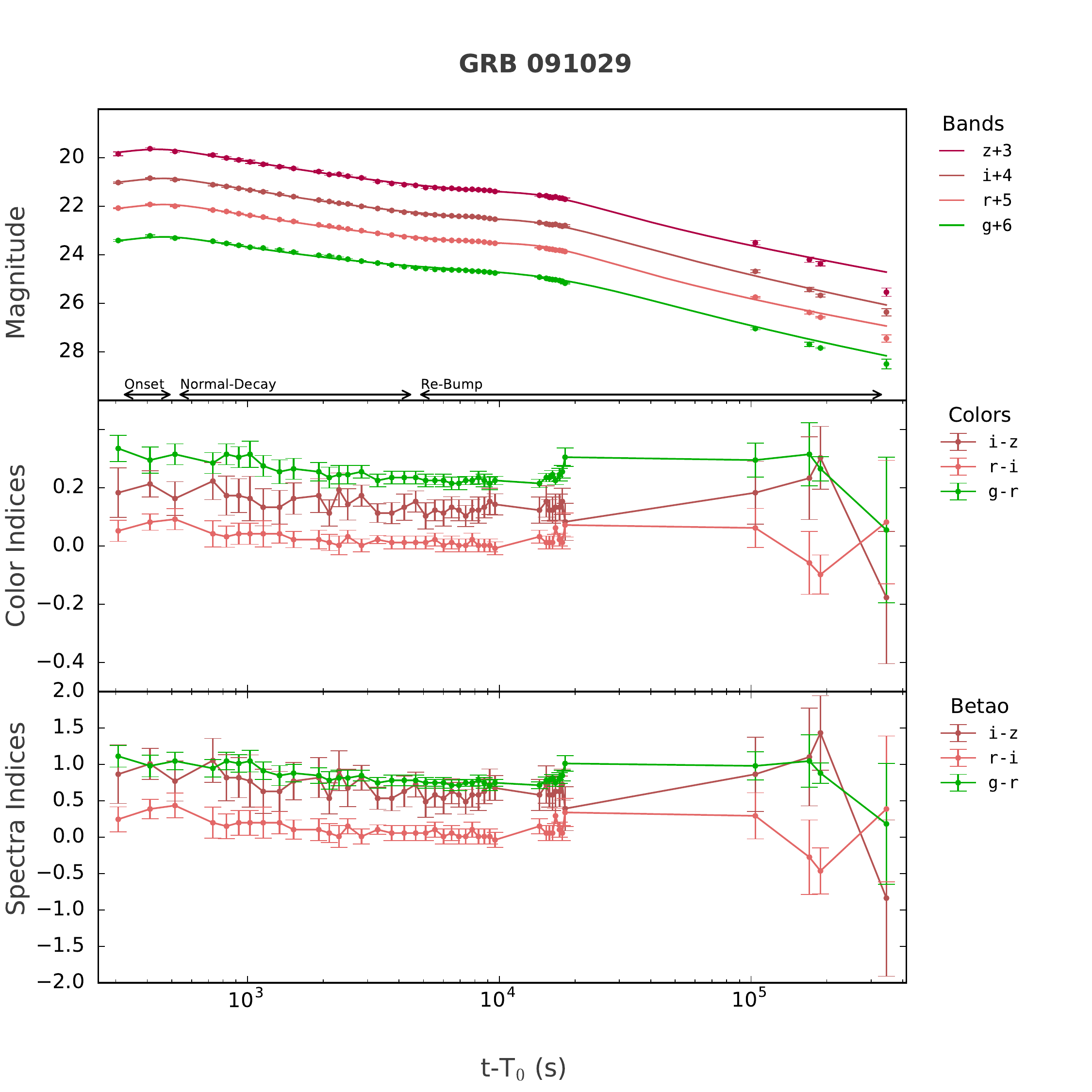}
\center{Fig. \ref{GoldenLCs}--- Continued}
\end{figure*}
\begin{figure*}
\includegraphics[angle=0,scale=0.40]{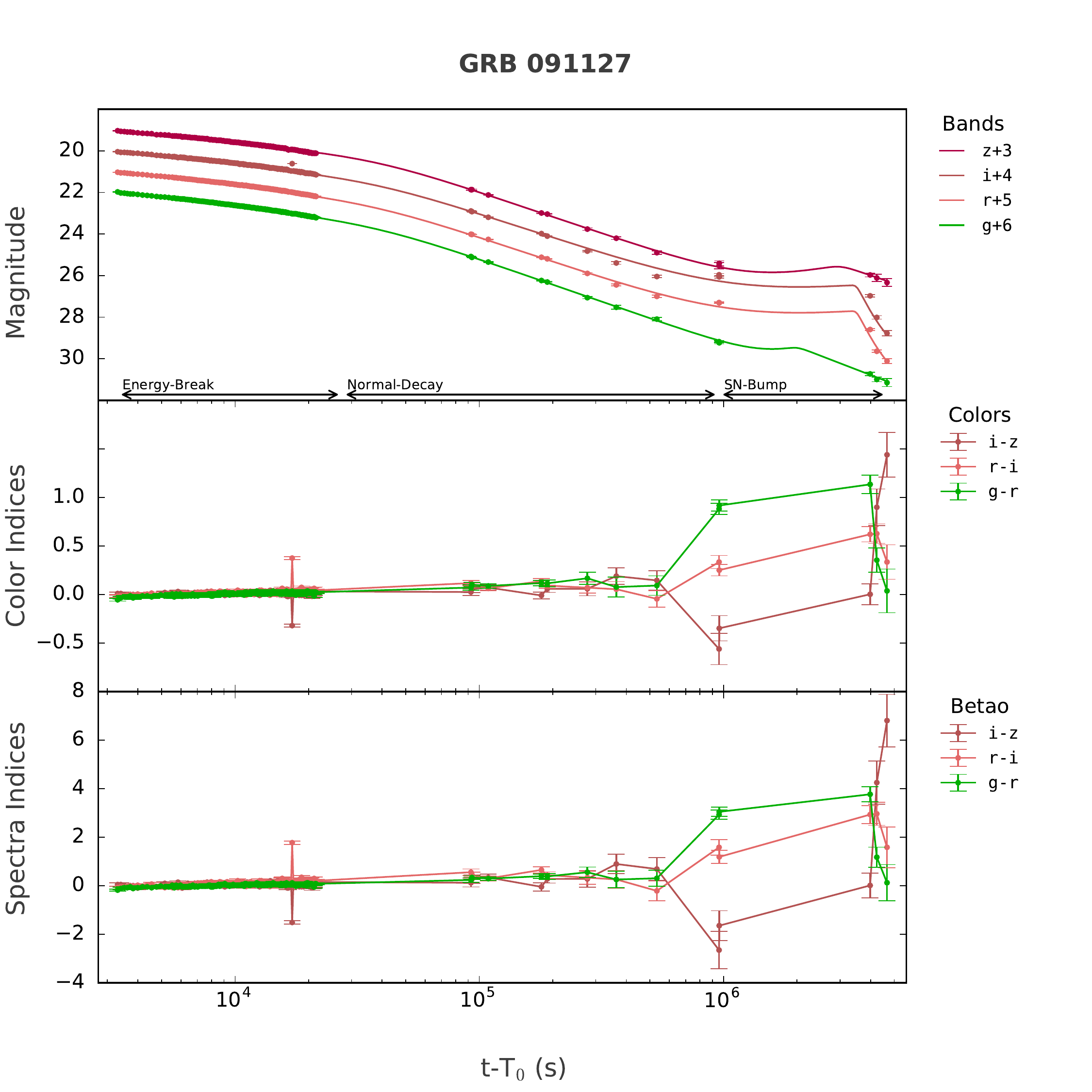}
\includegraphics[angle=0,scale=0.40]{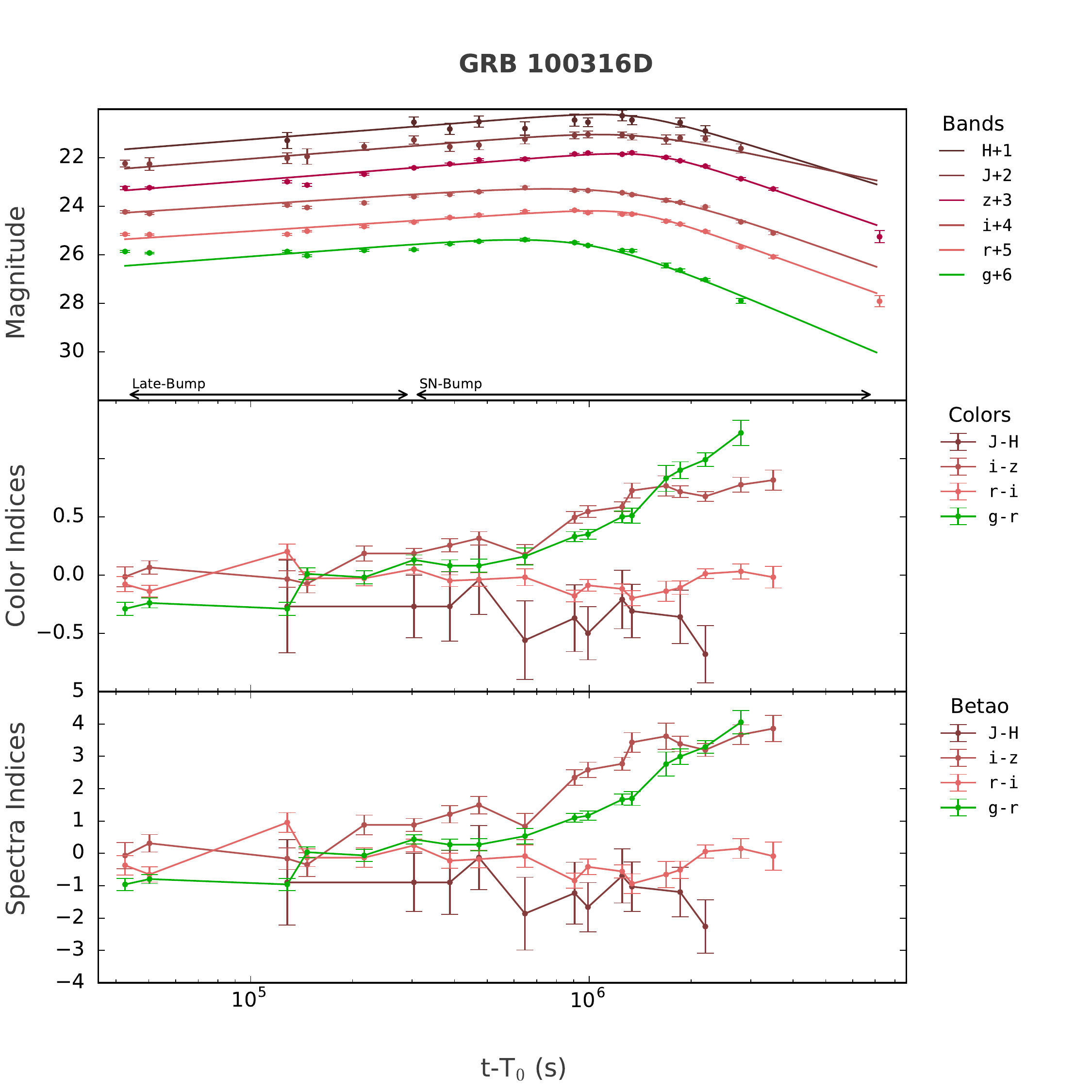}
\includegraphics[angle=0,scale=0.40]{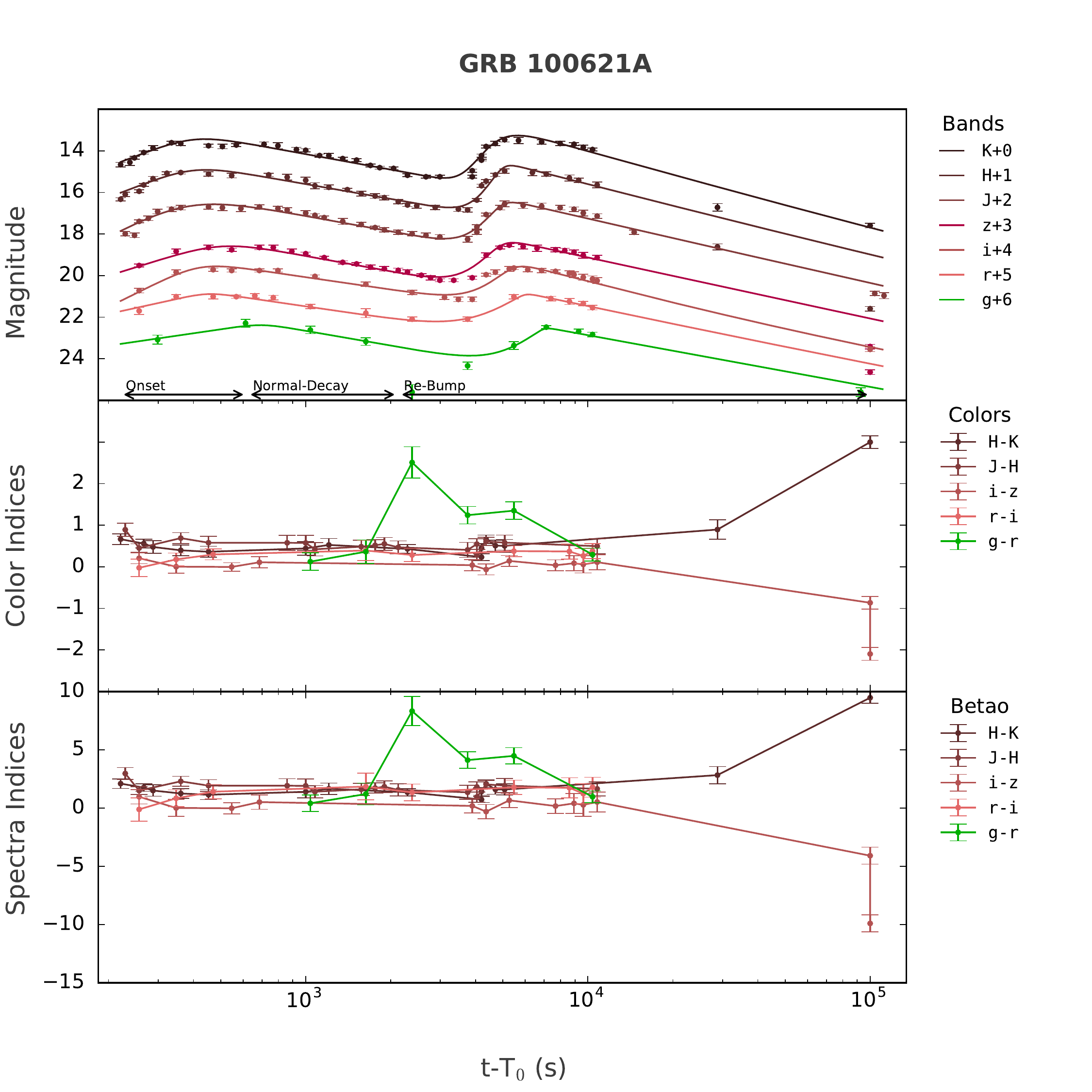}
\includegraphics[angle=0,scale=0.40]{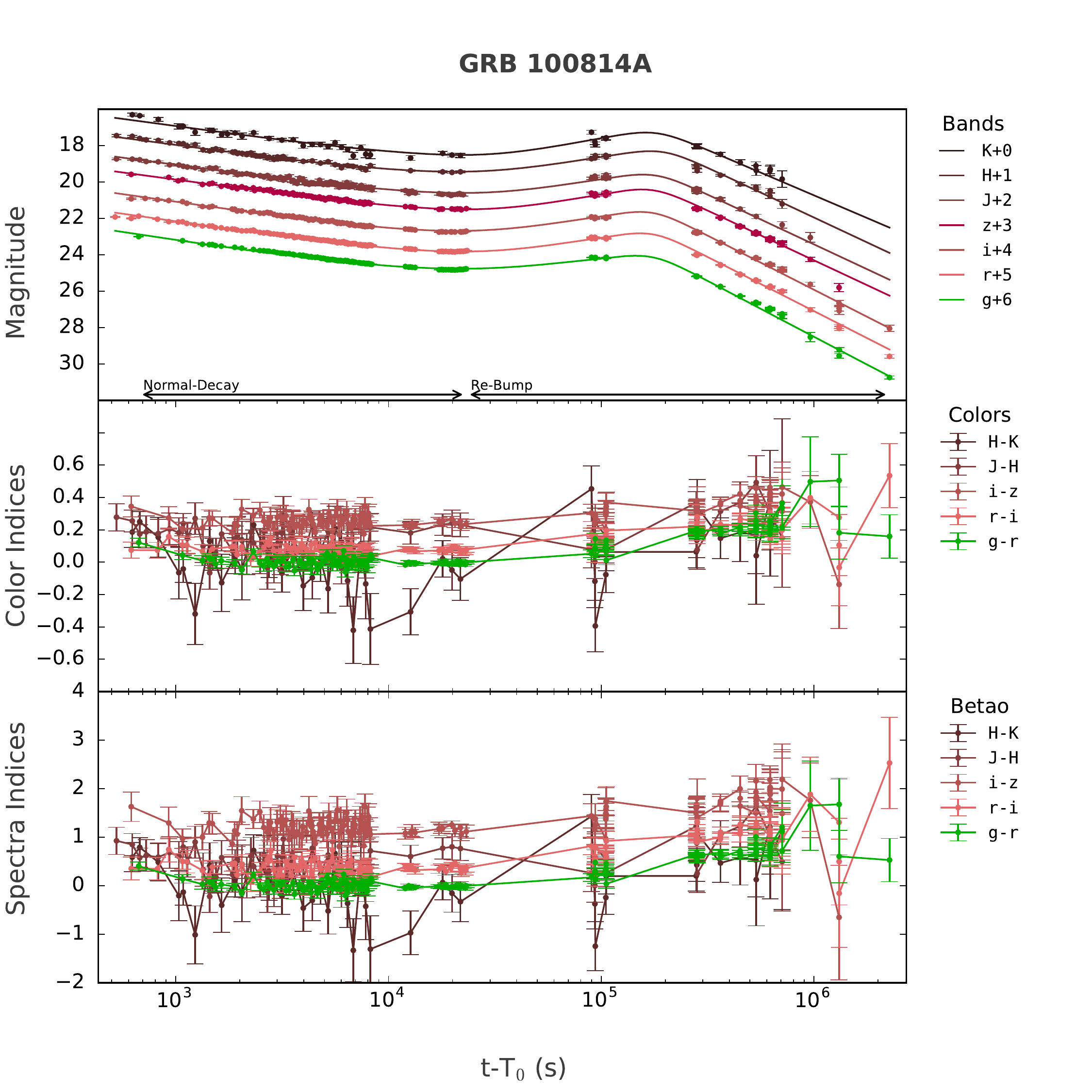}
\center{Fig. \ref{GoldenLCs}--- Continued}
\end{figure*}
\begin{figure*}
\includegraphics[angle=0,scale=0.40]{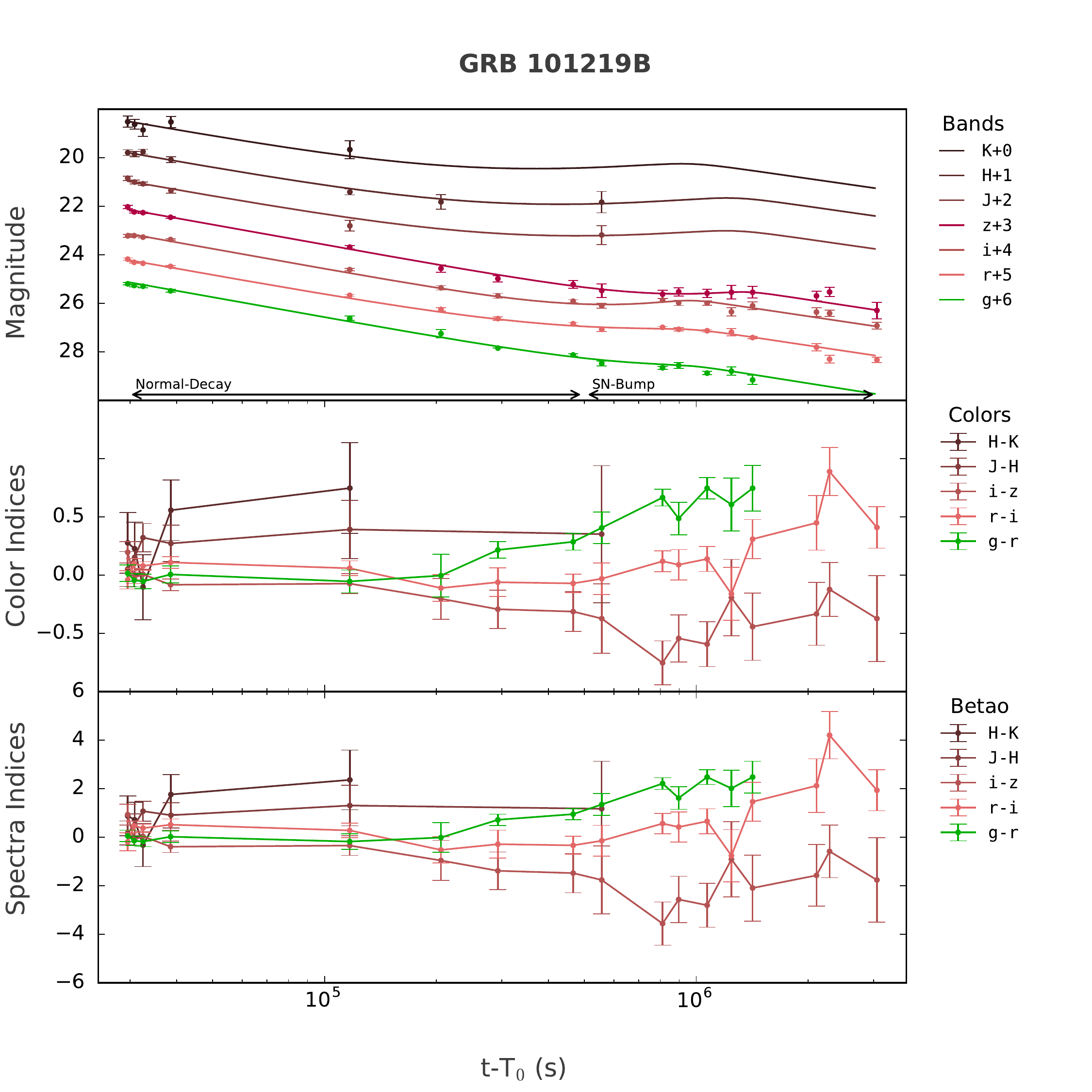}
\includegraphics[angle=0,scale=0.40]{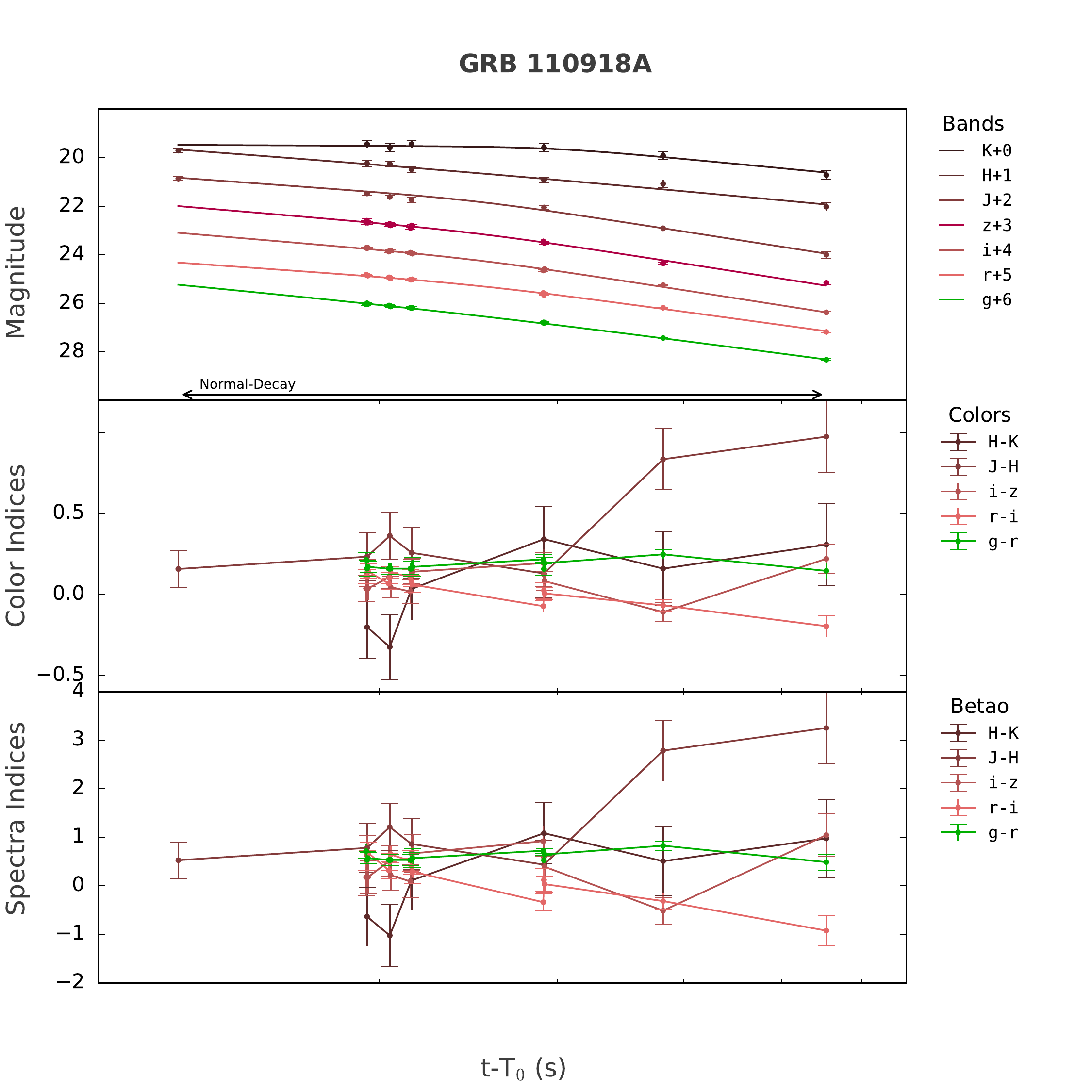}
\includegraphics[angle=0,scale=0.40]{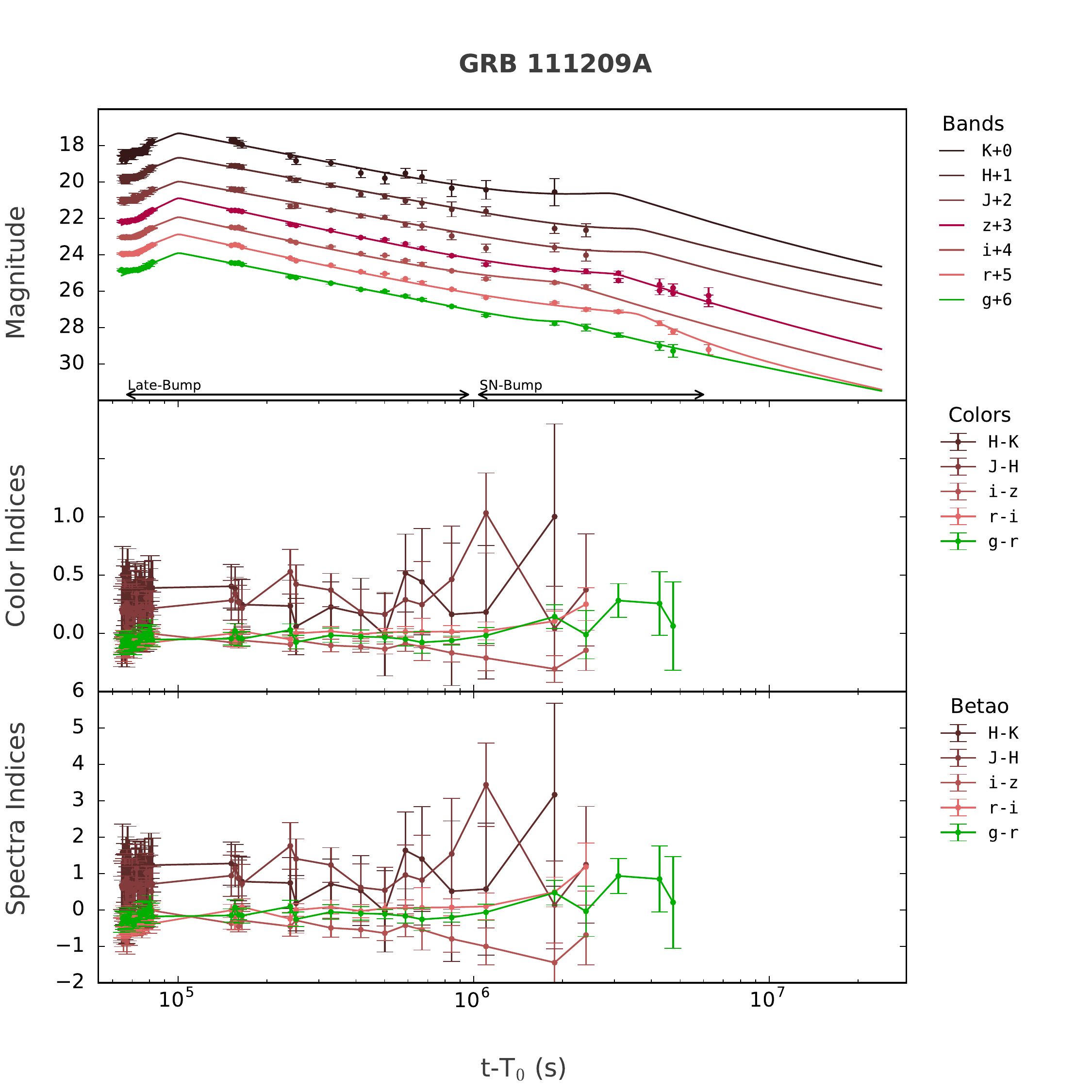}
\includegraphics[angle=0,scale=0.40]{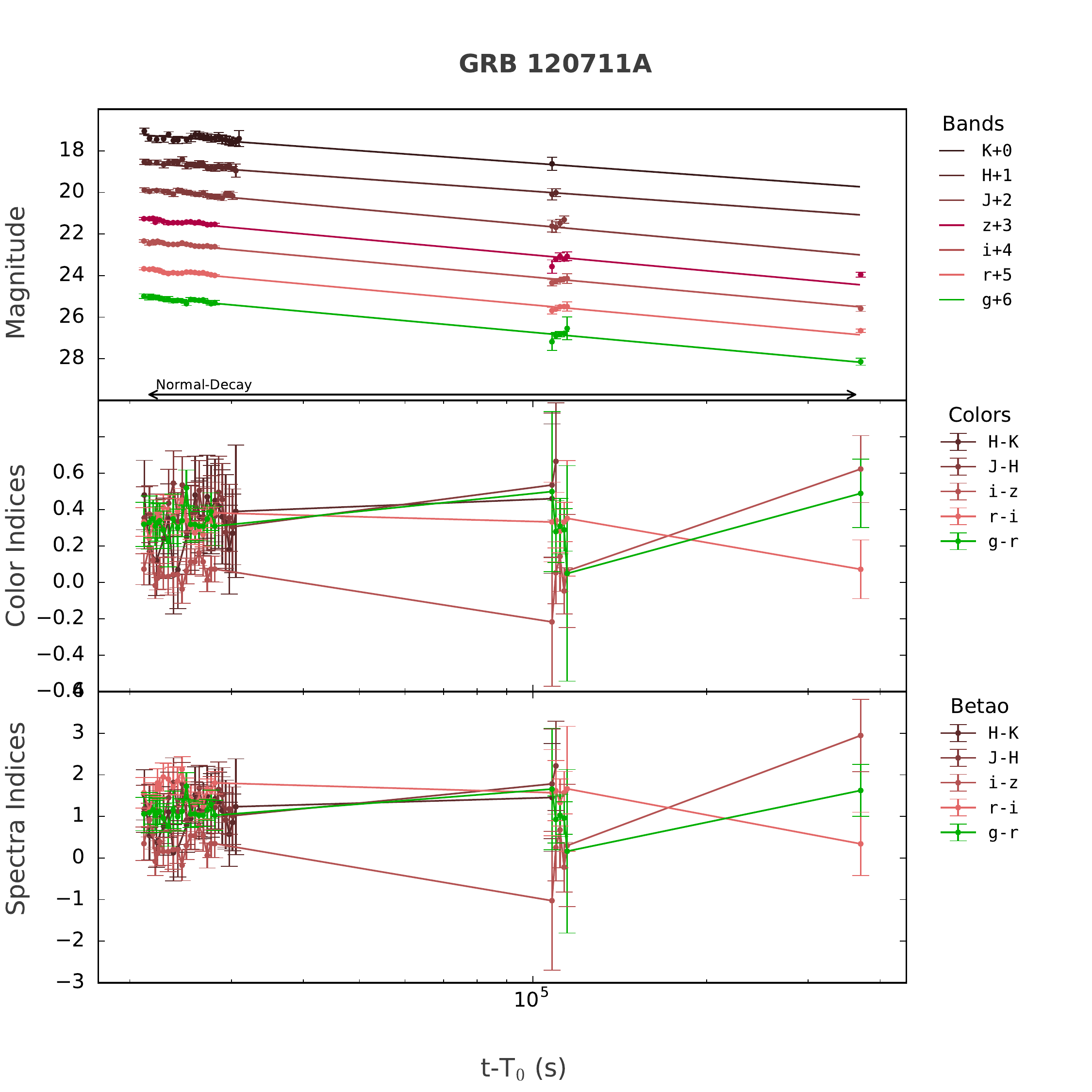}
\center{Fig. \ref{GoldenLCs}--- Continued}
\end{figure*}
\begin{figure*}
\includegraphics[angle=0,scale=0.40]{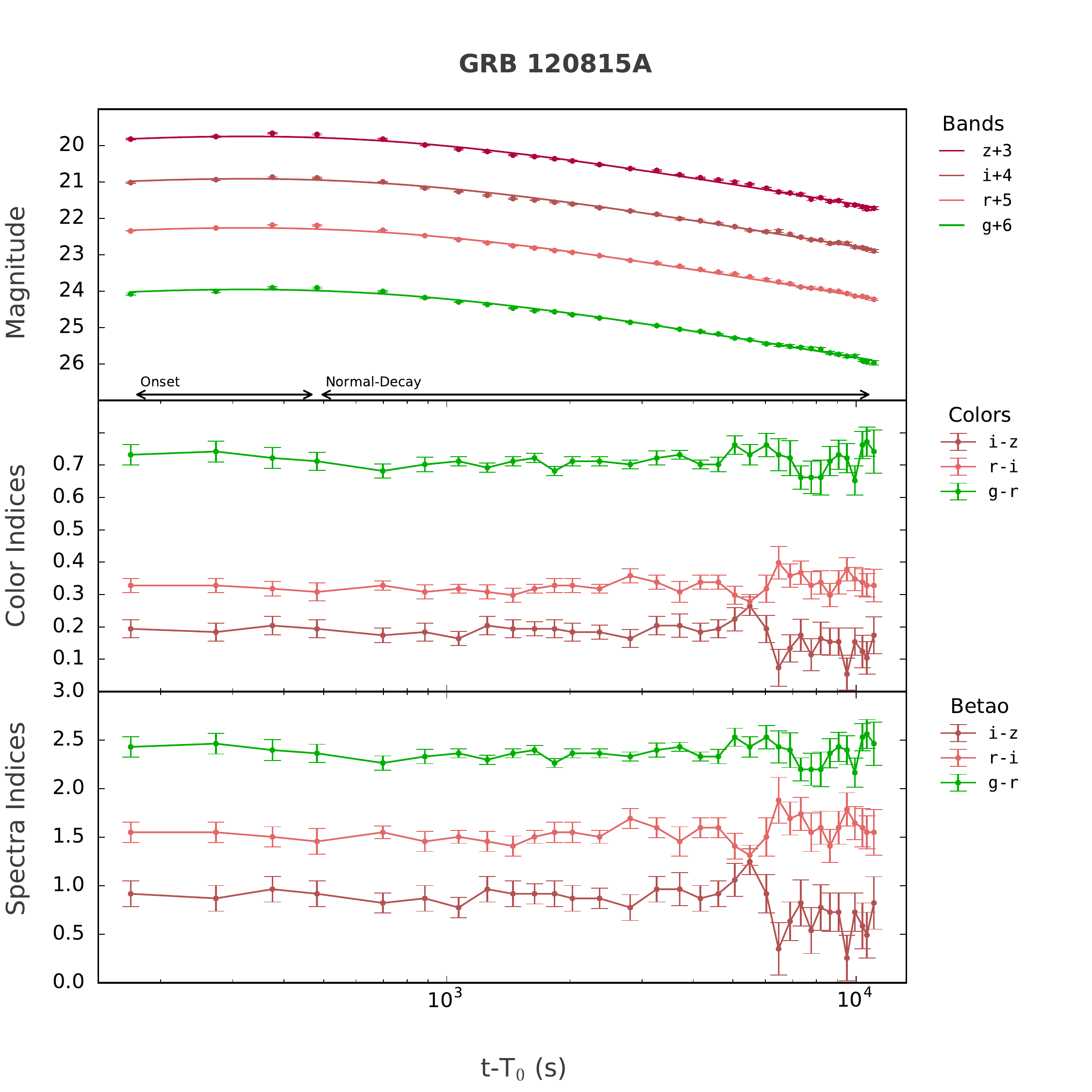}
\includegraphics[angle=0,scale=0.40]{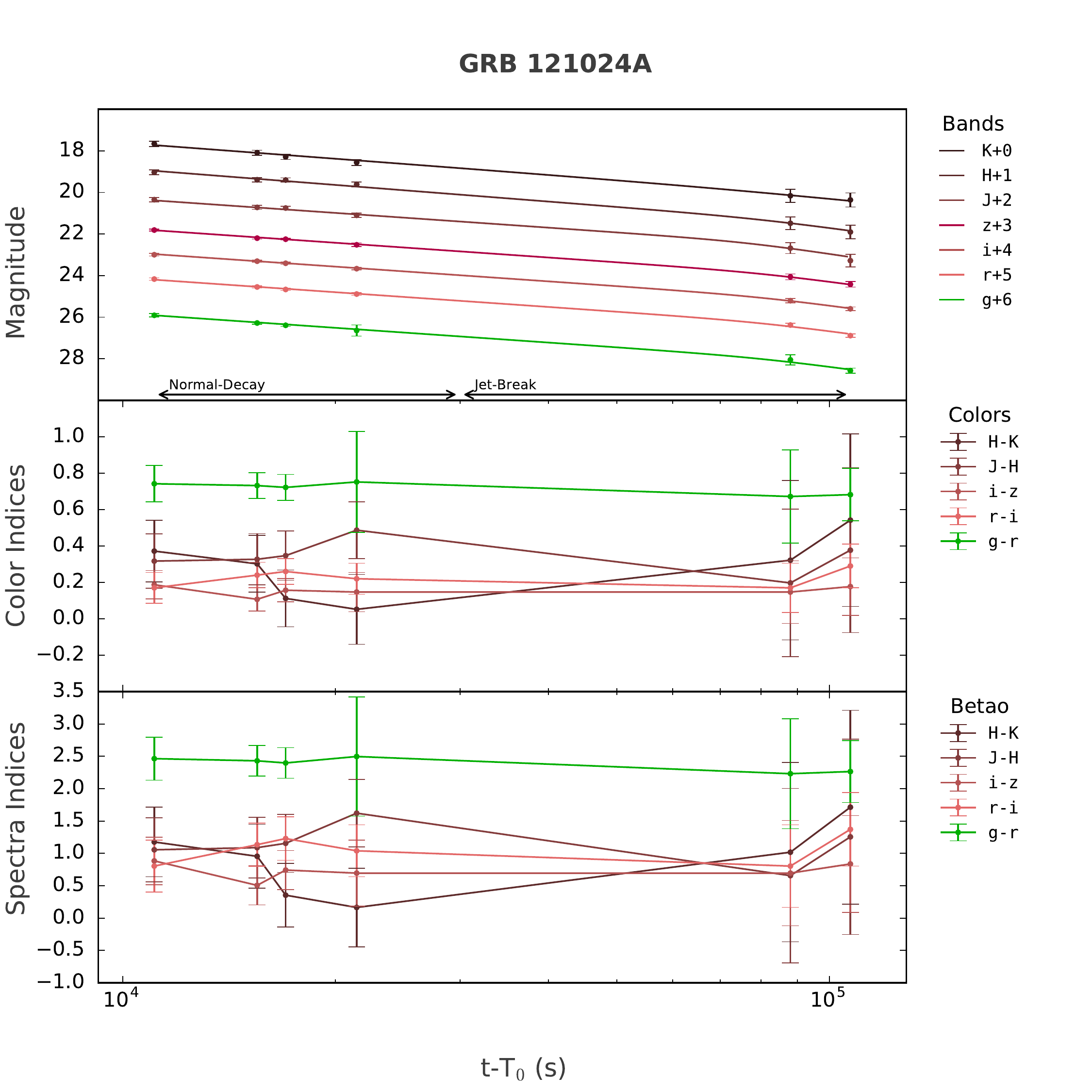}
\includegraphics[angle=0,scale=0.40]{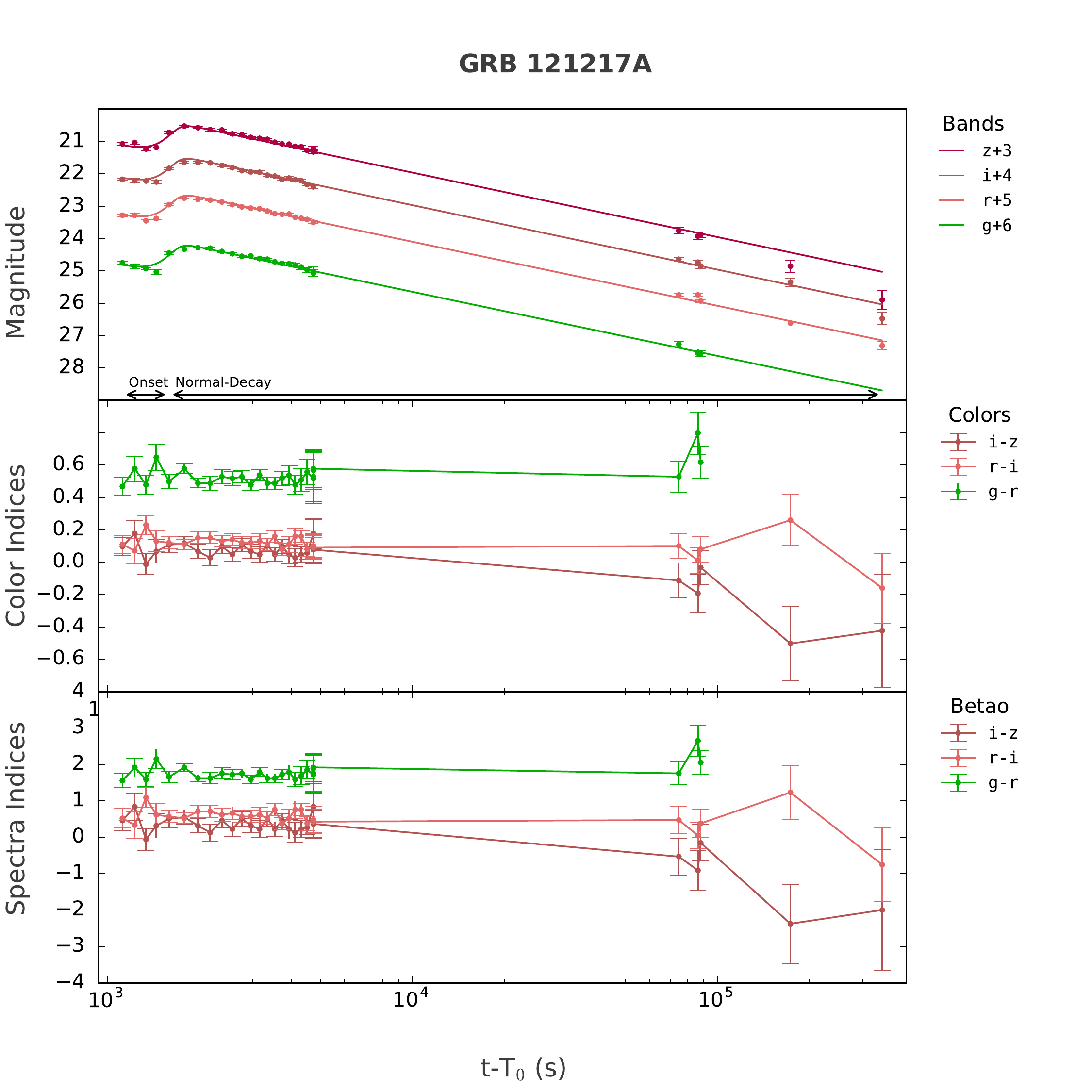}
\includegraphics[angle=0,scale=0.40]{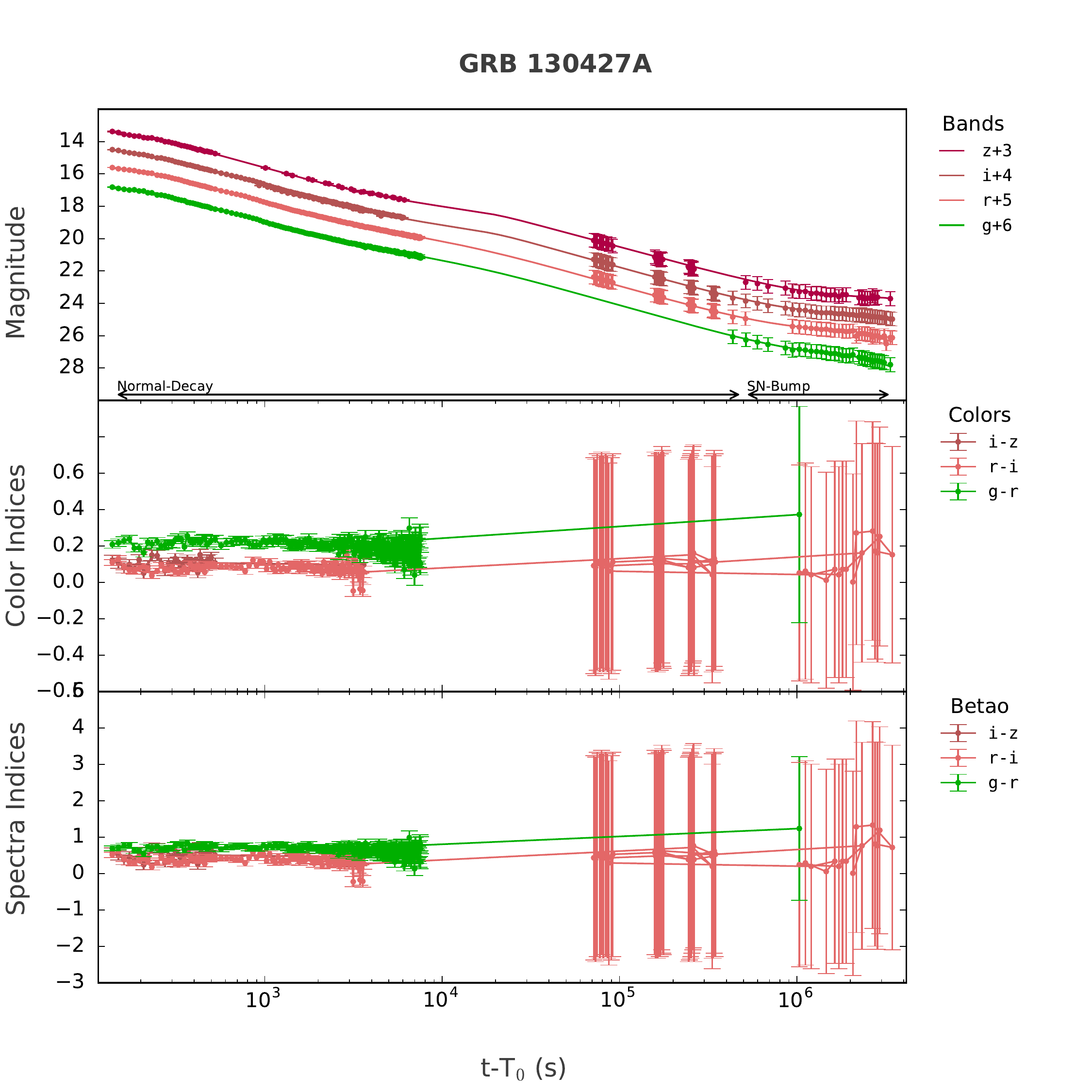}
\center{Fig. \ref{GoldenLCs}--- Continued}
\end{figure*}
\begin{figure*}
\includegraphics[angle=0,scale=0.40]{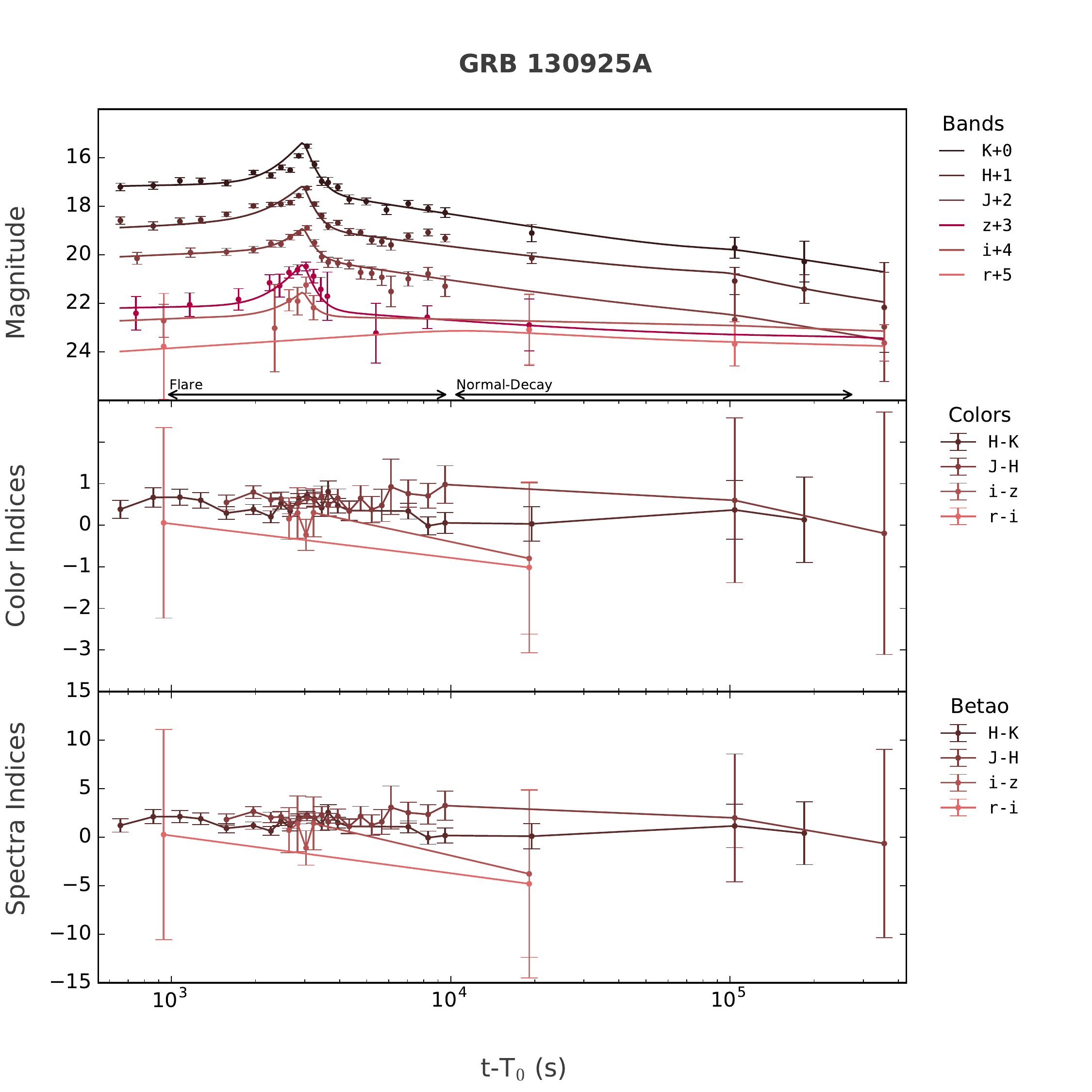}
\center{Fig. \ref{GoldenLCs}--- Continued}
\end{figure*}

\clearpage
\thispagestyle{empty}
\setlength{\voffset}{-18mm}
\begin{figure*}
\includegraphics[angle=0,scale=0.40]{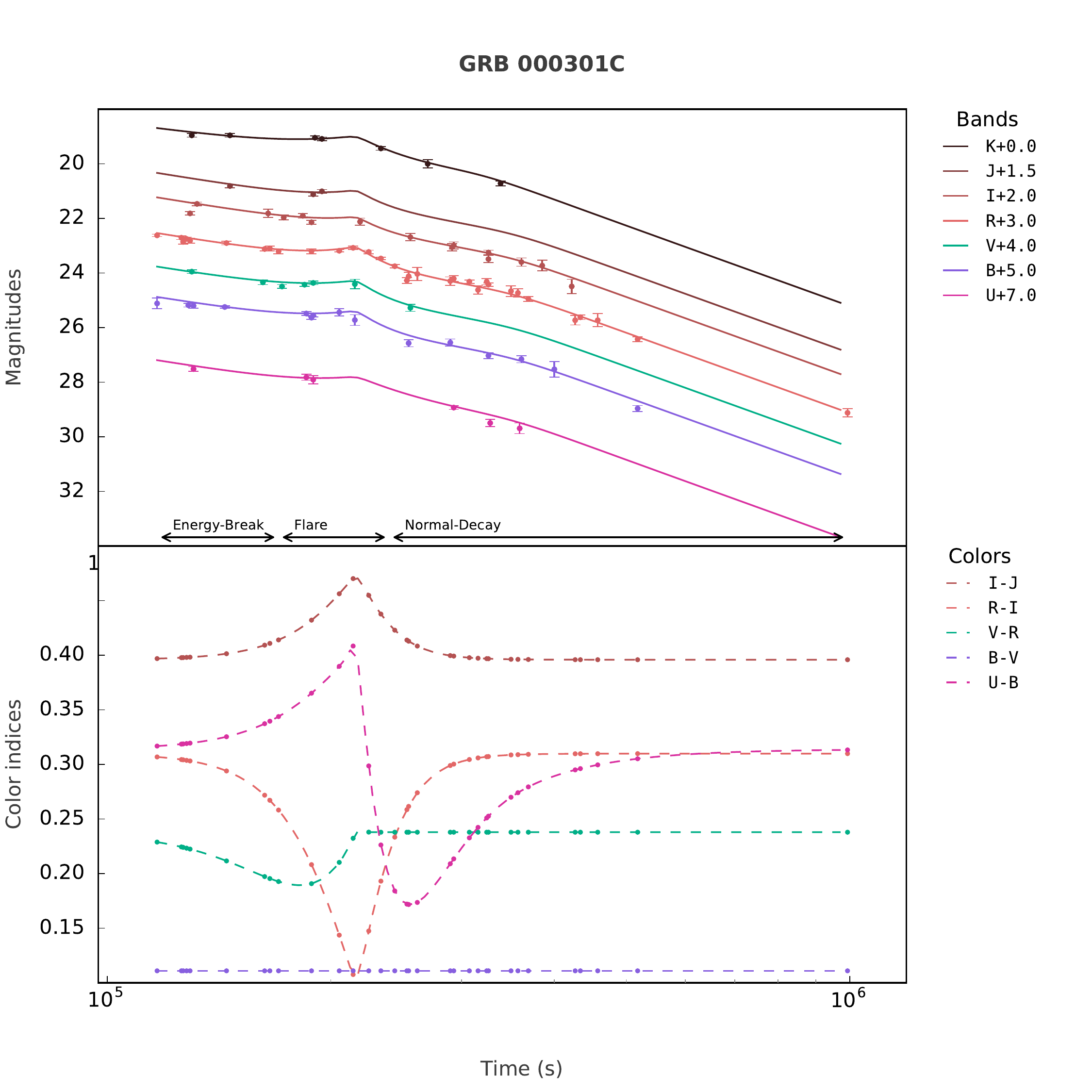}
\includegraphics[angle=0,scale=0.40]{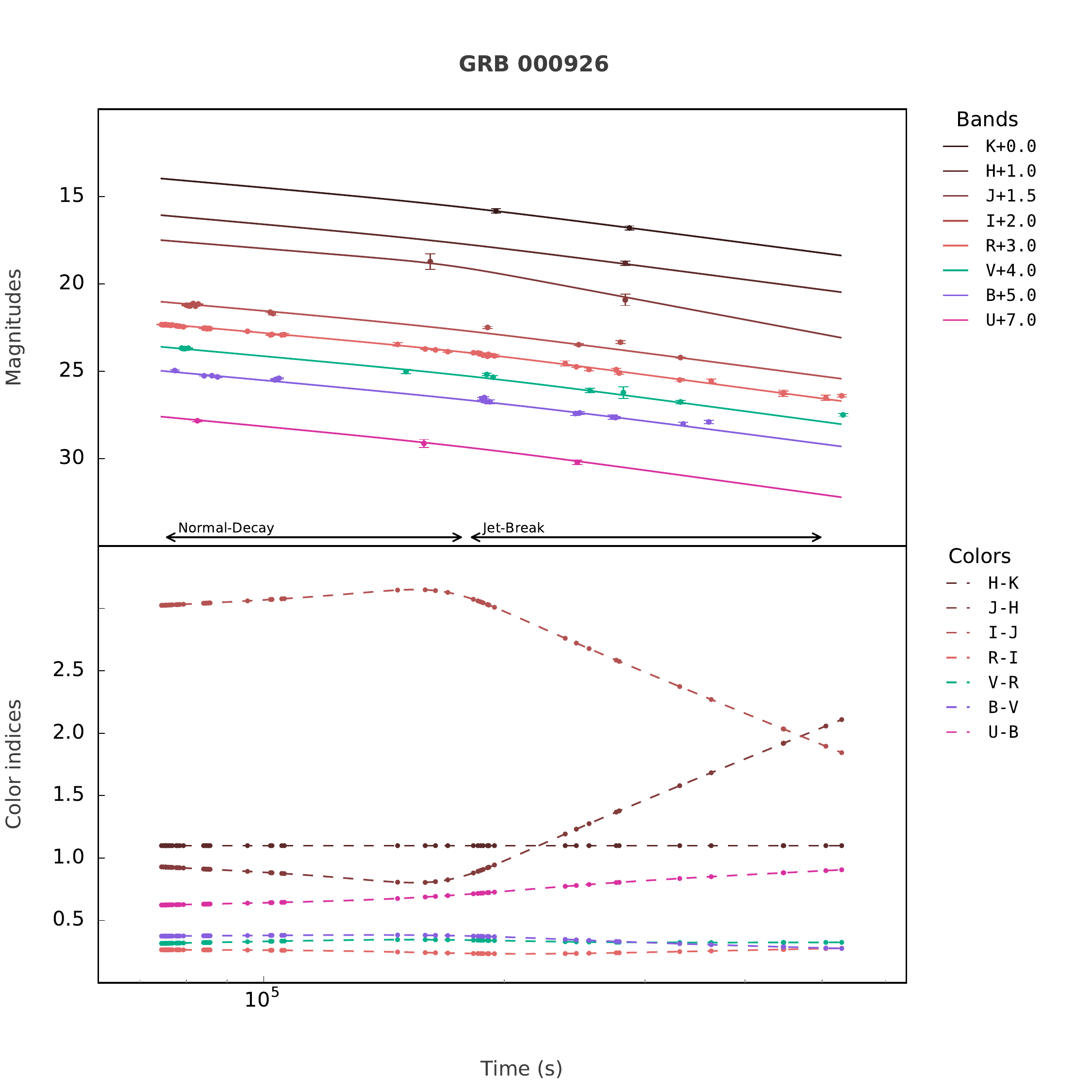}
\includegraphics[angle=0,scale=0.40]{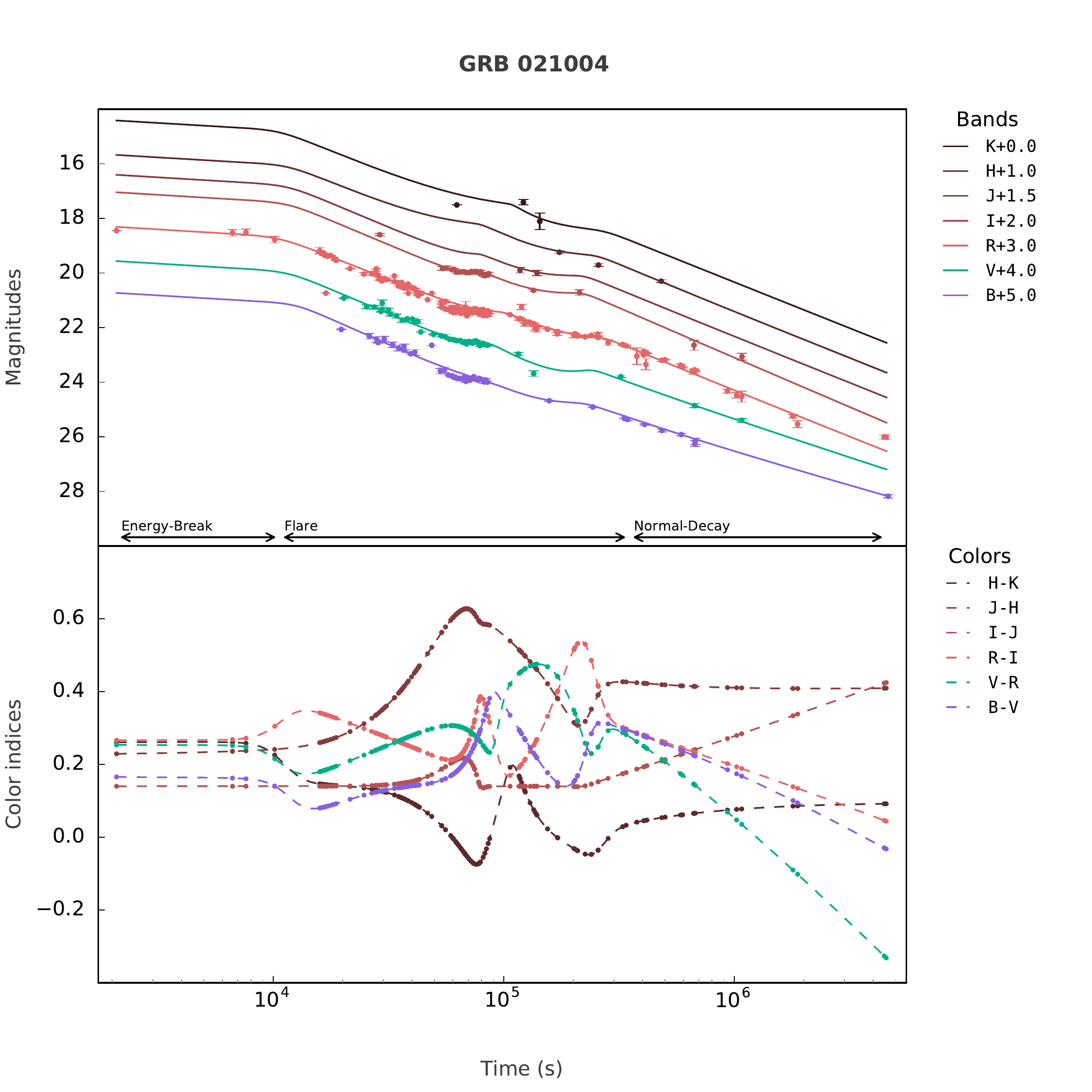}
\includegraphics[angle=0,scale=0.40]{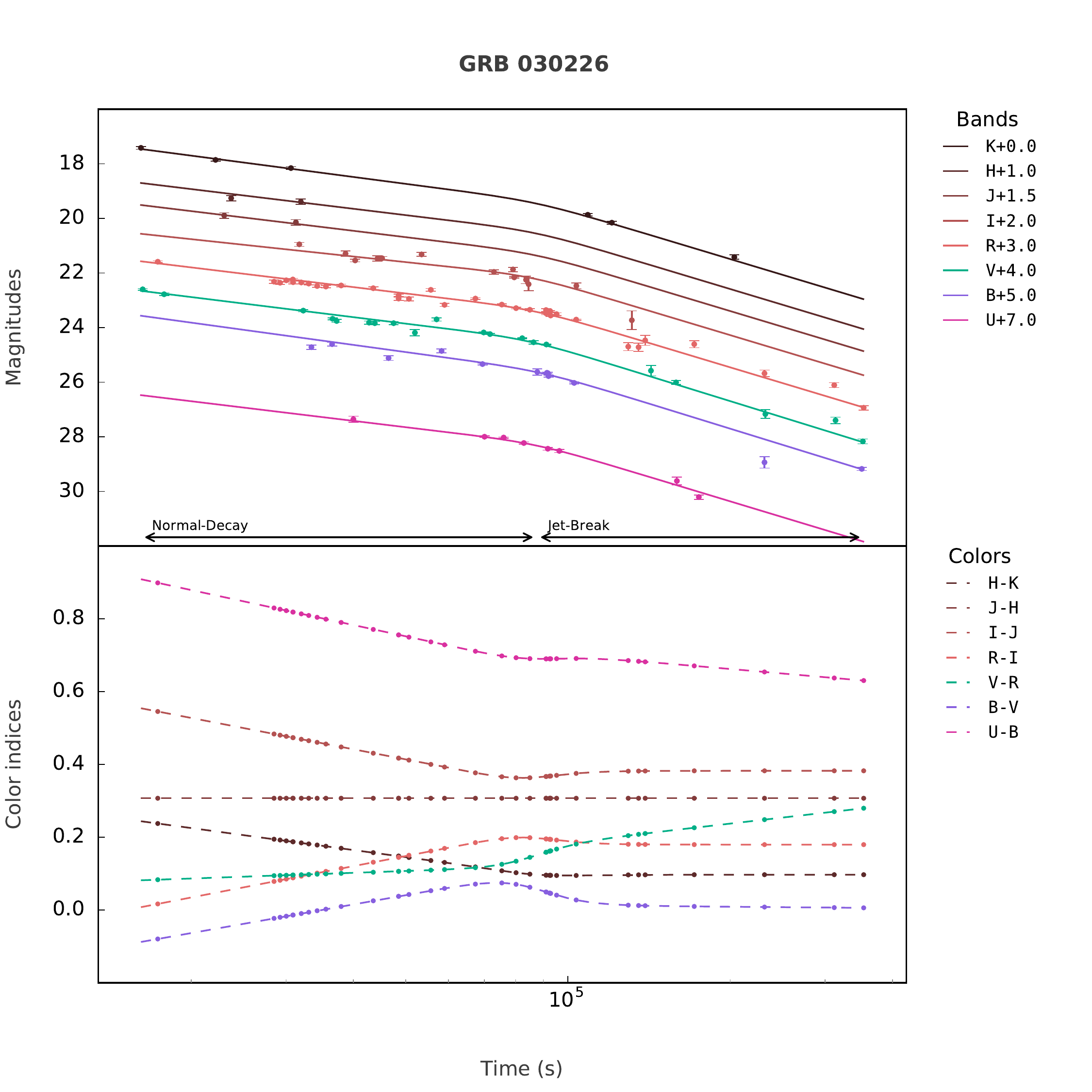}
\center{Fig. \ref{SilverLCs}--- Continued}
\end{figure*}
\begin{figure*}
\includegraphics[angle=0,scale=0.40]{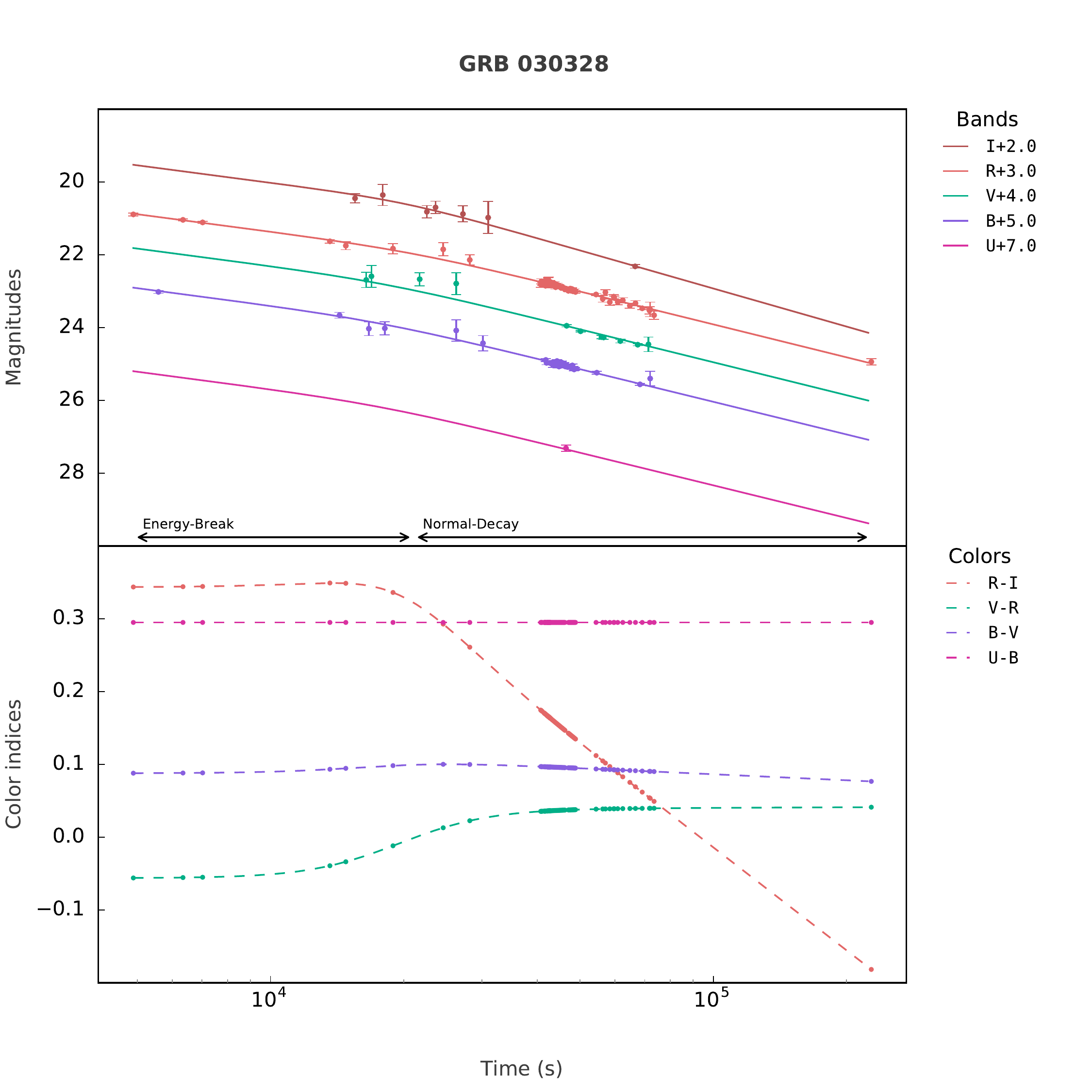}
\includegraphics[angle=0,scale=0.40]{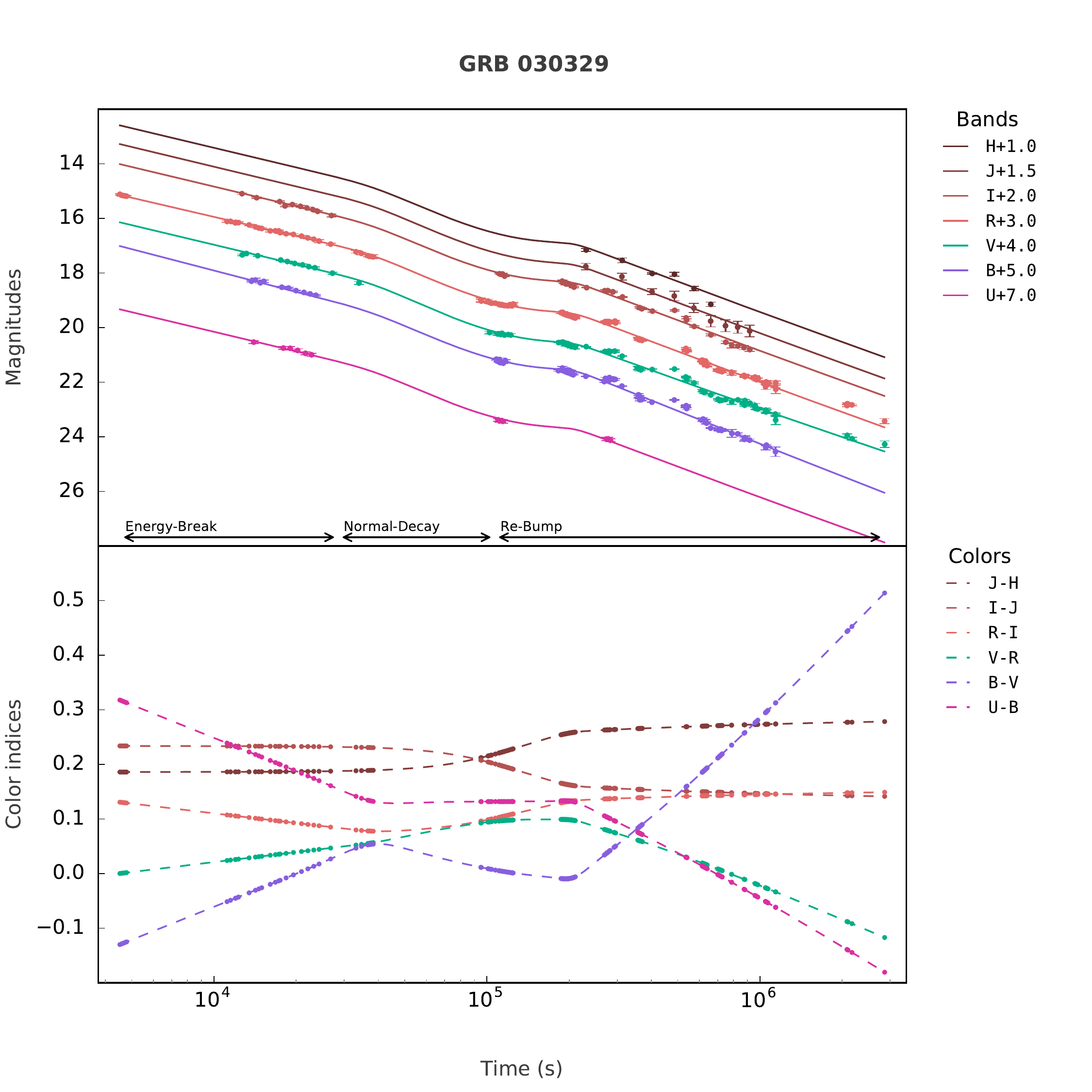}
\includegraphics[angle=0,scale=0.40]{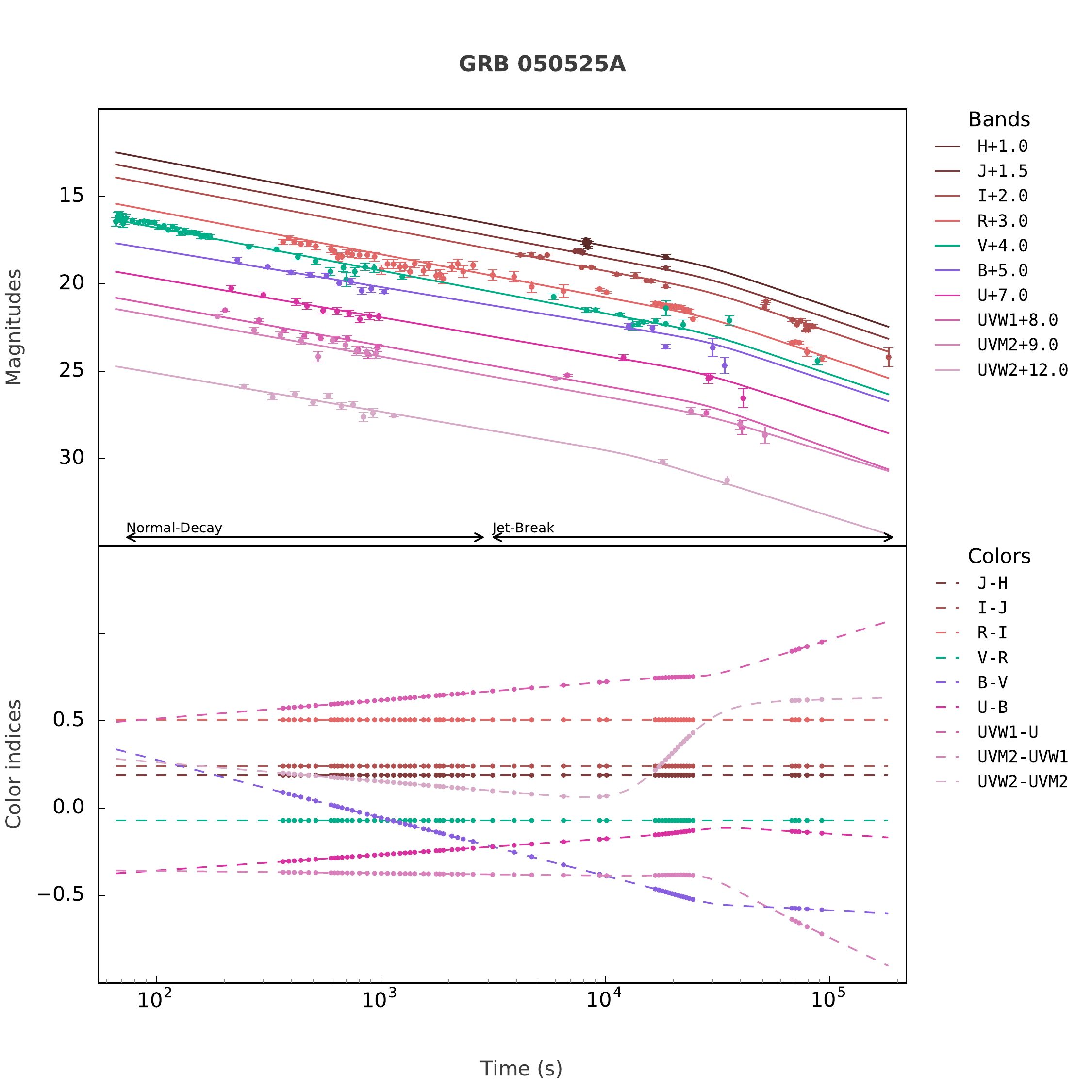}
\includegraphics[angle=0,scale=0.40]{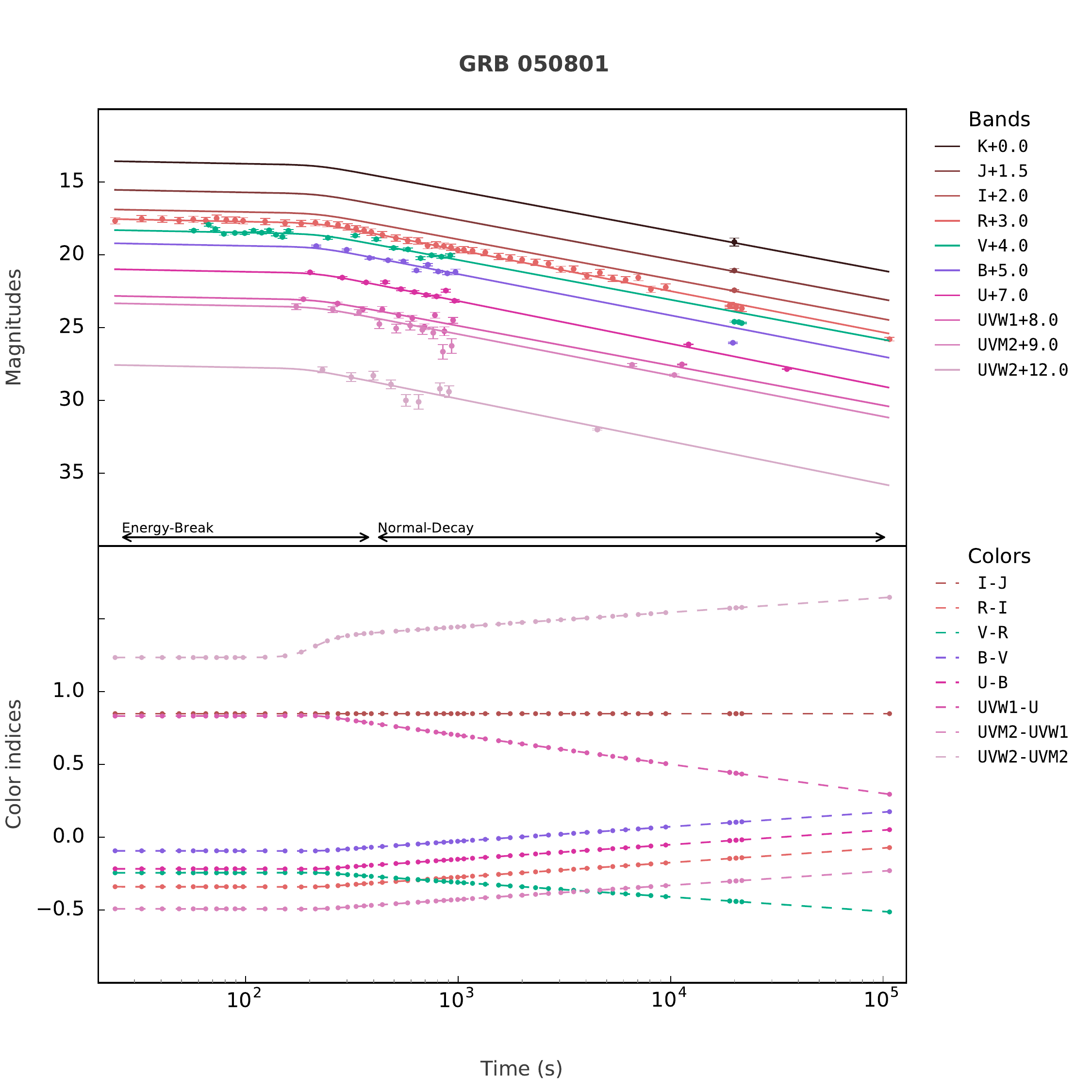}
\center{Fig. \ref{SilverLCs}--- Continued}
\end{figure*}
\begin{figure*}
\includegraphics[angle=0,scale=0.40]{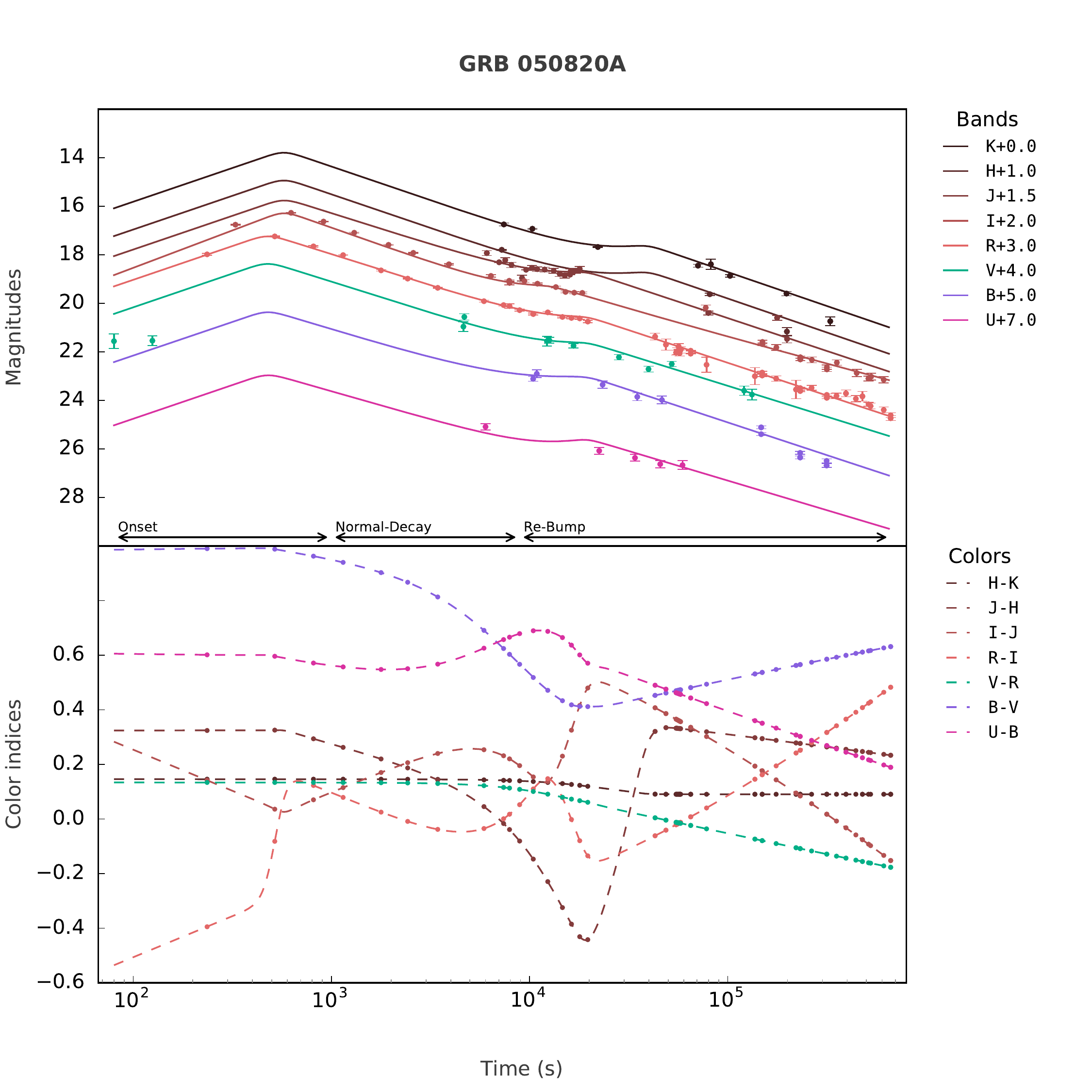}
\includegraphics[angle=0,scale=0.40]{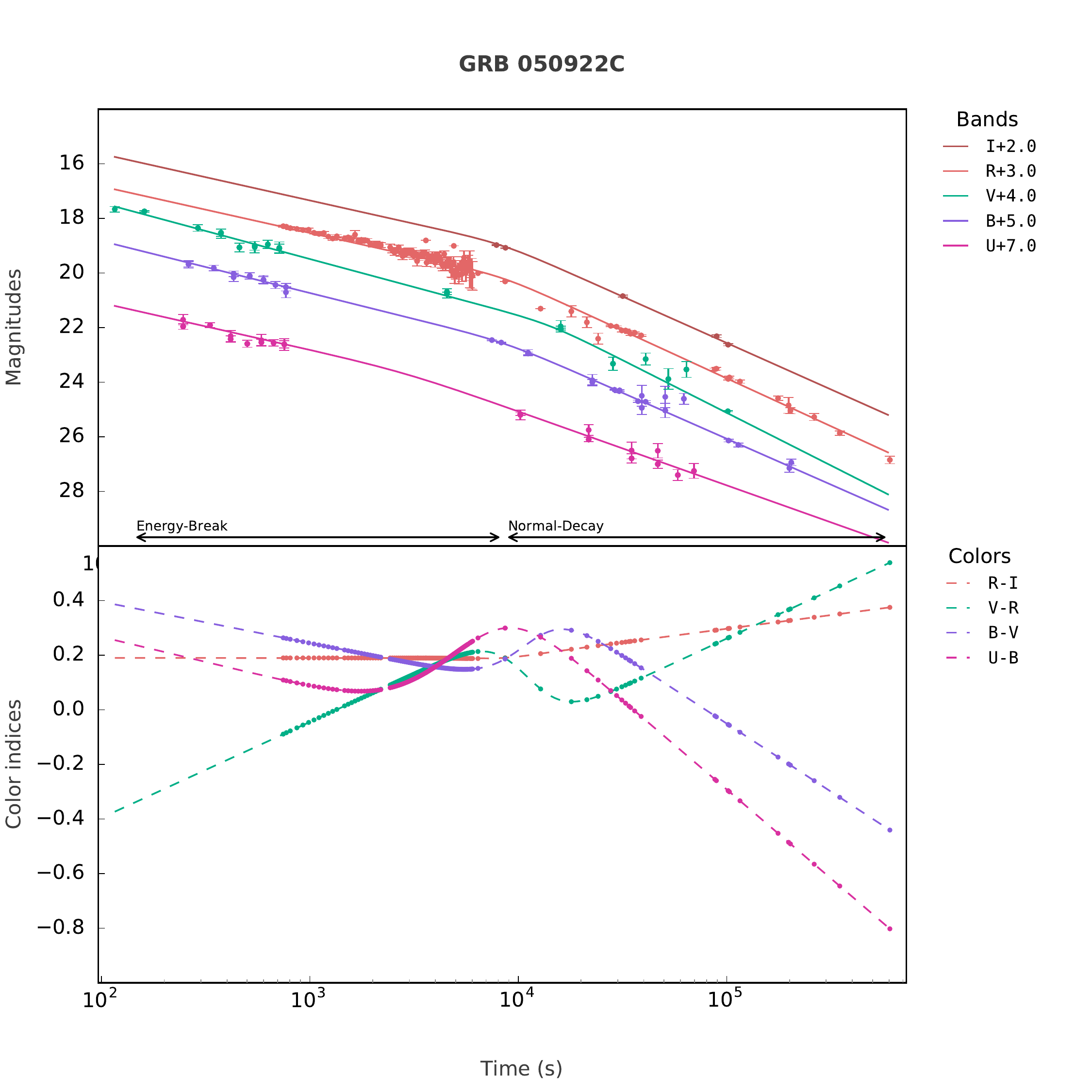}
\includegraphics[angle=0,scale=0.40]{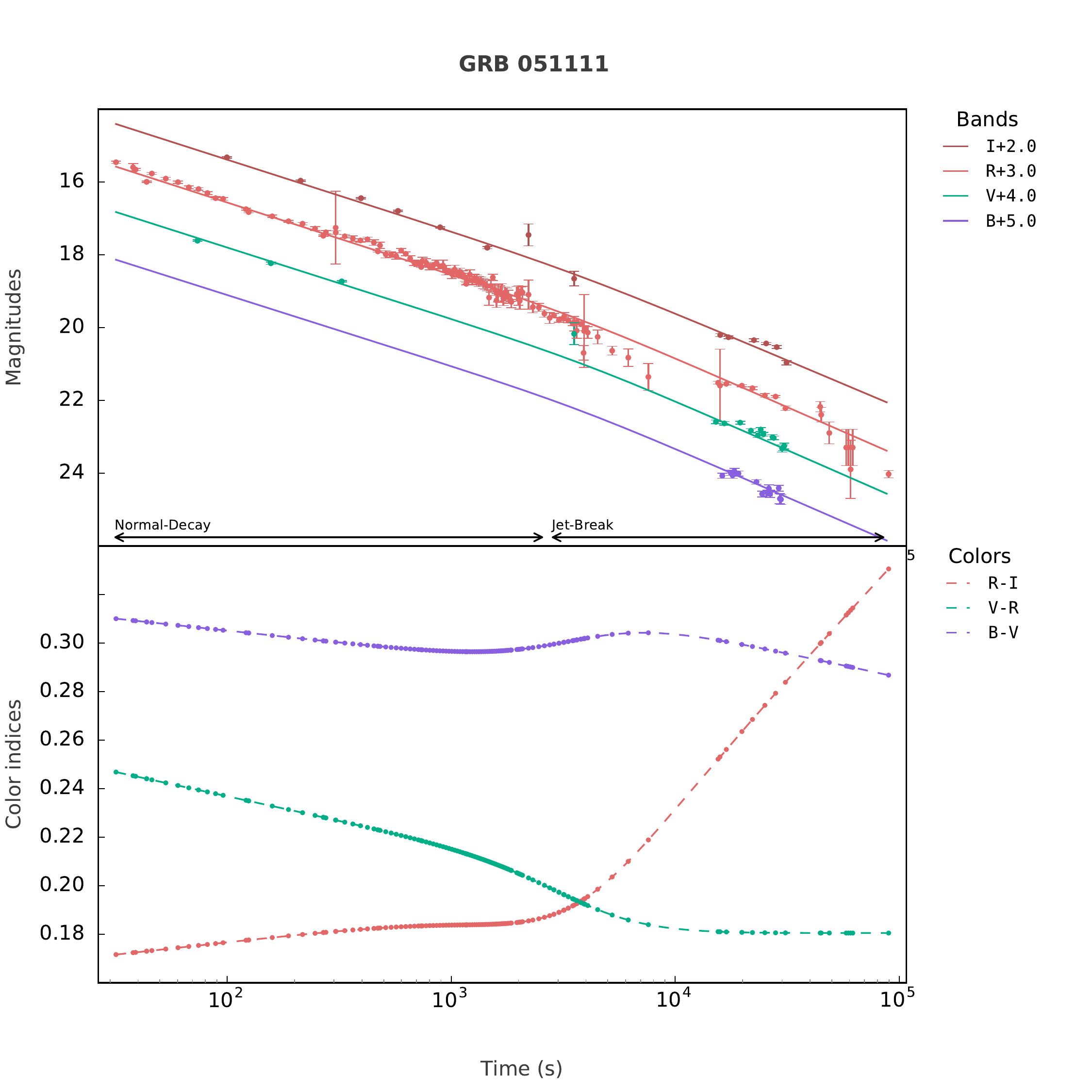}
\includegraphics[angle=0,scale=0.40]{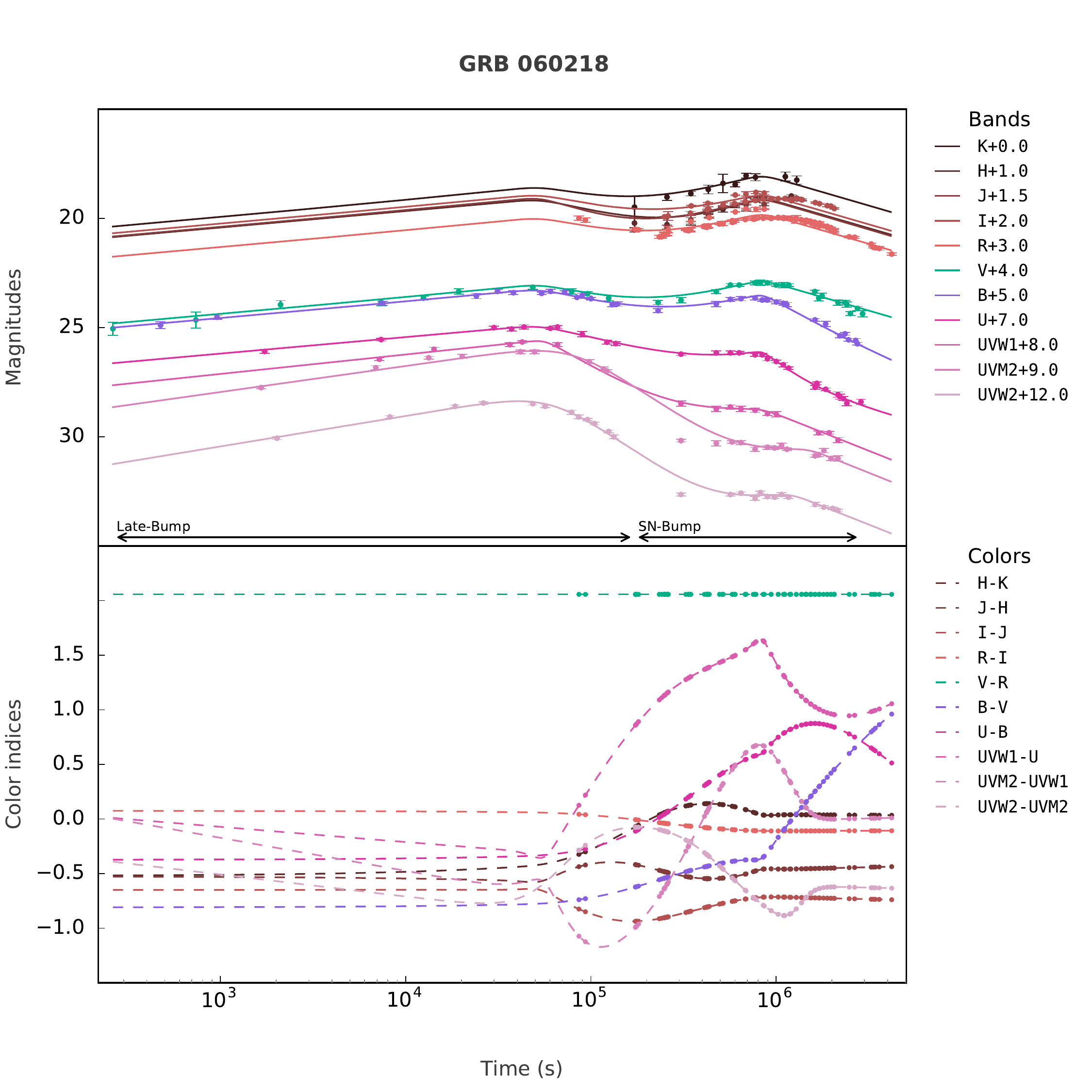}
\center{Fig. \ref{SilverLCs}--- Continued}
\end{figure*}
\begin{figure*}
\includegraphics[angle=0,scale=0.40]{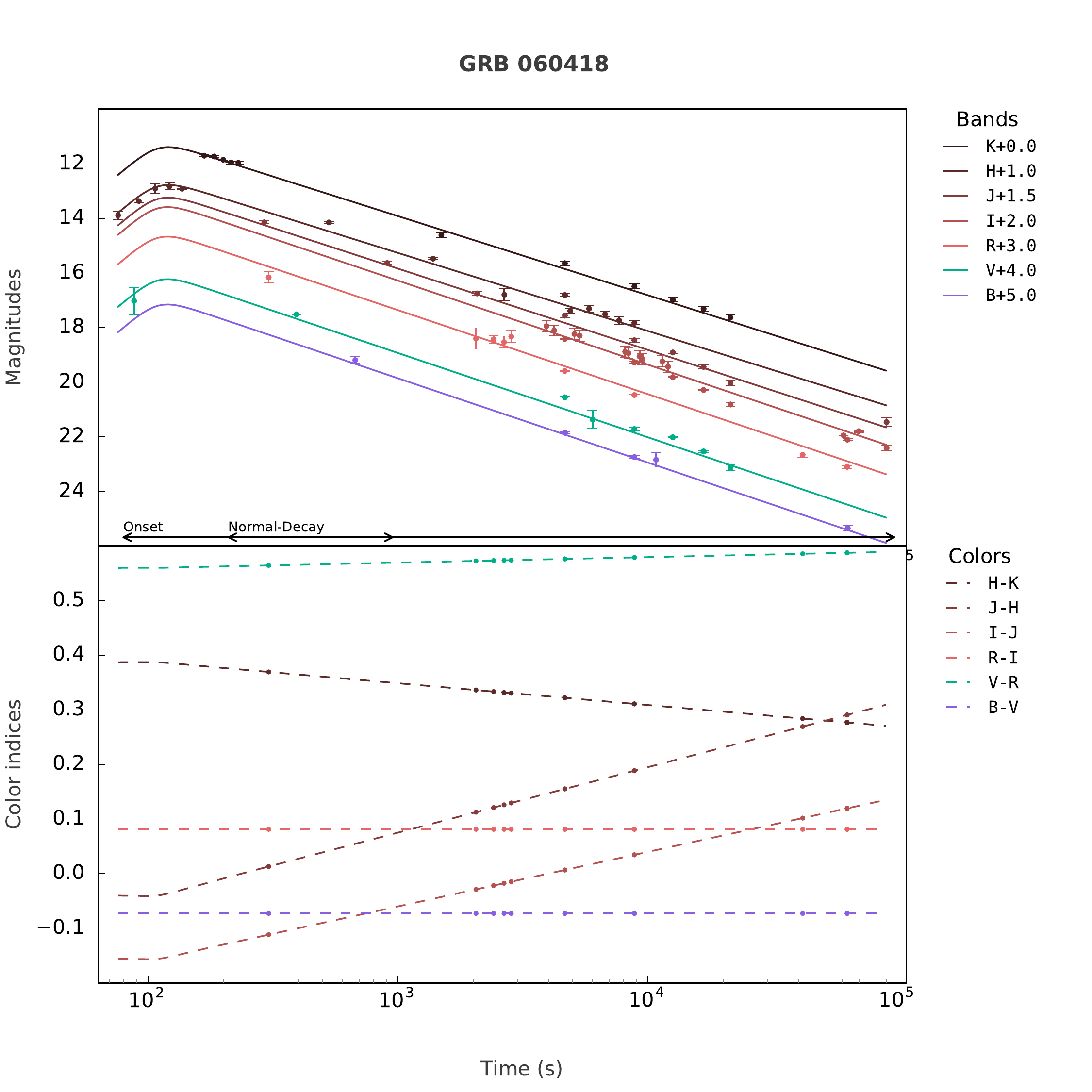}
\includegraphics[angle=0,scale=0.40]{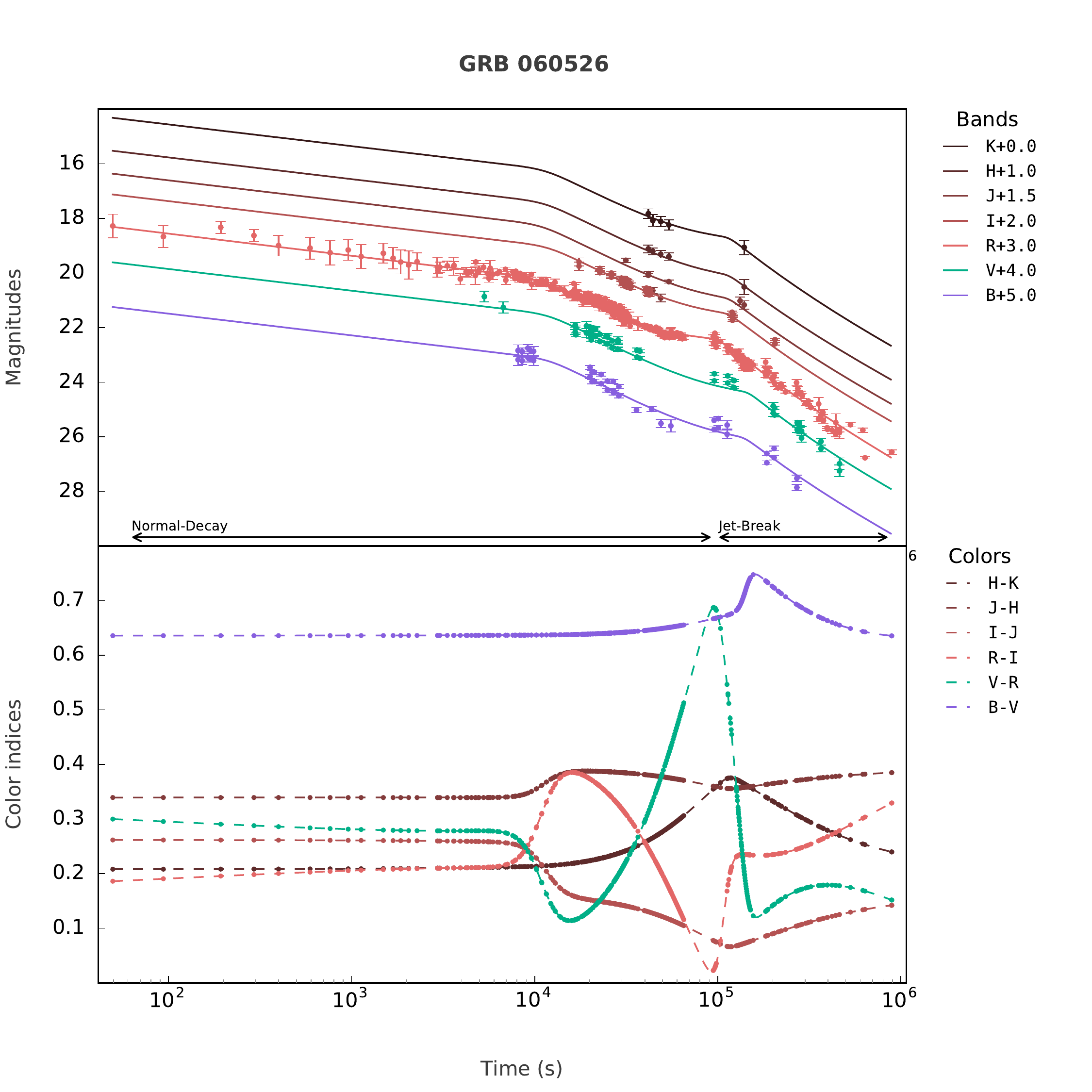}
\includegraphics[angle=0,scale=0.40]{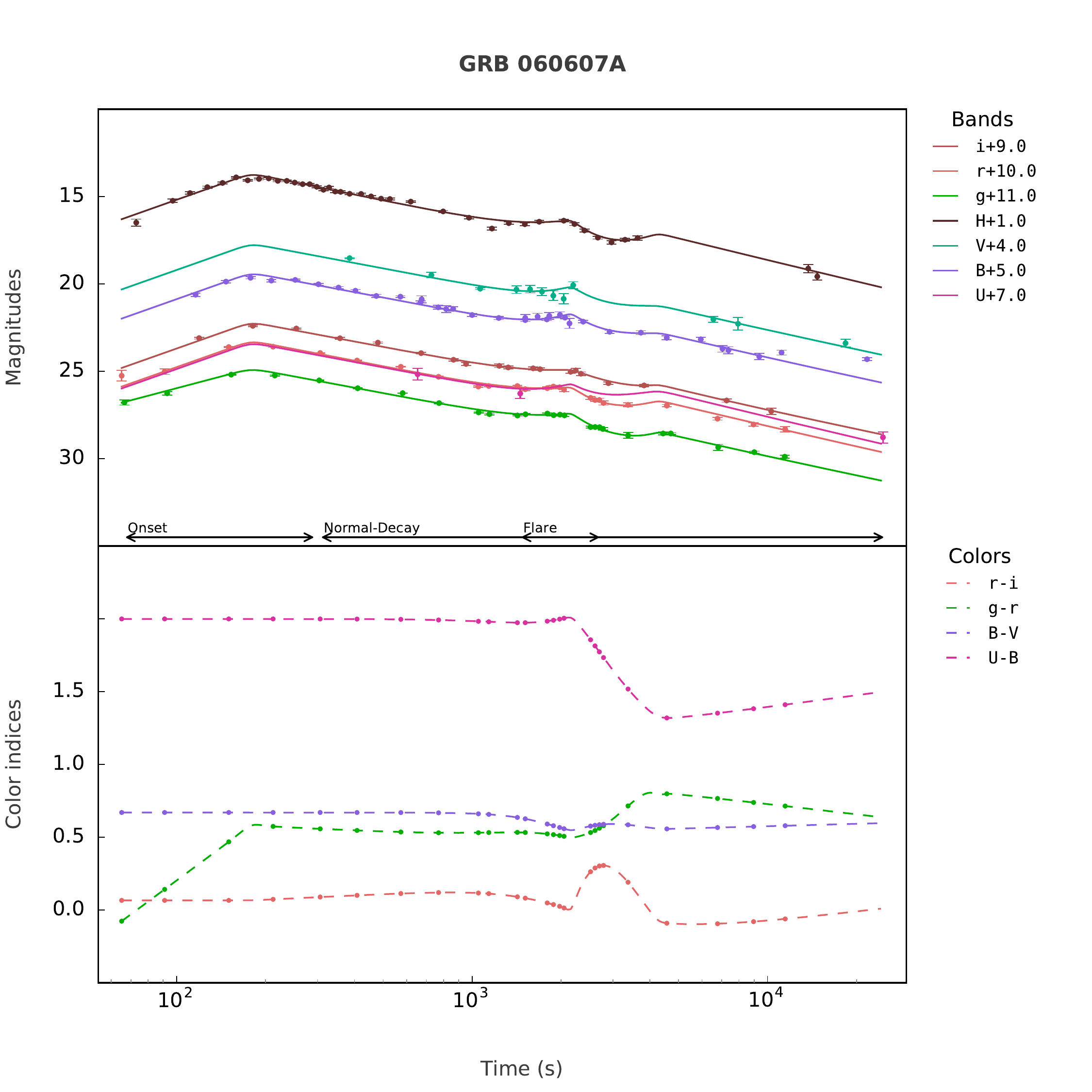}
\includegraphics[angle=0,scale=0.40]{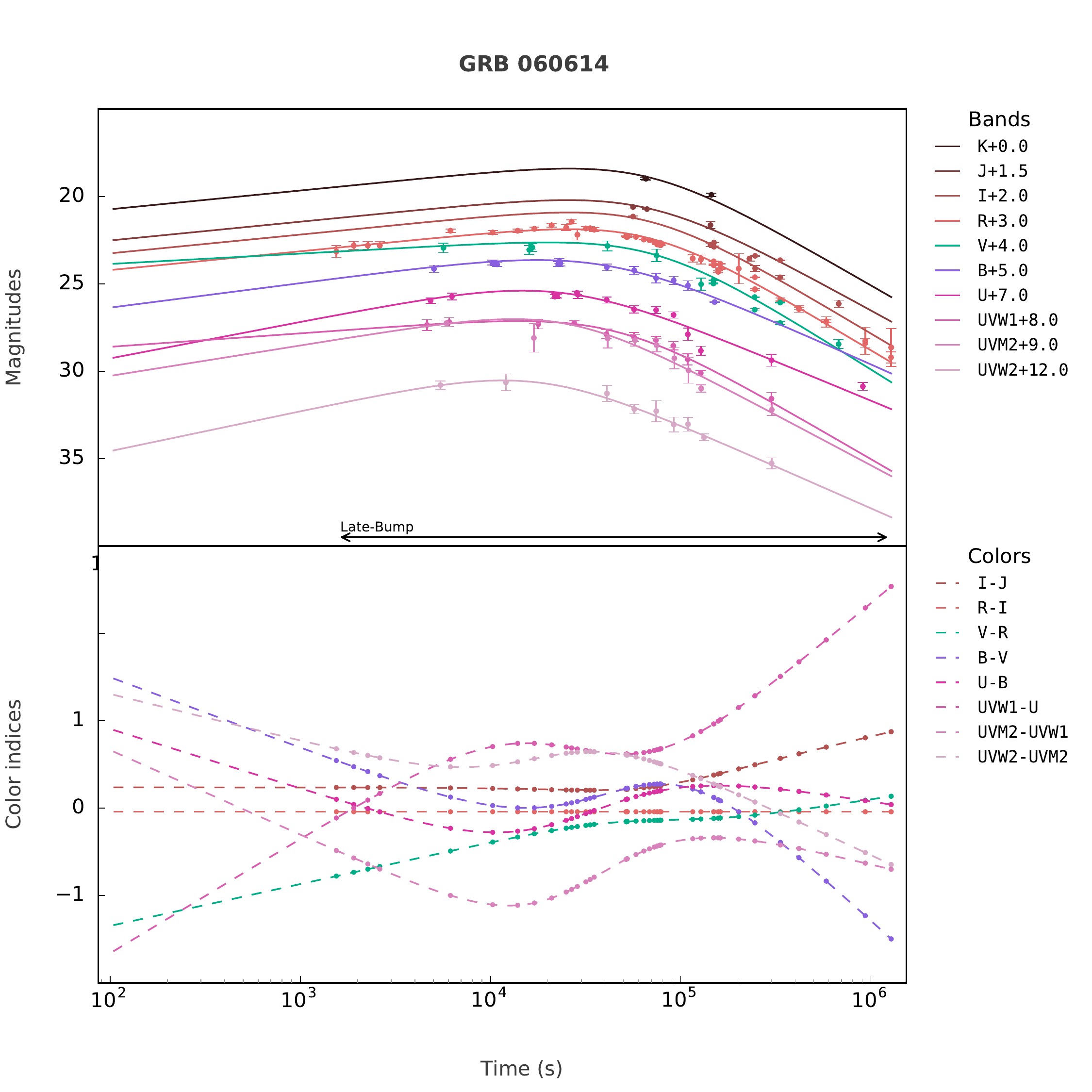}
\center{Fig. \ref{SilverLCs}--- Continued}
\end{figure*}
\begin{figure*}
\includegraphics[angle=0,scale=0.40]{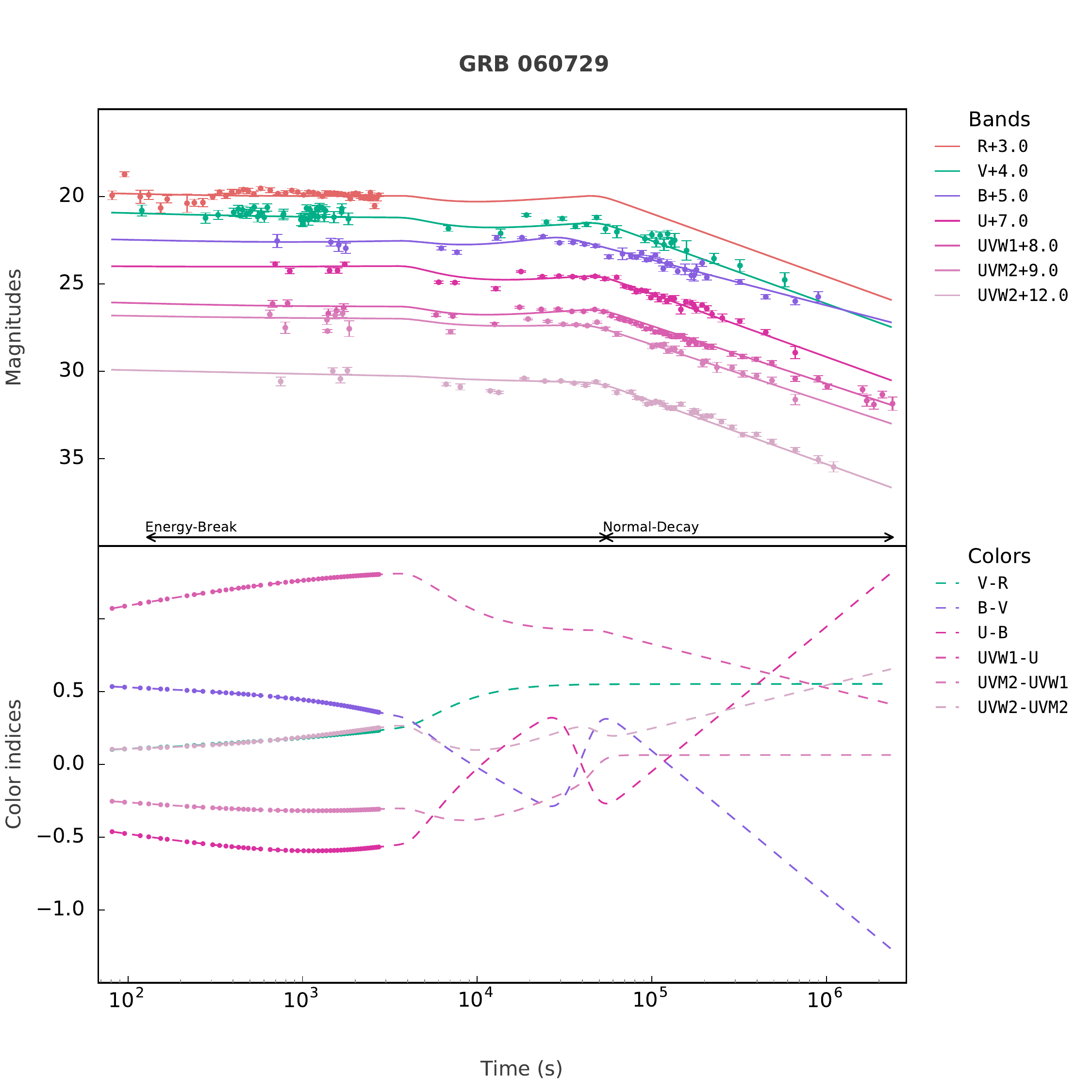}
\includegraphics[angle=0,scale=0.40]{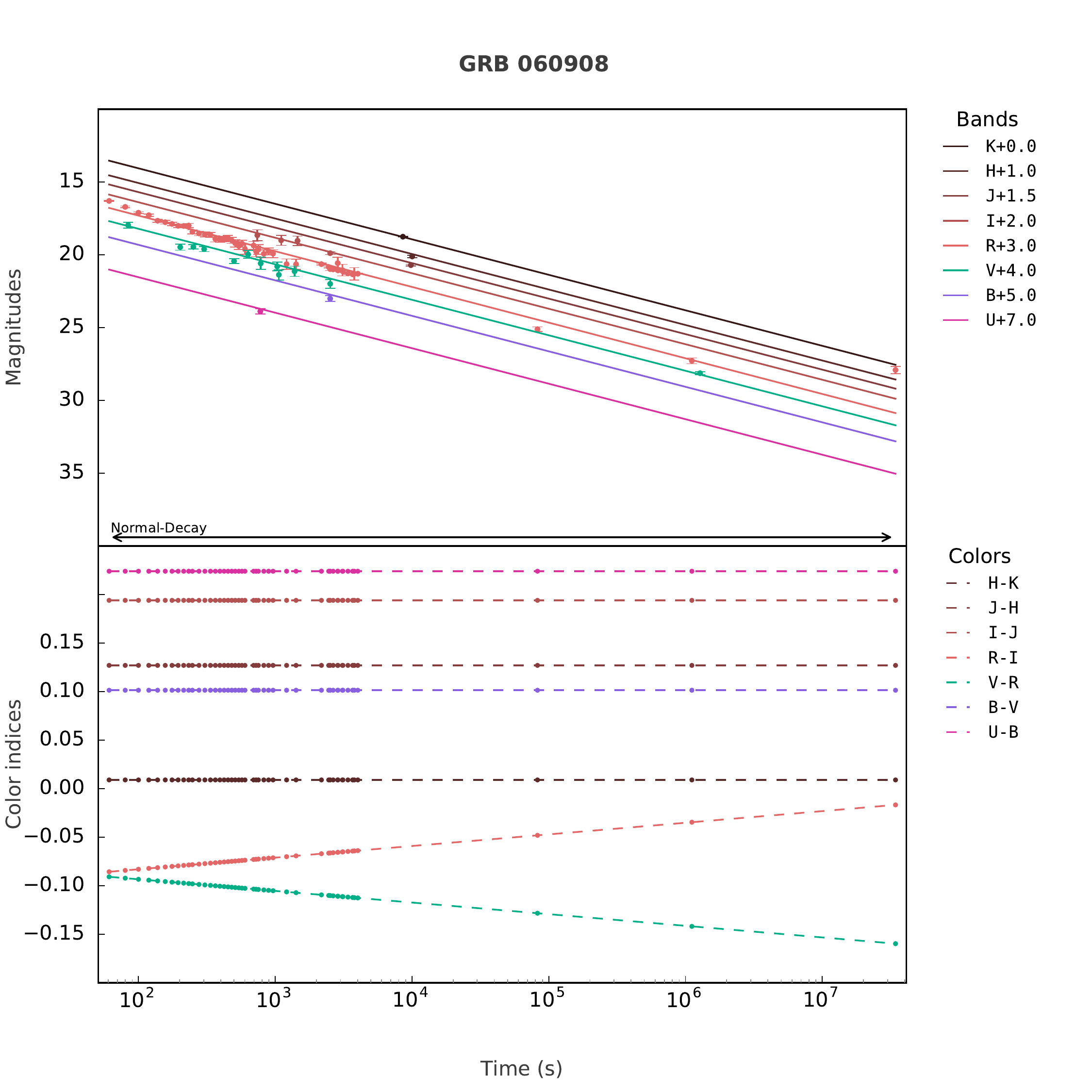}
\includegraphics[angle=0,scale=0.40]{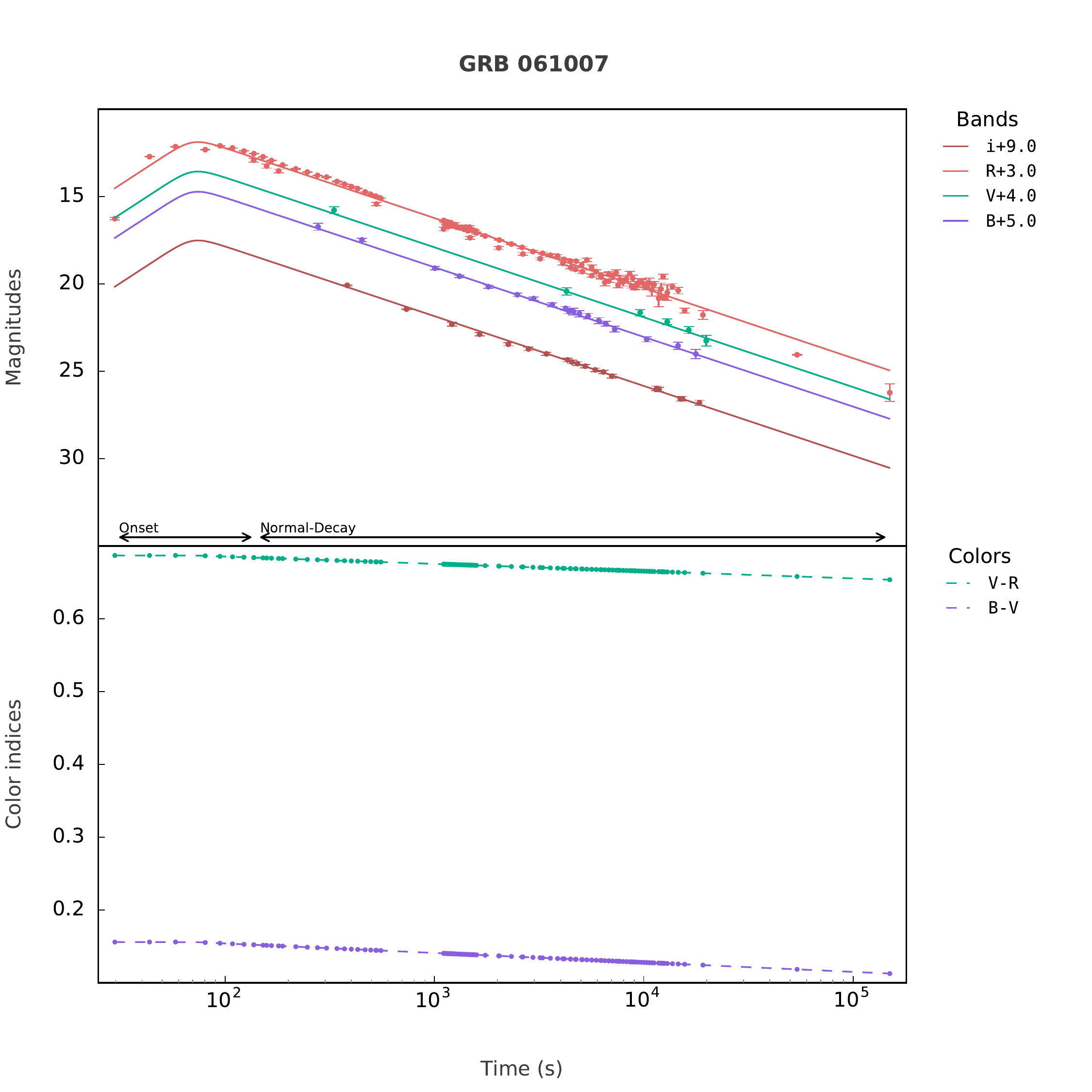}
\includegraphics[angle=0,scale=0.40]{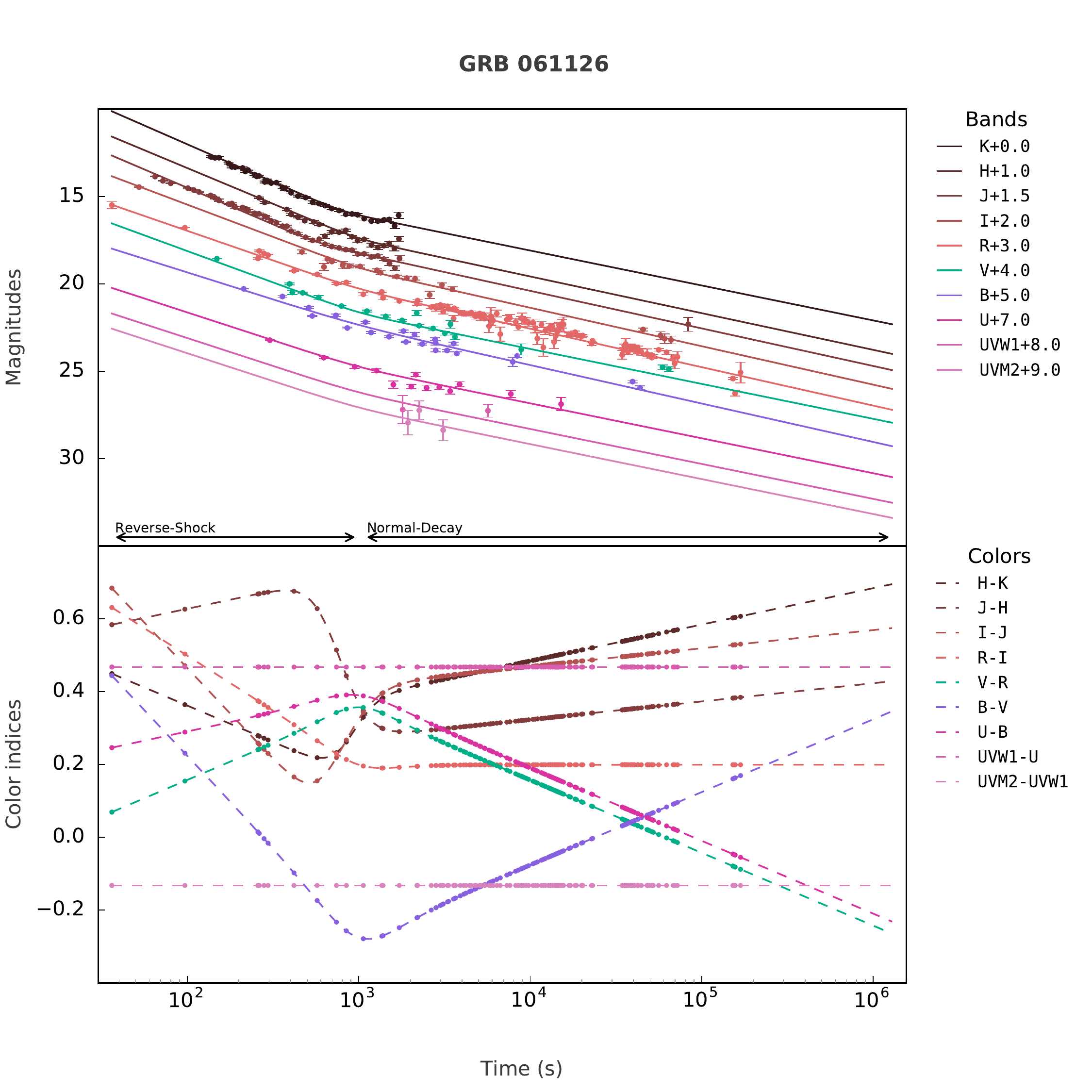}
\center{Fig. \ref{SilverLCs}--- Continued}
\end{figure*}
\begin{figure*}
\includegraphics[angle=0,scale=0.40]{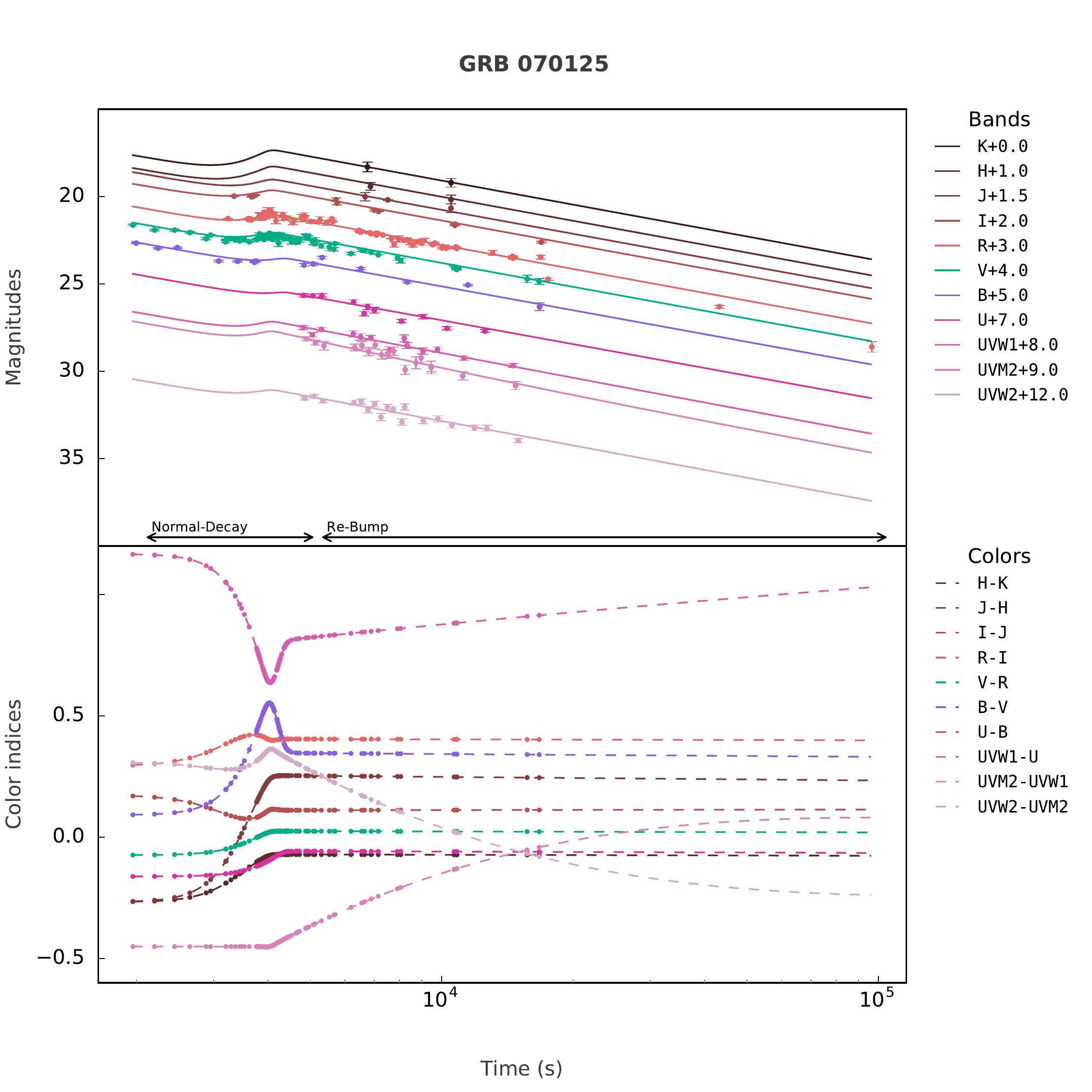}
\includegraphics[angle=0,scale=0.40]{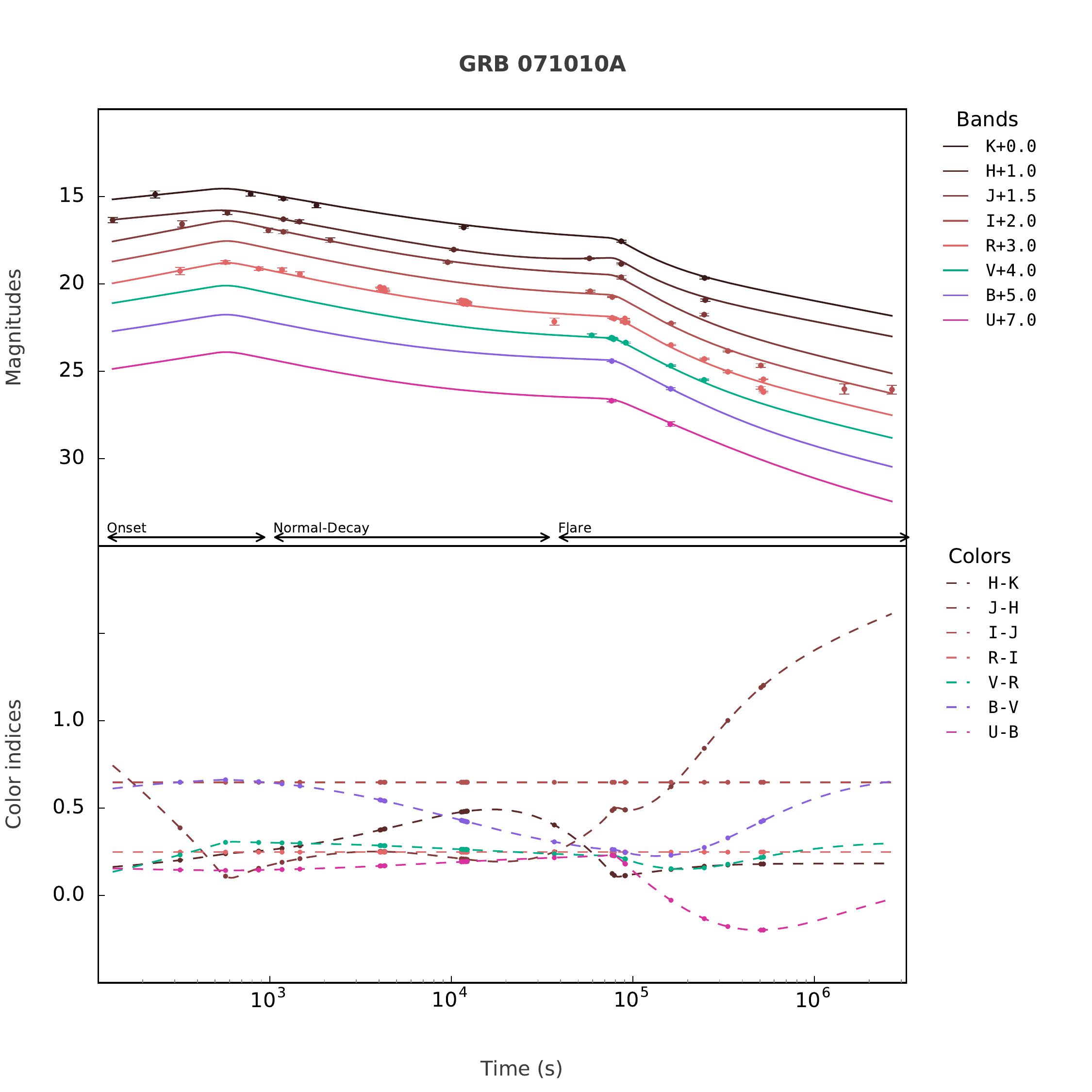}
\includegraphics[angle=0,scale=0.40]{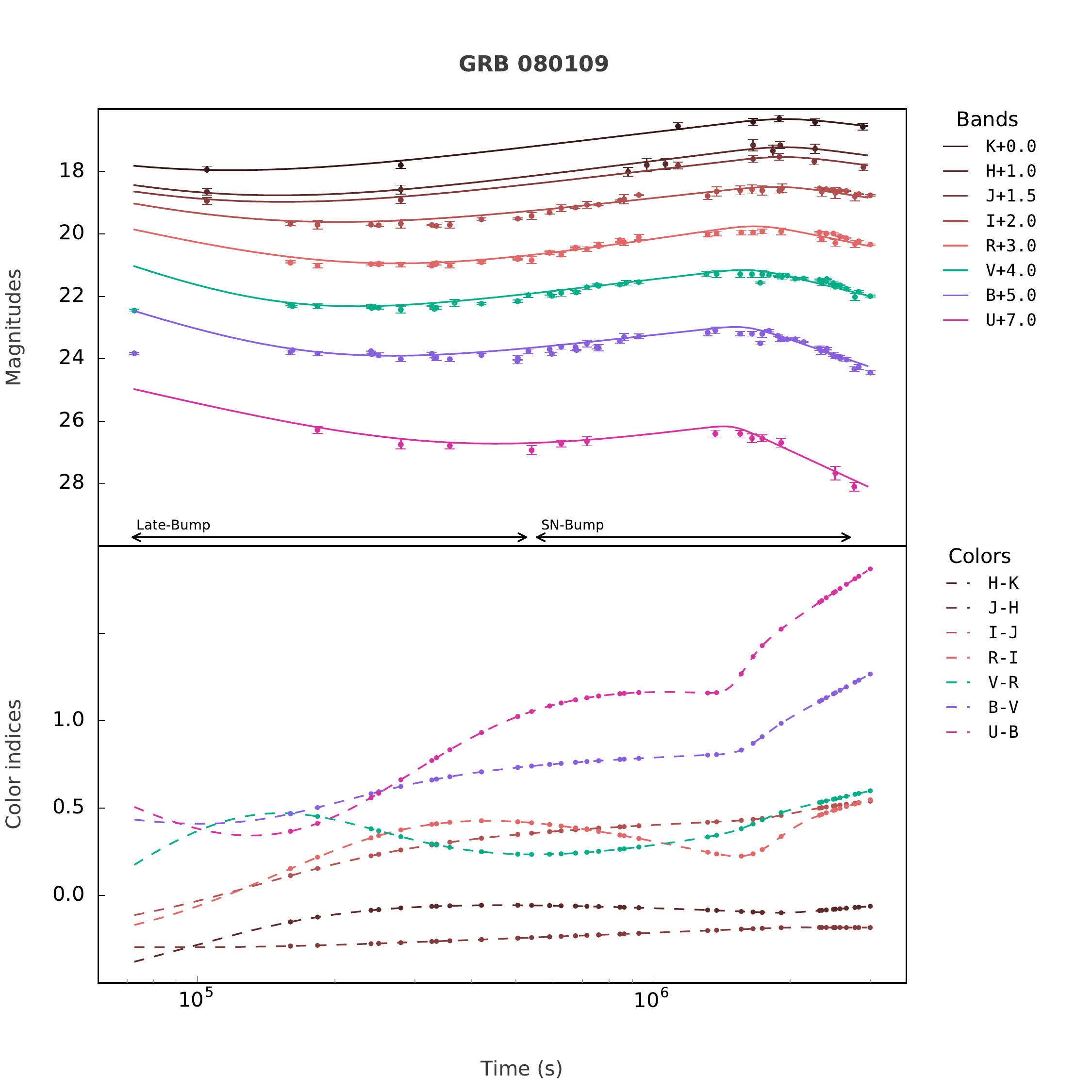}
\includegraphics[angle=0,scale=0.40]{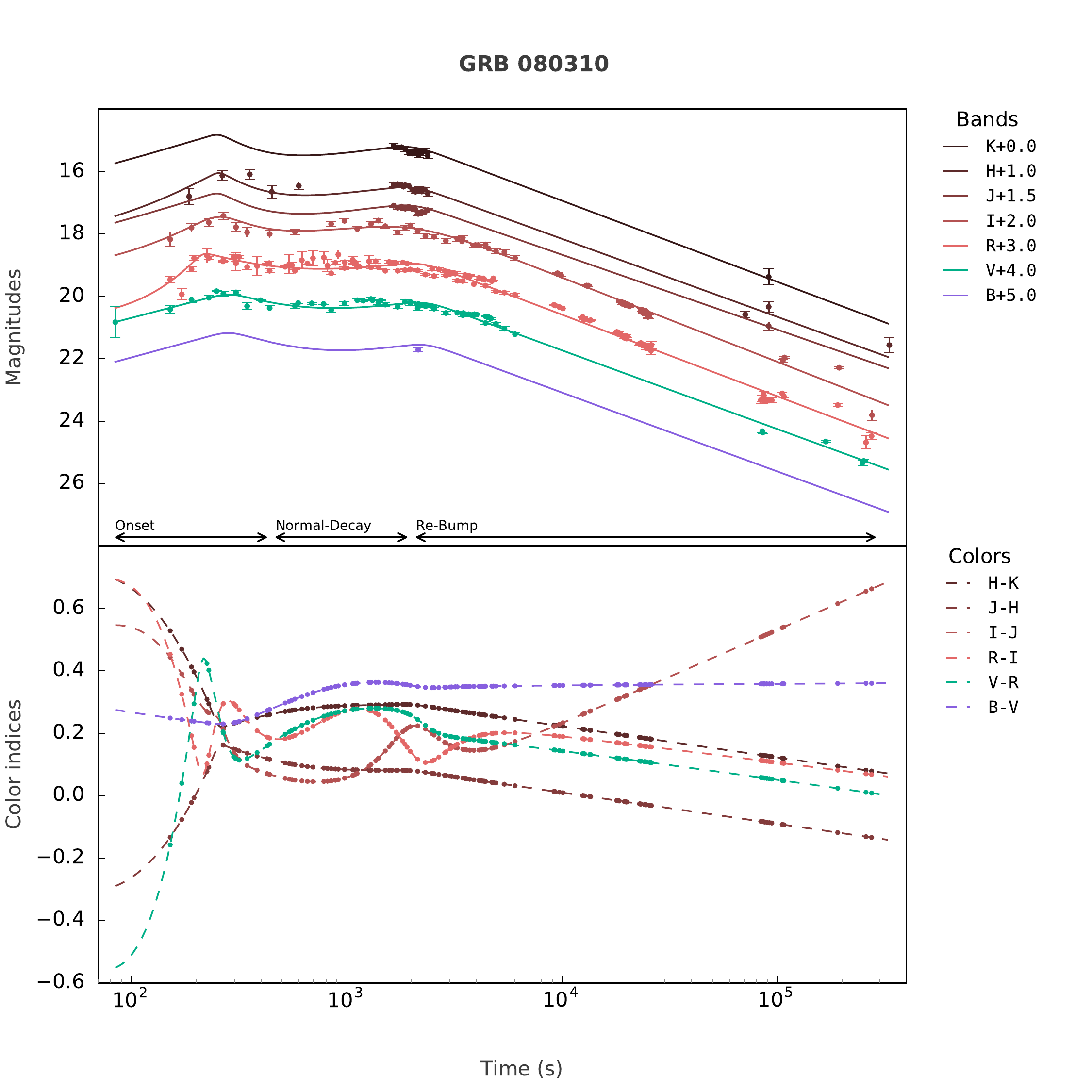}
\center{Fig. \ref{SilverLCs}--- Continued}
\end{figure*}
\begin{figure*}
\includegraphics[angle=0,scale=0.40]{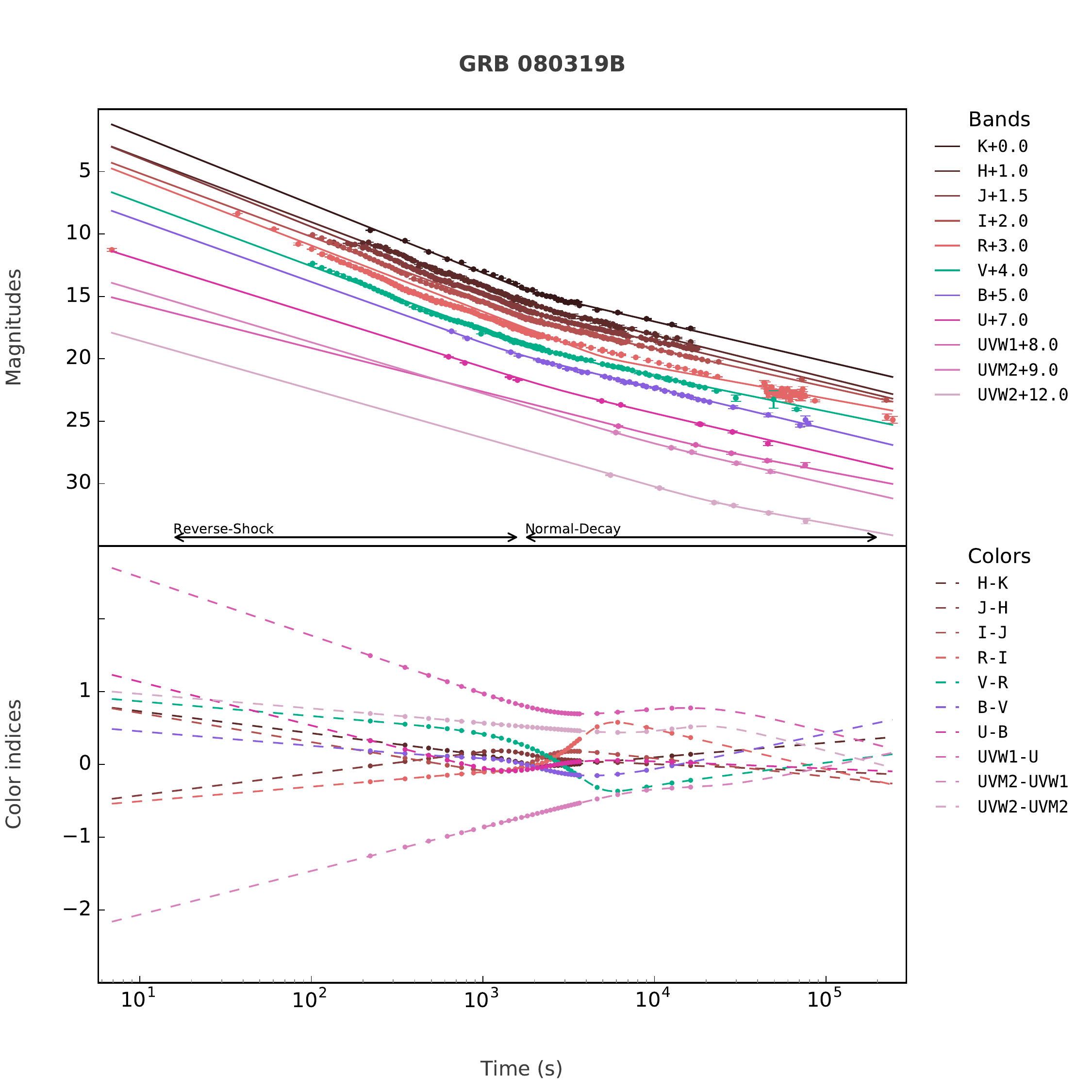}
\includegraphics[angle=0,scale=0.40]{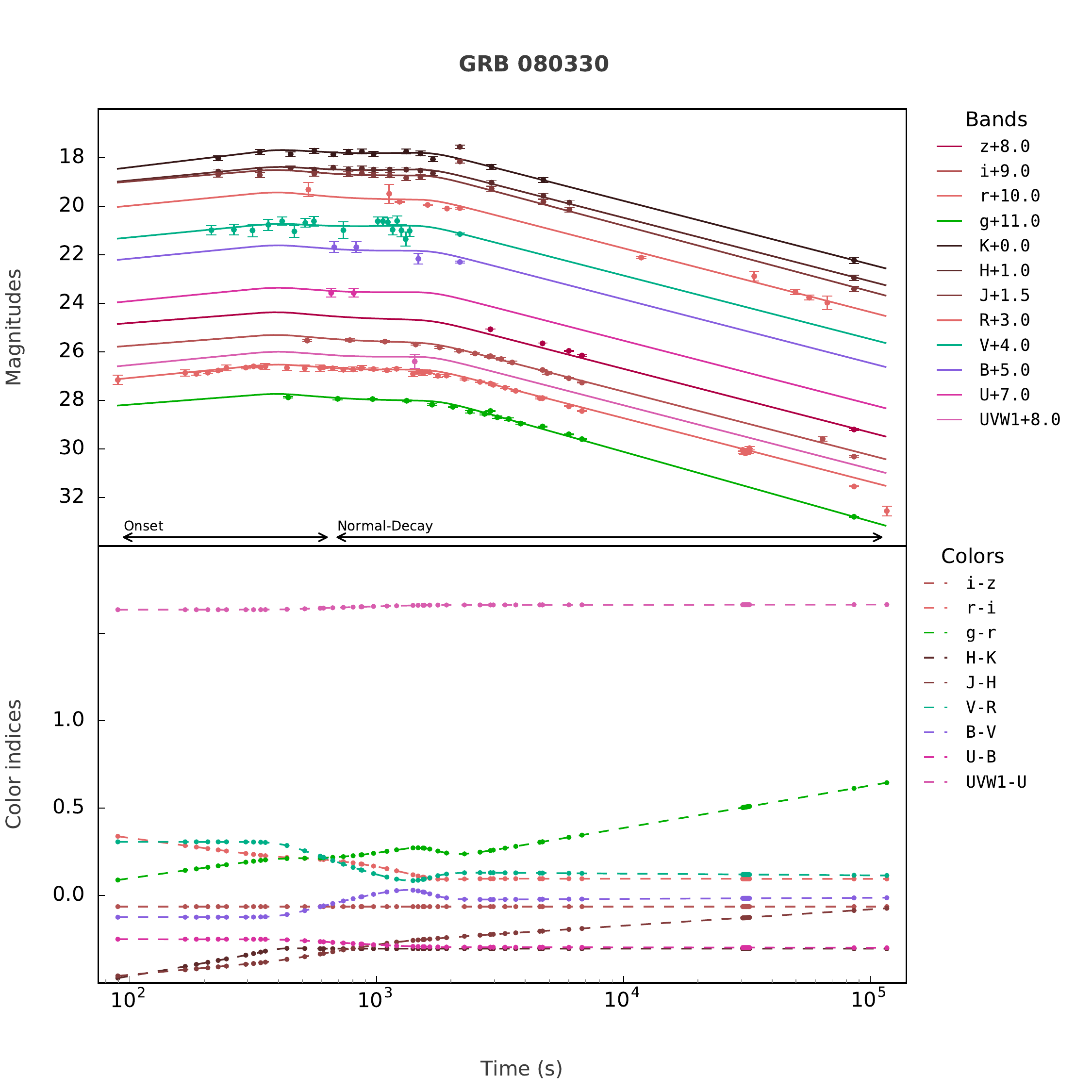}
\includegraphics[angle=0,scale=0.40]{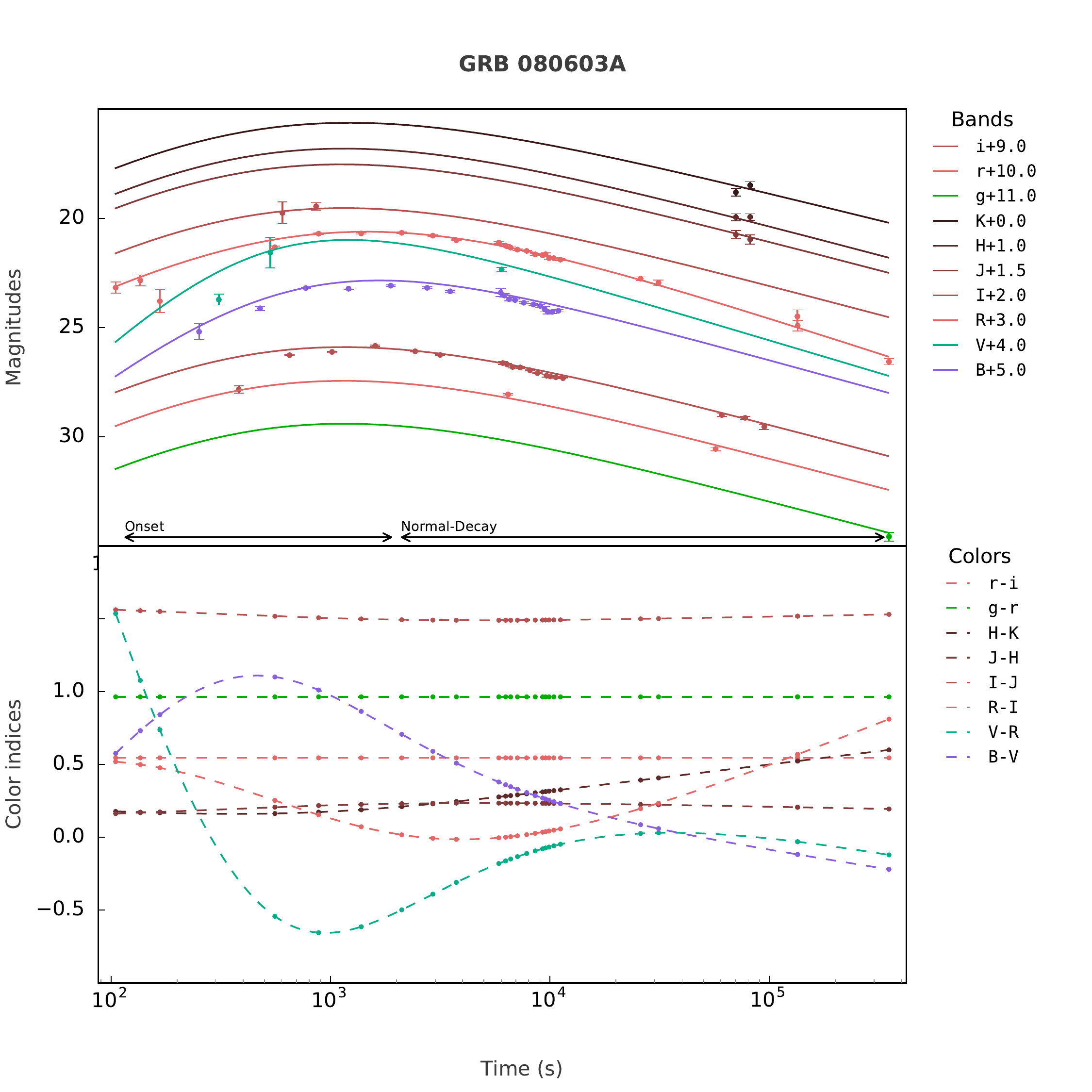}
\includegraphics[angle=0,scale=0.40]{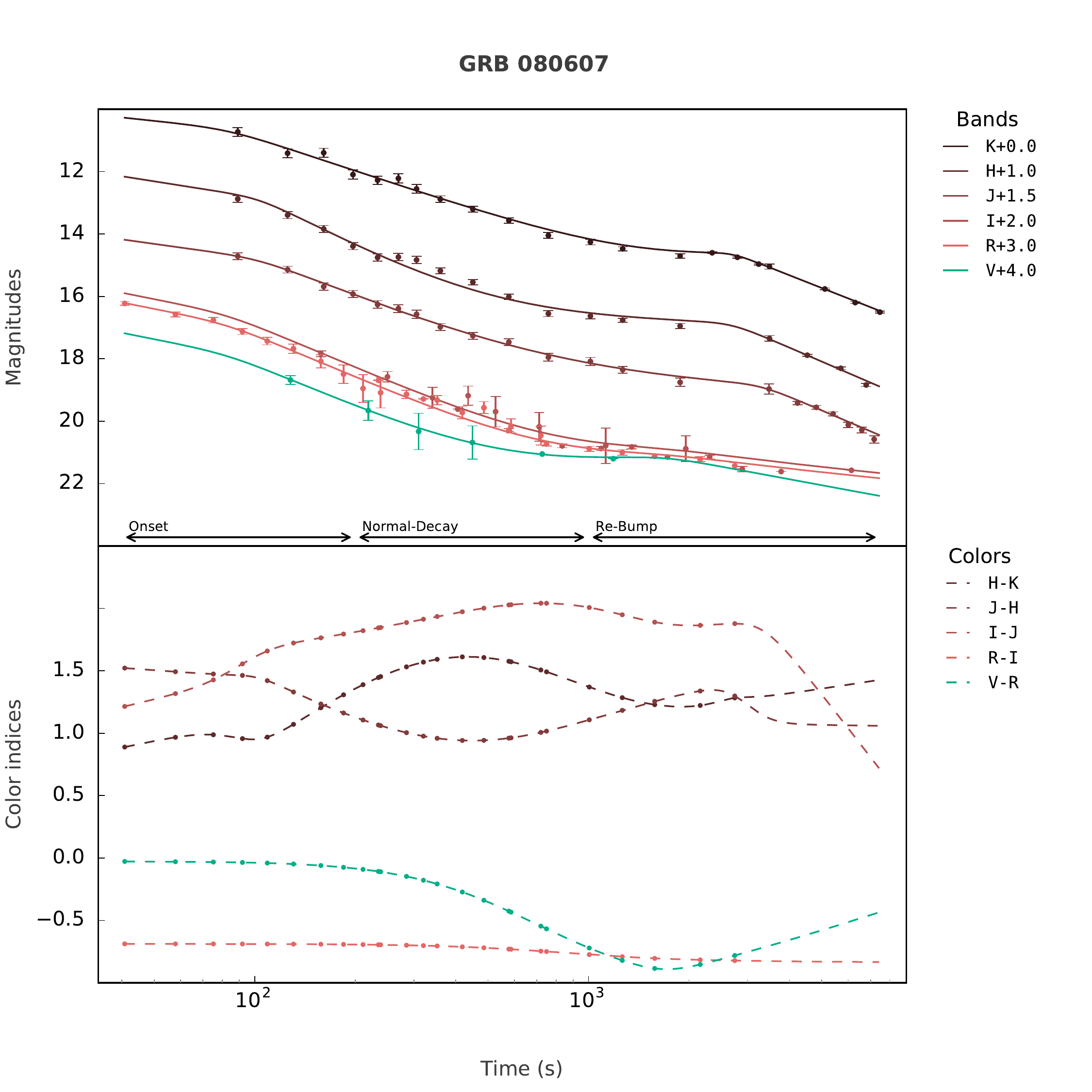}
\center{Fig. \ref{SilverLCs}--- Continued}
\end{figure*}
\begin{figure*}
\includegraphics[angle=0,scale=0.40]{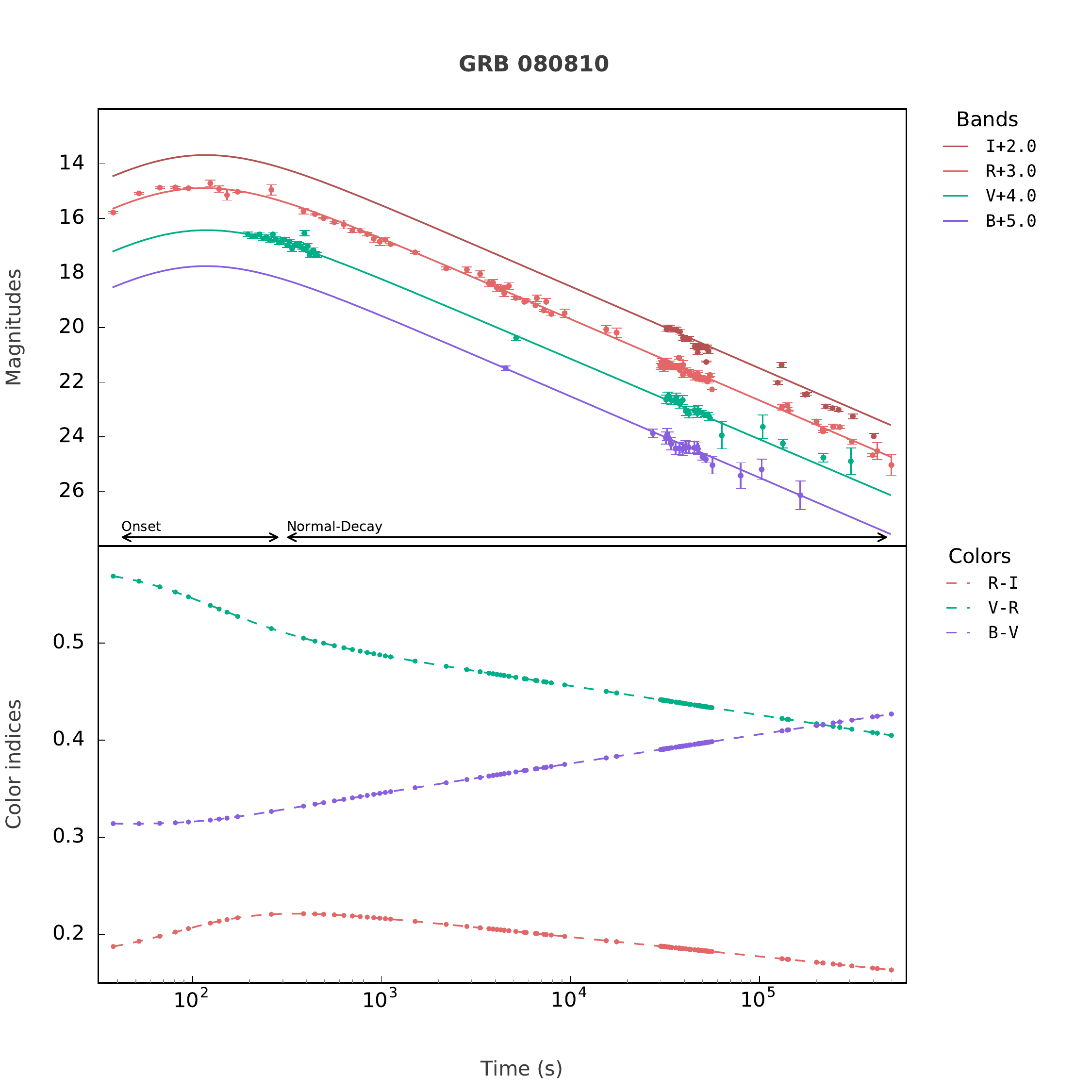}
\includegraphics[angle=0,scale=0.40]{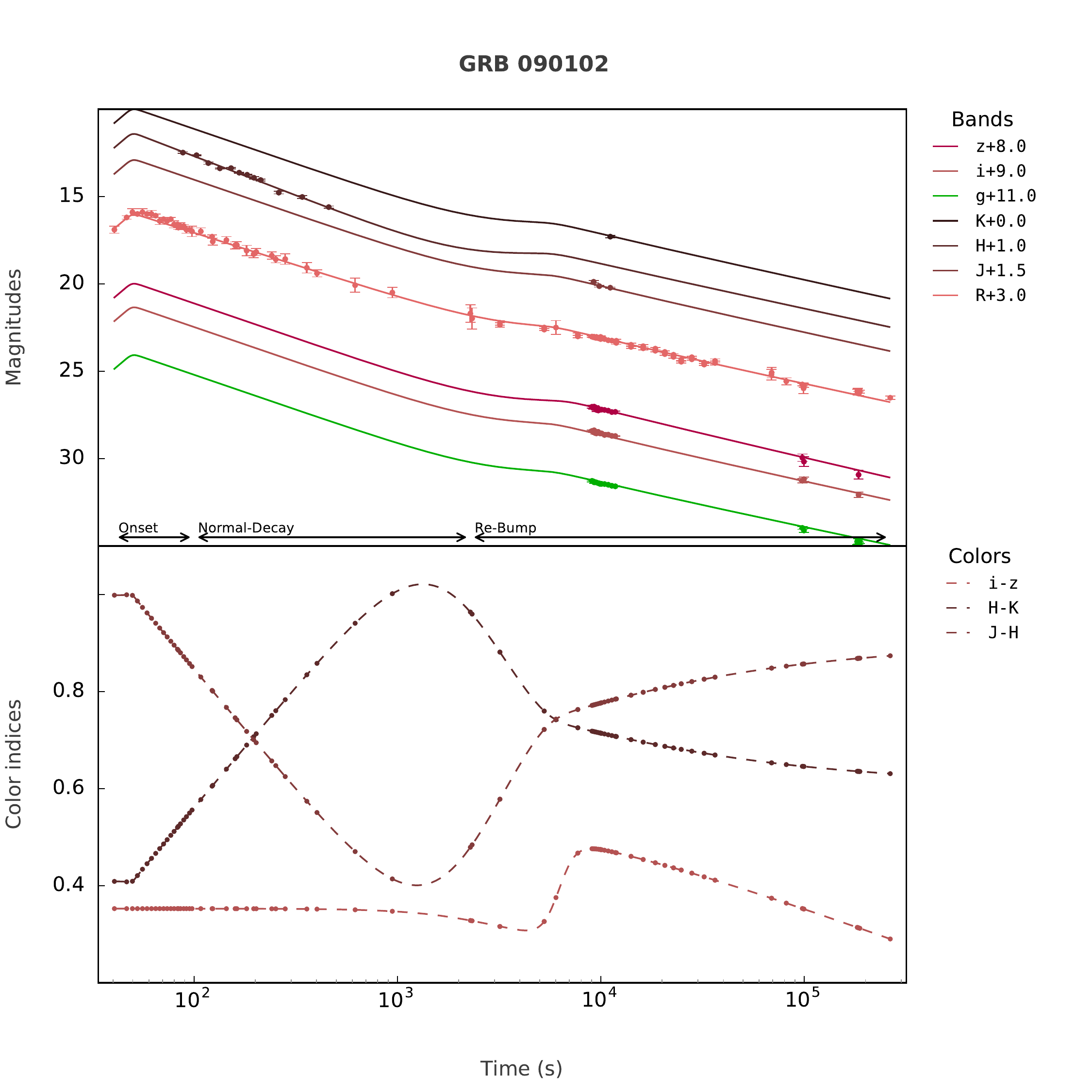}
\includegraphics[angle=0,scale=0.40]{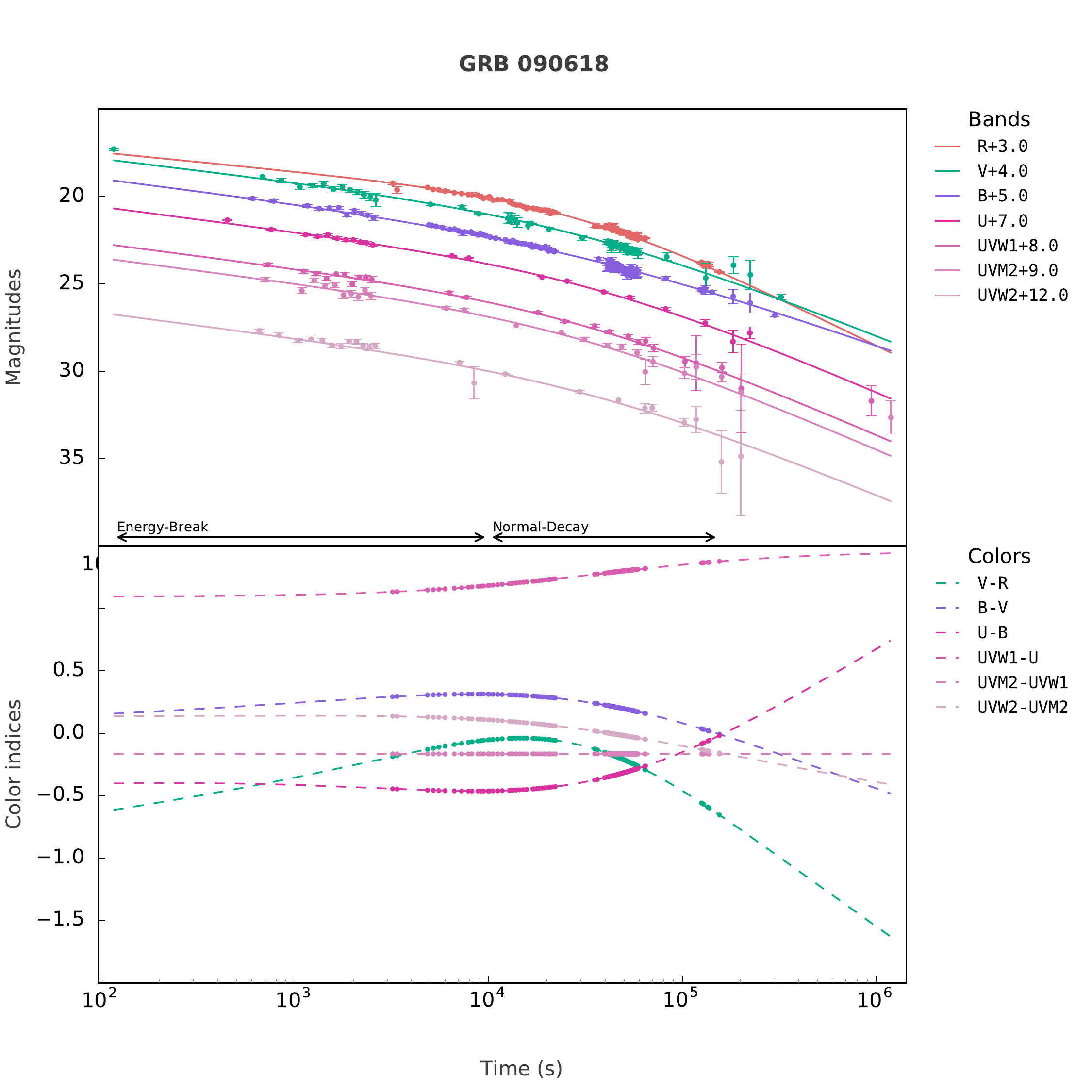}
\includegraphics[angle=0,scale=0.40]{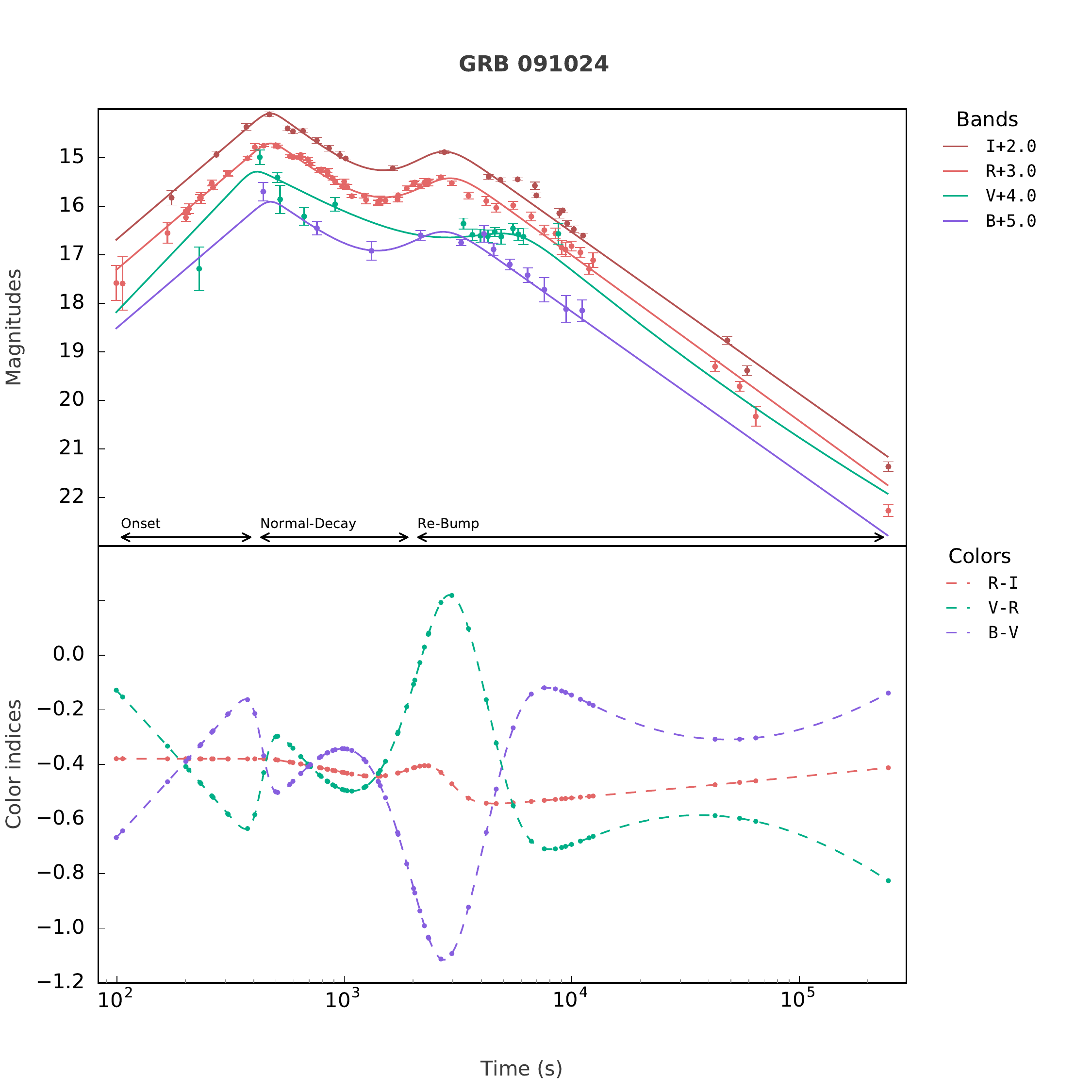}
\center{Fig. \ref{SilverLCs}--- Continued}
\end{figure*}
\begin{figure*}
\includegraphics[angle=0,scale=0.40]{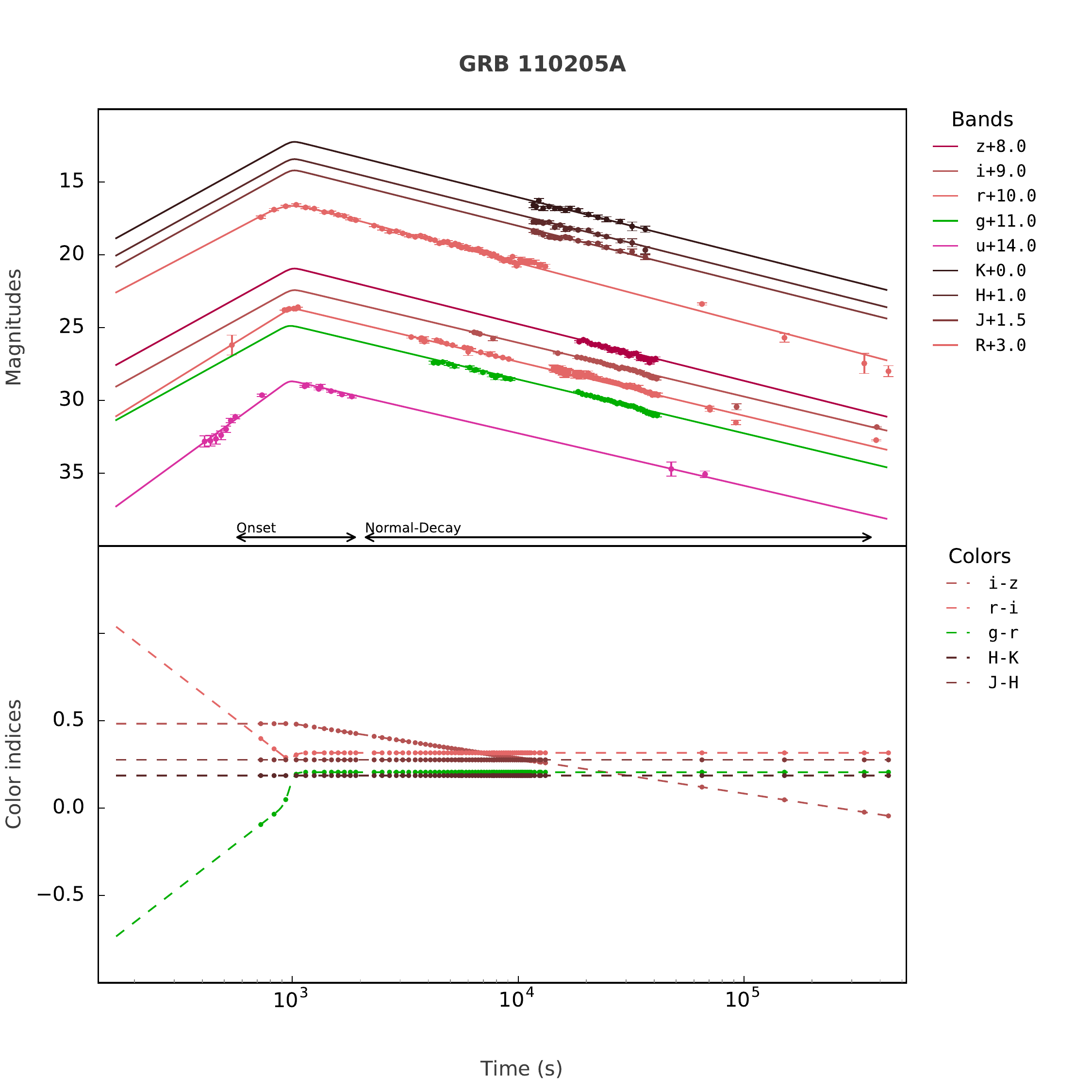}
\includegraphics[angle=0,scale=0.40]{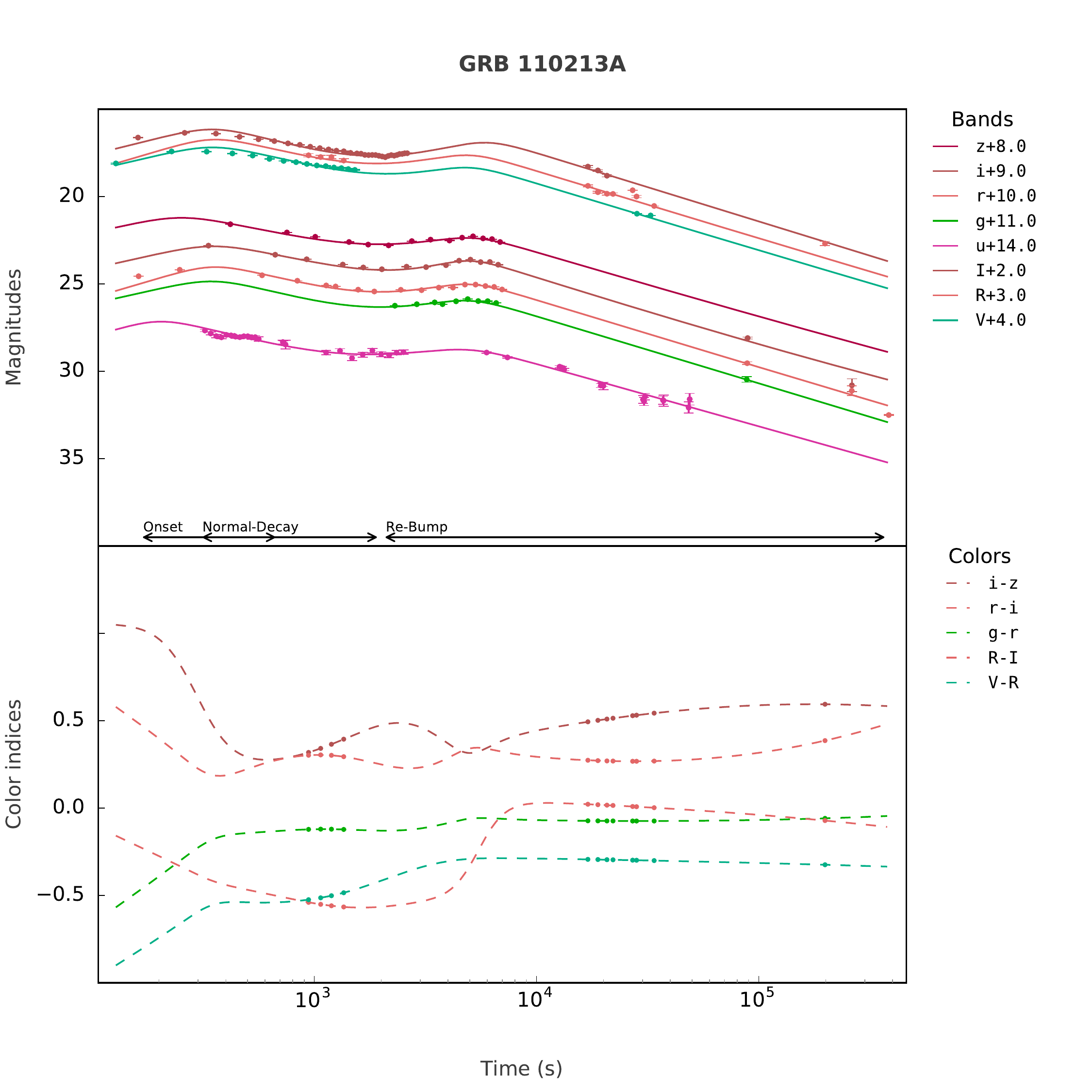}
\includegraphics[angle=0,scale=0.40]{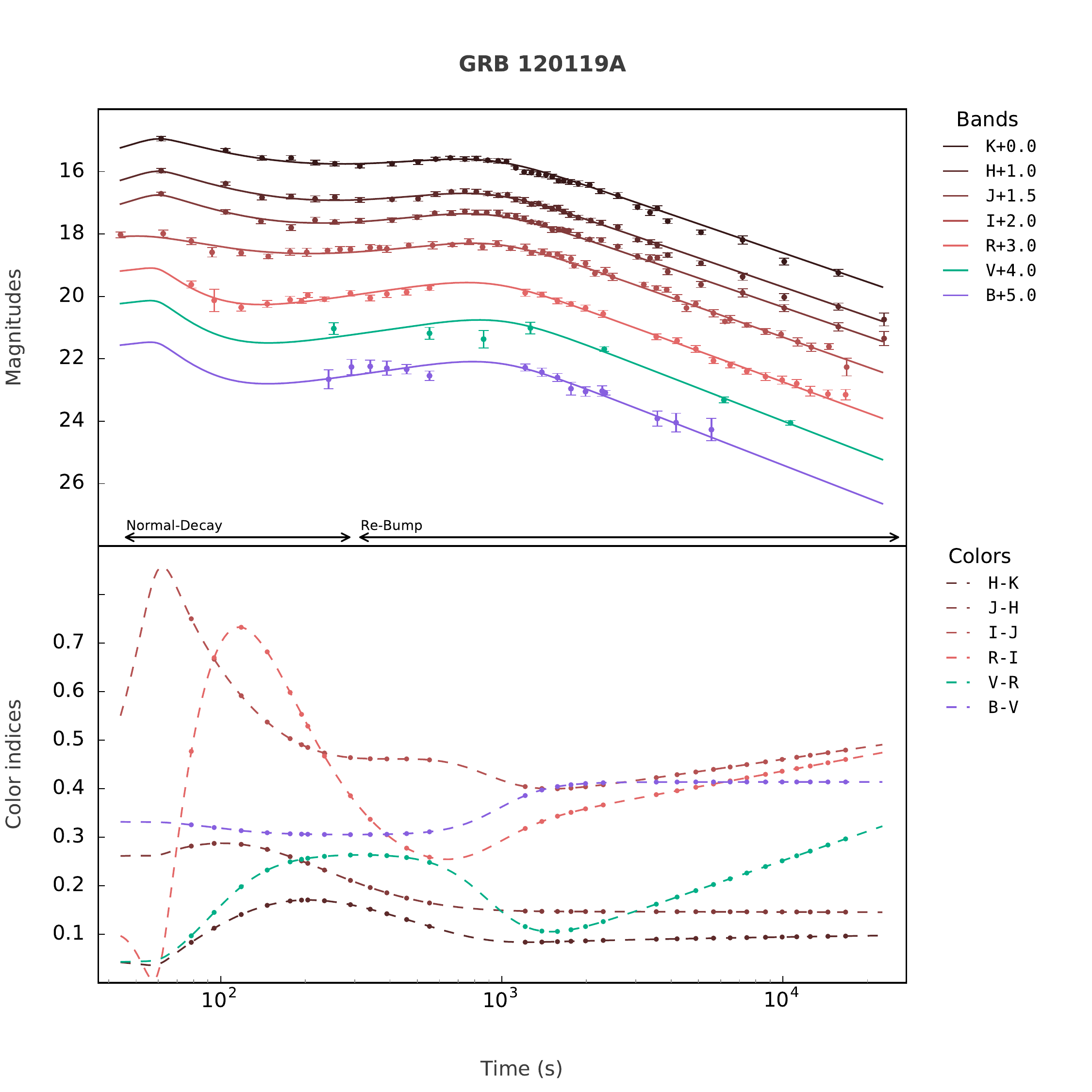}
\includegraphics[angle=0,scale=0.40]{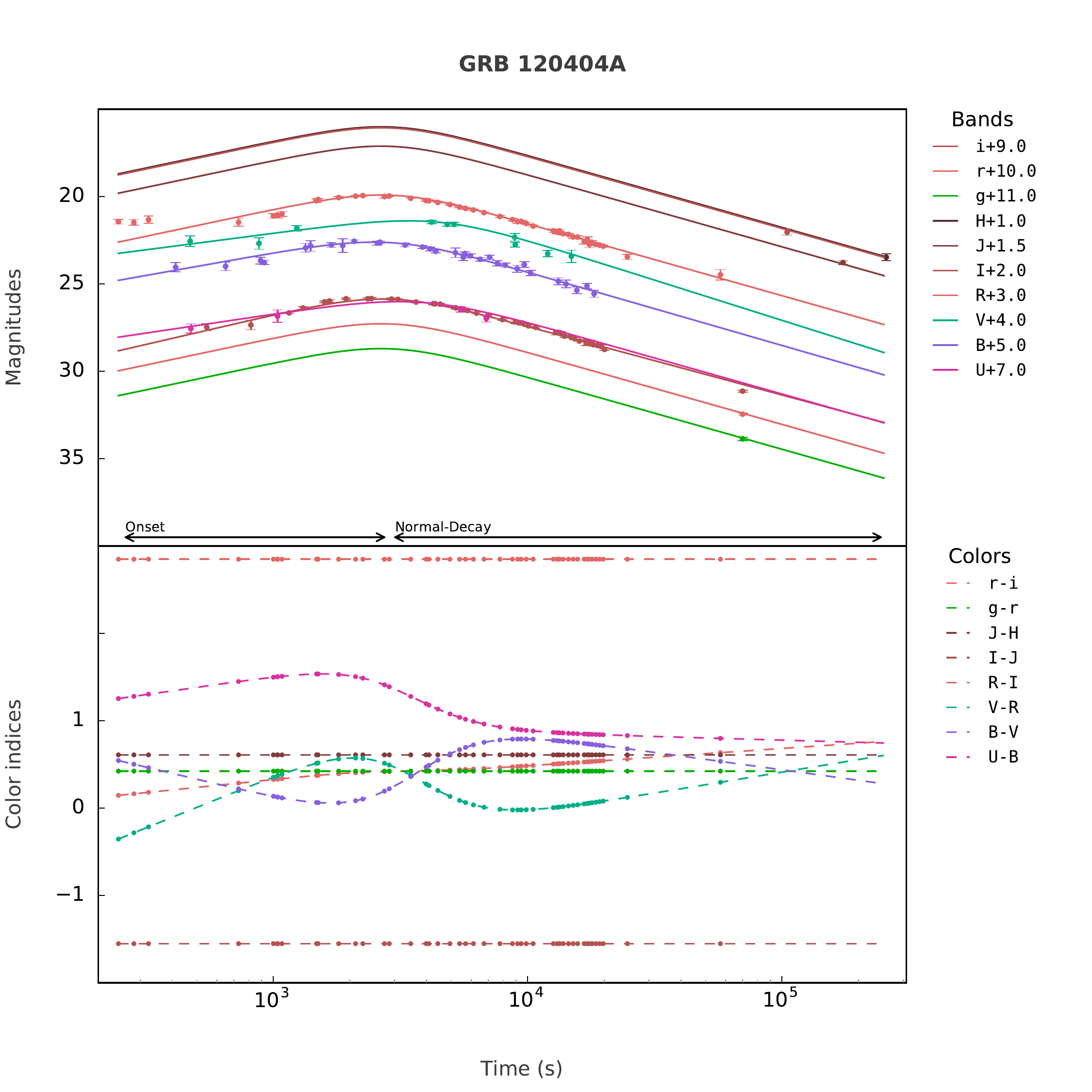}
\center{Fig. \ref{SilverLCs}--- Continued}
\end{figure*}
\begin{figure*}
\includegraphics[angle=0,scale=0.40]{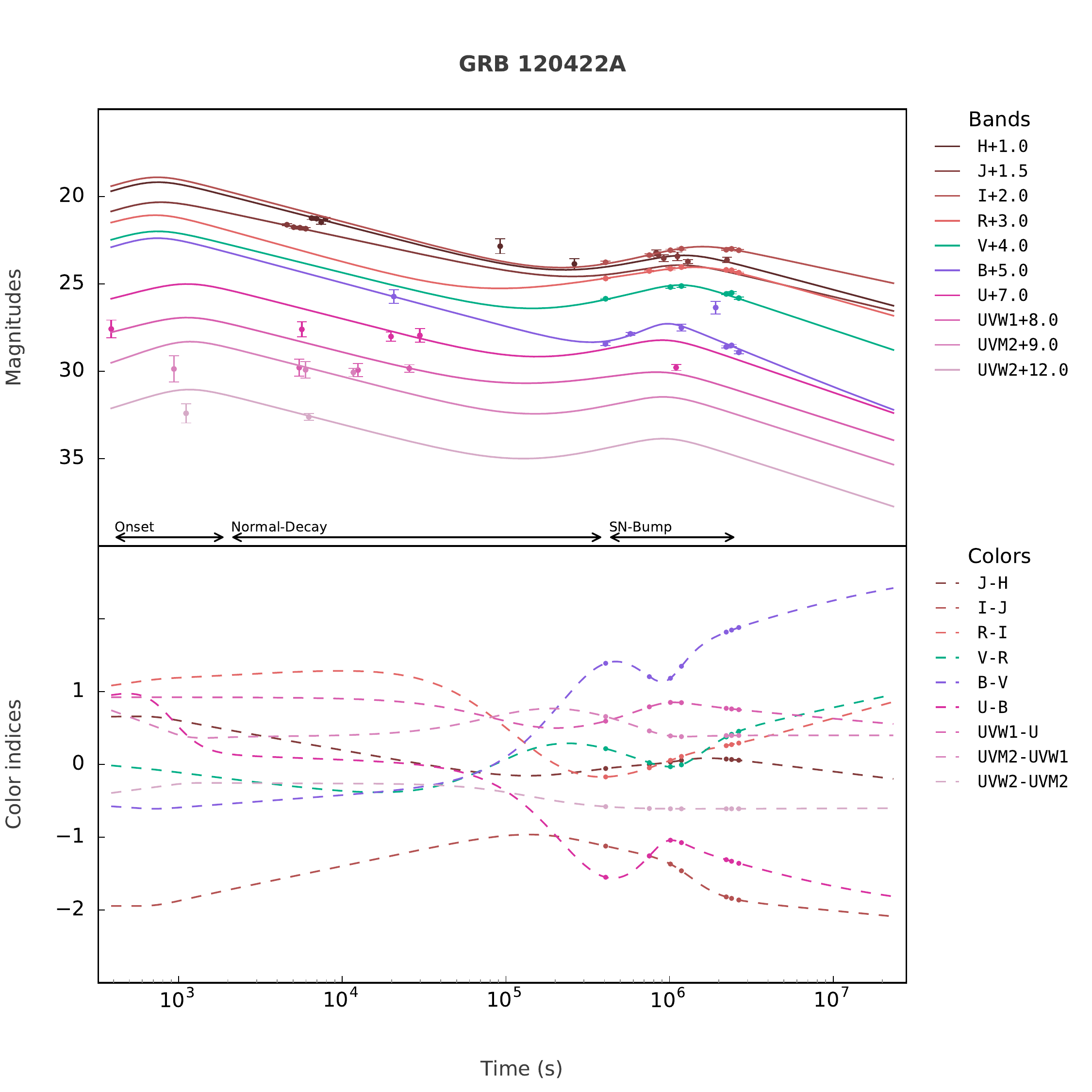}
\includegraphics[angle=0,scale=0.40]{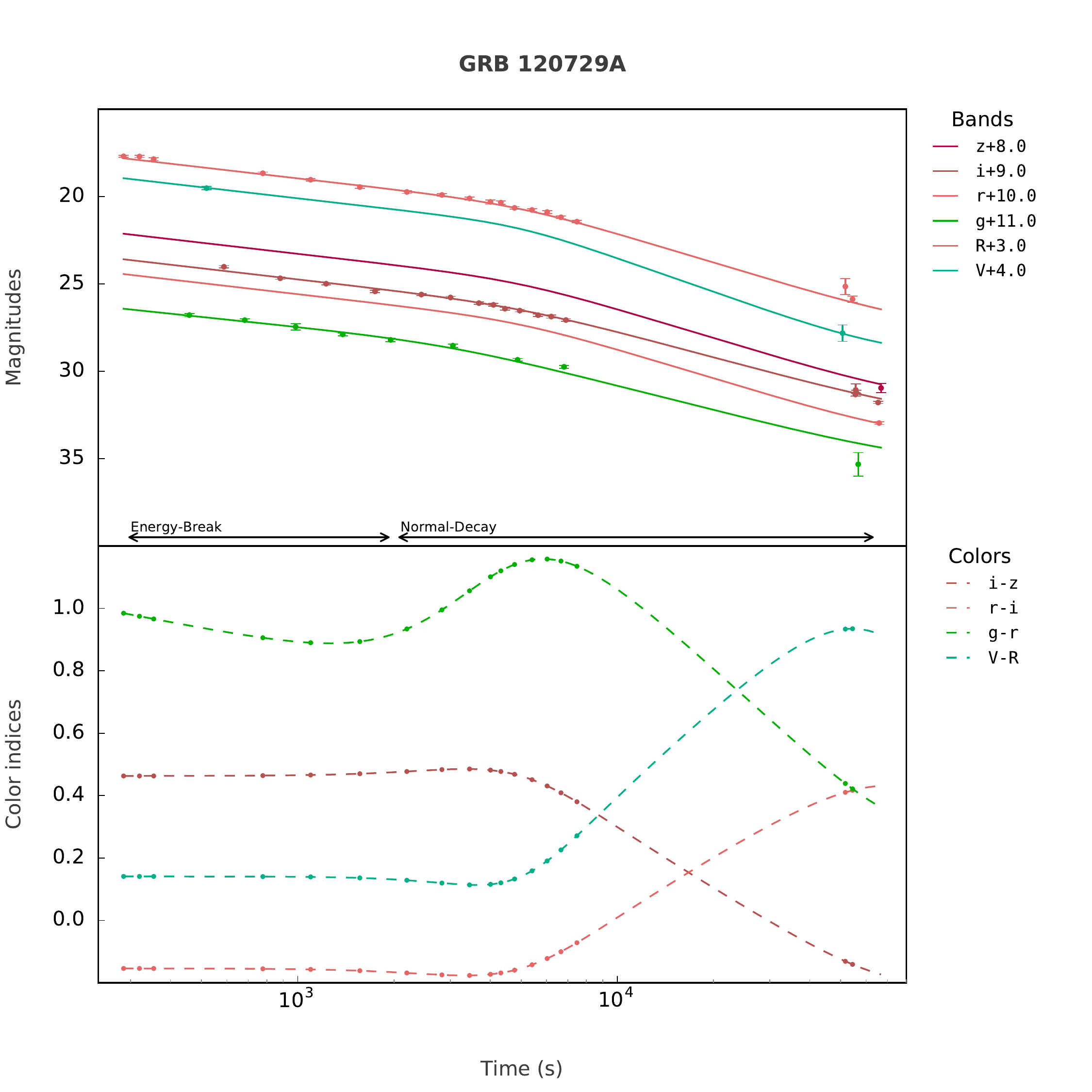}
\includegraphics[angle=0,scale=0.40]{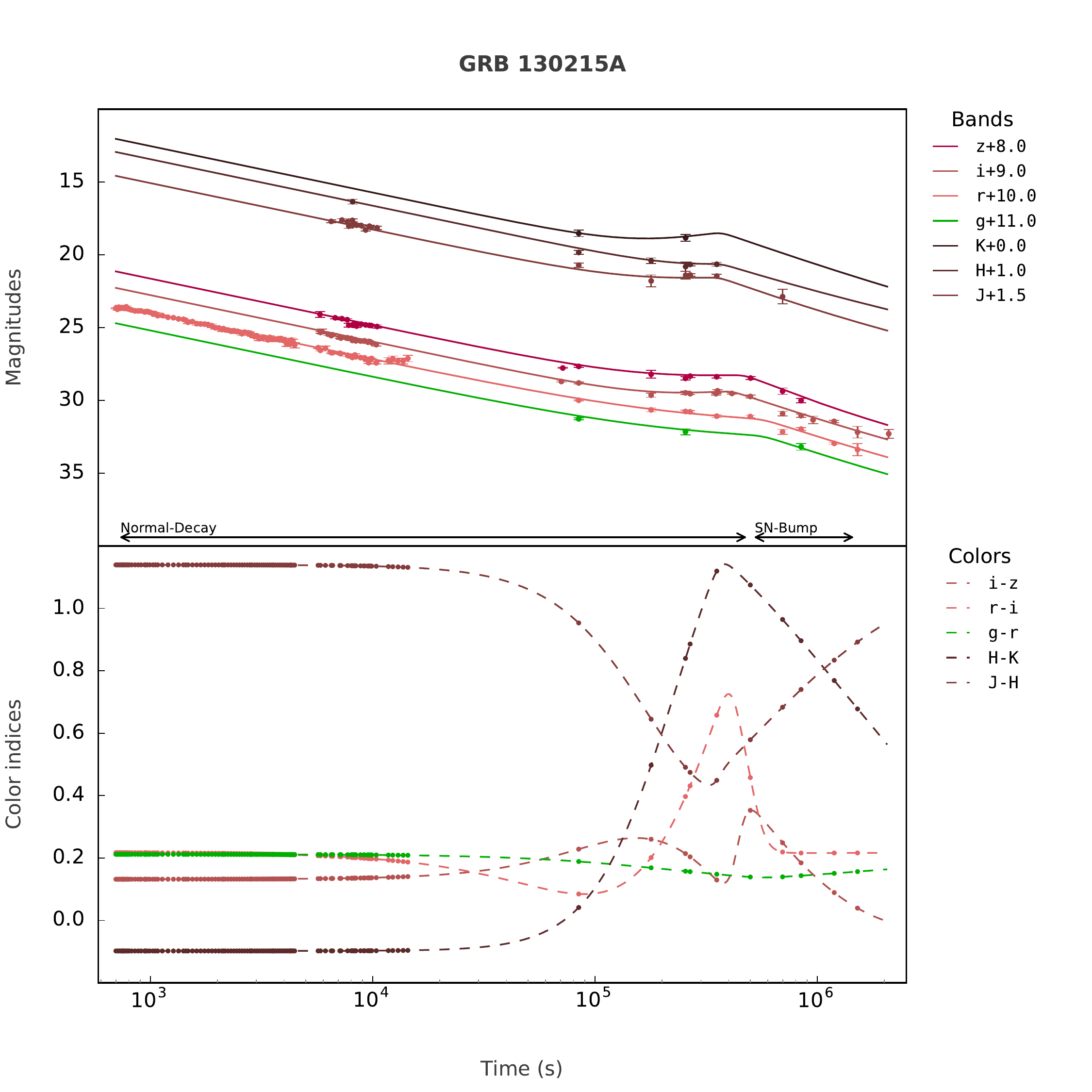}
\includegraphics[angle=0,scale=0.40]{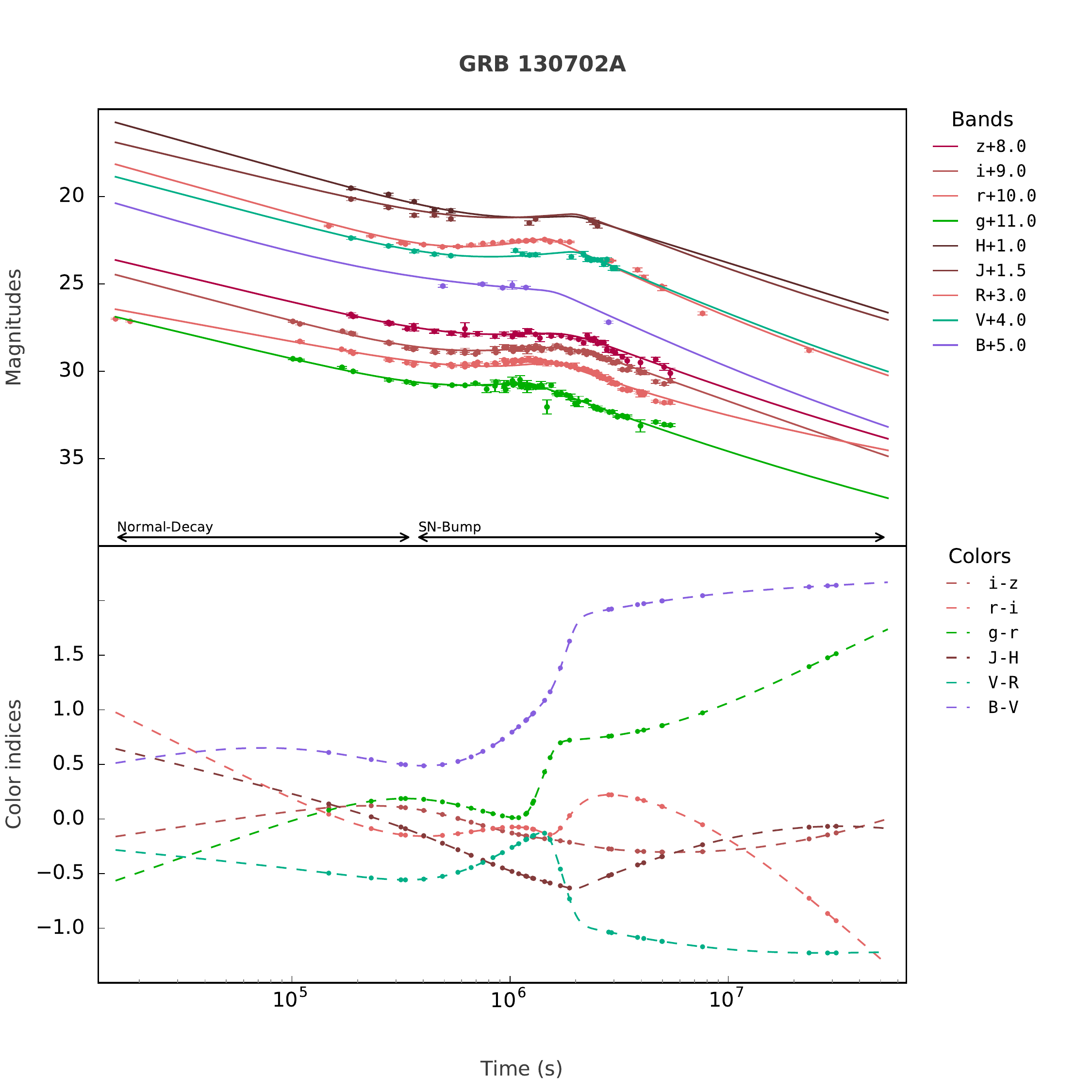}
\center{Fig. \ref{SilverLCs}--- Continued}
\end{figure*}
\begin{figure*}
\includegraphics[angle=0,scale=0.40]{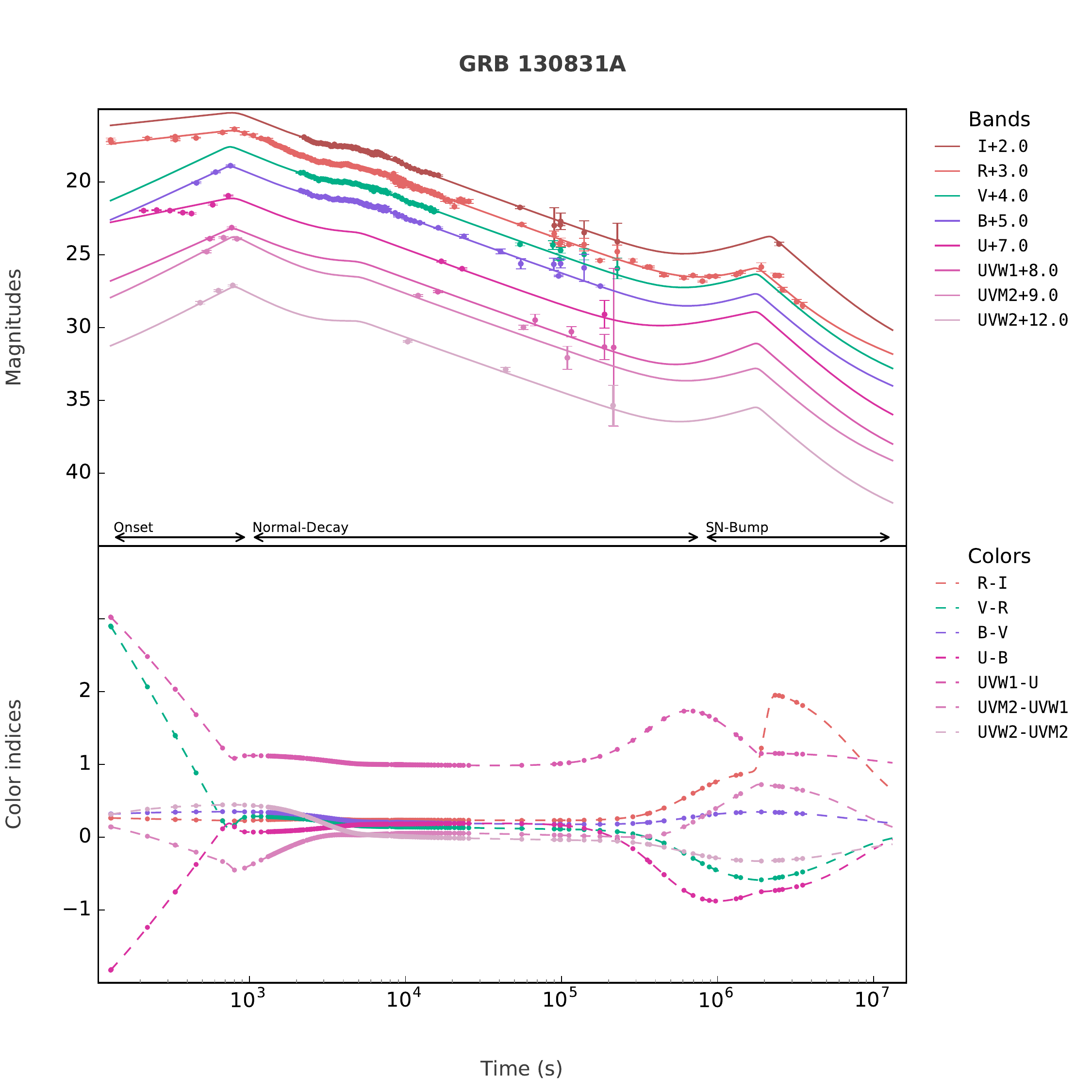}
\center{Fig. \ref{SilverLCs}--- Continued}
\end{figure*}

\clearpage
\thispagestyle{empty}
\setlength{\voffset}{-18mm}

\clearpage
\thispagestyle{empty}
\setlength{\voffset}{-18mm}

\clearpage
\thispagestyle{empty}
\setlength{\voffset}{-18mm}

\clearpage
\thispagestyle{empty}
\setlength{\voffset}{-18mm}

\clearpage
\thispagestyle{empty}
\setlength{\voffset}{-18mm}

\clearpage
\thispagestyle{empty}
\setlength{\voffset}{-18mm}

\clearpage
\thispagestyle{empty}
\setlength{\voffset}{-18mm}

\clearpage
\thispagestyle{empty}
\setlength{\voffset}{-18mm}

\clearpage
\thispagestyle{empty}
\setlength{\voffset}{-18mm}

\clearpage
\thispagestyle{empty}
\setlength{\voffset}{-18mm}

\clearpage
\thispagestyle{empty}
\setlength{\voffset}{-18mm}

\clearpage
\thispagestyle{empty}
\setlength{\voffset}{-18mm}

\clearpage
\thispagestyle{empty}
\setlength{\voffset}{-18mm}

\clearpage
\thispagestyle{empty}
\setlength{\voffset}{-18mm}

\clearpage
\vspace{5mm}
\facilities{{\it Swift}/UVOT}
\software{Python}
\bibliography{./my_citation_C.bib}

\clearpage
\appendix

\begin{longrotatetable}
\setlength{\tabcolsep}{0.35em}
\begin{deluxetable}{ccccccccccccccccccccccccccccc}
\tabletypesize{\tiny}
\tablewidth{0pt}
\rotate
\tablecaption{Color Indices and Their Relation Spectral Indices for the Golden Sample}
\tablehead{
\colhead{GRB}&
\colhead{Time}&
\colhead{g-r}&
\colhead{$\beta_{o}^{\rm CI}$}&
\colhead{Time}&
\colhead{r-i}&
\colhead{$\beta_{o}^{\rm CI}$}&
\colhead{Time}&
\colhead{i-z}&
\colhead{$\beta_{o}^{\rm CI}$}&
\colhead{Time}&
\colhead{J-H}&
\colhead{$\beta_{o}^{\rm CI}$}&
\colhead{Time}&
\colhead{H-K$_{\rm s}$}&
\colhead{$\beta_{o}^{\rm CI}$}&\\
&
\colhead{(ks)}&
&
\colhead{(g-r)}&
\colhead{(ks)}&
&
\colhead{(r-i)}&
\colhead{(ks)}&
&
\colhead{(i-z)}&
\colhead{(ks)}&
&
\colhead{(J-H)}&
\colhead{(ks)}&
&
\colhead{(H-Ks)}&
}
\startdata
071025&...&...  &...  &...&...  &...  &...&...  &...  &0.175&-0.38$\pm$0.89&-1.27$\pm$2.97&0.175&1.29$\pm$0.87&4.02$\pm$2.73\\
071025&...&...    &...    &...&...    &...    &...&...    &...    &0.210&-0.06$\pm$0.25&-0.19$\pm$0.83&0.210&0.60$\pm$0.28&1.87$\pm$0.87\\
071025&...&...  &...  &...&...  &...  &...&...  &...  &0.246&0.29$\pm$0.22&0.96$\pm$0.72&0.246&0.64$\pm$0.20&2.01$\pm$0.62\\
071025&...&...  &...  &...&...  &...  &...&...  &...  &0.282&0.39$\pm$0.17&1.31$\pm$0.58&0.282&0.49$\pm$0.16&1.54$\pm$0.49\\
071025&...&...  &...  &...&...  &...  &...&...  &...  &0.318&0.40$\pm$0.19&1.34$\pm$0.63&0.318&0.45$\pm$0.18&1.42$\pm$0.56\\
071025&...&...  &...  &...&...  &...  &...&...  &...  &0.355&0.22$\pm$0.17&0.73$\pm$0.56&0.355&0.53$\pm$0.18&1.67$\pm$0.55\\
071025&...&...  &...  &...&...  &...  &...&...  &...  &0.391&0.23$\pm$0.16&0.78$\pm$0.54&0.391&0.52$\pm$0.16&1.64$\pm$0.49\\
071025&...&...    &...    &...&...    &...    &...&...    &...    &0.427&0.39$\pm$0.17&1.28$\pm$0.58&0.427&0.66$\pm$0.16&2.07$\pm$0.51\\
\enddata
\end{deluxetable}
\end{longrotatetable}

\end{document}